\documentclass[12pt]{article}
\pdfoutput = 1
\usepackage{cite}
\usepackage{pstricks}
\usepackage{color}

\usepackage{array}
\usepackage{epsfig}
\usepackage{amssymb}
\usepackage{graphics,graphpap}
\usepackage{amssymb}
\usepackage{amsmath}
\usepackage{slashed}
\usepackage{dsfont}

\def\no{\nonumber}

\newcommand{\bsi}{B_6^{(1/2)}}
\newcommand{\bei}{B_8^{(3/2)}}

\def\eps{\varepsilon}
\def\epe{\varepsilon'/\varepsilon}
\newcommand{\imlt}{\IM \lambda_t}
\newcommand{\relt}{\RE \lambda_t}

\newcommand{\RE}{{\rm Re}}
\newcommand{\IM}{{\rm Im}}

\newcommand{\tev}{\, {\rm TeV}}
\newcommand{\gev}{\, {\rm GeV}}
\newcommand{\mev}{\, {\rm MeV}}

\newcommand{\Heff}{{\cal H}_\text{ eff}}

\newcommand{\mt}{m_{\rm t}}
\newcommand{\mtb}{\overline{m}_{\rm t}}

\newcommand{\be}{\begin{equation}}
\newcommand{\ee}{\end{equation}}
\newcommand{\bea}{\begin{eqnarray}}
\newcommand{\eea}{\end{eqnarray}}

\newcommand{\bi}{\begin{itemize}}
\newcommand{\ei}{\end{itemize}}
\newcommand{\ord}{{\cal O}}
\newcommand{\vtb}{|V_{tb}|}
\newcommand{\vcb}{|V_{cb}|}
\newcommand{\vtd}{|V_{td}|}
\newcommand{\vub}{|V_{ub}|}
\newcommand{\vts}{|V_{ts}|}
\newcommand{\vus}{|V_{us}|}

\newcommand{\bbs}{\ensuremath{B^0_s\!-\!\overline{\!B}^0_s\,}}

\newcommand{\bbd}{\ensuremath{B^0_d\!-\!\overline{\!B}^0_d\,}}
\newcommand{\bbmd}{\bbd\ mixing}
\newcommand{\kk}{\ensuremath{K^0\!-\!\overline{\!K}^0\,}}
\newcommand{\kkm}{\kk\ mixing}

\def\kpn{K^+\rightarrow\pi^+\nu\bar\nu}

\def\klpn{K_{L}\rightarrow\pi^0\nu\bar\nu}

\usepackage{graphicx}

\medskipamount 

\usepackage{fancyhdr}
\advance \headheight by 3.0truept       

\newlength{\textlength}
\newlength{\overlinelength}

 \def\s#1{\setbox0=\hbox{$#1$}%
   \rlap{\ifdim\wd0>.7em\kern.22\wd0\else\kern.1\wd0\fi /}#1}


\begin{document}

\begin{titlepage}
\begin{flushright}
{FLAVOUR(267104)-ERC-45}
\end{flushright}
\vskip1.2cm
\begin{center}
{\Large \bf \boldmath
Towards the Identification of New Physics through Quark Flavour Violating Processes*\let\thefootnote\relax\footnote{*Review article
to be submitted for
publication in Reports on Progress in Physics}
}
\vskip1.0cm
{\bf
Andrzej J. Buras and 
Jennifer Girrbach}
\vskip0.3cm
TUM-IAS, Lichtenbergstr. 2a, D-85748 Garching, Germany\\
Physik Department, TUM, D-85748 Garching, Germany
\\
\vskip0.51cm


\vskip0.35cm

{\large\bf Abstract\\[10pt]} \parbox[t]{\textwidth}{
We outline a systematic strategy  which should help in this decade to 
identify New Physics (NP) beyond the Standard Model (SM) 
by means of quark flavour violating processes 
and thereby to extend the picture of short distance physics 
down to the scales as short as $10^{-20}$ m and even shorter distance 
scales corresponding to energies of $100\tev$. Rather than using all 
possible flavour violating observables that will be measured in the 
coming years at the LHC, SuperKEKB 
and in Kaon physics 
dedicated experiments at CERN, J-PARC and Fermilab, we concentrate on those
observables that are theoretically clean and very sensitive to NP.
Assuming that the data on the selected  observables by us will be very precise, we stress  the importance of correlations between these
observables as well as of 
future precise calculations of non-perturbative parameters
by means of lattice QCD simulations with dynamical fermions. Our strategy 
consists of {\it twelve} steps which we will discuss in detail illustrating possible outcomes with the help of the SM, models with constrained 
Minimal Flavour Violation (CMFV), MFV at large and models with tree-level 
FCNCs mediated by neutral gauge bosons and scalars.  We also briefly summarize the status of a number of concrete models. We propose
{\it DNA-charts} 
that exhibit correlations between 
flavour observables in different NP scenarios. The models with new left-handed 
and/or right-handed currents and  non-MFV interactions can be distinguished 
transparently in this manner. We emphasize the important role of the stringent CMFV relations between various observables as {\it standard 
candles of flavour physics.}  The pattern of deviations from these relations  may help 
in identifying the correct NP scenario.
The success of this
program will be very much facilitated through direct signals of NP 
at the LHC even if LHC will not be able to probe the physics at scales 
shorter than  $4\times 10^{-20}$~m. We also emphasize the importance 
of lepton flavour violation, electric dipole moments and $(g-2)_{e,\mu}$ 
in these studies.

}

\vfill

\end{center}
\end{titlepage}

\setcounter{footnote}{0}

\clearpage
\pagestyle{empty}
\tableofcontents
\clearpage
\pagestyle{fancyplain}

\section{Overture}
\label{sec:intro}
The main goal of elementary particle physics is to search for fundamental laws
at very short distance scales. From the Heisenberg uncertainty principle 
we
know that to test scales of order $10^{-18}{\rm m}$ we need the energy of 
approximately $200\gev$. Therefore the LHC we will be able to probe distances as short as $4\cdot 10^{-20}{\rm m}$.  Unfortunately, it will take some time before 
we can reach a higher resolution using high energy processes. On the other
hand flavour-violating and CP-violating processes are very strongly suppressed
and are governed by quantum fluctuations that allow us to test energy scales 
as high as  $200\tev$ corresponding to short distances in the ballpark of
 $10^{-21}{\rm m}$. Even shorter distance scales
can be tested, albeit indirectly, in this manner. Consequently frontiers
in testing ultrashort distance scales belong to flavour physics or more
concretely to very rare processes like particle-antiparticle mixing, rare decays of mesons, CP violation
and lepton flavour violation. Also electric dipole moments and $(g-2)_\mu$ 
belong to these
frontiers even if they are flavour conserving.
While such tests are not limited by the available energy, they are limited
by the available precision. The latter has to be very high as the 
Standard Model (SM) has been until now very successful and finding 
departures from its predictions in the quark sector 
has become a real challenge. This 
precision applies both to experiments and theoretical calculations. Among 
the latter higher order renormalization group improved 
perturbative QCD calculations and in particular calculations of non-perturbative parameters by means of QCD lattice simulations with dynamical fermions 
play prominent roles in the search for NP at very short distance 
scales.

Flavour physics developed over the last two decades into a very broad field.
In addition to $K$, $D$ and $B_d$ decays and $K^0-\bar K^0$ and $B^0_d-\bar B^0_d$
mixings that were with us for quite some time, $B_s^0-\bar B_s^0$ mixing, 
$B_s$ decays and $D^0-\bar D^0$
mixing belong these days to the standard repertoire of any flavour 
workshop. Similarly lepton flavour violation (LFV) 
gained in importance after the
discovery of neutrino oscillations and related non-vanishing neutrino 
masses even if within the SM  enriched with { tiny} neutrino masses LFV is basically unmeasurable. The recent 
precise measurement of the parameter $\theta_{13}$ resulting in a much higher 
value than expected by many theorists enhanced the importance of this field.
Simultaneously new ideas for the explanation of the quark and lepton
mass spectra and the related weak mixings, summarized by the
CKM \cite{Cabibbo:1963yz,Kobayashi:1973fv} and PMNS 
\cite{Pontecorvo:1957qd,Maki:1962mu}
matrices, developed significantly in last two decades. Moreover the analyses
of electric dipole moments (EDMs), of the $(g-2)_\mu$ anomaly and of 
flavour changing neutral current (FCNC) processes in top quark decays 
intensified during the last years in
view of the related experimental progress that is expected to take place
in this decade.

The correlations between all these observables and the interplay of
flavour physics with direct searches for NP
 and electroweak precision
studies will tell us hopefully one day which is the proper extension of the
SM. In writing this paper we have been guided by the impressive success of the CKM
picture of flavour changing interactions 
\cite{Cabibbo:1963yz,Kobayashi:1973fv} accompanied by the 
GIM mechanism \cite{Glashow:1970gm}
and also by several tensions between the
flavour data and the SM that possibly are the first signals of NP. Fortunately,
there is still a lot of room for NP contributions, in particular in rare 
decays of mesons and charged leptons, in CP-violating transitions and in
electric dipole moments of leptons, of the neutron and of other particles. 
There
is also a multitude of models that attempt to explain the existing tensions
and to predict what experimentalists should find in this decade. 

The main goal of this writing is to have still another look at this fascinating 
field.
However, we should strongly emphasize that we do not intend to present 
here a review 
of flavour physics. Comprehensive reviews, written by a hundred of 
flavour experts are already present on the market
\cite{Buchalla:2008jp,Raidal:2008jk,Antonelli:2009ws} and moreover,
extensive studies of the physics at future flavour machines and other visions
can be found in 
\cite{Bona:2007qt,Browder:2008em,Adeva:2009ny,Buras:2009if,Isidori:2010kg,Fleischer:2010qb,Nir:2010jr,Hurth:2010tk,Buras:2010wr,Ciuchini:2011ca,Meadows:2011bk,Buras:2012ts,Borissov:2013yha,Bediaga:2012py,Hewett:2012ns,Hurth:2012vp,Stone:2012yr,Isidori:2013ez,Kronfeld:2013uoa,Cirigliano:2013lpa,Charles:2013aka,Butler:2013kdw}.

Even if this overture follows closely the one in \cite{Buras:2009if} 
and some goals listed there will be encountered below, our presentation 
is more explicit and is meant as a strategy which we hope we can execute 
systematically in the coming years. Undoubtedly several ideas presented below 
appeared already in the literature including those present in our papers. 
But the collection 
of these ideas at one place, various correlations between them and in 
particular new proposals and observations will hopefully facilitate to monitor 
the coming advances of our experimental colleagues who are searching 
for
the footprints of NP directly at
the LHC and indirectly through flavour and CP-violating processes and other 
rare processes in this decade.

However, in contrast to  \cite{Buras:2009if} we will not confine 
our discussion to scales explored by the ATLAS and CMS but also consider 
much shorter distance scales.

Our paper is organized as follows. In Section~\ref{sec:1} we set the scene for 
our strategy stressing the importance of correlations between observables.
 In Section~\ref{sec:2} we summarize briefly the theoretical framework for weak decays and briefly present 
a number of simplest models which will be used to illustrate our ideas.
These are in  particular models with 
MFV and  models with tree-level FCNCs mediated by neutral gauge bosons and 
scalars that exhibit transparently non-MFV interactions and the effects 
of right-handed currents.
In Section~\ref{sec:correlations}, as a preparation for the subsequent main { section} of our paper, we 
present a classification of various correlations between various processes 
that depend on the NP scenario considered. 

Section~\ref{sec:4}, a very long section,  is 
devoted 
to the presentation of our strategy that consists  of twelve steps and except for Step 12 involves only quark flavour physics. In the
course of
this presentation we will frequently refer to models of 
Section~\ref{sec:2}, 
illustrating our ideas by means of them. In 
 Section~\ref{sec:5} we collect the lessons gained in Section~\ref{sec:4} 
and  propose DNA-charts with the goal to transparently exhibit correlations between various observables that are characteristic for a given
NP scenario.
Finally we briefly review a number of concrete extensions of the SM, investigating how they face the most recent LHCb data.
In Section~\ref{sec:sum} we close this report with a  shopping list for 
this decade.

\section{Strategy}\label{sec:1}
\subsection{Setting the Scene}

 In order to illustrate the basic spirit of our strategy for the identification 
of NP through flavour violating processes we recall here a few deviations from 
SM expectations which could be some signs of NP at work but require further 
investigations. For non-experts the appearance of several observables not 
familiar to them already at the start could be some challenge. The detailed 
table of context should then allow them to quickly find out what 
a given observable means. In particular various definitions of observables, 
like $\varepsilon_K$ and $S_{\psi K_S}$, that are related to $\Delta F=2$ 
transitions, can be found in Section~\ref{Step3}, that is in Step 3 of our strategy for the search for NP. It is also a fact that many observables discussed in this review were at the basis of the construction of the SM and appear in the textbooks \cite{Branco:1999fs,Bigi:2000yz} already, so that the general strategy outlined here should not be difficult to follow. While at first sight the experts could in principle skip this section we would like to ask them not to do it as our strategy for the identification of NP through quark flavour violating processes differs significantly from 
other strategies found in the literature.

We begin then by recalling a visible 
tension between the CP-violating observables $\varepsilon_K$ and $S_{\psi K_S}$
within the SM first emphasized in \cite{Lunghi:2008aa,Buras:2008nn}. 
The nature of this tension depends sensitively on the value of the CKM 
element $\vub$ for which the exclusive semileptonic decays imply significantly 
lower value than the inclusive ones. While the latter problem will hopefully 
be solved in the coming years, it is instructive to consider presently 
two scenarios for $\vub$:
 \begin{itemize}
\item
{\bf Exclusive (small) $\vub$ Scenario 1:}
$|\varepsilon_K|$ is smaller than its experimental determination,
while $S_{\psi K_S}$ is  close to its central experimental value.
\item
{\bf Inclusive (large) $\vub$ Scenario 2:}
$\varepsilon_K$ is consistent with its experimental determination,
while $S_{\psi K_S}$ is significantly higher than its  experimental value.
\end{itemize}
The actual size of discrepancies will be considered in Step 3 of our 
strategy but the message is clear:
dependently which scenario is considered we need either  
{\it constructive} NP contributions to $|\varepsilon_K|$ (Scenario 1) 
or {\it destructive} NP contributions to  $S_{\psi K_S}$ (Scenario 2). 
However this  NP should not spoil the agreement with the data 
for $S_{\psi K_S}$ (Scenario 1) and for $|\varepsilon_K|$ (Scenario 2).

In view of the fact that the theoretical precision on $S_{\psi K_S}$ 
is significantly larger than in the case of $\varepsilon_K$, one may wonder 
whether removing $1-2\sigma$ anomaly in $\varepsilon_K$ 
by generating a $2-3\sigma$ anomaly in $S_{\psi K_S}$ is a reasonable 
strategy. However, we will proceed in this manner as this will 
teach us how different NP scenarios deal with this problematic. Definitely 
in order to resolve this puzzle we need not only precise determination 
of $\vub$ not polluted by NP but also precise values of non-perturbative 
parameters relevant for SM predictions in this case.

Until 2012 there was another significant tension between  SM branching ratio for $B^+\to\tau^+\nu_\tau$ and the data, with the experimental value 
being by a factor of two larger than the theory. This would favour strongly 
large $\vub$ scenario. However, presently after the data from BELLE this 
discrepancy, as discussed in Step 5 of our strategy, is practically absent. 
Yet, the agreement of the SM with the data still depends on the chosen 
value of $\vub$ which enters this branching ratio quadratically. In turn 
the kind of NP which would improve the agreement of the theory with the 
data depends on the chosen value of $\vub$. Other modest tensions 
between the SM and the data will be discussed as we proceed.

Now models with many new parameters can face successfully both scenarios 
for $\vub$
removing the deviations from the data for certain ranges of their parameters 
but as we will see below in simpler models often 
only one scenario can be admitted as only in that scenario for $\vub$ a given 
model has a chance to fit $\varepsilon_K$ and $S_{\psi K_S}$ simultaneously. 
For instance as we will see in the course of our presentation
 models with constrained Minimal Flavour Violation (CMFV) select Scenario 1, while the 2HDM with MFV and flavour blind phases,
${\rm 2HDM_{\overline{MFV}}}$, favours  Scenario 2 for $\vub$.
What is interesting is that the future precise determination of $\vub$ through 
tree level decays will 
be able to distinguish between these two NP scenarios. We will see that there 
are other models which can be distinguished in this simple manner.

Clearly, in order to get the full picture many more observables have 
to be considered. For instance in Table~\ref{tab:SMpred}, that can be found in Step 3, we illustrate the SM predictions for additional observables, 
in particular the mass differences  $\Delta M_s$  and $\Delta M_d$ in 
the $B_{s,d}-\bar B_{s,d}$ systems.
What is 
striking in this table is that with the present lattice input in Table~\ref{tab:input} the predicted central values of $\Delta M_s$  and $\Delta M_d$ 
 are both in a good agreement with the latter 
when hadronic uncertainties are taken into account. In particular 
the central value of the ratio $\Delta M_s/\Delta M_d$ is 
 very close to  the data. 
These results depend strongly on the lattice input and in the case 
of $\Delta M_d$ on the value of $\gamma$. Therefore to get a better insight 
both lattice input and the tree level determination of $\gamma$ 
have to improve.  Moreover the situation changes with time. 
While one year ago  lattice input was such that
models providing  $10\%$ {\it suppression} 
of {\it both} $\Delta M_s$ and $\Delta M_d$ were favoured, this is no longer 
the case as can be seen  in Table~\ref{tab:SMpred}. 

However, for the purpose of presenting our strategy, it will be
 useful to keep the old central values from lattice that are consistent 
within $1\sigma$ with the present ones but imply certain deviations from 
SM expectations. This will allow to illustrate how NP can remove these 
deviations. In doing this we will keep 
in mind that the pattern 
of deviations from SM expectations could be modified in the future. 
This is in particular the case of observables, like $\Delta M_{s,d}$, that 
still suffer from non-perturbative uncertainties. 
It could turn out that suppressions (enhancements) of some observables 
 required  in our examples from NP 
will be modified to enhancements (suppressions) in the future and it will 
be of interest to see whether a given model could cope with such changes. 
Having this in mind will lead us eventually in Section~\ref{sec:5} 
to a  proposal of {\it DNA-charts}, primarily with the goal to exhibit 
transparently  the pattern of enhancements and suppressions of flavour 
observables  in a given 
NP scenario and the correlations between them. Of course also this pattern 
will include situations in which no modifications in a given observable 
relative to the SM will take place.

\subsection{Towards New Standard Model in 12 Steps}
\begin{figure}[!tb]
\centerline{\includegraphics[width=0.65\textwidth]{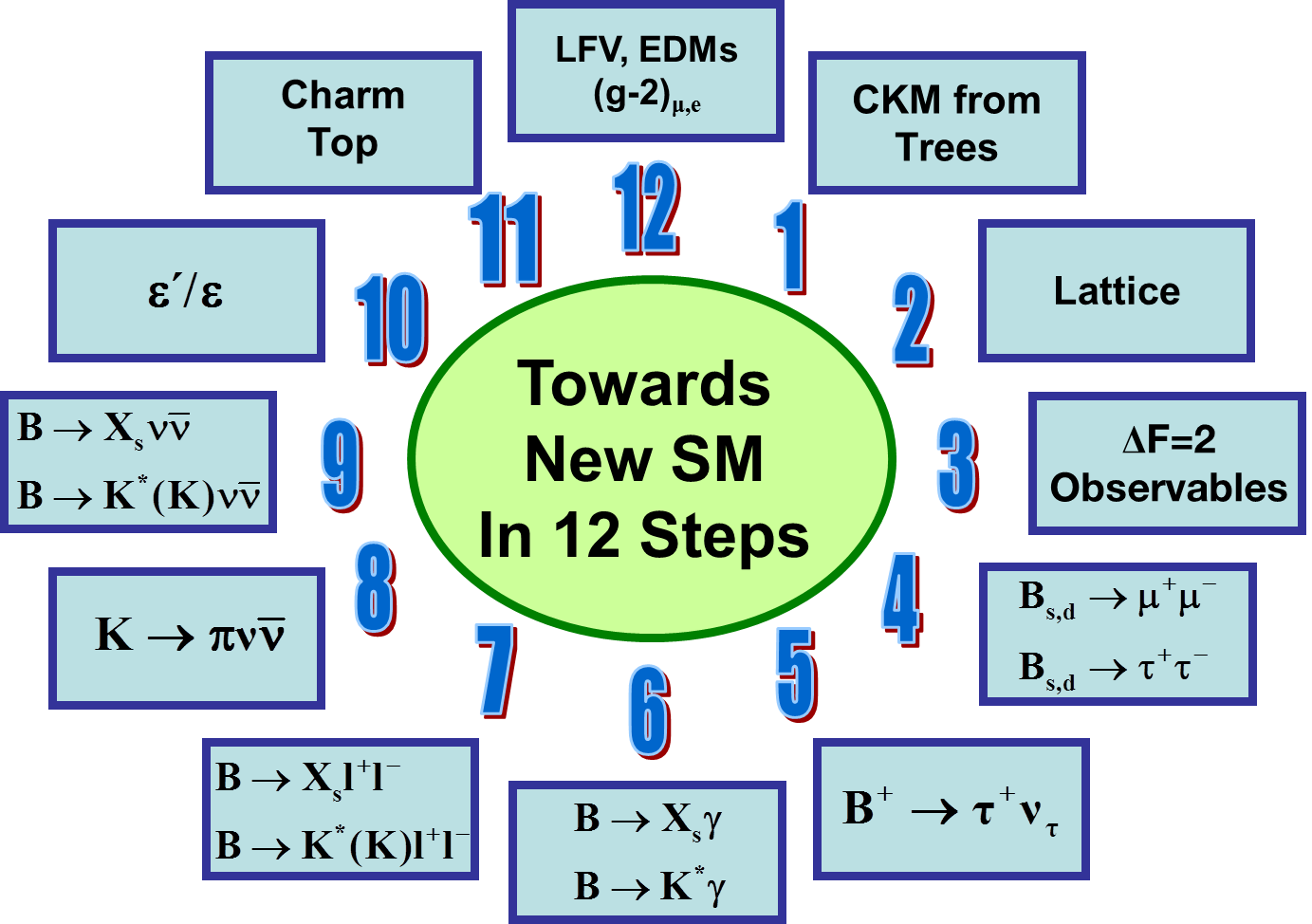}}
\caption{\it Towards New Standard Model in 12 Steps.}\label{Fig:1}~\\[-2mm]\hrule
\end{figure}

Our strategy involves twelve steps that 
we present in detail in Section~\ref{sec:4}. These steps involve a number 
of decays and transitions as shown in Fig.~\ref{Fig:1} and can be
 properly adjusted in case 
the pattern of deviations from the SM will be modified.

For the time being
 assuming that the present tensions will 
be strengthened with time, when the data improve, the specific 
questions that arise are:
\begin{itemize}
\item
Which model is capable of removing the $\varepsilon_K$-$S_{\psi K_S}$ 
tension and simultaneously providing modifications in 
 $B^+\to\tau^+\nu_\tau$ and
$\Delta M_{s,d}$ if they are required?
\item
What are the predictions of this model for:
\be\label{OBS1}
S_{\psi\phi},\quad B_{s,d}\to\mu^+\mu^-,\quad B\to K^*\ell^+\ell^-,\quad B\to X_s\ell^+\ell^-,
\ee
\be
B\to X_s\nu\bar\nu,\quad, B\to K^*\nu\bar\nu, \quad B\to K\nu\bar\nu,
\ee
\be\label{OBS3}
\kpn, \quad \klpn, \quad \frac{\varepsilon^\prime}{\varepsilon}, \quad K_L\to\mu^+\mu^-
\ee
and how are these predictions correlated with  $S_{\psi K_S}$ 
and $\varepsilon_K$?
\end{itemize}

 The comparison of processes and observables listed here with those 
appearing in Fig.~\ref{Fig:1} should not be understood that the ones 
missing in (\ref{OBS1})-(\ref{OBS3}), like lepton flavour violation 
and electric dipole moments, are less important. But as we discuss 
these topics in our review only in general terms. They will in fact remain 
under the shadow of the processes listed above.

\subsection{Correlations between Observables}
In order to reach our goal we need a strategy for uncovering new physics 
responsible for the observed anomalies and possible anomalies hopefully 
found in the future. One line of attack chosen by several authors are 
model independent studies of the Wilson coefficients with the goal to find 
out how much room for NP contributions is still left in each coefficient. 
In this context correlations between various Wilson coefficients are studied. 
While such studies are certainly useful and give some insight into the room 
left for new physics, one should keep in mind that Wilson Coefficients are 
scale and renormalization scheme dependent and correlations between them 
generally depend on the scale at which they are evaluated and the renormalization scheme used.

Therefore it is our strong believe that searching for correlations 
between the measured observables is more powerful. Extensive studies of 
correlations between various observables in concrete models illustrate 
very clearly the power of this strategy. Quite often only a qualitative 
behaviour of these correlations is sufficient to eliminate the model 
as a solution to observed anomalies or to select models as candidates 
for a new Standard Model. 
A detailed review of such explicit studies can be found in 
\cite{Buras:2010wr,Buras:2012ts}.
These studies allowed to construct various classifications of NP contributions 
in the form of ''DNA'' tables \cite{Altmannshofer:2009ne} and {\it flavour codes}  \cite{Buras:2010wr}
  as well as provided 
some insight into the physics behind resulting correlations in specific 
models \cite{Blanke:2009pq}.  Detailed analyses in this spirit  have been
 subsequently performed in \cite{Altmannshofer:2011gn,Altmannshofer:2012ir}.
With improved data all these results will be increasingly useful. 

In the present paper we will take a slightly different route. Instead of 
investigating explicit models 
we will illustrate the search for new Standard Model using very simple 
models being aware of the fact that in more complicated models certain patterns 
of flavour violations and correlations between various observables could 
be washed out and be less transparent. { This strategy has been used 
by us in our most recent papers \cite{Buras:2012sd,Buras:2012jb,Buras:2012dp,Buras:2013td,Buras:2013uqa,Buras:2013rqa,Buras:2013raa,Buras:2013qja,Buras:2013dea}.} 
In this context a prominent role 
will be played by new tree-level contributions to FCNC processes mediated 
either by heavy neutral gauge bosons or neutral heavy scalars. These contributions are governed in particular by the couplings $\Delta_{L,R}^{ij}(Z^\prime)$ and 
$\Delta_{L,R}^{ij}(H^0)$ for gauge bosons and scalars to quarks, respectively.
Here $(i,j)$ denote quark flavours. As we 
will see in addition to a general form of these couplings 
it will be instructive to consider the following four scenarios for them 
keeping the pair $(i,j)$ fixed:
\begin{enumerate}
\item
Left-handed Scenario (LHS) with complex $\Delta_L^{bq}\not=0$  and $\Delta_R^{bq}=0$,
\item
Right-handed Scenario (RHS) with complex $\Delta_R^{bq}\not=0$  and $\Delta_L^{bq}=0$,
\item
Left-Right symmetric Scenario (LRS) with complex  
$\Delta_L^{bq}=\Delta_R^{bq}\not=0$,
\item
Left-Right asymmetric Scenario (ALRS) with complex
$\Delta_L^{bq}=-\Delta_R^{bq}\not=0$,
\end{enumerate}
with analogous scenarios for the pair $(s,d)$. These ideas can also be 
extended to charged gauge boson ($W^{\prime +}$) and charged Higgs ($H^+)$ exchanges. 
We will see that these simple scenarios will give us a profound insight 
into the flavour structure of models in which NP is dominated by left-handed 
currents or right-handed currents or left-handed and right-handed currents 
of approximately the same size. 

The idea of looking at such  NP scenarios is not new and has been in 
particular motivated by a detailed study of supersymmetric flavour models 
with NP dominated by LH currents, RH currents or equal amount of LH and RH 
currents \cite{Altmannshofer:2009ne}. Moreover, it has been found in several studies of non-supersymmetric 
 frameworks 
like LHT model \cite{Blanke:2006eb} or Randall-Sundrum scenario with custodial protection (RSc) 
\cite{Blanke:2008yr}
that models with the dominance of LH or RH currents exhibit quite different 
patterns of flavour violation. Our simple models will demonstrate it in 
a transparent manner.

There is another point we would like to make. In several papers predictions 
for various observables in given extensions of the SM are made using presently 
available loop processes to determine CKM parameters. As we will emphasize 
in Step 1 below, in our view this is not the optimal time to proceed in this 
manner. As last years have shown  such predictions have
rather  a short life-time. It appears to us that it is more useful at this stage 
to develop transparent formulae which will allow to monitor the future events 
in flavour physics in the SM and its extensions when the experimental data 
improve and the uncertainties in lattice calculations decrease.

Our strategy will also be complementary to analyses in which allover fits 
using sophisticated computer machinery are made. We will start with a subset 
of observables which have simple theoretical structure ignoring first 
constraints 
from more complicated observables. In subsequent steps we will gradually  include  more observables in our analysis which necessarily will modify our 
insights gained in the first steps thereby teaching us something. Only  in Section~\ref{sec:5}  we will look at all observables
simultaneously and the grand 
view of simple models and the grand view of more complicated models should 
hopefully
allow us to monitor efficiently flavour events in this decade.

 With this general strategy in mind we can now enter the details recalling first 
briefly the theoretical framework for weak decays.

\section{Theoretical Framework}\label{sec:2}
\subsection{Preliminaries}
The field of weak decays is based on effective Hamiltonians with the generic 
form given as follows
\be\label{Heff-general}
\Heff^{\rm Process} =\kappa
\sum_{i}C_i(\mu)Q_i + h.c\,.
\ee
Here $Q_i$ are local operators and $C_i(\mu)$ their Wilson coefficients 
that can be evaluated in renormalization group improved perturbation theory. 
Details on the calculations of these coefficients and the related technology 
including QCD corrections at the NLO and NNLO level can be found in 
\cite{Buchalla:1995vs,Buras:1998raa,Buras:2011we}.

The overall factor $\kappa$  can be chosen at will in accordance with 
the overall normalization of Wilson coefficients and operators. Sometimes 
it is useful to set $\kappa$ to its value in the SM but this is not always 
the case as we will see below.
The scale $\mu$ can be the low 
energy scale $\mu_L$ at which actual lattice calculations 
are performed or any other 
scale, in particular the matching scale $\mu_\text{in}$, the border line between a given  full and corresponding effective theory.

The 
matrix elements of the effective Hamiltonian are directly 
related to decay amplitudes and can  be 
written generally as follows:
\be\label{Heff-general1}
\langle\Heff^{\rm Process}\rangle =\kappa
\sum_{i}C_i(\mu_L)\langle Q_i(\mu_L)\rangle\,
\ee
or
\be\label{Heff-general2}
\langle\Heff^{\rm Process}\rangle =\kappa
\sum_{i}C_i(\mu_\text{in})\langle Q_i(\mu_\text{in})\rangle\,.
\ee

These two expressions are equal to each other and the Wilson coefficients in 
them are connected through 
\be\label{RGevolution}
\vec{C}(\mu_L)=\hat{U}(\mu_L,\,\mu_{\rm in})\vec{C}(\mu_{\rm in}),
\ee
where $\hat{U}$ is the renormalization group evolution matrix and $\vec{C}$ a
column vector. Which of the formulations is more useful depends on the process and model 
considered. 

Now the Wilson coefficients depend directly on the couplings present in 
the fundamental theory. In our paper the quark-gauge boson and quark-scalar 
couplings will play the prominent role and it is useful to introduce 
a general notation for them so that they can be used in the context of 
any model considered.

\begin{figure}[!tb]
\includegraphics[width = 0.6\textwidth]{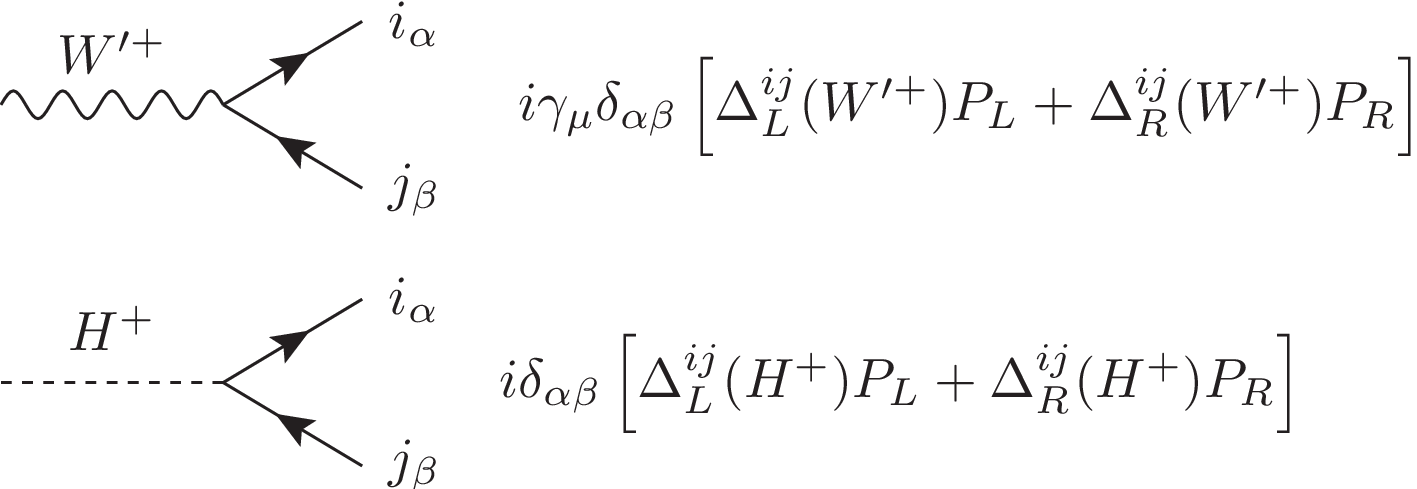}
\caption{\it Feynman rules for colourless charged gauge boson $W^{\prime +}$ with mass $M_{W^\prime}$, 
and charged colourless scalar particle $H^+$
with mass $M_H$, where $i\,(j)$ denotes an up-type (down-type) quark flavour with charge $+\frac{2}{3}$ ($-\frac{1}{3}$)  and
$\alpha,\,\beta$  are colour indices. $P_{L,R}=(1\mp\gamma_5)/2$.}\label{charged}
\end{figure}

\begin{figure}[!tb]
\includegraphics[width = 0.6\textwidth]{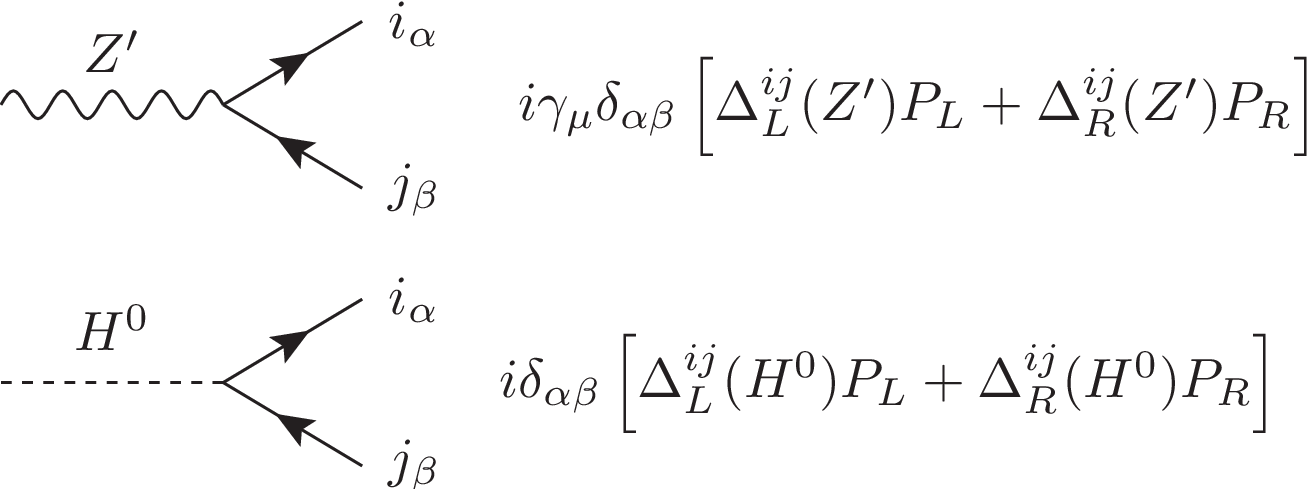}
\caption{\it Feynman rules for colourless neutral gauge boson $Z^\prime$ with mass $M_{Z^\prime}$, 
and neutral colourless scalar particle $H^0$
with mass $M_H$, where $i,\,j$ denote different
quark flavours and $\alpha,\,\beta$ the colours. $P_{L,R}=(1\mp\gamma_5)/2$.}\label{neutral}
\end{figure}

Quite generally we can consider  
the basic interactions of charged gauge bosons $W^{\prime +}$, 
charged scalars $H^+$, neutral gauge bosons $Z^\prime$ and neutral scalars $H^0$ 
with quarks that are shown as vertices 
in Figs.~\ref{charged} and \ref{neutral}. The gauge bosons shown there are 
all colourless but this notation could be easily extended to coloured 
gauge bosons and scalars. They can also be extended to heavy quarks 
interacting with SM quarks and to interactions of bosons with leptons. 
It should be emphasized that all the fields in these vertices are in 
the mass eigenstate basis. In the course of our presentation we will give 
the expressions for various coefficients in terms of these couplings.

 In Figs.~\ref{charged} and \ref{neutral} the couplings $\Delta_{L,R}$ 
are $3\times 3$ complex matrices in the flavour space with  $i,j$  denoting 
different quark flavours. In the case of charged boson exchanges the 
first flavour index in  Fig.~\ref{charged} denotes an up-type quark and the 
second  a down-type quark.

In models in which FCNC processes take place first at one-loop 
level, it is useful to work with (\ref{Heff-general2}) and express 
$C_i(\mu_\text{in})$ in terms of a
 set of gauge independent master functions which result from calculations of penguin and box diagrams and which govern
the FCNC processes. In particular this is the case for 
those models in 
which the operator structure is the same as in the SM. 
We will discuss such models soon.

On the other hand in models in which new operators with 
right-handed currents and scalar and pseudoscalar currents are present, 
it is necessary to exhibit these new structures explicitly by introducing 
new loop functions. This is also the case for  models with 
tree-level FCNC processes mediated by gauge bosons and scalars as such 
exchanges bring in necessarily new operators beyond the ones present in the 
SM.

We will next introduce a number of simple extensions of the SM that will 
serve to illustrate our strategy.
\subsection{Constrained Minimal Flavour Violation (CMFV)} 
This is possibly the simplest class of BSM scenarios. It is defined 
pragmatically as follows \cite{Buras:2000dm}:
\begin{itemize}
\item
The only source of flavour and CP violation is the CKM matrix. This 
implies that the only CP-violating phase is the KM phase and that
CP-violating flavour blind phases are assumed to be absent.
\item
The only relevant operators in the effective Hamiltonian below 
the electroweak scale are the ones present within the SM.
\end{itemize}
Detailed expositions of phenomenological consequences of this NP 
scenario has been given in \cite{Buras:2003jf,Blanke:2006ig} and recently 
in \cite{Buras:2012ts}.

In CMFV models it is useful to work with (\ref{Heff-general2}) and express 
$C_i(\mu_\text{in})$ in terms of a
 set of gauge independent master functions which result from calculations of penguin and box diagrams and which govern
the FCNC processes.
One has then
seven one-loop functions that are 
denoted by\footnote{The first calculation of these functions within the SM 
is due to  Inami and Lim \cite{Inami:1980fz}. The gauge independent form of these functions 
 as used presently in the 
literature has been introduced in the SM in  \cite{Buchalla:1990qz} and 
in CMFV models in \cite{Buras:2003jf}.}
\be\label{masterf}
S(v),~X(v),~Y(v),~Z(v),~E(v),~D'(v),~E'(v),
\ee
where the variable $v$ collects the parameters of a given model. It is often 
useful to keep the CKM factors outside these functions. Then in models with 
MFV without flavour blind phases these functions are real valued and 
universal with respect to different meson systems 
implying various stringent correlations between various decays and related 
observables. In models with MFV and flavour blind CPV phases and genuine 
non-MFV frameworks
these functions become complex valued and the universality 
between various meson systems is violated implying corrections to 
correlations present in models with MFV but no flavour blind phases.

Generally, several master functions contribute to a given decay,
although decays exist which depend only on a single function.
We have the following correspondence between the most interesting FCNC
processes and the master functions in the MFV models 
in which the operator structure is the same as in the SM:
\begin{center}
\begin{tabular}{lcl}
$K^0-\bar K^0$-mixing ($\varepsilon_K$) 
&\qquad\qquad& $S(v)$ \\
$B_{d,s}^0-\bar B_{d,s}^0$-mixing ($\Delta M_{s,d}$) 
&\qquad\qquad& $S(v)$ \\
$K \to \pi \nu \bar\nu$, $B \to X_{d,s} \nu \bar\nu$ 
&\qquad\qquad& $X(v)$ \\
$K_{\rm L}\to \mu \bar\mu$, $B_{d,s} \to \ell^+\ell^-$ &\qquad\qquad& $Y(v)$ \\
$K_{\rm L} \to \pi^0 e^+ e^-$ &\qquad\qquad& $Y(v)$, $Z(v)$, 
$E(v)$ \\
$\varepsilon'$, Nonleptonic $\Delta B=1$, $\Delta S=1$ &\qquad\qquad& $X(v)$,
$Y(v)$, $Z(v)$,
$E(v)$ \\
$B \to X_s \gamma$ &\qquad\qquad& $D'(v)$, $E'(v)$ \\
$B \to X_s~{\rm gluon}$ &\qquad\qquad& $E'(v)$ \\
$B \to X_s \ell^+ \ell^-$ &\qquad\qquad&
$Y(v)$, $Z(v)$, $E(v)$, $D'(v)$, $E'(v)$
\end{tabular}
\end{center}

This table means that the observables like branching ratios, mass differences
$\Delta M_{d,s}$ in $B_{d,s}^0-\bar B_{d,s}^0$-mixing and the CP violation 
parameters $\varepsilon$ and $\varepsilon'$, all can be to a very good 
approximation entirely expressed in
terms of the corresponding master functions and the relevant CKM factors. 

\subsection{CMFV Relations as Standard Candles of Flavour Physics}
The implications of this framework are so stringent that it appears to us 
to consider them as standard candles of flavour physics. Even if some of 
these relations will appear again in the context of our presentation 
it is useful to collect the most important ones at one place here. 
 A review of these relations is 
given in \cite{Buras:2003jf}. As NP effects in FCNC processes appear  smaller 
than anticipated in the past the importance of these relations increased 
in 2013.

We have:
\begin{enumerate}
  \item  $S_{\psi K_S}$ and $S_{\psi\phi}$ are as in the SM and therefore given 
 by 
 \be
 S_{\psi K_S} = \sin(2\beta)\,,\qquad S_{\psi\phi} =  \sin(2|\beta_s|)\,,
 \label{CMFV1}
 \ee
 where $\beta$ and $\beta_s$ are defined  in (\ref{vtdvts}).
 \item 
  While $\Delta M_d$ and $\Delta M_s$ can differ from the SM 
 values their ratio is as in the SM
 \be\label{CMFV2}
 \left(\frac{\Delta M_d}{\Delta M_s}\right)_{\rm CMFV}=
 \left(\frac{\Delta M_d}{\Delta M_s}\right)_{\rm SM}.
 \ee
 Moreover, this ratio is given entirely in terms of CKM parameters 
 and non perturbative parameter $\xi$:
 \be\label{CMFV3}
 \frac{\Delta M_d}{\Delta M_s}
 =\frac{m_{B_d}}{m_{B_s}}
 \frac{1}{\xi^2}
 \left|\frac{V_{td}}{V_{ts}}\right|^2r(\Delta M), 
 \qquad \xi^2=\frac{\hat B_{s}}{\hat B_{d}}\frac{F^2_{B_s}}{F^2_{B_d}}
 \end{equation}
 where we have introduced the quantity $r(\Delta M)$,  
 that is equal unity in models with CMFV. It 
 parametrizes the deviations from these relations found in several models 
 discussed by us below.
 \item These two properties allow the construction of the {\it Universal 
 Unitarity Triangle} (UUT) of models of CMFV that uses as inputs the measured 
 values of $S_{\psi K_S}$ and $\Delta M_s/\Delta M_d$ \cite{Buras:2000dm}.
 \item The flavour universality of  $S(v)$ allows to derive 
          universal expressions for $S_{\psi K_S}$ and the angle $\gamma$ in 
          the UUT that depend only on
           $\vus$, $\vcb$, known from tree-level decays, and 
           non-perturbative parameters entering the evaluation of 
           $\varepsilon_K$ and $\Delta M_{s,d}$ 
             \cite{Buras:1994ec,Buras:2000xq,Blanke:2006ig}. They 
            are valid for all CMFV models. We will present an
  update of these formulae in Step 3 of our strategy. Therefore, once 
 the data on $\vus$, $\vcb$, $\varepsilon_K$ and $\Delta M_{s,d}$ are 
 taken into account one is able in this framework to predict not only 
 $S_{\psi\phi}$ but also $\vub$.
 \item For fixed CKM parameters determined in tree-level decays, $|\varepsilon_K|$, 
 $\Delta M_s$ and $\Delta M_d$, if modified,  can only be {\it enhanced} 
 relative to SM predictions  \cite{Blanke:2006yh}. Moreover this 
 happens in a correlated manner \cite{Buras:2000xq}.
 \item  Two other interesting universal  relations in models with CMFV 
 are
 \begin{equation}\label{CMFV4}
 \frac{\mathcal{B}(B\to X_d\nu\bar\nu)}{\mathcal{B}(B\to X_s\nu\bar\nu)}=
 \left|\frac{V_{td}}{V_{ts}}\right|^2r(\nu\bar\nu)
 \end{equation}
 \begin{equation}\label{CMFV5}
 \frac{\mathcal{B}(B_d\to\mu^+\mu^-)}{\mathcal{B}(B_s\to\mu^+\mu^-)}=
 \frac{\tau({B_d})}{\tau({B_s})}\frac{m_{B_d}}{m_{B_s}}
 \frac{F^2_{B_d}}{F^2_{B_s}}
 \left|\frac{V_{td}}{V_{ts}}\right|^2 r(\mu^+\mu^-),
 \end{equation}
 where we have again introduced the quantities $r(\nu\bar\nu)$ 
 and $r(\mu^+\mu^-)$ that are all equal unity in CMFV models.
 
 \item 
 Eliminating $|V_{td}/V_{ts}|$ from (\ref{CMFV3}) and (\ref{CMFV5}) 
 allows 
 to obtain another   universal relation within the
 CMFV models
 \cite{Buras:2003td}  
 \be\label{CMFV6}
 \frac{\mathcal{B}(B_{s}\to\mu^+\mu^-)}{\mathcal{B}(B_{d}\to\mu^+\mu^-)}
 =\frac{\hat B_{d}}{\hat B_{s}}
 \frac{\tau( B_{s})}{\tau( B_{d})} 
 \frac{\Delta M_{s}}{\Delta M_{d}}r, \quad r=\frac{r(\Delta M)}{r(\mu^+\mu^-)}
 \ee
 that does not 
 involve $F_{B_q}$ and CKM parameters and consequently contains 
 smaller hadronic and parametric uncertainties than the formulae considered 
 above. It involves
 only measurable quantities except for the ratio $\hat B_{s}/\hat B_{d}$
 that is now known already from  lattice calculations with impressive 
accuracy of $\pm 2-3\%$ \cite{Carrasco:2013zta} and this 
 precision should be even improved. Therefore the relation (\ref{CMFV6}) should allow
 a precision test of CMFV even if the branching ratios $\mathcal{B}(B_{s,d}\to\mu^+\mu^-)$ would turn out to deviate from SM 
predictions by $10-20\%$.
 \item 
  All amplitudes for FCNC processes within the CMFV framework can be expressed 
  in terms of seven {\it real} and {\it universal} master loop functions 
  listed in (\ref{masterf}). The implications of this property are numerous 
  correlations between various observables 
  that are discussed more explicitly in Section~\ref{sec:correlations}.

\end{enumerate}

\subsection{Minimal Flavour Violation at Large (MFV)}
In the more general case of MFV the formulation  
with the help of global symmetries present in the limit of vanishing 
Yukawa couplings  as formulated in 
\cite{D'Ambrosio:2002ex} is elegant and useful. See also \cite{Feldmann:2006jk}
for a similar formulation that goes beyond the MFV. 
Other profound discussions of various aspects of MFV 
can be found in 
\cite{Colangelo:2008qp,Paradisi:2008qh,Mercolli:2009ns,Feldmann:2009dc,Kagan:2009bn,Paradisi:2009ey}.
An excellent  compact formulation of MFV as effective theory has been given by
Gino Isidori \cite{Isidori:2010gz}.  We also recommend the reviews in 
\cite{Hurth:2008jc,Isidori:2012ts}, where phenomenological aspects of MFV 
are summarized.

In short the hypothesis of MFV amounts to assuming that the Yukawas are the only sources of the breakdown of flavour and CP-violation.
The phenomenological implications of the MFV hypothesis formulated in this 
more grander manner than the CMFV formulation given above can be found 
model independently 
by using an effective field theory approach (EFT) \cite{D'Ambrosio:2002ex}. 
In this framework the SM Lagrangian is supplemented by all higher dimension
 operators consistent with the MFV hypothesis, built using the Yukawa 
couplings as spurion fields. The NP effects in this 
framework
are then parametrized in terms of a few {\it flavour-blind} 
free parameters and SM Yukawa couplings that are solely responsible for 
 flavour violation  and also CP violation if these flavour-blind parameters 
are chosen as {\it real} quantities as done in \cite{D'Ambrosio:2002ex}. 
This approach naturally 
suppresses FCNC processes to the level observed experimentally even in the 
presence of new particles with masses of a few hundreds GeV. It also implies 
specific correlations between various observables, which are not as  stringent 
as in the CMFV but are still very powerful.

Yet, it should be stressed  that the MFV symmetry principle in itself does 
not forbid the presence of
{\it flavour blind} CP violating sources~\cite{Baek:1998yn,Baek:1999qy,Bartl:2001wc,Paradisi:2009ey,Ellis:2007kb,Colangelo:2008qp,Altmannshofer:2008hc,Mercolli:2009ns,Feldmann:2009dc,Kagan:2009bn} that make effectively the flavour blind free parameters 
{\it complex} quantities having flavour-blind phases (FBPs). These phases can in 
turn enhance the electric dipole moments (EDMs) of various particles and 
atoms and in the interplay with the CKM matrix can have also profound 
impact on flavour violating observables, in particular the CP-violating ones. 
In the context of the so-called aligned 2HDM model such effects have also been 
emphasized 
in \cite{Pich:2009sp}.

The 
introduction of flavour-blind 
CPV phases compatible with the MFV symmetry principle turns out to 
be a very interesting set-up~\cite{Kagan:2009bn,Colangelo:2008qp,Mercolli:2009ns,Paradisi:2009ey,Ellis:2007kb}.
In particular, as noted in \cite{Kagan:2009bn}, 
a large new phase in $B^0_s$--$\bar B^0_s$ mixing could
in principle be obtained in the MFV framework if additional FBPs
are present. This idea cannot be realized in the ordinary MSSM
with MFV, as shown in~\cite{Altmannshofer:2009ne,Blum:2010mj}. The difficulty of realizing this scenario in the MSSM 
is due to the suppression in the MSSM 
of effective operators with several Yukawa insertions. 
Sizable couplings for these operators are necessary both to have an 
effective large CP-violating phase in $B^0_s$--$\bar B^0_s$ mixing and, at the same time,
 to evade bounds from 
other observables, such as $B_s\to \mu^+\mu^-$ and $B \to X_s \gamma$.
However, it could be realized in different underlying models, 
such as the up-lifted MSSM, as pointed out in \cite{Dobrescu:2010rh}, 
in the so-called
beyond-MSSM scenarios~\cite{Altmannshofer:2011iv,Altmannshofer:2011rm}
and in the 2HDM with MFV and FBPs, the so-called ${\rm 2HDM_{\overline{MFV}}}$ \cite{Buras:2010mh} to which we will return at various places in this writing. 
 An excellent review of 2HDMs at large can be found in \cite{Branco:2011iw}. 

 As we will see in Step 3 of our strategy the present data from the LHCb 
 show that the new phases in  $B^0_s$--$\bar B^0_s$ mixing, if present, must
 be rather small. Consequently also the role of flavour blind phases in 
 describing  data decreased significantly relatively to the one they 
played in the studies 
summarized above. However, the full assessment of the importance of these 
phases will only be possible when the CP-violation in  $B^0_s$--$\bar B^0_s$ mixing will be precisely measured and also the bounds on electric dipole moments 
improve.

\subsection{Simplest Models with non-MFV Sources}
In models with new sources of flavour and CP violation in which the 
operator structure is not modified, the formulation of FCNC processes 
in terms of seven one-loop functions is useful as well 
but when the CKM factors are the only ones kept explicit as 
overall factors, 
these functions become complex and are different for different meson systems.
We have then
 ($i=K,d,s$):
\begin{equation}\label{eq31}
S_i\equiv|S_i|e^{i\theta_S^i},
\quad
X_i \equiv \left|X_i\right| e^{i\theta_X^{i}}, \quad
Y_i \equiv \left|Y_i\right| e^{i\theta_Y^i}, \quad
Z_i \equiv \left|Z_i\right| e^{i\theta_Z^i}\,,
\end{equation}

\begin{equation} \label{eq32}
E_i \equiv \left|E_i\right| e^{i\theta_{E}^i}\,,\quad
D'_i \equiv \left|D_i'\right| e^{i\theta_{D'}^i}\,, \quad
E'_i \equiv \left|E_i'\right| e^{i\theta_{E'}^i}\,.
\end{equation}

As now the property of the universality of these
functions
is lost, the usual CMFV relations between $K$, $B_d$ and $B_s$
systems listed above can be violated and the parameters $r(k)$ introduced 
in the context of our discussion of CMFV models are generally different 
from unity and can be complex.  A known example is the Littlest Higgs 
Model with T-parity (LHT) \cite{Blanke:2006eb}.

\boldmath
\subsection{The $U(2)^3$ Models}
\unboldmath
Probably the simplest models with new sources of flavour violation are 
models in which the $U(3)^3$ symmetry of MFV models is reduced to 
$U(2)^3$ symmetry \cite{Barbieri:2011ci,Barbieri:2011fc,Barbieri:2012uh,Barbieri:2012bh,Crivellin:2011fb,Crivellin:2011sj,Crivellin:2008mq}.
As pointed out in \cite{Buras:2012sd} a number of properties of CMFV models 
remains in this class of models,  in particular the relation (\ref{CMFV6}) is still valid. On the other hand there are profound 
differences due 
to the presence of new CP-phases which we will discuss in the course of our 
presentation.

\subsection{Tree-Level Gauge Boson and Scalar Exchanges}\label{toy}
In a number of BSM scenarios NP can enter already at tree level, both 
in charged current processes and in FCNC processes.  

In the case of charged current processes prominent examples are the right-handed $W^{\pm\prime}$ bosons in left-right symmetric models and charged Higgs ($H^\pm$) particles in models with extended scalar sector like  two Higgs doublet models and supersymmetric models. In these models new operators are present, 
the simplest example being $(V+A)\times(V+A)$ operators originating 
in the exchange of $W^{\pm\prime}$ gauge bosons in the left-right symmetric 
models.   In these models also  $(V-A)\times(V+A)$ operators 
contribute. These operators generate in turn  through QCD corrections 
 $(S-P)\times(S+P)$ operators present also in models with $H^\pm$ 
particles. In the latter models also $(S\pm P)\times(S\pm P)$ operators 
are present. Needless to say all these statements also apply to neutral 
gauge bosons and scalars mediating $\Delta F=1$ transitions.
It should also be stressed that anomalous right-handed couplings of SM gauge 
bosons $W^\pm$ to quarks can be generated through the mixing with heavy 
vectorial fermions.

Concerning FCNC processes, tree-level transitions are present in any 
model in which GIM mechanism is absent in some sectors of a given model. 
This is the case of numerous 
 $Z^\prime$ models, gauged flavour models with new very heavy neutral 
gauge bosons 
and Left-Right symmetric models with 
heavy neutral scalars. 
They can also be generated at one loop in models 
having GIM at the fundamental level and Minimal Flavour Violation of which 
Two-Higgs Doublet models with and without supersymmetry are the best known 
examples. 
Tree-level  $Z^0$ and SM neutral Higgs $H^0$ contributions to $\Delta F=2$ 
processes are also possible
in models containing vectorial heavy fermions that mix with the standard chiral quarks. This is also the case of
 models in which $Z^0$ and SM neutral Higgs $H^0$ mix with new heavy 
gauge bosons and scalars in the process of electroweak symmetry breaking. 
Recently two very detailed analyses of FCNCs within models with tree-level 
gauge boson and neutral scalar and pseudoscalar exchanges have been performed in
\cite{Buras:2012jb,Buras:2013rqa} and we will include the highlights from 
these two papers in our discussion.

In the previous section we defined in Figs.~\ref{charged} and \ref{neutral}
the basic interactions of charged gauge bosons $W^{\prime +}$, 
charged scalars $H^+$, 
with quarks. In the flavour precision era also QCD corrections to tree-level exchanges have to be taken into account. They depend  
on whether a gauge 
boson or scalar is exchanged and of course on the process considered. 
Fortunately, the NLO matching conditions for tree-level 
 neutral gauge bosons $Z^\prime$ and neutral scalars $H^0$ exchanges  
have been calculated recently in \cite{Buras:2012fs,Buras:2012gm}. Combining them
 with previously calculated two-loop anomalous dimensions of four-quark 
operators, it is possible to perform complete NLO renormalization group 
analysis in this case.

Finally, we would like to make a general comment on the expressions for 
various observables in this class of models that we will encounter below.
They are very general and 
apply also to models in which the FCNC processes enter first at the 
one-loop level. Indeed they contain very general operator structure and 
general new flavour violating and CP-violating interactions. However, 
having simpler coupling structure than in the case of models in which NP is 
dominated by loop contributions, allows us to have an {\it analytic look} 
at various correlations between various observables as we will see below.

\section{Classifying Correlations between various Observables}\label{sec:correlations}
As we have seen in preceding sections, in the SM and in models with CMFV the 
observables measured in the processes 
shown in Fig.~\ref{Fig:1} depend on selected number of basic universal functions that are the same for $K$ and $B_{s,d}$ decays. In particular $\Delta F=2$ 
processes depend only on the function $S(v)$, while the most important 
rare $K$ and $B_{s,d}$ decays depend on three universal functions $X(v)$, $Y(v)$, $Z(v)$. 
Consequently, a number of correlations exist between various observables
not only within the $K$ and $B$ systems but also between $K$ 
and $B$ systems. In particular the latter correlations are very interesting
as they are characteristic for this class of models. A review of these
correlations is given in \cite{Buras:2003jf}. These correlations
are violated in several extensions of the SM either through the presence 
of new source of flavour violation or the presence of new operators. However, 
as the SM constitutes the main bulk of most branching ratios, the CMFV 
correlations can be considered as {\it standard candles of flavour physics} with 
help of which new sources of flavour violation or effects of new operators 
could be identified. It is for the latter reason that we prefer to use CMFV 
correlations  as 
standard
flavour candles and not those present in MFV at large, but models with MFV and 
one Higgs doublet give the same results.

In \cite{Blanke:2008yr} a classification of correlations following from 
CMFV has been presented. In what follows we will somewhat modify this classification so that it fits better to our presentation in the next section that 
considers a number of models in contrast to \cite{Blanke:2008yr} where only the RSc 
model has been studied.

We distinguish the following classes of correlations in CMFV models\footnote{In this list we do not include a known model independent correlation 
between the asymmetries $S_{\psi\phi}$ and $A^s_\text{SL}$
 \cite{Ligeti:2006pm} that has to be satisfied basically in any 
 extension of the SM.}:

{\bf Class 1:}

 Correlations implied by the universality of the real function $X$. They
 involve rare $K$ and $B$ decays with $\nu\bar\nu$ in the final state. 
 These are:
\be\label{Class1decays} 
 \kpn, \quad \klpn, \quad B\to X_{s,d}\nu\bar\nu, \quad B\to K^*(K)\nu\bar\nu.
 \ee
 
 {\bf Class 2:} 

 Correlations implied by the universality of the real function $Y$. They
 involve rare $K$ and $B$ decays with $\mu^+\mu^-$ in the final state. These 
 are
\be\label{Class2decays}
B_{s,d}\to\mu^+\mu^-,\quad K_L\to\mu^+\mu^-, \quad
K_L\to \pi^0 \mu^+\mu^-,\quad  K_L\to \pi^0 e^+e^-.
\ee

{\bf Class 3:} 

 In models with CMFV NP contributions enter the functions $X$ and $Y$
 approximately in the same manner as at least in the Feynman gauge
 they come dominantly from $Z^0$-penguin diagrams. This implies 
 correlations between
 rare decays with $\mu^+\mu^-$ and $\nu\bar\nu$ in the final state.
 It should be emphasized that this is a separate class as NP can generally
 have {a} different impact on decays with $\nu\bar\nu$ {and} $\mu^+\mu^-$ in the
 final state. This class involves simply the decays of Class 1 and Class 2.

 {\bf Class 4:}

 Here we group correlations between $\Delta F=2$ and $\Delta F=1$ transitions
 in which the one-loop functions $S$ and $(X,Y)$, respectively, cancel out and
 the correlations follow from the fact that  the CKM parameters extracted 
from tree-level decays are universal.
One 
 known correlation of this type involves
\cite{Buchalla:1994tr,Buras:2001af} 
\be
\kpn,\quad \klpn \quad {\rm and} \quad S_{\psi K_S},
\ee
another one
\cite{Buras:2003td}
\be
 B_{s,d}\to\mu^+\mu^- \quad {\rm and} \quad \Delta M_{s,d}.
\ee
 As we will see in Section \ref{sec:4}, some of these correlations, in
 particular those between $K$ and $B$ decays are strongly violated in certain 
models, others
 are approximately satisfied. Clearly the full picture is only obtained by looking
 simultaneously at patterns of violations of the correlations in question in a
given NP scenario.

 At later stages of our presentation in Section~\ref{sec:4} we will study 
correlations in models with tree-level FCNCs mediated by neutral gauge bosons 
and scalars that go beyond the CMFV framework. In these models multi-correlations between various observables in a given meson system are predicted and it 
is useful 
to group these  processes in the following three classes. 
These are:

 {\bf Class 5:}

\be\label{Class5}
\varepsilon_K, \quad \kpn, \quad \klpn, \quad K_L\to\mu^+\mu^-, \quad 
K_L\to \pi^0 \ell^+\ell^-,\quad  \epe.
\ee

 {\bf Class 6:}

\be\label{Class6} 
\Delta M_d, \quad S_{\psi K_S}, \quad  B_d\to\mu^+\mu^-, \quad S_{\mu\mu}^d,
\ee
where the CP-violating asymmetry $S_{\mu\mu}^d$ can only be obtained from time-dependent rate of 
 $B_d\to\mu^+\mu^-$ and will remain in the realm of theory for 
the foreseeable future.

{\bf Class 7:}

\be\label{Class7} 
\Delta M_s, \quad S_{\psi \phi}, \quad B_s\to\mu^+\mu^-, \quad S_{\mu\mu}^s, \quad
B\to K \nu \bar \nu, \quad B\to K^* \nu \bar \nu, \quad B\to X_s \nu \bar \nu,
\ee
 where the measurement of $S_{\mu\mu}^s$ will require heroic efforts from 
experimentalists but apparently is not totally hopeless.

 {\bf Class 8:}

\be 
B\to X_s\gamma, \quad B\to K^*\gamma,\quad B^+\to \tau^+\nu_\tau
\ee
in which new charged gauge bosons and heavy scalars can play significant role. 
 The first two differ from previous decays as they are governed by dipole 
operators.

{\bf Class 9}

\be
B\to K \mu^+\mu^-, \quad B\to K^*\mu^+\mu^-,\quad  B\to X_s \mu^+\mu^- 
\ee
to which several operators contribute and for which multitude of 
observables can be defined. Moreover in the case of FCNCs mediated by 
tree-level neutral gauge boson exchanges interesting correlations between 
these observables and the ones of Class 7 exist.

{\bf Class 10:}

Correlations between $K$ and $D$ observables.

{\bf Class 11:}

 Correlations between quark flavour violation, lepton flavour violation, 
     electric dipole moments and $(g-2)_{e,\mu}$.

\section{Searching for New Physics in twelve Steps}\label{sec:4}

\subsection{Step 1: The CKM Matrix from tree level decays}
As the SM represents already the dominant part in very many 
flavour observables it is crucial to determine the CKM parameters 
as precise as possible independently of NP contributions. Here the 
tree-level decays governed by $W^\pm$ exchanges  play the prominent 
role. The charged current decays could be affected by heavy charged
new gauge boson exchanges and heavy 
charged Higgs boson exchanges that 
could contribute directly to tree level decays. Also non-standard 
$W^\pm$ couplings could be generated through mixing of $W^\pm$ with 
the new heavy gauge bosons in the process of electroweak symmetry breaking. 
 Moreover, the mixing of heavy fermions, both sequential 
like the case of fourth generation or vectorial ones present in various 
NP scenarios, could make the CKM matrix to be non-unitary not allowing to 
use the well known unitarity relations of this matrix. This mixing would 
also generate non-standard $W^\pm$ couplings to SM quarks.

The non-observation of any convincing NP signals at the LHC until now 
gives some hints that the masses of new charged particles are shifted above the
$500\gev$ scale. Therefore NP effects in charged current decays are 
likely to be at most at the level of a few percent. While effects of 
this sort could play a role one day, it is a good strategy to assume 
in the first step that tree level charged current decays are fully dominated by
$W^\pm$ exchanges with SM couplings and consequently by the CKM matrix.

The goal of this first step is then a very precise determination of 
\be\label{CKMtree}
\vus\simeq s_{12}, \qquad \vub\simeq s_{13},\qquad \vcb\simeq s_{23}, \qquad \gamma=\delta,
\ee
where on the l.h.s we give the measured quantities and on the r.h.s the 
determined parameters of the CKM matrix given in the standard parametrization: 
\begin{equation}\label{2.72}
\hat V_{\rm CKM}=
\left(\begin{array}{ccc}
c_{12}c_{13}&s_{12}c_{13}&s_{13}e^{-i\delta}\\ -s_{12}c_{23}
-c_{12}s_{23}s_{13}e^{i\delta}&c_{12}c_{23}-s_{12}s_{23}s_{13}e^{i\delta}&
s_{23}c_{13}\\ s_{12}s_{23}-c_{12}c_{23}s_{13}e^{i\delta}&-s_{23}c_{12}
-s_{12}c_{23}s_{13}e^{i\delta}&c_{23}c_{13}
\end{array}\right)\,.
\end{equation}

\begin{figure}[!tb]
\vspace{0.10in}
\centerline{
\epsfysize=2.1in
\epsffile{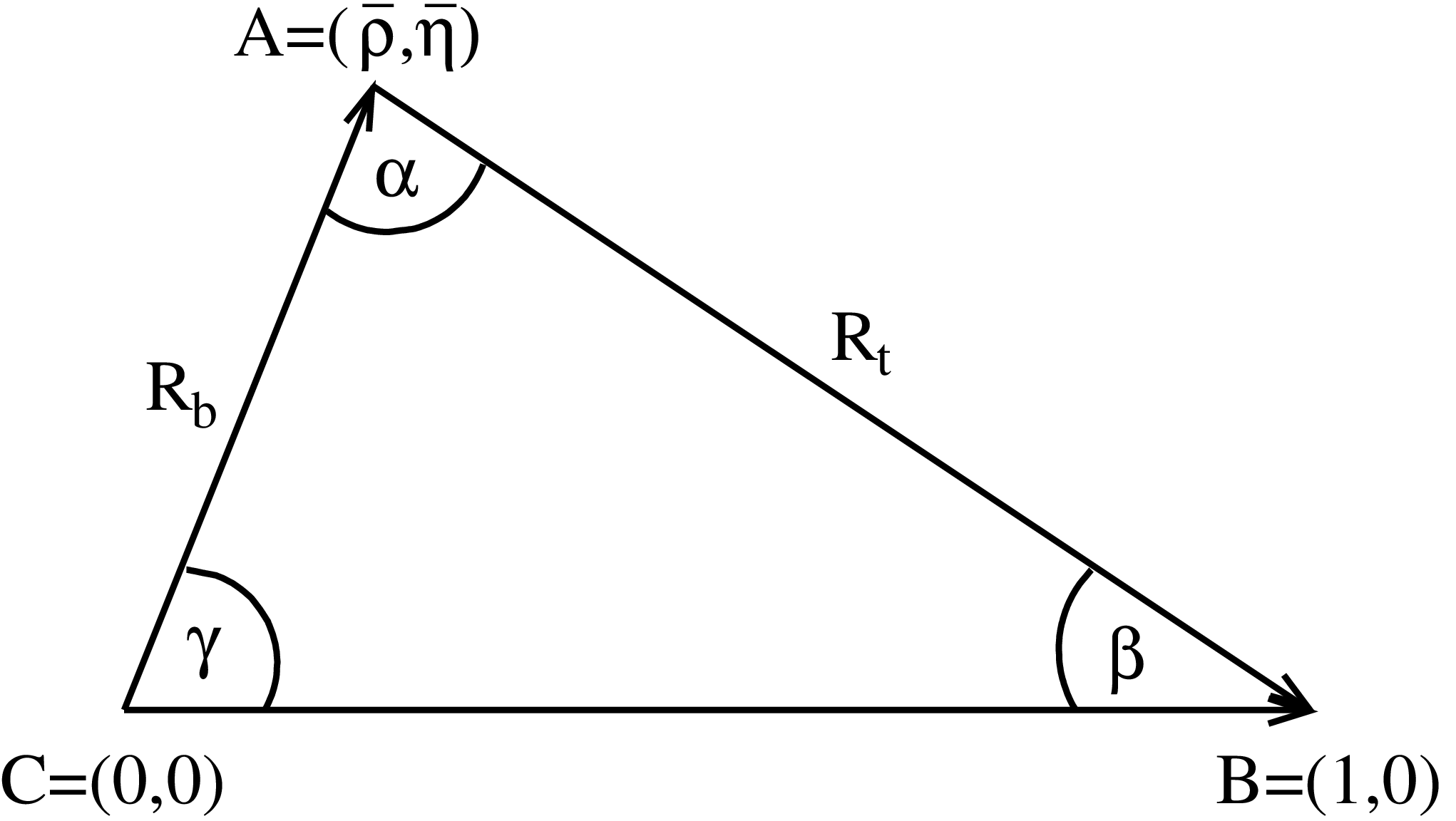}
}
\vspace{0.08in}
\caption{\it Unitarity Triangle.}\label{fig:utriangle}~\\[-2mm]\hrule
\end{figure}

The phase $\gamma$ is one of the angles of the unitarity triangle shown 
 Fig.~\ref{fig:utriangle}. We 
emphasize that the relations in (\ref{CKMtree}) are excellent 
approximations. Indeed $c_{13}$ and $c_{23}$ are very close to unity. 
The parameters $\bar\varrho$ and $\bar\eta$ are the 
 generalized Wolfenstein 
parameters \cite{Wolfenstein:1983yz,Buras:1994ec}.  Extensive analyses of 
the unitarity triangle have been performed for many years by CKMfitter \cite{Charles:2004jd} and UTfit \cite{Bona:2005eu} collaborations and
recently by SCAN-Method collaboration \cite{Eigen:2013cv}.

\begin{table}[!tb]
\center{\begin{tabular}{|l|l|}
\hline
$G_F = 1.16637(1)\times 10^{-5}\gev^{-2}$\hfill\cite{Nakamura:2010zzi} 	&  $m_{B_d}= m_{B^+}=5279.2(2)\mev$\hfill\cite{Beringer:1900zz}\\
$M_W = 80.385(15) \gev$\hfill\cite{Nakamura:2010zzi}  								&	$m_{B_s} =
5366.8(2)\mev$\hfill\cite{Beringer:1900zz}\\
$\sin^2\theta_W = 0.23116(13)$\hfill\cite{Nakamura:2010zzi} 				& 	$F_{B_d} =
(190.5\pm4.2)\mev$\hfill \cite{Aoki:2013ldr}\\
$\alpha(M_Z) = 1/127.9$\hfill\cite{Nakamura:2010zzi}									& 	$F_{B_s} =
(227.7\pm4.5)\mev$\hfill \cite{Aoki:2013ldr}\\
$\alpha_s(M_Z)= 0.1184(7) $\hfill\cite{Nakamura:2010zzi}			&  $F_{B^+} =(185\pm3)\mev$\hfill \cite{Dowdall:2013tga}
\\\cline{1-1}
$m_u(2\gev)=2.16(11)\mev $ 	\hfill\cite{Aoki:2013ldr}						&  $\hat B_{B_d} =1.27(10)$,  $\hat
B_{B_s} =
1.33(6)$\hfill\cite{Aoki:2013ldr}\\
$m_d(2\gev)=4.68(0.15)\mev$	\hfill\cite{Aoki:2013ldr}							& $\hat B_{B_s}/\hat B_{B_d}
= 1.01(2)$ \hfill \cite{Carrasco:2013zta} \\
$m_s(2\gev)=93.8(24) \mev$	\hfill\cite{Aoki:2013ldr}				&
$F_{B_d} \sqrt{\hat
B_{B_d}} = 216(15)\mev$\hfill\cite{Aoki:2013ldr} \\
$m_c(m_c) = (1.279\pm 0.013) \gev$ \hfill\cite{Chetyrkin:2009fv}					&
$F_{B_s} \sqrt{\hat B_{B_s}} =
266(18)\mev$\hfill\cite{Aoki:2013ldr} \\
$m_b(m_b)=4.19^{+0.18}_{-0.06}\gev$\hfill\cite{Nakamura:2010zzi} 			& $\xi =
1.268(63)$\hfill\cite{Aoki:2013ldr}
\\
$m_t(m_t) = 163(1)\gev$\hfill\cite{Laiho:2009eu,Allison:2008xk} &  $\eta_B=0.55(1)$\hfill\cite{Buras:1990fn,Urban:1997gw}  \\
$M_t=173.2\pm0.9 \gev$\hfill\cite{Aaltonen:2012ra}						&  $\Delta M_d = 0.507(4)
\,\text{ps}^{-1}$\hfill\cite{Amhis:2012bh}\\\cline{1-1}
$m_K= 497.614(24)\mev$	\hfill\cite{Nakamura:2010zzi}								&  $\Delta M_s = 17.72(4)
\,\text{ps}^{-1}$\hfill\cite{Amhis:2012bh}
\\	
$F_K = 156.1(11)\mev$\hfill\cite{Laiho:2009eu}												&
$S_{\psi K_S}= 0.68(2)$\hfill\cite{Amhis:2012bh}\\
$\hat B_K= 0.766(10)$\hfill\cite{Aoki:2013ldr}											&
$S_{\psi\phi}= 0.00(7)$\hfill\cite{Amhis:2012bh}\\
$\kappa_\epsilon=0.94(2)$\hfill\cite{Buras:2008nn,Buras:2010pza}				& $\Delta\Gamma_s/\Gamma_s=0.123(17)$\hfill\cite{Amhis:2012bh}
\\	
$\eta_{cc}=1.87(76)$\hfill\cite{Brod:2011ty}												
	& $\tau_{B_s}= 1.509(11)\,\text{ps}$\hfill\cite{Amhis:2012bh}\\		
$\eta_{tt}=0.5765(65)$\hfill\cite{Buras:1990fn}												
& $\tau_{B_d}= 1.519(7) \,\text{ps}$\hfill\cite{Amhis:2012bh}\\
$\eta_{ct}= 0.496(47)$\hfill\cite{Brod:2010mj}												
& $\tau_{B^\pm}= 1.642(8)\,\text{ps}$\hfill\cite{Amhis:2012bh}    \\
$\Delta M_K= 0.5292(9)\times 10^{-2} \,\text{ps}^{-1}$\hfill\cite{Nakamura:2010zzi}	&
$|V_{us}|=0.2252(9)$\hfill\cite{Amhis:2012bh}\\
$|\eps_K|= 2.228(11)\times 10^{-3}$\hfill\cite{Nakamura:2010zzi}					& $|V_{cb}|=(40.9\pm1.1)\times
10^{-3}$\hfill\cite{Beringer:1900zz}\\
  $|V^\text{incl.}_{cb}|=42.4(9)\times10^{-3}$\hfill\cite{Gambino:2013rza}  &
$|V^\text{incl.}_{ub}|=4.40(25)\times10^{-3}$\hfill\cite{Aoki:2013ldr}\\
$|V^\text{excl.}_{cb}|=39.4(6)\times10^{-3}$\hfill\cite{Aoki:2013ldr}	&
$|V^\text{excl.}_{ub}|=3.42(31)\times10^{-3}$\hfill\cite{Aoki:2013ldr}\\
\hline
\end{tabular}  }
\caption {\textit{Values of the experimental and theoretical
    quantities used as input parameters  as of April 2014. For future 
updates see PDG \cite{Beringer:1900zz}, FLAG \cite{Aoki:2013ldr} and HFAG \cite{Amhis:2012bh}. }}
\label{tab:input}~\\[-2mm]\hrule
\end{table}

Under the assumption made above this determination would give us the values of the elements of the CKM 
matrix without NP pollution. From the present perspective most important are 
the determinations of
$\vub$ and $\gamma$ because as seen in Table~\ref{tab:input}
they are presently not as well known as $\vcb$ and $\vus$. 
In this table we give other most recent values of the relevant parameters  to which we will return 
in the course of our review.

Looking 
at Table~\ref{tab:input} we make the following observations:
\begin{itemize}
\item
The element $\vus$ is already well measured.
\item
The accuracy of the determination of $\vcb$ is  quite good but 
{  disturbing is the  discrepancy between the inclusive and exclusive 
determinations \cite{Ricciardi:2013cda,Gambino:2013rza}, with the exclusive ones being visibly smaller \cite{Bailey:2014tva}. } We quote here 
only the average value provided by PDG. It should 
be recalled that the knowledge of this CKM matrix element is very 
important for rare decays and CP violation in the $K$-meson system. Indeed
$\varepsilon_K$, $\mathcal{B}(\kpn)$ and $\mathcal{B}(\klpn)$ are all roughly 
proportional to $\vcb^4$ and even a respectable accuracy of $2\%$ in $\vcb$ 
translates 
into $8\%$ parametric uncertainty in these observables. This is in particular 
disturbing for  $\mathcal{B}(\kpn)$ and $\mathcal{B}(\klpn)$ as these branching ratios are 
practically independent of any theoretical uncertainties. 
 Future $B$-facilities accompanied by improved 
theory should be able to determine $\vcb$ with precision of $1-2\%$.
\item
The case of $\vub$ is more disturbing with central values from 
inclusive determinations being by roughly $25\%$ higher than the corresponding 
value resulting from exclusive semi-leptonic decays. 
We will see below 
that dependently on which of these values is assumed, different 
 conclusions on the properties of  NP responsible 
for certain anomalies  seen in the data will  be reached. 
Again,  future $B$-facilities accompanied by improved 
theory should be able to determine $\vub$ with precision of
$1-2\%$.
\item
Finally, the only physical CP phase in the CKM matrix, $\gamma$, is still 
poorly known from tree-level decays. But LHCb should be able to determine 
this angle with an error of a few degrees, which would be a great achievement. 
Further improvements could come from SuperKEKB.
\end{itemize}

 The importance of precise determinations of $\vcb$, $\vub$ and $\gamma$ 
should not be underestimated. Table~3 and Fig~2 in \cite{Buras:2014sba} showing 
SM predictions for various combinations of  $\vcb$ and $\vub$
demonstrate this very clearly. Therefore 
the consequences of reaching our first goal would be profound.
Indeed, having determined precisely the four parameters of the CKM matrix 
 without influence from NP, will allow us
to reconstruct all its elements. In turn they could be used efficiently in 
the calculation of the SM predictions for all decays and in particular 
FCNC processes, both CP-conserving and CP-violating. Moreover, 
this would allow to calculate not only an important element $\vtd$ but 
also its phase $-\beta$, with $\beta$ denoting another, very important 
angle of the unitarity triangle in Fig.~\ref{fig:utriangle}.

In order to be prepared for these developments we collect here the most 
important formulae related to the unitarity triangle and CKM matrix. The 
phases of $V_{td}$ and $V_{ts}$ are defined by
\be\label{vtdvts}
V_{td}=\vtd e^{-i\beta}, \qquad V_{ts}=-\vts e^{-i\beta_s}.
\ee

Next, the lengths $CA$ and $BA$ in the unitarity triangle are given respectively by
\begin{equation}\label{2.94}
R_b \equiv \frac{| V_{ud}^{}V^*_{ub}|}{| V_{cd}^{}V^*_{cb}|}
= \sqrt{\bar\varrho^2 +\bar\eta^2}
= (1-\frac{\lambda^2}{2})\frac{1}{\lambda}
\left| \frac{V_{ub}}{V_{cb}} \right|,
\end{equation}
\begin{equation}\label{2.95}
R_t \equiv \frac{| V_{td}^{}V^*_{tb}|}{| V_{cd}^{}V^*_{cb}|} =
 \sqrt{(1-\bar\varrho)^2 +\bar\eta^2}
=\frac{1}{\lambda} \left| \frac{V_{td}}{V_{cb}} \right|.
\end{equation}
An important very accurate relation is 
\be
\sin 2\beta=2 \frac{\bar\eta(1-\bar\varrho)}{R^2_t}.
\ee
We also note that the knowledge of $(R_b,\gamma)$ from tree-level decays 
gives 
\begin{equation} \label{eq:Rt_beta}
\vtd=\vus\vcb R_t, \qquad
R_t=\sqrt{1+R_b^2-2 R_b\cos\gamma}, \qquad
\cot\beta=\frac{1-R_b\cos\gamma}{R_b\sin\gamma}~.
\end{equation}
Similarly  the knowledge of
$(R_t,\beta)$ allows to determine $(R_b,\gamma)$ through 
\be\label{VUBG}
R_b=\sqrt{1+R_t^2-2 R_t\cos\beta},\qquad
\cot\gamma=\frac{1-R_t\cos\beta}{R_t\sin\beta}
\ee
and consequently with known $\lambda=\vus$ and $\vcb$, one finds $\vub$ 
by means of (\ref{2.94}).
Similarly $V_{ts}$ can be calculated. $\vts$ is slightly below $\vcb$ but 
in the flavour precision era it is better to calculate its value numerically 
by using the standard parametrization. Then one also finds that
the value of $\beta_s$ is tiny: $\beta_s\approx -1^\circ$.

There is still another powerful route to the determination of the Unitarity 
Triangle. As pointed out in  \cite{Buras:2002yj} in addition to 
the determination of UT without any NP pollution through the determination 
of $(R_b,\gamma)$, in models with CMFV and MFV in which NP is absent in 
$S_{\psi K_S}$, the determination can proceed through $(\beta,\gamma)$. Then 
\be\label{BPS}
R_t=\frac{\sin\gamma}{\sin(\beta+\gamma)}, \qquad R_b=\frac{\sin\beta}{\sin(\beta+\gamma)}.
\ee
In fact as demonstrated in  \cite{Buras:2002yj} $(R_b,\gamma)$ and $(\beta,\gamma)$ are the two most powerful ways to determine UT in the 
sense that the accuracy on these two pairs does not have to be very high in 
order to determine  $(\bar\varrho,\bar\eta)$  with good precision. But as we have seen $\vub$ 
is not known very well and even if there are hopes to determine it within 
few $\%$ in the second half of this decade, it is more probable that 
$\gamma$ from tree-level decays will be known with this accuracy first 
and the $(\beta,\gamma)$ strategy will be leading one in getting  $(\bar\varrho,\bar\eta)$ within CMFV and MFV models.

The values of $\vtd$ and $\vts$ are crucial for the predictions of 
various rare decays but in particular for the mass differences 
$\Delta M_d$ and  $\Delta M_s$ and the phases $\beta$ and $\beta_s$ 
for the corresponding mixing induced 
CP-asymmetries $S_{\psi K_S}$ and $S_{\psi \phi}$, which are defined within the 
SM in 
(\ref{eq:3.43}).
Also the CP-violating parameter $\varepsilon_K$ depends crucially on 
$V_{td}$ and $V_{ts}$. 

Before making some statements about the present status of the first five 
super stars of flavour physics
\be\label{STARS1}
\Delta M_d,\qquad \Delta M_s, \qquad S_{\psi K_S}, \qquad S_{\psi\phi}, \qquad
\varepsilon_K
\ee
within the SM, we have to make the second very important step.

\subsection{Step 2: Improved Lattice Calculations of Hadronic Parameters}

Precise knowledge of  the meson decay constants $F_{B_s}$, $F_{B_d}$, $F_{B^+}$ and of various 
non-perturbative 
parameters $B_i$ related to hadronic matrix elements of SM operators and operators found in the extensions of the 
SM is very important. Indeed this would allow
in conjunction with Step 1 to perform
precise calculations of $\Delta M_s$, $\Delta M_d$, $\varepsilon_K$, 
$\mathcal{B}(B_{s,d}\to\mu^+\mu^-)$, $\mathcal{B}(B^+\to\tau^+\nu_\tau)$ and of other observables in the SM. We  could 
then directly see whether the SM is capable
of describing these observables or not. The recent unquenched lattice 
calculations allow
for optimism and in fact a very significant progress in the calculation of 
$\hat B_K$, that is relevant for $\varepsilon_K$, has been made recently. Also the weak  decay constants $F_{B_s}$, $F_{B_d}$ and  $F_{B^+}$ and some 
non-perturbative 
$B_i$ parameters are much better known than few years ago.

In Table~\ref{tab:input} we collect most relevant 
non-perturbative parameters relevant for $\Delta F=2$ observables that we 
extracted from the most  recent FLAG average \cite{Aoki:2013ldr}. It should be
remarked that these values are consistent with the ones presented in 
\cite{Laiho:2009eu,Dowdall:2013tga} but generally have larger errors 
as FLAG prefers to be conservative. In particular in the latter two 
papers one finds:
\be\label{oldf1}
F_{B_s} \sqrt{\hat B_{B_s}}=279 (13)\mev, \qquad F_{B_d} \sqrt{\hat B_{B_d}}=226 (13)\mev, \qquad \xi= 1.237(32),
\ee
\be\label{oldf2}
F_{B_s} =225(3)\mev, \qquad F_{B_d} =188(4)\mev, 
\ee
which contain smaller errors than quoted in \cite{Aoki:2013ldr}.

We should also mention recent results 
 from the Twisted Mass Collaboration \cite{Carrasco:2013zta}
\be\label{twist}
\sqrt{\hat B_{B_s}}F_{B_s} = 262(10)\mev,\qquad \sqrt{\hat B_{B_d}}F_{B_d} = 216(10)\mev,
\ee
which are not yet included in the FLAG average but having smaller errors 
are consistent with the latter

 Evidently there is a big progress in determining all these relevant parameters  but one would like to decrease 
the errors further and it appears that this should be possible in the coming 
years. Selected reviews about the status and prospects can be found in \cite{Tarantino:2012mq,Davies:2012qf,Gamiz:2013waa,Carrasco:2013zta,Sachrajda:2013fxa,Christ:2013lxa}.

\boldmath
\subsection{Step 3: $\Delta F=2$ Observables}\label{Step3}
\unboldmath
\subsubsection{Contributing operators}
In order to describe these processes in generality we begin by listing 
the operators which can contribute to $\Delta F=2$ observables in any 
extension of the SM. Specifying to the $K^0-\bar K^0$ system the full 
basis is given as follows
\cite{Buras:2001ra,Buras:2012fs}:
\begin{subequations}\label{equ:operatorsZ}
\bea
{Q}_1^\text{VLL}&=&\left(\bar s\gamma_\mu P_L d\right)\left(\bar s\gamma^\mu P_L d\right)\,,\\
{Q}_1^\text{VRR}&=&\left(\bar s\gamma_\mu P_R d\right)\left(\bar s\gamma^\mu P_R d\right)\,,\\
{Q}_1^\text{LR}&=&\left(\bar s\gamma_\mu P_L d\right)\left(\bar s\gamma^\mu P_R d\right)\,,\\
{Q}_2^\text{LR}&=&\left(\bar s P_L d\right)\left(\bar s P_R d\right)\,,
\eea
\end{subequations}
{\allowdisplaybreaks
\begin{subequations}\label{equ:operatorsHiggs}
 \bea
{Q}_1^\text{SLL}&=&\left(\bar s P_L d\right)\left(\bar s P_L d\right)\,,\\
{Q}_1^\text{SRR}&=&\left(\bar s P_R d\right)\left(\bar s P_R d\right)\,,\\
{Q}_2^\text{SLL}&=&\left(\bar s \sigma_{\mu\nu} P_L d\right)\left(\bar s\sigma^{\mu\nu}  P_L d\right)\,,\\
{Q}_2^\text{SRR}&=&\left(\bar s \sigma_{\mu\nu}  P_R d\right)\left(\bar s \sigma^{\mu\nu} P_R d\right)\,,
\eea
\end{subequations}}%
where $P_{R,L}=(1\pm\gamma_5)/2$ and 
we suppressed colour indices as they are summed up in each factor. For instance $\bar s\gamma_\mu P_L d$ stands for $\bar s_\alpha\gamma_\mu P_L d_\alpha$ and similarly for other factors.
For $B_q^0-\bar B_q^0$ mixing our conventions for operators are:
\begin{subequations}\label{equ:operatorsZb}
\bea
{Q}_1^\text{VLL}&=&\left(\bar b\gamma_\mu P_L q\right)\left(\bar b\gamma^\mu P_L q\right)\,,\\
{Q}_1^\text{VRR}&=&\left(\bar b\gamma_\mu P_R q\right)\left(\bar b\gamma^\mu P_R q\right)\,,\\
{Q}_1^\text{LR}&=&\left(\bar b\gamma_\mu P_L q\right)\left(\bar b\gamma^\mu P_R q\right)\,,\\
{Q}_2^\text{LR}&=&\left(\bar b P_L q\right)\left(\bar b P_R q\right)\,,
\eea
\end{subequations}
{\allowdisplaybreaks
\begin{subequations}\label{equ:operatorsHiggsb}
 \bea
{Q}_1^\text{SLL}&=&\left(\bar b P_L q\right)\left(\bar b P_L q\right)\,,\\
{Q}_1^\text{SRR}&=&\left(\bar b P_R q\right)\left(\bar b P_R q\right)\,,\\
{Q}_2^\text{SLL}&=&\left(\bar b \sigma_{\mu\nu} P_L q\right)\left(\bar b\sigma^{\mu\nu}  P_L q\right)\,,\\
{Q}_2^\text{SRR}&=&\left(\bar b \sigma_{\mu\nu}  P_R q\right)\left(\bar b \sigma^{\mu\nu} P_R q\right)\,,
\eea
\end{subequations}}%


\begin{table}[!tb]
{\renewcommand{\arraystretch}{1.3}
\begin{center}
\begin{tabular}{|c||c|c|c|c|}
\hline
&$\langle Q_1^\text{LR}(\mu_H)\rangle$& $\langle Q_2^\text{LR}(\mu_H)\rangle$&$\langle Q_1^\text{SLL}(\mu_H)\rangle$&$\langle
Q_2^\text{SLL}(\mu_H)\rangle$\\
\hline
\hline
$K^0$-$\bar K^0$ &$-0.14$ &$0.22$ & $-0.074$ & $-0.128$\\
\hline
$B_d^0$-$\bar B_d^0$& $-0.25$ &$0.34$  &$-0.11$ &$-0.22 $\\
\hline
$B_s^0$-$\bar B_s^0$& $-0.37$ &$ 0.51$ &$-0.17$  & $-0.33$\\
\hline
\end{tabular}
\end{center}}
\caption{\it Hadronic matrix elements $\langle Q_i^a(\mu_H)\rangle$  in units of GeV$^3$ at $\mu_H=1\tev$.
\label{tab:Qi}}~\\[-2mm]\hrule
\end{table}

\begin{table}[!tb]
{\renewcommand{\arraystretch}{1.3}
\begin{center}
\begin{tabular}{|c||c|c|c|c|}
\hline
&$\langle Q_1^\text{LR}(m_t)\rangle$& $\langle Q_2^\text{LR}(m_t)\rangle$&$\langle Q_1^\text{SLL}(m_t)\rangle$&$\langle
Q_2^\text{SLL}(m_t)\rangle$\\
\hline
\hline
$K^0$-$\bar K^0$ &$-0.11$ &$0.18$  & $-0.064$ & $-0.107$\\
\hline
$B_d^0$-$\bar B_d^0$& $-0.21$ &$0.27$  &$-0.095$ &$-0.191 $\\
\hline
$B_s^0$-$\bar B_s^0$& $-0.30$ &$ 0.40$ &$-0.14$  & $-0.29$\\
\hline
\end{tabular}
\end{center}}
\caption{\it Hadronic matrix elements $\langle Q_i^a(\mu_t)\rangle$  in units of GeV$^3$ at $m_t(m_t)$.
\label{tab:Qi1}}~\\[-2mm]\hrule
\end{table}

 As already mentioned in Step 2 the main theoretical uncertainties in 
$\Delta F=2$ transitions reside in the hadronic matrix elements of 
the contributing operators. These matrix elements are usually evaluated 
by lattice QCD at scales corresponding roughly to the scale of decaying 
hadron although in the case of $K$ meson decays, in order to improve the 
matching with the Wilson coefficients, the lattice calculations are performed 
these days at scales $\mu\approx 2\gev$. However, for the study of NP 
contributions it is useful, starting from their values at these low scales, 
to evaluate them at  scales where NP is at work. This can be done 
by means of renormalization group methods and the corresponding analytic 
formulae to achieve this goal can be found in  \cite{Buras:2001ra}.

The most recent values of the matrix elements of the operators at a high 
scale $\mu_H=1\tev$ are given in Table~\ref{tab:Qi}. The matrix elements of 
operators with L replaced by R are equal to the ones given in this 
table.
The values  in Table~\ref{tab:Qi} correspond to the 
$\overline{\text{MS}}$-NDR scheme and are based on lattice calculations
in \cite{Boyle:2012qb,Bertone:2012cu} for $K^0-\bar K^0$ system and  in
\cite{Bouchard:2011xj} for
$B_{d,s}^0-\bar B^0_{d,s}$ systems. For the $K^0-\bar K^0$ system we have just
used the average of the results in \cite{Boyle:2012qb,Bertone:2012cu} that
are consistent with each other\footnote{The recent results using staggered fermions from SWME collaboration in  $K^0-\bar K^0$ system 
\cite{Bae:2013tca}   are not included here. While for $Q_{1,2}^\text{SLL}$ this group obtains  results consistent with  
\cite{Boyle:2012qb,Bertone:2012cu}, the matrix elements of $Q_{1,2}^\text{LR}$ are 
by $50\%$ larger. Let us hope this difference will be clarified soon.}.
As the values of the relevant $B_i$ parameters in these papers have been
evaluated at $\mu=3\gev$ and $4.2\gev$, respectively, we have used the
formulae in  \cite{Buras:2001ra} to obtain the values of the matrix
elements in question at $\mu_{H}$. For simplicity we choose this scale to
be $M_{H}$ but any scale of this order would give the same results for
the physical quantities up to NNLO QCD corrections that are negligible
at these high scales.  The renormalization scheme dependence of the
matrix elements is canceled by the one of the Wilson coefficients as discussed 
below.

In the case of tree-level SM $Z$ and SM  Higgs exchanges we evaluate the matrix elements at $m_t(m_t)$ as the inclusion of NLO QCD corrections allows us to choose any scale of $\ord(M_H)$ without changing physical results. The values
of hadronic matrix elements at $m_t(m_t)$ in the
$\overline{\text{MS}}$-NDR scheme
are given in Table~\ref{tab:Qi1}.

The Wilson coefficients of these operators depend on the short distance 
properties of a given theory. They can be directly expressed in terms of 
the couplings $\Delta_{L,R}^{ij}(Z^\prime)$ and $\Delta_{L,R}^{ij}(H^0)$ in the case of tree-level gauge boson and scalar exchanges. In models with GIM mechanism at work 
they are given in terms of loop functions. Then couplings 
$\Delta_{L,R}^{ij}(W^{\prime +})$ and $\Delta_{L,R}^{ij}(H^+)$ enter the game.
\subsubsection{Standard Model Results}
In the SM only the operator ${Q}_1^\text{VLL}$ contributes to 
each meson system. 
With the information gained in Steps 1 and 2 at hand we are ready to calculate the SM values 
for the five super stars in (\ref{STARS1}). To this end we recall the 
formulae for $\Delta M_{d,s}$, $S_{\psi K_S}$, $S_{\psi \phi}$, and $\varepsilon_K$.

Defining 
\be
\lambda_i^{(K)} =V_{is}^*V_{id},\qquad 
\lambda_t^{(d)} =V_{tb}^*V_{td}, \qquad \lambda_t^{(s)} =V_{tb}^*V_{ts},
\ee
we have first
\be\label{DMs}
\Delta M_s =\frac{G_F^2}{6 \pi^2}M_W^2 m_{B_s}|\lambda_t^{(s)}|^2   F_{B_s}^2\hat B_{B_s} \eta_B S_0(x_t),
\ee

\be\label{DMd}
\Delta M_d =\frac{G_F^2}{6 \pi^2}M_W^2 m_{B_d}|\lambda_t^{(d)}|^2   F_{B_d}^2\hat B_{B_d} \eta_B S_0(x_t).
\ee
which result from ($q=d,s$)
\bea
\left(M_{12}^q\right)^*_\text{SM}&=&{\frac{G_F^2}{12\pi^2}F_{B_q}^2\hat
B_{B_q}m_{B_q}M_{W}^2
\left[
\left(\lambda_t^{(q)}\right)^2\eta_BS_0(x_t)
\right]}\,\label{eq:3.6}
\eea
and
\be
\Delta M_q=2|M_{12}^q|.
\ee
Here $x_t=m_t^2/M_W^2$, $\eta_B=0.55$ is a QCD factor  and
\begin{align}
S_0(x_t)  = \frac{4x_t - 11 x_t^2 + x_t^3}{4(1-x_t)^2}-\frac{3 x_t^2\log x_t}{2
(1-x_t)^3} = 2.31 \left[\frac{\mtb(\mt)}{163\gev}\right]^{1.52} ~.
\end{align}

We find then  three useful formulae ($|V_{tb}|=1$)
\begin{equation}\label{DMS}
\Delta M_{s}=
17.7/{\rm ps}\cdot\left[ 
\frac{\sqrt{\hat B_{B_s}}F_{B_s}}{267\mev}\right]^2
\left[\frac{S_0(x_t)}{2.31}\right] 
\left[\frac{\vts}{0.0402} \right]^2
\left[\frac{\eta_B}{0.55}\right] \,,
\end{equation}

\begin{equation}\label{DMD}
\Delta M_d=
0.51/{\rm ps}\cdot\left[ 
\frac{\sqrt{\hat B_{B_d}}F_{B_d}}{218\mev}\right]^2
\left[\frac{S_0(x_t)}{2.31}\right] 
\left[\frac{\vtd}{8.5\cdot10^{-3}} \right]^2 
\left[\frac{\eta_B}{0.55}\right]
\end{equation}
and 
\be\label{DetRt}
R_{\Delta M_B}=\frac{\Delta M_d}{\Delta M_s}=  \frac{m_{B_d}}{m_{B_s}}\frac{\hat B_d}{\hat
B_s}\frac{F_{B_d}^2}{F_{B_s}^2}\left|\frac{V_{td}}{V_{ts}}\right|^2 
\equiv  \frac{m_{B_d}}{m_{B_s}}\frac{1}{\xi^2} \left|\frac{V_{td}}{V_{ts}}\right|^2.
\ee

The mixing induced CP-asymmetries are given within the SM simply by
\be
S_{\psi K_S} = \sin(2\beta)\,,\qquad S_{\psi\phi} =  \sin(2|\beta_s|)\,. 
\label{eq:3.43}
\ee
 They are the coefficients of $\sin(\Delta M_d t)$ and $\sin(\Delta M_s t)$ 
in the 
time dependent asymmetries in $B_d^0\to\psi K_S$ and $B_s^0\to\psi\phi$, respectively.

For the CP-violating parameter $\varepsilon_K$ we have
\be
\varepsilon_K=\frac{\kappa_\eps e^{i\varphi_\eps}}{\sqrt{2}(\Delta M_K)_\text{exp}}\left[\Im\left(M_{12}^K\right)_\text{\rm SM}\right]\,,
\label{eq:3.35}
\ee
where $\varphi_\eps = (43.51\pm0.05)^\circ$ and $\kappa_\eps=0.94\pm0.02$ \cite{Buras:2008nn,Buras:2010pza} takes into account 
that $\varphi_\eps\ne \tfrac{\pi}{4}$ and includes long distance effects in $\Im( \Gamma_{12})$ and $\Im (M_{12})$. Moreover
\be\label{eq:3.4}
\left(M_{12}^K\right)^*_\text{SM}=\frac{G_F^2}{12\pi^2}F_K^2\hat
B_K m_K M_{W}^2\left[
\lambda_c^{2}\eta_{cc}S_0(x_c)+\lambda_t^{2}\eta_{tt}S_0(x_t)+2\lambda_c\lambda_t\eta_{ct}S_0(x_c,x_t)
\right],
\ee
where $\eta_i$ are QCD factors  given in Table~\ref{tab:input}
and $S_0(x_c,x_t)$ can be found in  \cite{Blanke:2011ry}.

\begin{table}[!tb]
\centering
\begin{tabular}{|c||c|c|c|c|c|c|}
\hline
 $|V_{ub}| \times 10^3$  & $3.1$ & $3.4$ & $3.7$&  $4.0$ & $4.3$  & Experiment\\
\hline
\hline
  \parbox[0pt][1.6em][c]{0cm}{} $|\varepsilon_K|\times 10^3$ & $1.76$ & $1.91$  & $2.05$ & $2.19$ & $2.33$ &$2.228(11)$\\
 \parbox[0pt][1.6em][c]{0cm}{}$\mathcal{B}(B^+\to \tau^+\nu_\tau)\times 10^4$&  $0.58$ & $0.70$ & $0.83$ & $0.97$ & $1.12$ & $1.14(22)$\\
 \parbox[0pt][1.6em][c]{0cm}{}$(\sin2\beta)_\text{true}$ & $0.619$ & $0.671$ & $0.720$ &  $0.766$  & 
$0.808$ & $0.679(20)$\\
\parbox[0pt][1.6em][c]{0cm}{}$S_{\psi\phi}$ & $0.032$ & $0.035$ & $0.038$ &  
$0.042$  & 
$0.046$ & $0.001(9)$\\
\parbox[0pt][1.6em][c]{0cm}{}$\Delta M_s\, [\text{ps}^{-1}]$ (I)& $17.5$ & $17.5$ & $17.5$ & $17.6$ & 
$17.6$ &$17.69(8)$ \\
 \parbox[0pt][1.6em][c]{0cm}{} $\Delta M_d\, [\text{ps}^{-1}]$ (I) & $0.52$ & $0.51 $ & $0.51$& $0.52$ & $0.52$   &  $0.507(4)$\\
\parbox[0pt][1.6em][c]{0cm}{}$\Delta M_s\, [\text{ps}^{-1}]$  (II)& $19.2$ & $19.2$ & $19.2$ & $19.3$ & 
$19.3$ &$17.72(4)$ \\
 \parbox[0pt][1.6em][c]{0cm}{} $\Delta M_d\, [\text{ps}^{-1}]$  (II)& $0.56$ & $0.56 $ & $0.56$& $0.57 $& $0.57$   &  $0.510(4)$\\
\parbox[0pt][1.6em][c]{0cm}{}$\vtd\times10^3$& $8.56$&  $8.54$& $8.54$ 
& $8.56$ & $8.57 $ & $--$\\
\parbox[0pt][1.6em][c]{0cm}{}$\vts\times10^3 $& $40.0$ &$40.0$ & $40.0$ &  
$40.0$&$40.0$ & $--$\\
 \hline
\end{tabular}
\caption{\it SM prediction for various observables as functions of 
$|V_{ub}|$ and $\gamma =
68^\circ$. The two results for $\Delta M_{s,d}$ correspond to two sets of 
the values of $F_{B_s} \sqrt{\hat B_{B_s}}$ and $F_{B_d} \sqrt{\hat B_{B_d}}$:   { central values in Table~\ref{tab:input} (I) and  older 
values in (\ref{oldf1}) (II).} 
}\label{tab:SMpred}~\\[-2mm]\hrule
\end{table}

In Table~\ref{tab:SMpred} we 
summarize the results for $|\varepsilon_K|$, 
$\mathcal{B}(B^+\to \tau^+\nu_\tau)$, $\Delta M_{s,d}$, 
$\left(\sin 2\beta\right)_\text{true}$, $\Delta M_{s,d}$, $\vtd$ and $\vts$ obtained from (\ref{eq:Rt_beta}), 
setting 
\be\label{fixed}
\vus=0.2252, \qquad \vcb=0.0409, \qquad \gamma=68^\circ,
\ee
and  choosing five values for $\vub$. Two of them 
correspond to two 
scenarios defined in Section~\ref{sec:1}.  The value of $\gamma$ is close 
to its most recent value from $B\to DK$ decays obtained by LHCb  using 3~fb$^{-1}$ and neglecting $D^0-\bar D^0$ mixing 
\cite{LHCb-CONF-2013-006} 
\be
\gamma=(67.2\pm 12)^\circ, \qquad  {\rm (LHCb)}
\ee
and to the extraction from U-spin analysis of $B_s\to K^+K^-$ and $B_d\to\pi^+\pi^-$ decays ($\gamma=(68.2\pm 7.1)^\circ$) 
\cite{Fleischer:2010ib}.  In \cite{Aaij:2013zfa} both $B\to DK$ and $B\to D\pi$ decays are used and furthermore $D^0-\bar D^0$ mixing 
fully included and the combination of results gives as best-fit value $\gamma = 72.6^\circ$ and the confidence interval $\gamma 
\in [55.4,82.3]^\circ$ at 68\% CL. 
We do not show the uncertainties in SM predictions but just quote rough estimate  of them: 
\be\label{errors}
 |\varepsilon_K|:~ \pm 11\%, \qquad  \mathcal{B}(B^+\to \tau^+\nu_\tau):~\pm 15\%, \qquad \Delta M_{s,d}:~ \pm 10\%, \qquad S_{\psi K_S}:~\pm 3.0\%.
\ee

In order to show the importance of precise values of the non-perturbative 
parameters we show the results for present central values of 
$F_{B_s} \sqrt{\hat B_{B_s}}$ and $F_{B_d} \sqrt{\hat B_{B_d}}$ in Table~\ref{tab:input} (I) and for the older values in (\ref{oldf1})
indicated by (II).

We observe that while  $\Delta M_{s,d}$, 
$\vtd$ and $\vts$, practically do not depend on $\vub$, this is not the case 
for the remaining observables, although the $\vub$ dependence in $S_{\psi\phi}$ is very weak. Clearly 
the data show that it is difficult to fit simultaneously $\varepsilon_K$ 
and $S_{\psi\phi}$ within the SM  but 
the character of the NP which could cure these tensions depends 
 on the choice of $\vub$. On the other hand the agreement of the SM with 
the data on $\Delta M_s$ and $\Delta M_d$ is very good. In particular
for the set (I) we find 
\be
\left(\frac{\Delta M_s}{\Delta M_d}\right)_{\rm SM}= 34.1\pm 3.0\qquad {\rm exp:~~ 34.7\pm 0.3}
\ee
in excellent  agreement with the data.

We learn the following lessons to be remembered 
when we start investigating models beyond the SM:

{\bf Lesson 1:} 

We learn that in the case of exclusive determination of $\vub$
any NP model that pretends to be able to remove or soften 
the observed departures from the data should
simultaneously:
\begin{itemize}
\item
Enhance $|\varepsilon_K|$ by roughly $20\%$ without affecting significantly the 
result for  $S_{\psi K_S}$.
\item
Suppress slightly $\Delta M_s$ and $\Delta M_d$ without affecting 
significantly their ratio in the case of the set (II).  This suppression is not  required if the set (I) is used.
\end{itemize}

{\bf Lesson 2:} 

We learn that  in the case of inclusive determination of $\vub$
any NP model that pretends to be able to remove or soften 
the observed departures from the data should
simultaneously:
\begin{itemize}
\item
Suppress $S_{\psi K_S}$ by roughly $20\%$ without affecting significantly the 
result for  $|\varepsilon_K|$
\item
Suppress slightly $\Delta M_s$ and $\Delta M_d$ without affecting 
significantly their ratio in the case of the set (II).  This suppression is not  required if the set (I) is used.
\end{itemize}

Clearly $\vub$ could have an intermediate value but we find that a more
transparent picture emerges for these two values.

{\bf Lesson 3:} 

 The next lesson comes from HQAG \cite{Amhis:2012bh}:
\be\label{LHCb1}
S_{\psi\phi}=-(0.04^{+0.10}_{-0.13}), \quad S^{\rm SM}_{\psi\phi}=0.038\pm 0.005,
\ee
where we have shown also SM prediction and 
the experimental 
error on $S_{\psi\phi}$ has been obtained by adding statistical and systematic 
errors in quadrature. Indeed it looks like the SM still survived another test: 
mixing induced CP-violation in $B_s$ decays is significantly lower than in 
$B_d$ decays as expected in the SM already for 25 years. However from the present perspective $S_{\psi\phi}$ could still be found in the range 
\be\label{Spsiphirange}
-0.20\le S_{\psi\phi}\le 0.20
\ee
and finding it to be negative would be a clear signal of NP. Moreover finding 
it above $0.1$ would also be a signal of NP but not as pronounced as the 
negative value. The question then arises whether this NP is somehow correlated 
with the one related to the anomalies identified above. We will return to 
this issue in the course of our presentation.

{\bf Lesson 4:} 

The final lesson comes from the recent analysis in \cite{Buras:2013dea} 
were the values $|V_{cb}|=(42.4(9))\times 10^{-3}$ \cite{Gambino:2013rza} and
$|V_{ub}|=(3.6\pm0.3)\times10^{-3}$\hfill\cite{Beringer:1900zz} have been 
used. For such values there is an acceptable simultaneous agreement of the 
SM with both $S_{\psi K_S}$ and $\varepsilon_K$ but then
\be\label{BFGNEW}
\Delta M_s = 18.8\,\text{ps}^{-1}, \qquad \Delta M_d = 0.530\,\text{ps}^{-1}~,
\ee
slightly above the data.

This discussion shows how important is the determinations of the CKM and 
and non-perturbative parameters if we want to identify NP indirectly through 
flavour violating processes. We will return to this point below and refer 
to  \cite{Buras:2013qja,Buras:2013dea}, where extensive numerical analysis of this issue has  been presented in the context of models with tree-level FCNC transitions.

\subsubsection{Going Beyond the Standard Model}
In view of NP contributions, required to remove the anomalies just 
discussed, we have to generalize the formulae of the SM.
First for $M_{12}^K$, $M_{12}^d$ and $M_{12}^s$, that govern the analysis of $\Delta F=2$ transitions in any extension of the SM we have
\be
M_{12}^i=\left(M_{12}^i\right)_\text{\rm SM}+\left(M_{12}^i\right)_\text{NP},
\qquad(i=K,d,s)
\,,
\label{eq:3.33}
\ee
with $\left(M_{12}^i\right)_\text{SM}$ given in (\ref{eq:3.6}) and (\ref{eq:3.4}).

For the mass differences in the $B_{d,s}^0-\bar B_{d,s}^0$ systems we have then
\be
\Delta M_q=2\left|\left(M_{12}^q\right)_\text{\rm SM}+\left(M_{12}^q\right)_\text{NP}\right|\qquad (q=d,s)\,.
\label{eq:3.36}
\ee
Now
\be
M_{12}^q=\left(M_{12}^q\right)_\text{\rm SM}+\left(M_{12}^q\right)_\text{NP}=\left(M_{12}^q\right)_\text{\rm SM}C_{B_q}e^{2i\varphi_{B_q}}\,,
\label{eq:3.37}
\ee
where
\be
\left(M_{12}^d\right)_\text{\rm SM}=\left|\left(M_{12}^d\right)_\text{\rm SM}\right|e^{2i\beta}\,,\qquad
\left(M_{12}^s\right)_\text{\rm SM}=\left|\left(M_{12}^s\right)_\text{\rm SM}\right|e^{2i\beta_s}.
\label{eq:3.39}
\ee
The phases $\beta$ and $\beta_s$ are defined in (\ref{vtdvts})  and 
one has approximately $\beta\approx (22\pm3)^\circ$ and $\beta_s\simeq -1^\circ$
with precise values depending on $\vub$.
We find then
\be
\Delta M_q=(\Delta M_q)_\text{\rm SM}C_{B_q}\,,
\label{eq:3.41}
\ee
and
\begin{equation}
S_{\psi K_S} = \sin(2\beta+2\varphi_{B_d})\,, \qquad
S_{\psi\phi} =  \sin(2|\beta_s|-2\varphi_{B_s})\,.
\label{eq:3.44}
\end{equation}
Thus in the presence of 
non-vanishing $\varphi_{B_d}$ and $\varphi_{B_s}$ these two asymmetries do not measure $\beta$ and $\beta_s$ 
but $(\beta+\varphi_{B_d})$ and $(|\beta_s|-\varphi_{B_s})$, respectively.

It should be remarked that
the experimental results are usually given for the phase
\be
\phi_s=2\beta_s+\phi^{\rm NP}
\ee
so that
\be
S_{\psi\phi}=-\sin(\phi_s), \qquad  2\varphi_{B_s}=\phi^{\rm NP}.
\ee
In particular the minus sign in this equation should be remembered when 
comparing our results with those quoted by the LHCb.

Next, the parameter $\varepsilon_K$ is given by
\be
\varepsilon_K=\frac{\kappa_\eps e^{i\varphi_\eps}}{\sqrt{2}(\Delta M_K)_\text{exp}}\left[\Im\left(M_{12}^K\right)_\text{\rm SM}+\Im\left(M_{12}^K\right)_\text{NP}\right]\,.
\label{eq:3.35a}
\ee
Finally, the ratio in (\ref{DetRt}) can be modified
\be\label{DetRtmod}
R_{\Delta M_B}=\frac{\Delta M_d}{\Delta M_s}=  \frac{m_{B_d}}{m_{B_s}}\frac{1}{\xi^2} \left|\frac{V_{td}}{V_{ts}}\right|^2 r(\Delta M),
\ee
where the departure of $r(\Delta M)$ from unity signals non-MFV sources at work.
 In this review we only rarely  consider $\Delta M_K$ as it is subject to large hadronic 
uncertainties. Moreover generally $\varepsilon_K$ gives  a stronger constraint 
on NP.

We will now investigate which of the models introduced in 
Section~\ref{sec:2}  could remove the anomalies just discussed 
dependently whether exclusive or inclusive value of
 $\vub$ has been chosen by nature and which models are put under significant 
pressure in both cases. In the latter case the hope is that the final 
value for $\vub$ will be some average of inclusive and exclusive determinations, that is in the ballpark of $\vub=3.7\times 10^{-3}$.
If this will turn out not to be the case
the latter models are then either close to being 
ruled out or are incomplete requiring new sources of flavour and/or CP 
violation in order to agree with the data. As we will soon see the simplest models 
considered by us  have a sufficiently low number of parameters that  
concrete answers about their ability to remove the anomalies in question 
can be given, in particular when subsequent steps will be considered.

\subsubsection{Constrained Minimal Flavour Violation (CMFV)}
 
The flavour structure in this class of models 
implies that the mixing induced CP asymmetries $S_{\psi K_S}$ 
and $S_{\psi\phi}$ are not modified with respect to the SM and the 
expressions in (\ref{eq:3.43}) still apply.

This structure also implies the flavour universality of loop functions 
contributing to various processes that is broken only by the CKM 
factors multiplying these functions. In the case of $\Delta F=2$ 
processes considered here this means that in this class of models 
NP can only modify the loop function $S_0(x_t)$ to some real valued
function $S(v)$ without modifying the values of the CKM parameters that 
have been determined in Step 1 without any influence of NP.

Now, it has been demonstrated diagrammatically in \cite{Blanke:2006yh}  that 
in the context of CMFV:
\be\label{BBbound}
S_0(x_t)\le S(v).
\ee
This simply implies that $|\varepsilon_K|$, $\Delta M_d$ and $\Delta M_s$ 
can only be enhanced in this class of models. Moreover, this happens 
in a correlated manner. A correlation between $|\varepsilon_K|$, $\Delta M_d$ 
and $S_{\psi K_S}$ within the SM has been pointed out in 
\cite{Buras:1994ec,Buras:2000xq} and 
generalized to all models with CMFV in \cite{Blanke:2006ig}. 
This 
correlation follows from the universality of $S(v)$ and the fact that
in all 
CMFV models considered, only the term in $\varepsilon_K$ involving 
$(V_{ts}^*V_{td})^2$ is affected visibly by NP with the remaining terms 
described by the SM. 

Here we want to look at this correlation from a bit different point 
of view. In fact 
eliminating the one-loop function $S(v)$ in 
$\varepsilon_K$ in favour of $\Delta M_d$ and using 
also $\Delta M_s$ one can find 
 universal expressions for $S_{\psi K_S}$ and the angle $\gamma$ in 
         the UUT that depend only on
          $\vus$, $\vcb$, known from tree-level decays, and 
          non-perturbative parameters entering the evaluation of 
          $\varepsilon_K$ and $\Delta M_{s,d}$.  They 
           are valid for all CMFV models. Therefore, once 
the data on $\vus$, $\vcb$, $\varepsilon_K$ and $\Delta M_{s,d}$ are 
taken into account one is able in this framework to predict not only 
$S_{\psi K_S}$ and $\gamma$ but also $\vub$.

Explicitly we find first
\be\label{RobertB}
S_{\psi K_S}=\sin 2\beta=\frac{1}{b\Delta M_d} \left[\frac{|\varepsilon_K|}{\vcb^2\hat B_K}
-a\right],
\ee
where
\be
a=r_\varepsilon R_t\sin\beta\left[\eta_{ct} S_0(x_t,x_c)-\eta_{cc} x_c\right], \quad
b=\frac{\eta_{tt}}{\eta_B}\frac{r_\varepsilon}{2r_d\vus^2}
\frac{1}{F_{B_d}^2\hat B_{B_d}},
\ee
with 
\be
r_\varepsilon=\kappa_\varepsilon \vus^2\frac{G_F^2 F_K^2 m_K M_W^2}{6\sqrt{2}\pi^2\Delta M_K}, \quad r_d=\frac{G_F^2}{6\pi^2}M_W^2 m_{B_d}.
\ee

The following remarks should be made
\begin{itemize}
\item
The second term $a$ in the parenthesis in (\ref{RobertB})
is roughly by a factor of  4-5 smaller than the first term. 
It depends on $\beta$ through $\sin\beta$ and ($\lambda=\vus$)
\be
R_t=\eta_R\frac{\xi}{\vus}\sqrt{\frac{\Delta M_d}{\Delta M_s}}
    \sqrt{\frac{m_{B_s}}{m_{B_d}}}, \quad \eta_R=1 -\vus\xi\sqrt{\frac{\Delta M_d}{\Delta M_s}}\sqrt{\frac{m_{B_s}}{m_{B_d}}}\cos\beta+\frac{\lambda^2}{2}+\ord(\lambda^4),
\ee
 but this dependence is very weak  and  $0.34\le a \le0.41$  in the full range 
of parameters considered.
\item
The ratio of $\eta_{tt}/\eta_B$ is independent of NP.
\item
With $R_t$ and $\beta$ determined in this manner one can calculate $\gamma$ 
and $\vub$ by means of (\ref{2.94}) and (\ref{VUBG}).
\item
The element $\vcb$ appears only as square in these expressions and not 
as $\vcb^4$ in $\varepsilon_K$, which improves the accuracy of 
the determination.
\end{itemize}

We should emphasize that in this determination the experimental input 
$\Delta M_{s,d}$ and $\varepsilon_K$ is very precise. $\vus$ is known very well
and $\vcb$ is better known than $\vub$ from tree-level decays. 

Setting then the experimental values of $\Delta M_{s,d}$, $\varepsilon_K$ 
and $\vcb$ as well as central values of the non-perturbative parameters in 
Table~\ref{tab:input}
into (\ref{RobertB}) we find 
\begin{align}
 & S_{\psi K_S} = 0.81~(0.87)\,\Rightarrow \beta = 27~(30^\circ)\,,\qquad R_t = 0.92~(0.92)
\end{align}
and thus
\begin{align}
 &R_b = 0.46~(0.50)\,,\qquad |V_{ub}| = 0.0043~(0.0047)\,,\qquad \gamma = 67.2~(66.4^\circ)\,,
\end{align}
 where the values in parentheses correspond to the input in (\ref{oldf1}).
This demonstrates sensitivity  to the non-perturbative parameters.

 While a sophisticated analysis including all uncertainties would somewhat 
wash out these results, the message from this exercise is clear. The fact that $S_{\psi K_S}$ is much larger than the data 
requires the presence of new CP-violating phases,  although with the most recent lattice input these phases can be smaller. This
exercise is 
equivalent to the one performed in \cite{Lunghi:2008aa}, where  $\varepsilon_K$ has been set to its experimental value but $\sin 2\beta$ was free.
On the other hand setting $S_{\psi K_S}$ to its experimental value but 
keeping $\varepsilon_K$ free as done in \cite{Buras:2008nn} one finds that
$|\varepsilon_K|$ is significantly below the data. Yet, this difficulty 
can be resolved in CMFV models by increasing the value of $S(v)$. 
While, the latter approach is clearly legitimate, it hides possible problems 
of CMFV as it assumes that this NP scenario can describe the data on 
$\Delta M_{s,d}$ and $\varepsilon_K$ simultaneously, which as we will 
now show is not really the case.

Indeed, with respect to the anomalies discussed above we note that 
\begin{itemize}
\item CMFV models
favour the exclusive determination of $\vub$ as only then 
they are capable of reproducing the experimental value of $S_{\psi K_S}$.
\item
$|\varepsilon_K|$ can be naturally enhanced by increasing the value of $S(v)$
thereby solving the $|\varepsilon_K|$-$S_{\psi K_S}$ tension.
\item
 $\Delta M_{s,d}$ are enhanced simultaneously with the ratio $\Delta M_s/\Delta M_d$ unchanged with respect to the SM ($r(\Delta M)=1$).
While the latter property is certainly 
good news, the enhancements of $\Delta M_s$ and  $\Delta M_d$ 
are clearly problematic.
Therefore the present values of hadronic matrix elements imply  
 new tensions, namely the $|\varepsilon_K|$-$\Delta M_{s,d}$ tensions 
pointed out in \cite{Buras:2012ts,Buras:2011wi}.
\end{itemize}

\begin{figure}[!tb]
 \centering
\includegraphics[width = 0.6\textwidth]{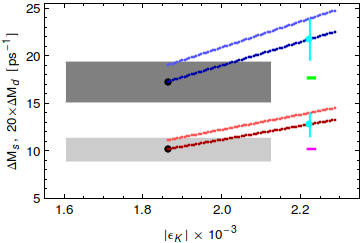}
\caption{\it $\Delta M_{s}$ (blue) and $20\cdot\Delta M_{d}$ (red) as functions of $|\varepsilon_K|$ in models 
with CMFV for Scenario 1 chosen by these models \cite{Buras:2012ts}. The short green and magenta  lines represent the data, while the large black and grey regions the SM 
predictions. For the light blue and light red line the old values from~(\ref{oldf1}) are used and for dark blue
and dark red the new ones from Table~\ref{tab:input}. More
information can be found in the text.}\label{fig:DeltaMvsepsK}~\\[-2mm]\hrule
\end{figure}

In Fig.~\ref{fig:DeltaMvsepsK}  we plot $\Delta M_s$ and $\Delta M_d$ as functions of 
$|\varepsilon_K|$.
In obtaining this plot we have simply varied the 
master one-loop  $\Delta F=2$ function $S$ keeping CKM parameters and other input parameters 
fixed. The value of $S$ at which central experimental value of $|\varepsilon_K|$
is reproduced 
turns out to be $S=2.9$  to        
be compared 
with $S_{\rm SM}=2.31$.
At this value the central values of $\Delta M_{s,d}$ read 
\be\label{BESTCMFV}
\Delta M_d=0.64(6)~\text{ps}^{-1} \quad (0.69(6)~\text{ps}^{-1}),\quad  \Delta M_s=21.7(2.1)~\text{ps}^{-1}\quad(23.9(2.1)~\text{ps}^{-1})~.
\ee
They both differ from experimental values by $3\sigma$.
The error on  $|\varepsilon_K|$ coming dominantly from the 
error of $\vcb$ and the error of the QCD factor $\eta_{cc}$ in the charm contribution 
\cite{Brod:2011ty} is however disturbing. Clearly this plot gives only some 
indication for possible  difficulties of the CMFV models and we need 
a significant decrease of theoretical errors in order to see how solid this 
result is.

In summary, we 
observe  
that simultaneous good agreement for  $\varepsilon_K$ and $\Delta M_{s,d}$ with 
the data is difficult to achieve in this NP scenario.
It 
also implies that to improve the agreement with data 
we need at least one of the following four ingredients:
\begin{itemize}
\item
Modification of the values of 
\be\label{par}
\vcb,\qquad F_{B_s}\sqrt{\hat B_{B_s}},\qquad  F_{B_d} \sqrt{\hat B_{B_d}}
\ee
\item
New CP phases, flavour violating and/or flavour blind,
\item
New flavour violating contributions beyond the CKM matrix,
\item
New local operators which could originate in tree-level heavy gauge boson 
or scalar exchanges. They could also be generated at one-loop level.
\end{itemize}

The first possibility has been addressed in \cite{Buras:2013raa}, where 
the experimental values of $\Delta M_{s,d}$, $\varepsilon_K$, $\vus$ and 
$S_{\psi K_S}$ 
have been used as input and $\hat B_K$ has been set to $0.75$ in
perfect agreement with the lattice results and the large $N$ approach 
\cite{Buras:1985yx,Bardeen:1987vg,Gerard:2010jt,Buras:2014maa}. Subsequently the 
parameters in (\ref{par}) have been calculated as  functions of 
$S(v)$ and $\gamma$ in order to see whether there is any hope for  removing  all 
the tensions in CMFV simultaneously in case the future more 
precise determinations of  $F_{B_s}\sqrt{\hat B_{B_s}}$, $F_{B_d} \sqrt{\hat B_{B_d}}$ and $\vcb$  would result in different 
values than the present ones. The results of 
\cite{Buras:2013raa} can be summarized briefly as follows:

\begin{itemize}
\item
The tension between $\varepsilon_K$ and $\Delta M_{s,d}$ in CMFV models 
accompanied with $|\varepsilon_K|$ being smaller than the data within the 
SM, cannot be removed by varying $S(v)$ when the present input parameters
in Table~\ref{tab:input} are used.
\item
Rather the value of $\vcb$ has to be increased and the values of $F_{B_s} \sqrt{\hat B_{B_s}}$ and $F_{B_d} \sqrt{\hat B_{B_d}}$ decreased relatively to the 
presently quoted lattice values. These enhancements and suppressions are correlated 
with each other and depend on $\gamma$.  Setting the QCD corrections 
$\eta_{ij}$ at their central values one finds the results in 
Table~\ref{tab:CMFVpred}.
\item
However, the present significant uncertainty in $\eta_{cc}$ softens 
these problems. Yet, it turns out that
 the knowledge of long distance contributions to $\Delta M_K$ accompanied 
by the very precise experimental value of the latter allows a 
significant reduction of the present uncertainty in the value of 
the QCD factor $\eta_{cc}$ under the plausible assumption that $\Delta M_K$ 
in CMFV models is fully dominated by the SM contribution. Indeed, using 
the large $N$ estimate of long distance contribution to 
$\Delta M_K$ \cite{Buras:2014maa} we 
find
\be\label{etaBG}
\eta_{cc}\approx 1.70 \pm 0.21,
\ee
which  implies
the reduction of the theoretical 
error in $\varepsilon_K$ and in turn the reduction of the error 
in the extraction of the favoured 
value of  $\vcb$ in the CMFV framework. 
\end{itemize}

We should remark that the reduction of the error in $\eta_{cc}$ by a factor 
of more than $3.5$ relatively to the one resulting from direct calculation 
\cite{Brod:2011ty} is significant as the uncertainty in  $\varepsilon_K$ from $\eta_{cc}$ alone is reduced from roughly $7\%$ to $2\%$ and is consequently lower 
than the present uncertainty of $3\%$ from $\eta_{ct}$. Yet, this reduction 
cannot be appreciated at present as by far the dominant uncertainty in 
$\varepsilon_K$ comes from $\vcb$.

\begin{figure}[!tb]
 
\centering
\includegraphics[width = 0.45\textwidth]{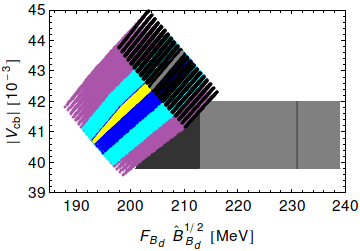}
\includegraphics[width = 0.45\textwidth]{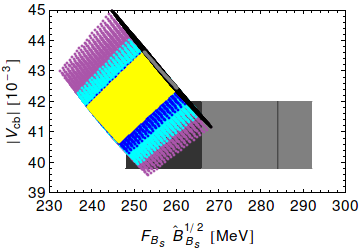}

 \caption{\it $\vcb$ versus $F_{B_d} \sqrt{\hat B_{B_d}}$ and $F_{B_s} \sqrt{\hat B_{B_s}}$ for
 $\gamma\in[63^\circ,71^\circ]$. The yellow region corresponds to $S(v)\in[2.31,2.8]$, $\eta_{cc} =
1.87$, $\eta_{ct} = 0.496$. In the purple region we include the errors in $\eta_{cc,ct}$:  $S(v)\in[2.31,2.8]$,
$\eta_{cc} \in [1.10,2.64]$,
$\eta_{ct} \in [0.451,0.541]$. In the cyan region we use instead the reduced error of $\eta_{cc}$ as in Eq.~(\ref{etaBG}): 
$S(v)\in[2.31,2.8]$,
$\eta_{cc} \in [1.49,1.91]$,
$\eta_{ct} \in [0.451,0.541]$. In the blue region we fix $\eta_{ct}$ to its central value: $S(v)\in[2.31,2.8]$,
$\eta_{cc} \in [1.49,1.91]$,
$\eta_{ct} =0.496$. To test the SM we include the black region for fixed $S(v) = S_0(x_t) = 2.31$ and $\eta_{cc,ct}$ as in the purple
region. The gray line within the black SM region corresponds to   $\eta_{cc} =
1.87$ and $\eta_{ct} = 0.496$. Dark (light) gray box: $1\sigma$ range of  $F_{B_d} \sqrt{\hat B_{B_d}}$, $F_{B_s} \sqrt{\hat B_{B_s}}$
and $\vcb$ as given in Table~\ref{tab:input} and (\ref{oldf1}), respectively. The vertical dark gray lines indicate where the dark gray
boxes end.
}\label{fig:VcbvsFBscan}~\\[-2mm]\hrule
\end{figure}

\begin{figure}[!tb]
 
\centering
\includegraphics[width = 0.45\textwidth]{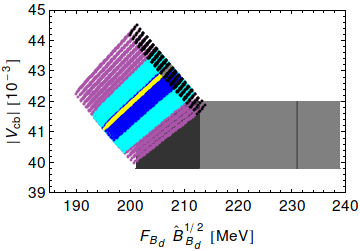}
\includegraphics[width = 0.45\textwidth]{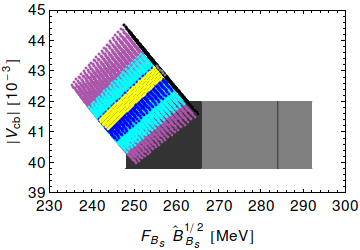}

 \caption{\it $\vcb$ versus $F_{B_d} \sqrt{\hat B_{B_d}}$ and $F_{B_s} \sqrt{\hat B_{B_s}}$ as in Fig.~\ref{fig:VcbvsFBscan} but for 
 $\gamma = (67\pm1)^\circ$. 
}\label{fig:VcbvsFBscan2}~\\[-2mm]\hrule
\end{figure}

In Fig.~\ref{fig:VcbvsFBscan} 
 on the left hand side we show the correlation between $F_{B_d} \sqrt{\hat B_{B_d}}$ and $\vcb$ for $\gamma\in[63^\circ,71^\circ]$.
 Analogous correlation 
between  $F_{B_s} \sqrt{\hat B_{B_s}}$ and $\vcb$ is shown on the right hand side.   The dark gray boxes represent the present values of the
parameters 
as given in Table~\ref{tab:input}, while the light gray the ones from 
(\ref{oldf1}). The vertical dark gray lines show where the dark gray boxes end, respectively. In these plots
we show the anatomy of various uncertainties  
with different ranges described in the figure caption. We observe that the 
reduced error on $\eta_{cc}$ corresponding to the cyan region 
decreased the allowed region  which with future 
lattice calculations could be decreased further. Comparing the blue 
and cyan regions we note that the reduction in the error on $\eta_{ct}$ 
would be welcomed as well. It should also be stressed that in a given 
CMFV model with fixed $S(v)$ the uncertainties are reduced further. 
This is illustrated with the black range for the case of the SM. 
 Finally an  impact on  Fig.~\ref{fig:VcbvsFBscan} will have a
precise measurement of $\gamma$ or equivalently precise lattice determination 
of 
$\xi$.  We illustrate this impact in Fig.~\ref{fig:VcbvsFBscan2} by setting in the plots of  Fig.~\ref{fig:VcbvsFBscan} $\gamma=(67\pm1)^\circ$. Further 
details can be found in \cite{Buras:2013raa}.

 We note that the most recent values of  $F_{B_s}\sqrt{\hat B_{B_s}}$ and $F_{B_d} \sqrt{\hat B_{B_d}}$ softened significantly
the problems of CMFV in question, even if 
still an enhanced value of $\vcb$ is required. For instance,  in accordance with  the lesson 4 above, if one would ignore 
the present exclusive determination of $\vcb$ and used the most recent 
inclusive determination \cite{Gambino:2013rza}
\be
\vcb=(42.42\pm0.86)\times 10^{-3}
\ee
 CMFV would be in a much better shape but also the SM-like values for $S(v)$ 
would be favoured. We are looking forward 
to the improved lattice calculations and improved determinations of $\vcb$ 
in order to see whether CMFV will survive flavour precision tests.

\begin{table}[!tb]
\centering
\begin{tabular}{|c|c||c|c|c|c|c|c|c|c|}
\hline
 $S(v)$  & $\gamma$ & $\vcb$ & $\vub$ & $\vtd$&  $\vts$ & $F_{B_s}\sqrt{\hat B_{B_s}}$ & 
$F_{B_d} \sqrt{\hat B_{B_d}}$ & $\xi$ &  $\mathcal{B}(B^+\to \tau^+\nu)$\\
\hline
\hline
  \parbox[0pt][1.6em][c]{0cm}{} $2.31$ & $63^\circ$ & $43.6$ & $3.69$  & $8.79$ & $42.8$ & $252.7$ &$210.0$ & $1.204$ &  $0.822$\\
 \parbox[0pt][1.6em][c]{0cm}{}$2.5$ & $63^\circ$ & $42.8$& $3.63$ & $8.64$ & $42.1$ & $247.1$ & $205.3$ &$1.204$ &  $0.794$\\
 \parbox[0pt][1.6em][c]{0cm}{}$2.7$ &$63^\circ$ & $42.1$ & $3.56$ & $8.49$ &  $41.4$  & 
$241.8$ & $200.9$ & $1.204$ &  $0.768$\\
\hline
  \parbox[0pt][1.6em][c]{0cm}{} $2.31$ & $67^\circ$ & $42.9$ & $3.62$  & $8.90$ & $42.1$ & $256.8$ &$207.2$ &$1.240$ & $0.791$\\
 \parbox[0pt][1.6em][c]{0cm}{}$2.5$ & $67^\circ$ & $42.2$& $3.56$ & $8.75$ & $41.4$ & $251.1$ & $202.6$ &$1.240$ & $0.765$\\
 \parbox[0pt][1.6em][c]{0cm}{}$2.7$ &$67^\circ$ & $41.5$ & $3.50$ & $8.61$ &  $40.7$  & 
$245.7$ & $198.3$ &$1.240$ & $0.739$\\
\hline
  \parbox[0pt][1.6em][c]{0cm}{} $2.31$ & $71^\circ$ & $42.3$ & $3.57$  & $9.02$ & $41.5$ & $260.8$ &$204.5$ &$1.276$ & $0.770$\\
 \parbox[0pt][1.6em][c]{0cm}{}$2.5$ & $71^\circ$ & $41.6$& $3.51$ & $8.87$ & $40.8$ & $255.1$ & $200.0$ &$1.276$ & $0.744$\\
 \parbox[0pt][1.6em][c]{0cm}{}$2.7$ &$71^\circ$ & $40.9$ & $3.45$ & $8.72$ &  $40.1$  & 
$249.6$ & $195.7$ &$1.276$ & $0.719$\\
 \hline
\end{tabular}
\caption{\it CMFV predictions for various quantities as functions of 
$S(v)$ and $\gamma$. The four elements of the CKM matrix are in units of $10^{-3}$,  
 $F_{B_s} \sqrt{\hat B_{B_s}}$ and $F_{B_d} \sqrt{\hat B_{B_d}}$ in units of $\mev$ and  $\mathcal{B}(B^+\to \tau^+\nu)$ in units of $10^{-4}$.
}\label{tab:CMFVpred}~\\[-2mm]\hrule
\end{table}

\subsubsection{2HDM with MFV and Flavour Blind Phases (${\rm 2HDM_{\overline{MFV}}}$)}

In view of our discussion above, 
this model \cite{Buras:2010mh} has in principle
a better chance to remove simultaneously 
the anomalies in question than CMFV models but as we will soon see it 
approaches this problem in a different manner. The basic new features in 
 ${\rm 2HDM_{\overline{MFV}}}$ relative to CMFV are:
\begin{itemize}
\item
The presence of flavour blind phases (FBPs) in this MFV framework modifies through their interplay 
with the standard CKM flavour violation the usual characteristic 
relations of  the CMFV framework. In particular the mixing induced CP
asymmetries $S_{\psi K_S}$ and $S_{\psi\phi}$  take the form known 
from non-MFV frameworks like LHT, RSc and SM4 as given in (\ref{eq:3.44}).
\item
 The FBPs in the ${\rm 2HDM_{\overline{MFV}}}$ can appear  both 
in Yukawa interactions and  in the Higgs potential. While in  
\cite{Buras:2010mh} only the case of FBPs in Yukawa interactions has been 
considered, in \cite{Buras:2010zm} these considerations have been extended
to include also the FBPs in the Higgs potential.
The two flavour-blind CPV mechanisms can be distinguished
through the correlation between $S_{\psi K_S}$ and $S_{\psi\phi}$ that is
strikingly different if only one of them is relevant. 
In fact the 
relation between generated new phases are very different in each case:
\be\label{Phase1}
\varphi_{B_d}=\frac{m_d}{m_s}\varphi_{B_s}\quad\quad {\rm and}\quad \varphi_{B_d}=\varphi_{B_s}
\ee
for FBPs in Yukawa couplings and Higgs potential, respectively.
\item
New local operators are generated through the contributions of tree level 
heavy Higgs exchanges which also implies modified structure of flavour 
violation relatively to CMFV.
\item
Sizable FBPs, necessary to explain possible  sizable 
non-standard CPV effects in $B_{s}$ mixing could, in principle, be 
forbidden
by the upper bounds on 
EDMs of the neutron and the atoms. This question has been addressed 
in \cite{Buras:2010zm} and it has been shown that even for 
$S_{\psi\phi}=\ord(1)$, this model still satisfied  these bounds.
\end{itemize}

It is not our goal to describe the phenomenology of this model here in details 
as such details can be 
found in \cite{Buras:2010mh,Buras:2010zm}. Moreover a review appeared in \cite{Isidori:2012ts}.
We rather want to emphasize 
that the model addresses the anomalies in question in a manner which differs 
profoundly from CMFV and thus a distinction between these two models can 
be already made on the basis of the data on $\Delta F=2$ processes.

Indeed in this model
new 
contributions to $\varepsilon_K$ originating in tree level neutral Higgs 
exchanges are tiny being suppressed by small quark masses $m_{s,d}$. 
Consequently the correct value of   $\varepsilon_K$ can only be obtained 
by choosing sufficiently large value of $\sin 2\beta$ which corresponds 
to the large ({\it inclusive}) $\vub$ scenario. If the formula (\ref{eq:3.43}) is 
used this in turn implies, as seen in Table~\ref{tab:SMpred}, a 
value of $S_{\psi K_S}$ which is much larger than the data. However, in this 
model
the interplay of the CKM phase with the flavour blind phases in Yukawa 
couplings and Higgs potential  generates non-vanishing new phases 
$\varphi_{B_q}$ and the formulae in (\ref{eq:3.44}) instead of (\ref{eq:3.43})
 should be used.
The new phases can suppress  $S_{\psi K_S}$ simultaneously 
enhancing uniquely the asymmetry $S_{\psi \phi}$.

Now while the rate of the 
suppression of $S_{\psi K_S}$ for a given $S_{\psi\phi}$ is much stronger 
if  significant FBPs in the Higgs potential rather than in 
Yukawa couplings are at work, both mechanism share a very important 
property:

\begin{itemize}
\item
The necessary suppression of $S_{\psi K_S}$ necessarily implies uniquely the
enhancement 
of $S_{\psi\phi}$ so that this asymmetry is larger than in the 
SM and consequently has  positive sign. Finding  eventually $S_{\psi\phi}$ at the LHC to be negative 
would be a real problem for  the ${\rm 2HDM_{\overline{MFV}}}$.
\end{itemize} 

Now $\varepsilon_K$ can only be made consistent in this model by 
properly choosing $\gamma$ and in particular $\vub$ that has to be 
sufficiently large. The question then arises, whether simultaneously also 
$S_{\psi K_S}$, $S_{\psi\phi}$ and $\Delta M_{d,s}$ can be made consistent 
with the data. We find then \cite{Buras:2012xxx}: 

\begin{itemize}
\item
The removal of the $\varepsilon_K-S_{\psi K_S}$ anomaly, which proceeds through 
the  negative phase $\varphi_{B_d}$, is only possible with the help of 
FBPs in the Higgs potential. This is achieved in the case of
the full dominance of the 
 $Q^{\rm SLL}_{1,2}$ operators as far as CP-violating contributions are 
concerned. If these operators also dominate the CP-conserving contributions
two important properties follow:
\be\label{MAIN1}
\varphi_{B_d}=\varphi_{B_s}, \qquad C_{B_s}= C_{B_d}.
\ee
The second of the equalities implies
\be\label{MAIN2}
\left(\frac{\Delta M_s}{\Delta M_d}\right)_{{\rm 2HDM_{\overline{MFV}}}}=
\left(\frac{\Delta M_s}{\Delta M_d}\right)_{\rm SM}.
\ee
This relation is known from models with CMFV but there $C_{B_s}= C_{B_d}\ge 1$.
In  ${\rm 2HDM_{\overline{MFV}}}$ also $C_{B_s}= C_{B_d}\le 1$ is possible.
Moreover, the CMFV correlation between 
$\varepsilon_K$ and $\Delta M_{s,d}$ is absent and $\Delta M_{s,d}$ can 
be both suppressed and enhanced if necessary. 
\item
A significant contribution of the operators $Q_{1,2}^{\rm LR}$ is unwanted 
as it spoils the relation~(\ref{MAIN2}) having much larger effect 
on $\Delta M_s$  than $\Delta M_d$.  But as this contribution uniquely 
suppresses $\Delta M_s$ below its SM value, it could turn out relevant 
one day if the lattice results for hadronic matrix changed. 
This contribution cannot help in 
solving $\varepsilon_K-S_{\psi K_S}$ anomaly as its effect on 
the phase  $\varphi_{B_d}$ is very small.
\end{itemize}

\begin{figure}[!tb]
 \centering
\includegraphics[width = 0.6\textwidth]{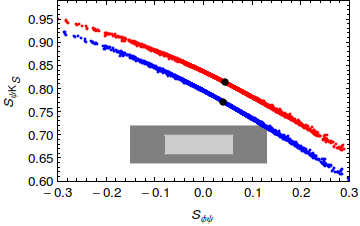}
\caption{ \it $S_{\psi K_S}$ vs. $S_{\psi \phi}$ in  ${\rm 2HDM_{\overline{MFV}}}$ 
for $\vub=4.0\cdot 10^{-3}$ (blue) and  $\vub=4.3\cdot 10^{-3}$ (red).
  SM is represented by black points while $1\sigma$ ($2\sigma$) experimental 
range by the grey (dark grey) area \cite{Buras:2012xxx}.}\label{fig:SvsSa}~\\[-2mm]\hrule
\end{figure}

Thus at first sight at the qualitative level this model provides 
a better description of $\Delta F=2$ data than the SM and models with 
CMFV. Yet, here comes a possible difficulty. As shown in Fig.~\ref{fig:SvsSa} the size of $\varphi_{B_d}$ that is necessary to obtain simultaneously 
good agreement with the data on $\varepsilon_K$ and $S_{\psi K_S}$ 
implies 
in turn $S_{\psi\phi}\ge0.15$ which is 
$2\sigma$ away from 
the LHCb central value in (\ref{LHCb1}). 

In summary ${\rm 2HDM_{\overline{MFV}}}$ is from the point of view of 
$\Delta F=2$ observables in a reasonable shape. Yet,
finding in the future that nature chooses a {\it negative} value of 
$S_{\psi\phi}$  and/or small (exclusive) value of $\vub$ would practically 
rule out  ${\rm 2HDM_{\overline{MFV}}}$. Also a decrease of the experimental 
error on $S_{\psi\phi}$ without the change of its central value would be 
problematic for this model.

We are looking forward to improved experimental data and improved lattice 
calculations to find out whether this simple model can satisfactorily describe 
the data on  $\Delta F=2$ observables.

\subsubsection{Tree-Level Gauge Boson Exchanges}

\begin{figure}[!tb]
 \centering
\includegraphics[width = 0.35\textwidth]{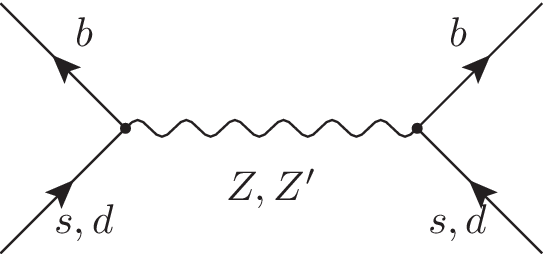}
\caption{\it Tree-level flavour-changing $Z,Z^\prime$ contribution to \bbd and \bbs mixing.}\label{fig:FD1}~\\[-2mm]\hrule
\end{figure}

We will next investigate what  a neutral gauge boson tree level exchange can contribute to this discussion. 
For the neutral gauge boson $Z^\prime$ contribution as shown in Fig.~\ref{fig:FD1} one has generally 
\cite{Buras:2012fs,Buras:2012jb}
\begin{align}\begin{split}\label{M12Z}
 \left(M_{12}^\star\right)^{bq}_{Z^\prime}   =&
 \frac{(\Delta_L^{bq}(Z^\prime))^2}{2M_{Z^\prime}^2}C_1^\text{VLL}(\mu_{Z^\prime})\langle Q_1^\text{VLL}(\mu_{Z^\prime})\rangle
 +\frac{(\Delta_R^{bq}(Z^\prime))^2}{2M_{Z^\prime}^2}
 C_1^\text{VRR}(\mu_{Z^\prime})\langle Q_1^\text{VLL}(\mu_{Z^\prime})\rangle \\
 &+\frac{\Delta_L^{bq}(Z^\prime)\Delta_R^{bq}(Z^\prime)}{
 M_{Z^\prime}^2} \left [ C_1^\text{LR}(\mu_{Z^\prime}) \langle Q_1^\text{LR}(\mu_{Z^\prime})\rangle +
 C_2^\text{LR}(\mu_{Z^\prime}) \langle Q_2^\text{LR}(\mu_{Z^\prime})\rangle \right]\,,\end{split}
 \end{align}
where including NLO QCD corrections \cite{Buras:2012fs}
\begin{align}\label{equ:WilsonZ}
\begin{split}
C_1^\text{VLL}(\mu_{Z^\prime})=C_1^\text{VRR}(\mu_{Z^\prime}) 
& = 1+\frac{\alpha_s}{4\pi}\left(-2\log\frac{M_{Z^\prime}^2}{\mu_{Z^\prime}^2}+\frac{11}{3}\right)\,,\end{split}\\
\begin{split}
 C_1^\text{LR}(\mu_{Z^\prime}) 
& =1+\frac{\alpha_s}{4\pi}
\left(-\log\frac{M_{Z^\prime}^2}{\mu_{Z^\prime}^2}-\frac{1}{6}\right)\,,\end{split}\\
C_2^\text{LR}(\mu_{Z^\prime}) &=\frac{\alpha_s}{4\pi}\left(-6\log\frac{M_{Z^\prime}^2}{\mu_{Z^\prime}^2}-1\right)\,.
\end{align}
Here $\langle Q^a_i(\mu_{Z^\prime})\rangle$
are the matrix elements of operators evaluated at the matching scale. 
 Their $\mu_{Z^\prime}$ dependence is canceled by the one of 
of $C^a_i(\mu_{Z^\prime})$ so that $M_{12}$ does not depend  on $\mu_{Z^\prime}$. The values of  $\langle Q^a_i(\mu_{Z^\prime})\rangle$ for
$\mu_H=\mu_{Z^\prime}=1\tev$ can be found 
in Table~\ref{tab:Qi}. In the case of the $K$ system the indices $bq$ should 
be replaced by $sd$. The Wilson coefficients listed above remain 
unchanged and the relevant hadronic matrix elements are also collected 
in Table~\ref{tab:Qi}.  If tree-level $Z$-boson exchanges are considered 
the matrix elements in Table~\ref{tab:Qi1} should be used, 
$M_{\mu_{Z^\prime}}\to M_Z$ and 
$\mu_{Z^\prime}\to m_t(m_t)$. 

In the case of VLL and VRR operators it is more convenient to incorporated 
NP effects as shifts in the one-loop functions $S(v)$. These shifts, 
denoted by 
$[\Delta S(M)]_{\rm VLL}$ and $[\Delta S(M)]_{\rm VRR}$ have been calculated 
in  \cite{Buras:2012dp} and are given as follows
\be\label{Zprime1}
[\Delta S(B_q)]_{\rm VLL}=
\left[\frac{\Delta_L^{bq}(Z^\prime)}{\lambda_t^{(q)}}\right]^2
\frac{4\tilde r}{M^2_{Z^\prime }g_{\text{SM}}^2}, \qquad
[\Delta S(K)]_{\rm VLL}=
\left[\frac{\Delta_L^{sd}(Z^\prime)}{\lambda_t^{(K)}}\right]^2
\frac{4\tilde r}{M^2_{Z^\prime}g_{\text{SM}}^2},
\ee
where
\be\label{gsm}
g_{\text{SM}}^2=4\frac{G_F}{\sqrt 2}\frac{\alpha}{2\pi\sin^2\theta_W}=1.78137\times 10^{-7} \gev^{-2}\,.
\ee
{ Here $\tilde r=0.985$, $\tilde r=0.965$, $\tilde r=0.953$ and $\tilde r = 0.925$ for $M_{Z^\prime} =1,~2,~3, ~10\tev$, respectively.}
$[\Delta S(M)]_{\rm VRR}$  is then
found from the formula above by simply replacing L by R. For the
case of tree-level $Z$ exchanges $\tilde r=1.068$.

For a qualitative discussion it is sufficient to set the Wilson coefficients 
to the LO values. Then
\be\label{M12Znew}
 \left(M_{12}^\star\right)_{Z^\prime}   =
 \left(\frac{(\Delta_L^{sd}(Z^\prime))^2}{2M_{Z^\prime}^2}
 +\frac{(\Delta_R^{sd}(Z^\prime))^2}{2M_{Z^\prime}^2}\right)
 \langle Q_1^\text{VLL}(\mu_{Z^\prime})\rangle 
 +\frac{\Delta_L^{sd}(Z^\prime)\Delta_R^{sd}(Z^\prime)}{
 M_{Z^\prime}^2} \langle Q_1^\text{LR}(\mu_{Z^\prime}) \rangle
 \ee
with analogous expressions for other meson systems. Now as seen in 
Table~\ref{tab:Qi}
model independently 
\be
 \langle Q_1^\text{VLL}(\mu_{Z^\prime})\rangle >0, \quad  
\langle Q_1^\text{LR}(\mu_{Z^\prime})\rangle<0, \quad 
 |\langle Q_1^\text{LR}(\mu_{Z^\prime})\rangle |\gg |\langle Q_1^\text{VLL}(\mu_{Z^\prime})\rangle|,
\ee
which has an impact on the signs and size of the couplings $\Delta_{L,R}(Z^\prime)$ 
if these contributions should remove the anomalies in the data.

The outcome for the phenomenology depends on whether $\Delta_L$ and $\Delta_R$ 
are of comparable size or if one of them is dominant and whether they are real or complex 
quantities. Moreover these properties can be different for different meson 
systems. Evidently we have here in mind the scenarios LHS, RHS, LRS and ALRS of 
Section~\ref{sec:1}.  Moreover, one has to distinguish between  the 
Scenario 1 (S1) and Scenario 2 (S2) for $\vub$, so that generally 
one deals with LHS1, LHS2 and similarly for RHS, LRS and ALRS.

As expected with these new contributions without any particular structure 
of the $\Delta_{L,R}$ couplings all tensions within the SM in the $\Delta F=2$ 
transitions can be removed in many ways and it will be important to investigate in the next 
steps which of them are also consistent with other constraints and which 
ones remove simultaneously other tensions, that are already present or 
will be generated when the data and lattice results improve in the future. 

In concrete BSM models the couplings  $\Delta^{ij}_{L,R}$,  corresponding to 
different meson systems, could be related to each other as they may depend 
on the same fundamental parameters of an underlying theory. For instance 
in the minimal 3-3-1 model, analyzed recently in \cite{Buras:2012dp,Buras:2013dea},  the flavour violating couplings 
$\Delta_L^{sd}(Z')$, $\Delta_L^{bd}(Z')$ and $\Delta_L^{bs}(Z')$ depend on 
two mixing angles and two complex phases, instead of six parameters, which 
implies correlations between observables in different meson systems (see also Sec.~\ref{sec:331}).

A very detailed analysis of $B^0_{d,s}-\bar B^0_{d,s}$  and $K^0-\bar K^0$ 
systems has been presented in \cite{Buras:2012jb} setting the CKM parameters 
as in (\ref{fixed}) and all the other input at the central values in 
Table~\ref{tab:input}  except that in \cite{Buras:2012jb} the input 
in (\ref{oldf1}) has been used. As the latter values are consistent with 
the present ones, in order to take partially hadronic
and experimental uncertainties into account we will still present here 
the results of  \cite{Buras:2012jb}. Moreover as in 
the latter paper we  
require  that values of observables in question satisfy the following
constraints 
\be\label{C1}
16.9/{\rm ps}\le \Delta M_s\le 18.7/{\rm ps},
\quad  -0.20\le S_{\psi\phi}\le 0.20,
\ee
\be\label{C2}
0.48/{\rm ps}\le \Delta M_d\le 0.53/{\rm ps},\quad
0.64\le S_{\psi K_S}\le 0.72 .
\ee

\be\label{C3}
0.75\le \frac{\Delta M_K}{(\Delta M_K)_{\rm SM}}\le 1.25,\qquad
2.0\times 10^{-3}\le |\varepsilon_K|\le 2.5 \times 10^{-3}.
\ee

The larger uncertainty for  $\varepsilon_K$  than  $\Delta M_{s,d}$ 
signals its strong $\vcb^4$ dependence. $\Delta M_K$ has even larger 
uncertainty because of potential long distance uncertainties.
When using the constraint from $S_{\psi\phi}$ and  $S_{\psi K_S}$ we take into
account that only mixing phases 
close to their SM value are allowed by the data thereby removing some discrete ambiguities. 

Parametrizing the different flavour violating  couplings of $Z'$ to quarks 
as follows
\be\label{Zprimecouplings}
\Delta_L^{bs}(Z')=-\tilde s_{23} e^{-i\delta_{23}},\quad 
\Delta_L^{bd}(Z')=\tilde s_{13} e^{-i\delta_{13}},\quad 
\Delta_L^{sd}(Z')=-\tilde s_{12} e^{-i\delta_{12}},
\ee
it was possible to find the allowed oases in the spaces
$(\tilde s_{ij},\delta_{ij})$ used to describe $Z'$ effects in each 
system. The minus sign is introduced to cancel the one in $V_{ts}$. 

\begin{figure}[!tb]
\begin{center}
\includegraphics[width=0.45\textwidth] {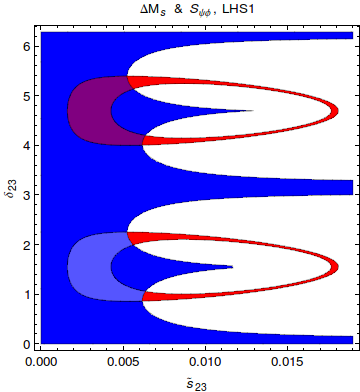}
\caption{\it  Ranges for $\Delta M_s$ (red region) and $S_{\psi \phi}$ (blue region) for $M_{Z^\prime}=1$~TeV in LHS1 satisfying the bounds
in Eq.~(\ref{C1}).
}\label{fig:oasesBsLHS1}~\\[-2mm]\hrule
\end{center}
\end{figure}

In the case of  $B^0_{s}-\bar B^0_{s}$ system 
the result of this search for $M_{Z'}=1\tev$ and LHS1 scenario 
is shown in Fig.~\ref{fig:oasesBsLHS1}.
The {\it red} regions correspond to the allowed ranges for $\Delta M_{s}$,
while the {\it blue} ones to the corresponding ranges for  $S_{\psi\phi}$. The overlap between red and blue regions (light blue and purple)
identifies the
oases we were looking for. We observe that the requirement of suppression
of $\Delta M_s$ implies $\tilde s_{23}\not=0$. As this system is immune 
to the value of $\vub$ the same results are obtained for LHS2.

\begin{figure}[!tb]
\begin{center}
\includegraphics[width=0.45\textwidth] {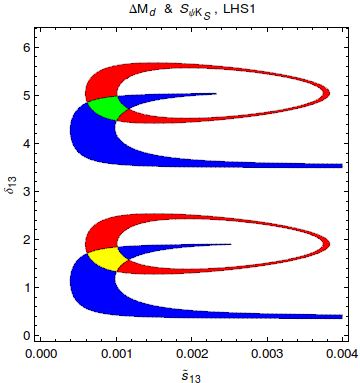}
\includegraphics[width=0.45\textwidth] {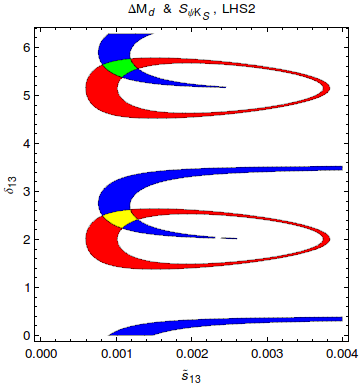}
\caption{\it  Ranges for $\Delta M_d$ (red region) and $S_{\psi K_S}$ (blue region) for $M_{Z^\prime}=1$ TeV in LHS1 (left) and
LHS2 (right) satisfying the bounds in Eq.~(\ref{C2}).
}\label{fig:oasesBdLHS}~\\[-2mm]\hrule
\end{center}
\end{figure}

\begin{figure}[!tb]
\begin{center}
\includegraphics[width=0.45\textwidth] {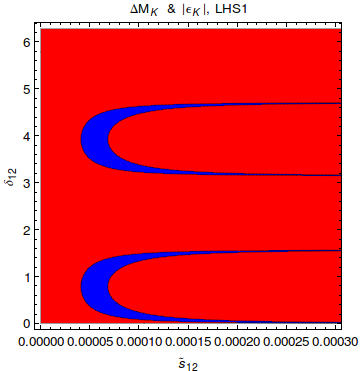}
\includegraphics[width=0.45\textwidth] {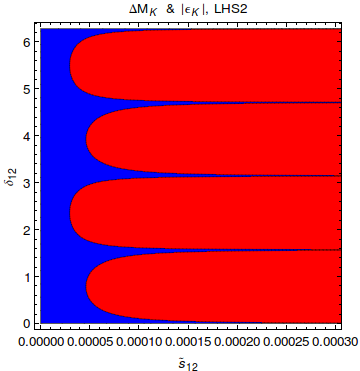}
\caption{\it  Ranges for $\Delta M_K$ (red region) and $\varepsilon_K$ (blue region) (LHS1: left, LHS2: right) for
$M_{Z^\prime}=1$ TeV  satisfying the bounds in Eq.~(\ref{C3}).
}\label{fig:oasesKLHS}~\\[-2mm]\hrule
\end{center}
\end{figure}

We note that
for each oasis with a given $\delta_{23}$ there is another oasis with $\delta_{23}$ shifted
by $180^\circ$ but the range for  $\tilde s_{23}$ is unchanged. This discrete
ambiguity results from the fact that $\Delta M_s$ and $S_{\psi\phi}$ are
governed by  $2\delta_{23}$. This ambiguity can be resolved by other 
observables discussed in the next steps.
The colour coding for the allowed oases, {\it blue} and {\it purple} 
for oasis with small and large $\delta_{23}$, respectively, will be useful 
in this context. 

The corresponding oases for 
$B^0_{d}-\bar B^0_{d}$  and $K^0-\bar K^0$ systems are shown in Figs.~\ref{fig:oasesBdLHS} and \ref{fig:oasesKLHS}, respectively. We note that now the results 
depend on whether LHS1 or LHS2 considered. Moreover in accordance with 
the quality of the constraints in (\ref{C1})-(\ref{C3}), the allowed oases 
in the $B^0_{d}-\bar B^0_{d}$ system  are smaller than in the $B^0_{s}-\bar B^0_{s}$ 
system, while they are larger in the $K^0-\bar K^0$ system. The colour coding for 
allowed oases in these figures will be useful to monitor the following steps 
in which rare $B_d$ and $K$ decays will be discussed and the distinction 
between the two allowed oases in each case will be possible.

\begin{figure}[!tb]
\begin{center}
\includegraphics[width=0.46\textwidth] {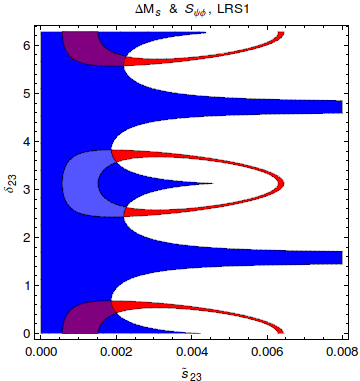}
\includegraphics[width=0.45\textwidth] {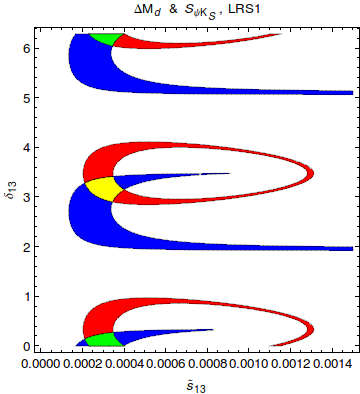}
\caption{\it   Ranges for $\Delta M_s$ and $S_{\psi \phi}$ (left) and  $\Delta M_d$  and $S_{\psi K_S}$ (right) for $M_{Z^\prime}=1$~TeV
in LRS1 satisfying the bounds
in Eq.~(\ref{C1}) and Eq.~(\ref{C2}).
}\label{fig:oasesBdBsLRS}~\\[-2mm]\hrule
\end{center}
\end{figure}

In \cite{Buras:2012jb} also the allowed oases in 
scenarios RHS, LRS and ALRS have been considered. We summarize here the main 
results and refer for details to this paper:
\begin{itemize}
\item
In the case of RHS scenarios the oases in the space of
parameters related to RH currents are precisely the same as those just
discussed for LHS scenarios, except that the parameters $\tilde s_{ij}$
and $\delta_{ij}$ parametrize now RH and not LH currents. 
Yet, as we will see in the next steps in the case of $\Delta F=1$ observables 
 some distinction between LH and RH currents will be possible.
\item
In the LRS scenarios 
NP contributions to $\Delta F=2$ observables are dominated by new LR
operators, whose contributions are enhanced through renormalization group
effects relative to LL and RR operators and in the case of $\varepsilon_K$
also through chirally enhanced hadronic matrix elements. Consequently
the oases will differ from the previous ones and typically the corresponding 
$\tilde s_{ij}$ will be smaller in order to obtain agreement with the data. 
The results can be found in Figs.~13-15 of \cite{Buras:2012jb}.
In order to understand these plots one should recall
 that the matrix element of the dominant $Q_1^{\rm LR}$
operator has the sign opposite to SM operators. Therefore, in the case 
of $B^0_{s,d}-\bar B^0_{s,d}$ systems
this operator
naturally suppresses $\Delta M_s$  and $\Delta M_d$ 
with the phase $\delta_{23}$ and  $\delta_{13}$
shifted down by roughly
$90^\circ$ relatively to the LHS scenarios. We illustrate this in Fig.~\ref{fig:oasesBdBsLRS} 
for LRS1 scenario. These plots should be compared with the one in 
 Fig.~\ref{fig:oasesBsLHS1} and in the left panel of
 Fig.~\ref{fig:oasesBdLHS}, respectively.
\item
 The allowed oases in ALR scenarios have the same phase structure as in 
 LHS scenarios because the contributions of the dominant LR operators
 have the same sign as SM contributions. Only the allowed values of 
 $\tilde s_{ij}$ are smaller because of larger hadronic matrix elements
 than in the LHS case.
\end{itemize}

The implications of these results for rare decays will be presented in the 
next steps.

\subsubsection{Tree-Level Scalar Exchanges}

\begin{figure}[!tb]
 \centering
\includegraphics[width = 0.35\textwidth]{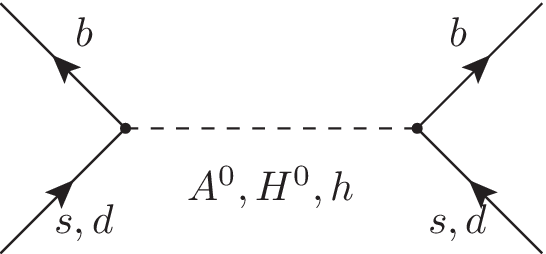}
\caption{\it Tree-level flavour-changing $A^0,H^0, h$ contribution to \bbd and \bbs mixing.}\label{fig:FD2}~\\[-2mm]\hrule
\end{figure}

We next turn our attention to  tree-level heavy scalar exchanges to 
$\Delta F=2$ transitions {(see Fig.~\ref{fig:FD2})}. Here one finds \cite{Buras:2012fs,Buras:2013rqa}
\allowdisplaybreaks{
  \begin{align}\begin{split}\label{M12H}
  \left(M_{12}^\star\right)_H =&
 -\frac{(\Delta_L^{sd}(H))^2}{2M_H^2}\left[C_1^\text{SLL}(\mu_H)
 \langle Q_1^\text{SLL}(\mu_H)\rangle +C_2^\text{SLL}(\mu_H)\langle Q_2^\text{SLL}(\mu_H)\rangle
 \right]\\
 &-\frac{(\Delta_R^{sd} (H))^2 } {2 M_H^2 }
 \left[C_1^\text{SRR}(\mu_H)\langle Q_1^\text{SRR}(\mu_H)\rangle 
 +C_2^\text{SRR}(\mu_H)\langle Q_2^\text{SRR}(\mu_H)\rangle \right]\\
 &-\frac{\Delta_L^{sd}(H)\Delta_R^{sd}(H)}{
 M_H^2} \left [ C_1^\text{LR}(\mu_H) \langle Q_1^\text{LR}(\mu_H)\rangle +
 C_2^\text{LR}(\mu_H) \langle Q_2^\text{LR}(\mu_H)\rangle \right]\,,\end{split}
 \end{align}}%
where including NLO QCD corrections \cite{Buras:2012fs}
\allowdisplaybreaks{
\begin{align}\label{equ:WilsonH}
C_1^\text{SLL}(\mu)= C_1^\text{SRR}(\mu)&=
1+\frac{\alpha_s}{4\pi}\left(-3\log\frac{M_H^2}{\mu^2}+\frac{9}{2}\right)\,,\\
\begin{split}
C_2^\text{SLL}(\mu) =
C_2^\text{SRR}(\mu) 
&=\frac{\alpha_s}{4\pi}\left(-\frac{1}{12}\log\frac{M_H^2}{\mu^2}+\frac{1}{8}
\right)\,,\end{split}\\
 C_1^\text{LR}(\mu)& =-\frac{3}{2}\frac{\alpha_s}{4\pi}\,,\\
C_2^\text{LR}(\mu) &=  1-\frac{\alpha_s}{4\pi}\,.
\end{align}}%

Note that the scalar contributions to $C_{1,2}^\text{LR}$ differ from the ones from gauge bosons. The relevant matrix elements can again be found 
in Tables~\ref{tab:Qi} and \ref{tab:Qi1} for tree-level heavy scalar and 
SM Higgs contributions. In the later case $M_H=M_h$ with $h$ standing for 
the SM Higgs.

For our qualitative discussion it is sufficient to set the Wilson coefficients 
to the LO values. Then
  \be\label{M12Hnew}
  \left(M_{12}^\star\right)_H =
-\left(\frac{(\Delta_L^{sd}(H))^2}{2M_H^2}
 +\frac{(\Delta_R^{sd}(H))^2}{2M_H^2}\right)
 \langle Q_1^\text{SLL}(\mu_H)\rangle 
 -\frac{\Delta_L^{sd}(H)\Delta_R^{sd}(H)}{
 M_H^2} \langle Q_2^\text{LR}(\mu_H)\rangle
 \ee
with analogous expressions for other meson systems. Now as seen in 
Table~\ref{tab:Qi}
model independently 
\be\label{PiH}
 \langle Q_1^\text{SLL}(\mu_H)\rangle <0, \quad   
\langle Q_2^\text{LR}(\mu_H)\rangle >0, \quad 
 | \langle Q_2^\text{LR}(\mu_H)\rangle|\gg |\langle Q_1^\text{VLL}(\mu_H)\rangle|,
\ee
which has an impact on the signs and size of the couplings $\Delta_{L,R}(H)$ 
if these contributions should remove the anomalies in the data.

Interestingly the signs of  $ \langle Q^a_i\rangle$ that are relevant in gauge 
boson and scalar cases are such that at the end it is not possible 
to distinguish these two cases on the basis of the signs of the couplings alone. On 
the other hand  $\langle Q^{\rm SLL}_i\rangle$ are absent in the case of 
gauge boson exchanges and $\Delta_{L,R}(Z^\prime)$  and $\Delta_{L,R}(H)$ are 
generally different from each other so that some distinction will be 
possible when other decays will be taken into account in later steps. 
Otherwise, the qualitative comments made in the context of tree-level 
gauge boson exchanges can be repeated in this case.

Indeed as analyzed recently in \cite{Buras:2013rqa} the phase structure 
of the allowed oases is identical to the one of the gauge boson case. 
As seen in the plots presented in this paper only the values of 
$\tilde s_{ij}$ change.

\boldmath
\subsubsection{Implications of $U(2)^3$ Symmetry}
\unboldmath

Possibly the simplest solution to the problems of various models with MFV 
is to reduce the flavour symmetry $U(3)^3$  to
$U(2)^3$  \cite{Barbieri:2011ci,Barbieri:2011fc,Barbieri:2012uh,Barbieri:2012bh,Crivellin:2011fb,Crivellin:2011sj,Crivellin:2008mq}.
 As pointed out in \cite{Buras:2012sd} in this case NP effects in $\varepsilon_K$ and $B^0_{s,d}-\bar B^0_{s,d}$ are not correlated with
each other so that the 
enhancement of $\varepsilon_K$ and suppression of $\Delta M_{s,d}$  
can be achieved if necessary in principle for
the values of  $\vcb$, $F_{B_s} \sqrt{\hat B_{B_s}}$ and $F_{B_d} \sqrt{\hat B_{B_d}}$ in
Table~\ref{tab:input}  or (\ref{oldf1}).

\begin{figure}[!tb]
 \centering
\includegraphics[width = 0.6\textwidth]{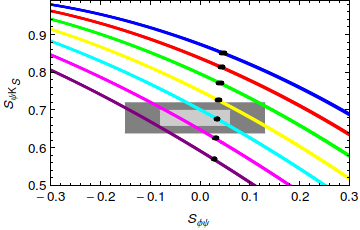}
\caption{ \it $S_{\psi K_S}$ vs. $S_{\psi \phi}$ in  models with 
$U(2)^3$ symmetry for different values of $\vub$ and $\gamma\in[58^\circ,78^\circ]$. From top to bottom: $\vub =$ $0.0046$ (blue), $0.0043$
(red), $0.0040$ (green),
$0.0037$ (yellow), $0.0034$ (cyan), $0.0031$ (magenta), $0.0028$ (purple). Light/dark gray: experimental $1\sigma/2\sigma$ region.
}\label{fig:SvsS}~\\[-2mm]\hrule
\end{figure}

 In particular,
\begin{itemize}
\item
NP effects in $\varepsilon_K$ are of CMFV type and $\varepsilon_K$ 
can only be enhanced.  But  because of the reduced flavour symmetry from 
$U(3)^3$ to $U(2)^3$ there is no correlation between
$\varepsilon_K$ and $\Delta M_{s,d}$ which was problematic for CMFV models.
 \item
In  $B^0_{s,d}-\bar B^0_{s,d}$ system, the ratio $\Delta M_s/\Delta M_d$ is 
equal to the one in the SM and in good agreement with the data. But 
in view of new CP-violating phases $\varphi_{B_d}$ and $\varphi_{B_s}$ even 
in the presence of only SM operators, $\Delta M_{s,d}$ can be suppressed. 
But the  $U(2)^3$ symmetry implies 
$\varphi_{B_d}=\varphi_{B_s}$ and consequently 
a  triple $S_{\psi K_S}-S_{\psi\phi}-|V_{ub}|$ correlation which constitutes 
an important test of this NP scenario \cite{Buras:2012sd}. We 
show this correlation in Fig.~\ref{fig:SvsS}  for  $\gamma$ between $58^\circ$ and $78^\circ$. Note that this correlation is independent of the values of 
 $F_{B_s} \sqrt{\hat B_{B_s}}$ and $F_{B_d} \sqrt{\hat B_{B_d}}$.
\item
As seen in this figure the important advantage  of $U(2)^3$ models over ${\rm 2HDM_{\overline{MFV}}}$ is that in the 
case of $S_{\psi\phi}$ being very small or even having opposite sign to 
SM prediction, this framework can survive with concrete prediction for 
$\vub$.
\end{itemize}

It is of interest to see how the parameter space in tree-level gauge boson 
or scalar $\Delta F=2$ transitions is further constrained when
the flavour $U(2)^3$ symmetry is imposed on the $Z^\prime$ or $H$ quark couplings. Indeed now
the observables in $B_d$ and $B_s$ systems are correlated with each other
due to the relations:
\be\label{equ:U23relation}
\frac{\tilde s_{13}}{\vtd}=\frac{\tilde s_{23}}{\vts}, \qquad
\delta_{13}-\delta_{23}=\beta-\beta_s.
\ee
Thus, once the allowed oases in the $B_d$ system are fixed, the oases in $B_s$
system are determined. Moreover, all observables in both systems are described
by only one real positive parameter and one phase, e.g. $({\tilde s}_{23},\delta_{23})$.

\begin{figure}[!tb]
\centering
\includegraphics[width = 0.45\textwidth]{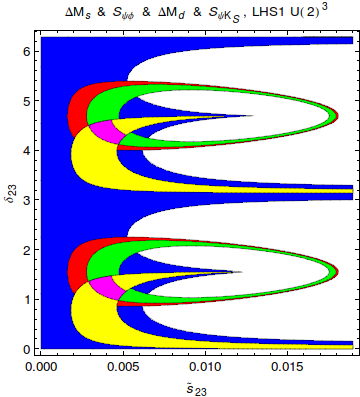}
\includegraphics[width = 0.45\textwidth]{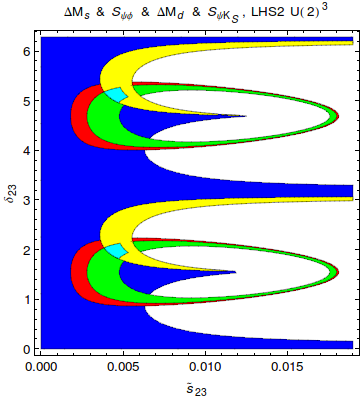}
\caption{\it Ranges for $\Delta M_s$ (red region), $S_{\psi \phi}$ (blue region), $\Delta M_d$ (green region) and $S_{\psi K_S}$
(yellow region) for $M_{Z^\prime}=1$~TeV in LHS1 (left) and LHS2 (right) in the $U(2)^3$ limit satisfying the bounds
in Eq.~(\ref{C1}) and ~(\ref{C2}). The overlap region of LHS1 (LHS2) is shown in magenta (cyan). }
 \label{fig:oasesU2}~\\[-2mm]\hrule
\end{figure}

The impact of $U(2)^3$ symmetry on tree level FCNCs due to gauge boson and 
scalar exchanges has been analyzed in \cite{Buras:2012jb} and \cite{Buras:2013rqa}, respectively. Again the phase structure in both cases is
the same. 
Fig.~\ref{fig:oasesU2} results from the combination of Figs.~\ref{fig:oasesBsLHS1} and~\ref{fig:oasesBdLHS} using the $U(2)^3$ symmetry
relations in (\ref{equ:U23relation}). 
We observe that in particular the 
$({\tilde s}_{23},\delta_{23})$ oases are significantly reduced.
Moreover the fact that
the results in the $B_d$ system depend on whether LHS1 or LHS2 is considered
is now transfered through the relations in (\ref{equ:U23relation}) into
the $B_s$ system. This is clearly seen in Fig.~\ref{fig:oasesU2}, in particular
the final oases (cyan) in LHS2 are  smaller than in LHS1 (magenta) due to the required
shift of $S_{\psi K_S}$. The corresponding results in the scalar case can be 
found in Fig.~15 of \cite{Buras:2013rqa}. It will be interesting to see 
what is the impact of  the $U(2)^3$ symmetry on rare decays in the 
next steps.

\boldmath
\subsection{Step 4: $\mathcal{B}(B_{s,d}\to\mu^+\mu^-)$ and 
$\mathcal{B}(B_{s,d}\to\tau^+\tau^-)$}
\unboldmath

\subsubsection{Preliminaries}
We now move to consider two superstars of rare $B$ decays: the 
decays $B_{s,d}\to\mu^+\mu^-$. We will also discuss $B_{s,d}\to\tau^+\tau^-$ which could become superstars in the future.  The particular
interest in 
 $B_{s,d}\to\mu^+\mu^-$ is related to the fact that in the SM their branching ratios  are not 
only loop and GIM suppressed as  other rare decays in the SM. As
 the final state is purely leptonic and the initial state is 
a pseudoscalar the decays in question are strongly helicity suppressed in view of the smallness of $m_\mu$ and equally importantly do not receive photon-mediated one-loop 
contributions.  As all these properties can be violated beyond the SM, these 
two decays are particularly suited for searching for NP being in addition 
theoretically very clean.

In the SM and in several of its extensions  $\mathcal{B}(B_{s}\to\mu^+\mu^-)$ 
is found in the ballpark of $(2-6)\cdot 10^{-9}$.
As several model studies show this is the case of models in which 
these decays proceed through $Z$-penguin diagrams and tree-level neutral 
gauge boson exchanges. Larger values 
can be obtained in the presence of  neutral heavy scalar and pseudoscalar  exchanges in 2HDM 
models and Supersymmetry. Here these decays are governed by scalar and 
pseudoscalar   
penguins
when the value of $\tan\beta$ is large. In certain models contributions 
from  tree-level scalars and pseudoscalars  can arise already at the fundamental level.
Therefore
a discovery of 
$\mathcal{B}(B_{s}\to\mu^+\mu^-)$ at $\ord (10^{-8})$ would be a clear signal of
NP, possibly related to such scalar and pseudoscalar exchanges 
\cite{Altmannshofer:2011gn}.
Unfortunately, as we will see below, the most recent data from   LHCb and  CMS tell us that the nature does 
not allow us for a clear distinction  between scalar, pseudoscalar and gauge boson contributions at least on the basis of the 
$\mathcal{B}(B_{s}\to\mu^+\mu^-)$  alone. 
Either other observables related  to the time-dependent rate of this decay
have to be studied \cite{Buras:2013uqa} or/and correlations with other 
observables have to be investigated.  We will see explicit examples below.
We refer also to \cite{Guadagnoli:2013mru,Altmannshofer:2013oia} where 
various virtues of these decays have been reviewed.

In order to discuss these issues 
we have to present the fundamental effective Hamiltonian relevant for 
these decays and  other
$b\to s\ell^+\ell^-$ transitions, like $B\to K^*\ell^+\ell^-$, $B\to K\ell^+\ell^-$ and $B\to X_s\ell^+\ell^-$, which we will consider in Step 7.
\boldmath
\subsubsection{Basic Formulae}\label{sec:bqll}
\unboldmath
There are different conventions for operators \cite{Isidori:2002qe,Bobeth:2001jm,Dedes:2008iw} relevant for $b\to s\ell^+\ell^-$ transitions and one has to be careful when using them along with the expressions for the branching ratios present in the literature. The effective Hamiltonian used here and in several 
recent papers is given as follows:
\be\label{eq:Heffqll}
 \Heff(b\to s \ell\bar\ell)
= \Heff(b\to s\gamma)
-  \frac{4 G_{\rm F}}{\sqrt{2}} \frac{\alpha}{4\pi}V_{ts}^* V_{tb} \sum_{i = 9,10,S,P} [C_i(\mu)Q_i(\mu)+C^\prime_i(\mu)Q^\prime_i(\mu)]
\end{equation}
where
\begin{subequations}
\begin{align}
Q_9 & = (\bar s\gamma_\mu P_L b)(\bar \ell\gamma^\mu\ell),&  &Q_9^\prime  =  (\bar s\gamma_\mu P_R b)(\bar \ell\gamma^\mu\ell),\\
Q_{10} & = (\bar s\gamma_\mu P_L b)(\bar \ell\gamma^\mu\gamma_5\ell),&  &Q_{10}^\prime  =  (\bar s\gamma_\mu P_R b)(\bar \ell\gamma^\mu\gamma_5\ell),\\
Q_S &= m_b(\bar s P_R b)(\bar \ell\ell),& & Q_S^\prime = m_b(\bar s P_L b)(\bar \ell\ell),\\
Q_P & = m_b(\bar s P_R b)(\bar \ell\gamma_5\ell),& & Q_P^\prime = m_b(\bar s P_L b)(\bar \ell\gamma_5\ell).
\end{align}
\end{subequations}
Here $\Heff(b\to s\gamma)$ stands for the effective Hamiltonian for the
$b\to s\gamma$ transition that involves the dipole operators (see Step 6).
 While we do not show explicitly the four-quark operators in (\ref{eq:Heffqll}) 
they are very important for decays considered in this step, in particular 
as far as QCD and electroweak corrections are concerned.

One should note the
difference of ordering of flavours relatively to $\Delta F=2$ operators 
considered in the previous step. This will play a role as we discuss below  (for example the relations of the couplings
in~(\ref{dictionary}) are useful when comparing $\Delta F = 1$ and $\Delta F = 2$ transitions). 
We neglect effects proportional to $m_s$  but keep $m_s$ and $m_d$ different from zero when they are shown explicitly. Analogous operators govern the 
$b\to d\ell^+\ell^-$ transitions, in particular the $B_d\to\mu^+\mu^-$ decay.

Concentrating first on $B_s\to\mu^+\mu^-$, there are three observables 
which can be used to search for NP in these decays. These are 
\be\label{trio}
\overline{\mathcal{B}}(B_{s}\to\mu^+\mu^-), \qquad \mathcal{A}^{\mu\mu}_{\Delta\Gamma}, \qquad S^s_{\mu\mu}.
\ee
Here $\overline{\mathcal{B}}(B_{s}\to\mu^+\mu^-)$ is the usual branching ratio 
which includes $\Delta\Gamma_s$ effects pointed out in 
in \cite{DescotesGenon:2011pb,deBruyn:2012wj,deBruyn:2012wk}. Following \cite{Buras:2013uqa} we 
will denote this branching ratio with a {\it bar} while the one without these effects without it.
These two branching ratios are related through
\cite{DescotesGenon:2011pb,deBruyn:2012wj,deBruyn:2012wk}
\be
\label{Fleischer1}
\mathcal{B}(B_{s}\to\mu^+\mu^-) =
r(y_s)~\overline{\mathcal{B}}(B_{s}\to\mu^+\mu^-),
\ee
where 
\be\label{rys}
r(y_s)\equiv\frac{1-y_s^2}{1+\mathcal{A}^{\mu^+\mu^-}_{\Delta\Gamma} y_s}.
\ee
with \cite{Amhis:2012bh} 
\begin{equation}\label{defys}
	y_s\equiv\tau_{B_s}\frac{\Delta\Gamma_s}{2}
=0.062\pm0.009.
\end{equation}

The observables $\mathcal{A}^{\mu\mu}_{\Delta\Gamma}$ and $S^s_{\mu\mu}$ 
can only be measured through time-dependent studies 
and appear in the time-dependent rate asymmetry as follows
\begin{align}
\frac{\Gamma(B^0_s(t)\to \mu^+\mu^-)-
\Gamma(\bar B^0_s(t)\to \mu^+\mu^-)}{\Gamma(B^0_s(t)\to \mu^+\mu^-)+
\Gamma(\bar B^0_s(t)\to \mu^+\mu^-)}
=\frac{ S_{\mu\mu}^s\sin(\Delta M_st)}{\cosh(y_st/ \tau_{B_s}) + 
{\mathcal{ A}}^{\mu\mu}_{\Delta\Gamma} \sinh(y_st/ \tau_{B_s})}.
\end{align}
$\mathcal{A}^{\mu\mu}_{\Delta\Gamma}$ can be extracted from the untagged data sample,
namely from the measurement of the effective lifetime, for which no distinction is made between initially present $B^0_s$ or 
$\bar B^0_s$ mesons. If tagging information is included, requiring the distinction  between initially present $B^0_s$ or $\bar B^0_s$ mesons, 
a CP-violating asymmetry $S^s_{\mu\mu}$ can also be measured.
Presently only $\overline{\mathcal{B}}(B_{s}\to\mu^+\mu^-)$ is known experimentally but once
$\mathcal{A}^{\mu\mu}_{\Delta\Gamma}$ will be extracted from time-dependent measurements, we will be able 
to obtain $\mathcal{B}(B_{s}\to\mu^+\mu^-)$ directly from experiment as well.
As emphasized and demonstrated in \cite{Buras:2013uqa}  $\mathcal{A}^{\mu\mu}_{\Delta\Gamma}$ and $S^s_{\mu\mu}$  provide additional information about possible NP 
which cannot be obtained on the basis of the branching ratio alone. In 
order to present the results for the trio in (\ref{trio}) in various 
models we have to express these observables in terms of the Wilson coefficients 
in the effective Hamiltonian in (\ref{eq:Heffqll}). 

To this end
one  introduces first
\begin{align}
P&\equiv \frac{C_{10}-C_{10}^\prime}{C_{10}^{\rm SM}}+
\frac{m^2_{B_s}}{2m_\mu}\frac{m_b}{m_b+m_s} \frac{C_P-C_P^\prime}{C_{10}^{\rm SM}}
\equiv |P|e^{i\varphi_P}\label{PP}\\
S&\equiv \sqrt{1-\frac{4m_\mu^2}{m_{B_s}^2}}\frac{m^2_{B_s}}{2m_\mu}
\frac{m_b}{m_b+m_s}
\frac{C_S-C_S^\prime}{C_{10}^{\rm SM}}
\equiv |S|e^{i\varphi_S},
\label{SS}
\end{align}
which carry the full information about dynamics in the decay. However, 
due to effects from $B_s^0-\bar B_s^0$ mixing, represented here by $y_s$,
also the new phase $\varphi_{B_s}$ in
$B_s^0-\bar B_s^0$ mixing will enter the expressions below.

One finds then three fundamental formulae \cite{deBruyn:2012wk,Fleischer:2012fy,Buras:2013uqa}
\begin{align}\label{Rdef}
	   &  \frac{\overline{\mathcal{B}}(B_{s}\to\mu^+\mu^-)}{\overline{\mathcal{B}}(B_{s}\to\mu^+\mu^-)_{\rm SM}}
	= \left[\frac{1+{\cal A}^{\mu\mu}_{\Delta\Gamma}\,y_s}{1+y_s} \right] \times (|P|^2 + |S|^2)\notag\\
	&= \left[\frac{1+y_s\cos(2\varphi_P-2\varphi_{B_s})}{1+y_s} \right] |P|^2 + \left[\frac{1-y_s\cos(2\varphi_S-2\varphi_{B_s})}{1+y_s} \right] |S|^2,
\end{align}
\begin{align}
{\cal A}^{\mu\mu}_{\Delta\Gamma} &= \frac{|P|^2\cos(2\varphi_P-2\varphi_{B_s}) - |S|^2\cos(2\varphi_S-2\varphi_{B_s})}{|P|^2 + |S|^2},\label{ADG}\\
	S^s_{\mu\mu}
	&=\frac{|P|^2\sin(2\varphi_P-2\varphi_{B_s})-|S|^2\sin(2\varphi_S-2\varphi_{B_s})}{|P|^2+|S|^2}.
	\label{Ssmu}
\end{align}
where 
\begin{align}
 &\overline{\mathcal{B}}(B_s\to\mu^+\mu^-)_\text{SM} = \frac{1}{1-y_s}\mathcal{B}(B_s\to\mu^+\mu^-)_\text{SM}\,,\\
&\mathcal{B}(B_s\to\mu^+\mu^-)_\text{SM} = \tau_{B_s}\frac{G_F^2}{\pi}\left(\frac{\alpha}{4\pi \sin^2\theta_W}\right)^2F_{B_s}^2m_\mu^2 
m_{B_s}\sqrt{1-\frac{4m_\mu^2}{m_{B_s}^2}}\left|V_{tb}^*V_{ts}\right|^2\eta_\text{eff}^2Y_0(x_t)^2\,
\end{align}
with $\eta_\text{eff}$ and $Y_0(x_t)$ given below.

It follows that in any model the branching ratio without $\Delta\Gamma_s$ 
effect is related to the corresponding SM branching ratio through
\be\label{THBr}
\mathcal{B}(B_{s}\to\mu^+\mu^-)=\mathcal{B}(B_{s}\to\mu^+\mu^-)_{\rm SM}(|P|^2 + |S|^2),
\ee
 which is obtained from (\ref{Rdef}) by setting $y_s=0$. 

Finally, all the formulae given above can be used for $B_d\to\mu^+\mu^-$ with 
$s$ replaced by $d$ and $y_d\approx 0$ so that in this case there is no
distinction between $\overline{\mathcal{B}}(B_{d}\to\mu^+\mu^-)$ and 
${\mathcal{B}}(B_{s}\to\mu^+\mu^-)$.  Still the CP asymmetry $S^d_{\mu\mu}$ 
can be considered, although measuring it would be a heroic effort.

These formulae are very general and can be used to study these 
observables model independently using as variables
\be\label{unknowns}
|P|,\qquad \varphi_P, \qquad |S|, \qquad \varphi_S. 
\ee
Such an analysis has been performed  in \cite{Buras:2013uqa}.
The classification of popular NP in various scenarios characterized  by 
the vanishing or non-vanishing values of the variables in (\ref{unknowns}) 
and of  the new phase $\varphi_{B_s}$ in  $B_s^0-\bar B_s^0$ mixing 
should help in  monitoring the improved data in the future. While some of 
the results of this paper and also of related analysis of tree-level 
gauge boson and scalar contributions in \cite{Buras:2013rqa} will be 
presented below, we collect already in  Table~\ref{tab:Models} the properties of the selected models discussed in these two papers with
respect to the basic phenomenological parameters listed in (\ref{unknowns})
and the classes defined in \cite{Buras:2013uqa} they belong to.

\begin{table}[t]
\centering
\begin{tabular}{|c||c||c|c|c|c|c|}
\hline
 Model & Scenario & $|P|$ & $\varphi_P$ & $|S|$ &  $\varphi_S$   &  $\varphi_{B_s}$\\
\hline
\hline
  \parbox[0pt][1.6em][c]{0cm}{} CMFV & A & $|P|$  &  $0$ & $0$ & $0$  & $0$ \\
 \parbox[0pt][1.6em][c]{0cm}{} MFV & D & $|P|$  &  $0$ & $|S|$ & $0$  & $0$ \\
 \parbox[0pt][1.6em][c]{0cm}{} LHT,~4G,~RSc,~$Z'$ & A & $|P|$  &  $\varphi_P$ & $0$ & $0$  &  $\varphi_{B_s}$ \\
 \parbox[0pt][1.6em][c]{0cm}{} 2HDM (Decoupling) & C & $|1\mp S|$  &   ${\rm arg}(1\mp S)$ & $|S|$
 & $\varphi_S$  &  $\varphi_{B_s}$ \\
\parbox[0pt][1.6em][c]{0cm}{} 2HDM (A Dominance) & A & $|P|$  &   $\varphi_P$ & $0$
 & $0$  &  $\varphi_{B_s}$  \\
\parbox[0pt][1.6em][c]{0cm}{} 2HDM (H Dominance) & B & $1$  &   $0$ & $|S|$
 & $\varphi_S$  &  $\varphi_{B_s}$ \\
\hline
\end{tabular}
\caption{\it General structure of basic variables in different NP models. The last three cases apply also to the MSSM. From
\cite{Buras:2013uqa}.}
\label{tab:Models}~\\[-2mm]\hrule
\end{table}

After these general introduction we will discuss the results in the SM 
and its simplest extensions.

\subsubsection{Standard Model Results and the Data}
In the SM $B_{s,d}\to\mu^+\mu^-$ are governed by $Z^0$-penguin diagrams 
and $\Delta F=1$ box diagrams which depend on the top-quark mass. The 
internal charm contribution can be safely neglected.

The only relevant Wilson coefficients in the SM are $C_9$ and $C_{10}$  
given by
\begin{align}\label{C9SM}
 \sin^2\theta_W C^{\rm SM}_9 &=\sin^2\theta_W P_0^{\rm NDR}+ [\eta_{\rm eff} Y_0(x_t)-4\sin^2\theta_W Z_0(x_t)],\\
   \sin^2\theta_W C^{\rm SM}_{10} &= -\eta_{\rm eff} Y_0(x_t) \label{Yeff}
 \end{align}
with all the entries given in \cite{Buras:2012jb,Buras:2013qja} except
for $\eta_{\rm eff}$ which is discussed below. With $m_s\ll m_b$ we have $C_9^\prime=C_{10}^\prime=0$.

Here  $Y_0(x_t)$ and  $Z_0(x_t)$ are SM one-loop functions given by
\be\label{YSM}
Y_0(x_t)=\frac{x_t}{8}\left(\frac{x_t-4}{x_t-1} + \frac{3 x_t \log x_t}{(x_t-1)^2}\right),
\ee
\begin{align}\label{ZSM}
  Z_0 (x) & = -\frac{1}{9} \log x + \frac{18 x^4 - 163 x^3 + 259 x^2 - 108 x}{144 (x-1)^3} + \frac{32 x^4 - 38 x^3 - 15 x^2 + 18 x}{72
(x-1)^4}\log x \,.
\end{align}
We have then
\be\label{CSM910}
C^{\rm SM}_9\approx 4.1, \qquad C^{\rm SM}_{10}\approx -4.1~.
\ee
The coefficient $\eta_{\rm eff}$ was until recently denoted by  $\eta_{Y}$ and 
included only NLO QCD corrections. For $m_t=m_t(m_t)$ one had $\eta_{Y}=1.012$
\cite{Buchalla:1998ba,Misiak:1999yg}.

Over several years  electroweak 
corrections to the branching ratios have been calculated \cite{Buchalla:1997kz,Bobeth:2003at,Huber:2005ig,Misiak:2011bf} but they were incomplete implying
dependence on renormalization scheme used for electroweak 
parameters as analysed in  detail in \cite{Buras:2012ru}.  
Recently complete NLO electroweak corrections \cite{Bobeth:2013tba} and QCD corrections up to NNLO \cite{Hermann:2013kca} have been calculated.
The inclusion of these new
higher order corrections that were missing until now reduced significantly
various scale uncertainties so that non-parametric uncertainties in both branching ratios are below $2\%$.

The calculations performed in \cite{Bobeth:2013tba,Hermann:2013kca}
are very involved and in analogy to the QCD factors, like $\eta_B$ and $\eta_{1-3}$ in $\Delta F=2$ processes, we find it 
 useful to include all QCD and electroweak 
corrections into  $\eta_{\rm eff}$ introduced in (\ref{Yeff}) that without 
these corrections would be equal to unity.
Inspecting the analytic formulae in \cite{Bobeth:2013uxa} one finds then 
\cite{Buras:2013dea}
\be\label{etaeff}
\eta_{\rm eff}= 0.9882\pm 0.0024~.
\ee

The small departure of  $\eta_{\rm eff}$ from unity  was already
anticipated in \cite{Misiak:2011bf,Buras:2012ru} but only the calculations in
\cite{Bobeth:2013uxa,Bobeth:2013tba,Hermann:2013kca} could put these expectations and conjectures on firm footing.
Indeed, in order to end up with such a simple result it was crucial to
perform such  involved calculations as these small corrections are only
valid for particular definitions of the top-quark mass and of other electroweak
parameters involved. In particular one has to use in 
$Y_0(x_t)$ the $\overline{\rm MS}$-renormalized top-quark mass $m_t(m_t)$ with respect to QCD but on-shell with respect to electroweak interactions. This means
$m_t(m_t)=163.5\gev$  as calculated in  \cite{Bobeth:2013uxa}. Moreover, in 
using (\ref{etaeff}) to calculate observables like branching ratios it is 
important to have the same normalization of effective Hamiltonian as in 
the latter paper. There this normalization is expressed in terms of $G_F$ 
and $M_W$ only.  Needless to say one can also use directly the formulae in  \cite{Bobeth:2013uxa}. 

In the present review we follow the normalization of effective Hamiltonian in 
\cite{Buras:1998raa} which uses $G_F$, $\alpha(M_Z)$ and $\sin^2\theta_W$ 
and in order to be consistent with the calculation in \cite{Bobeth:2013uxa} 
our $\eta_{\rm eff}= 0.991$ with $m_t(m_t)$ unchanged \cite{Buras:2013dea}. Interestingly also in the case of $\kpn$ and $\klpn$ the analog 
of $\eta_{\rm eff}$, multiplying this time $X_0(x_t)$, is found  with the  normalizations of effective Hamiltonian in \cite{Buras:1998raa}  
and definition of $m_t$ as given above to be within $1\%$ from unity \cite{Brod:2010hi}. { It should be remarked 
that presently only in the case of the $B_{s,d}\to \mu^+\mu^-$ decays discussed here and  $\kpn$ and $\klpn$ decays considered in Step 8 one has to take such a care about the definition of  $m_t$ with respect to electroweak corrections as 
in most cases such corrections are not known or hadronic uncertainties are 
too large so that the value $m_t(m_t)=163.0\gev$  in Table~\ref{tab:input} used by us otherwise can easily be defended.}

In view of still significant parametric uncertainties
it is useful to show the dependence of the branching ratios on various input parameters involved. Such formulae have been already presented in
\cite{Buras:2012ru,Buras:2013uqa} and have been recently updated 
by the authors of \cite{Bobeth:2013tba} and  \cite{Hermann:2013kca}. They find  
 \cite{Bobeth:2013uxa}
\begin{equation}
 \overline{\mathcal{B}}(B_{s}\to\mu^+\mu^-)_{\rm SM} = (3.65\pm0.06)\times 10^{-9} \left(\frac{m_t(m_t)}{163.5 \gev}\right)^{3.02}\left(\frac{\alpha_s(M_Z)}{0.1184}\right)^{0.032} R_s
\label{BRtheoRpar}
\ee
where
\be
\label{Rs}
R_s=
\left(\frac{F_{B_s}}{227.7\mev}\right)^2
\left(\frac{\tau_{B_s}}{1.516 {\rm ps}}\right)\left(\frac{0.938}{r(y_s)}\right)
\left|\frac{V_{tb}^*V_{ts}}{0.0415}\right|^2,
\end{equation}
where precise definition of $m_t(m_t)$ is given below. 
  We caution the reader that the parametric expression in (\ref{Rs}), 
which 
is based on the results in \cite{Bobeth:2013uxa}, differs slightly from the one 
presented by these authors and consequently the quoted uncertainty is only 
an approximation but a very good one.

Proceeding in the same manner with $B_d\to\mu^+\mu^-$ one finds  \cite{Bobeth:2013uxa}
\begin{equation}
\mathcal{B}(B_{d}\to\mu^+\mu^-)_{\rm SM} = (1.06\pm 0.02)\times 10^{-10} 
\left(\frac{m_t(m_t)}{163.5 \gev}\right)^{3.02}\left(\frac{\alpha_s(M_Z)}{0.1184}\right)^{0.032} R_d
\label{BRtheoRpard}
\ee
where
\be
R_d=\left(\frac{F_{B_d}}{190.5\mev}\right)^2
\left(\frac{\tau_{B_d}}{1.519 {\rm ps}}\right)\left|\frac{V_{tb}^*V_{td}}{0.0088}\right|^2.
\end{equation}
We emphasize that the overall factors in (\ref{BRtheoRpar}) and (\ref{BRtheoRpard}) include all the corrections calculated in   \cite{Bobeth:2013tba} and  \cite{Hermann:2013kca} and 
we do not expect that these numbers will change in the near future.  On the 
other hand the central value of $\vts$ in (\ref{Rs}) corresponds to the 
inclusive determination of $\vcb\approx 0.0425$. With $\vcb\approx 0.039$, 
as extracted from exclusive decays, one would find the central value for 
the branching ratio in question to be rather close to $3.0\times 10^{-9}$.

Concerning the other two observables in (\ref{trio}), with
$P=1$ and $S=0$ in the SM we have
\be
\mathcal{A}^{\mu\mu}_{\Delta\Gamma}=1, \quad S_{\mu\mu}^s=0,
\quad r(y_s)=0.938\pm0.009 \qquad  ({\rm SM}). 
\ee

Taking the parametric uncertainties into account one  finds then  \cite{Bobeth:2013uxa}
\be\label{LHCb2}
 \overline{\mathcal{B}}(B_{s}\to\mu^+\mu^-)_{\rm SM}=(3.65\pm0.23)\times 10^{-9},\quad
\overline{\mathcal{B}}(B_{s}\to\mu^+\mu^-)_{\rm exp} = (2.9\pm0.7) \times 10^{-9}, 
\ee
\be\label{LHCb3}
\mathcal{B}(B_{d}\to\mu^+\mu^-)_{\rm SM}=(1.06\pm0.09)\times 10^{-10}, \quad
\mathcal{B}(B_{d}\to\mu^+\mu^-)_{\rm exp} =\left(3.6^{+1.6}_{-1.4}\right)\times 10^{-10}, \quad
\ee
where we have also shown the most recent average of the results from LHCb and CMS
\cite{Aaij:2013aka,Chatrchyan:2013bka,CMS-PAS-BPH-13-007}. 
The agreement of  the SM prediction with the data for $B_s\to\mu^+\mu^-$
in (\ref{LHCb2}) is remarkable, although the rather large 
experimental error still allows for sizable NP contributions. In  
$B_d\to\mu^+\mu^-$ much bigger room for NP contributions is left.

We close our discussion of the SM with the correlations of 
$\mathcal{B}(B_q\to\mu^+\mu^-)$ and   $\Delta M_{s,d}$ that are 
free from $F_{B_q}$  and the $|V_{tq}|$ dependence \cite{Buras:2003td}
\begin{align}\label{NonDirect}
& \mathcal{B}(B_q\to\mu^+\mu^-) = C \frac{\tau_{B_q}}{\hat B_q}\frac{(\eta_{\rm eff} Y_0(x_t))^2}{S_0(x_t)}\Delta M_q,\\
&\text{with}\quad C = 6\pi \frac{1}{\eta_B^2}\left(\frac{\alpha}{4\pi \sin^2\theta_W}\right)^2\frac{m_\mu^2}{M_W^2}= 4.291\cdot 10^{-10},
\end{align}
where ${\hat B_q}$, known from Step 2, enters linearly as opposed to quadratic 
dependence on  $F_{B_q}$. 

The results for branching ratios obtained in this 
manner have presently comparable  errors to the ones  obtained by direct calculations of branching ratios with their values close to the ones quoted above. 
Of interest are also the  relations (\ref{CMFV5}) and 
(\ref{CMFV6}) with $r(\mu^+\mu^-)=1$ and $r=1$ 
which hopefully will be tested one day.

Let us next see what the simple models introduced in Section~\ref{sec:2} can tell us about 
these decays.

\subsubsection{CMFV}

In this class of models there are no new CP-violating phases 
and no new operators. Therefore all the formulae of the SM given until now remain 
valid except for the following changes:
\begin{itemize}
\item
The master functions $S_0(x_t)$ and $Y_0(x_t)$ are replaced by new functions 
$S(v)$ and $Y(v)$, respectively. Here $v$ denotes all parameters present 
in a given CMFV model, that is coupling and masses of new particles including 
those of the SM.
\item
QCD corrections to $B_{s,d}\to\mu^+\mu^-$, represented by $\eta_Y$,
are expected in this class of models to be small and this is also expected 
for electroweak corrections. On the other hand $\eta_B$ could be
visibly different in these models if the mases of particles involved are larger 
than $1\tev$. Yet, due to relatively small anomalous dimension of the 
$(V-A)\times(V-A)$ operator this change is much smaller than in the case 
of LR operators encountered in more complicated models. Therefore in view
of new parameters present in $S(v)$, it is a good idea to use first just the 
SM value for $\eta_B$.
\end{itemize}

A more precise treatment would be to make the following replacement:
\be
S_0(x_t) \to S_0(x_t)+\frac{\eta_B^{\rm NP}}{\eta_B^{\rm SM}}\Delta S_0(v),
\ee
where $\eta_B^{\rm SM}$ equals $\eta_B$ in previous expressions and 
$\Delta S_0(v)$ is the modification of the loop functions by NP contributions.   The 
new $\eta_B^{\rm NP}$ can easily be calculated 
in the LO if the NP scale is known. Then the sign of the anomalous dimension
of the operator $Q_1^{\rm VLL}$ implies $\eta_B^{\rm NP}\le\eta_B^{\rm SM}$
for NP scales 
larger than the electroweak scale.

\begin{figure}[!tb]
 \centering
\includegraphics[width = 0.6\textwidth]{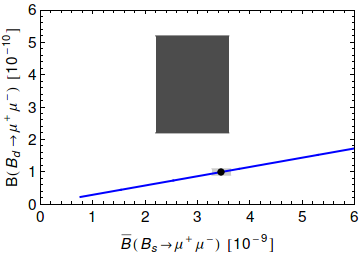}
\caption{\it $\mathcal{B}(B_d\to\mu^+\mu^-)$ vs $\overline{\mathcal{B}}(B_s\to\mu^+\mu^-)$ 
  in models 
with CMFV. SM is represented by the light grey area with black dot. Dark gray
 region: Combined exp 1$\sigma$
 range
 $\overline{\mathcal{B}}(B_s\to\mu^+\mu^-) = (2.9\pm0.7)\cdot 10^{-9}$ and $\mathcal{B}(B_d\to\mu^+\mu^-) = (3.6^{+1.6}_{-1.4})\cdot
10^{-10}$.}\label{fig:BdvsBs}~\\[-2mm]\hrule
\end{figure}

The branching ratios for $B_{s,d}\to\mu^+\mu^-$ will now be modified with 
respect to the SM but as seen in Fig.~\ref{fig:BdvsBs}  due to  relations in (\ref{CMFV5}) and (\ref{CMFV6}) with $r(\mu^+\mu^-)=1$ and $r=1$ 
strong correlation between these two branching ratios is predicted.  In Fig.~\ref{fig:BdvsBs} we included $\Delta\Gamma_s$ effects in 
$\mathcal{B}(B_{s}\to\mu^+\mu^-)$.

The calculations simplify considerably if CKM factors are fixed in Step 1. 
Then independently of $q$ we simply have 
\be\label{CMFV/SM}
\frac{\mathcal{B}(B_q\to\mu^+\mu^-)}{\mathcal{B}(B_q\to\mu^+\mu^-)^{\rm SM}}
=\left(\frac{Y(v)}{Y_0(x_t)}\right)^2
\ee
and {consequently 
\be\label{CMFVBS}
\left(\frac{\overline{\mathcal{B}}(B_s\to\mu^+\mu^-)}{\mathcal{B}(B_d\to\mu^+\mu^-)}\right)_{{\rm CMFV}}=
\left(\frac{\overline{\mathcal{B}}(B_s\to\mu^+\mu^-)}{\mathcal{B}(B_d\to\mu^+\mu^-)}\right)_{\rm SM} =34.4\pm 3.6,
\ee
where we have used the SM values in (\ref{LHCb2}) and (\ref{LHCb3}). Using 
(\ref{CMFV6}) with $r=1$ we would find $33.9\pm0.8$.   
Using (\ref{CMFVBS}) together with the measurement of
 $\overline{\mathcal{B}}(B_{s}\to\mu^+\mu^-)$  (\ref{LHCb2}) 
implies in turn in the context of these models 
\be\label{boundMFV1}
\mathcal{B}(B_{d}\to\mu^+\mu^-)=(0.84\pm 0.19)\times 10^{-10}, \qquad ({\rm CMFV}),
\ee
which is well be below the data in (\ref{LHCb3}). This  could then be 
an indication for new sources of flavour violation.  In fact as seen in 
Fig.~\ref{fig:BdvsBs} the present data differ from CMFV correlation between 
these two branching ratios by roughly 
$2\sigma$ but we have to wait for new improved data in order to claim NP at work.  Still it will be interesting to see what kind of NP could
bring the 
theory 
close to the present experimental central values for the branching ratios in 
this figure.

\boldmath
\subsubsection{${\rm 2HDM_{\overline{MFV}}}$}
\unboldmath

In ${\rm 2HDM_{\overline{MFV}}}$  scalar and pseudoscalar penguin diagrams generate new scalar and pseudoscalar
operators that can even dominate the decays 
$B_{s,d}\to\mu^+\mu^-$ at sufficiently high value of $\tan\beta$. However, 
due to recent LHCb  and CMS results such large enhancements are not possible 
for $B_{s}\to\mu^+\mu^-$ anymore and within this model the same applies 
to $B_{d}\to\mu^+\mu^-$. Indeed within an excellent approximation 
we have then similarly to (\ref{CMFVBS})
\cite{Buras:2010zm}
\be\label{MAIN3}
\left(\frac{\mathcal{B}(B_s\to\mu^+\mu^-)}{\mathcal{B}(B_d\to\mu^+\mu^-)}\right)_{{\rm 2HDM_{\overline{MFV}}}}=
\left(\frac{\mathcal{B}(B_s\to\mu^+\mu^-)}{\mathcal{B}(B_d\to\mu^+\mu^-)}\right)_{\rm SM}.
\ee
Combined with (\ref{MAIN2}) we then conclude that also 
(\ref{CMFV6}) with $r=1$ is well satisfied 
in this model. 
However, 
while the ratios in (\ref{MAIN2}) and (\ref{MAIN3}) are the same in 
${\rm 2HDM_{\overline{MFV}}}$ and the SM, the individual $\Delta M_{s,d}$ 
and $\mathcal{B}(B_{s,d}\to\mu^+\mu^-)$ can differ in these models.  
 Still the range for $\mathcal{B}(B_d\to\mu^+\mu^-)$ in (\ref{boundMFV1}) 
also applies and constitutes an important test of this model.

Finally in the limit $C_S=-C_P$  lower bounds on the two branching ratios can be derived ~\cite{Logan:2000iv,Buras:2013uqa}:
\be\label{LOWERB}
\mathcal{B}(B_{q}\to\mu^+\mu^-)_{{\rm 2HDM_{\overline{MFV}}}}\ge\frac{1}{2}(1-y_q)
\mathcal{B}(B_{q}\to\mu^+\mu^-)_{\rm SM},
\ee
which are also valid in the MSSM \cite{Altmannshofer:2012ks}.

\subsubsection{Tree-Level Gauge Boson Exchange}

\begin{figure}[!tb]
 \centering
\includegraphics[width = 0.35\textwidth]{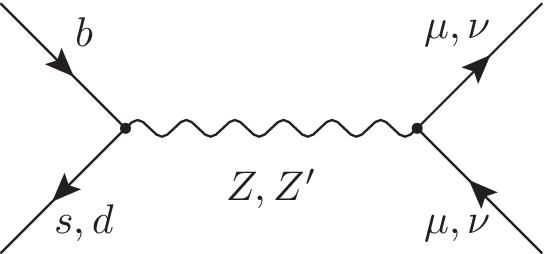}
\caption{\it Tree-level flavour-changing $Z$ and $Z^\prime$ contribution to $\Delta F = 1$ transitions.}\label{fig:FD3}~\\[-2mm]\hrule
\end{figure}

We will next consider the contributions of a tree-level gauge boson 
exchange to the Wilson coefficients of the operators involved {(see Fig.~\ref{fig:FD3})}. Including the 
SM contributions one has \cite{Buras:2012jb}
\begin{align}
 \sin^2\theta_W C_9 &=[\eta_Y Y_0(x_t)-4\sin^2\theta_W Z_0(x_t)]
-\frac{1}{g_{\text{SM}}^2}\frac{1}{M_{Z^\prime}^2}
\frac{\Delta_L^{sb}(Z^\prime)\Delta_V^{\mu\bar\mu}(Z^\prime)} {V_{ts}^* V_{tb}} ,\\
   \sin^2\theta_W C_{10} &= -\eta_Y Y_0(x_t) -\frac{1}{g_{\text{SM}}^2}\frac{1}{M_{Z^\prime}^2}
\frac{\Delta_L^{sb}(Z^\prime)\Delta_A^{\mu\bar\mu}(Z^\prime)}{V_{ts}^* V_{tb}},\\
  \sin^2\theta_W C^\prime_9         &=-\frac{1}{g_{\text{SM}}^2}\frac{1}{M_{Z^\prime}^2}
\frac{\Delta_R^{sb}(Z^\prime)\Delta_V^{\mu\bar\mu}(Z^\prime)}{V_{ts}^* V_{tb}},\\
  \sin^2\theta_W C_{10}^\prime   &= -\frac{1}{g_{\text{SM}}^2}\frac{1}{M_{Z^\prime}^2}
\frac{\Delta_R^{sb}(Z^\prime)\Delta_A^{\mu\bar\mu}(Z^\prime)}{V_{ts}^* V_{tb}},
 \end{align}
where we have defined 
\begin{align}
\begin{split}
 &\Delta_V^{\mu\bar\mu}(Z^\prime)= \Delta_R^{\mu\bar\mu}(Z^\prime)+\Delta_L^{\mu\bar\mu}(Z^\prime),\\
&\Delta_A^{\mu\bar\mu}(Z^\prime)= \Delta_R^{\mu\bar\mu}(Z^\prime)-\Delta_L^{\mu\bar\mu}(Z^\prime).
\end{split}
\end{align}
 In order to simplify the presentation we still work with $\eta_Y$  
  and $Y_0(x_t)$ which should be replaced by $Y_{\rm eff}$ in (\ref{Yeff}) if 
the future precision of experimental data will require it.

The vector Wilson coefficients $C_9,C_9^\prime$ do not contribute to decays in 
question but they will enter Step 7, where the decays  $B\to X_s\ell^+\ell^-$ and 
$B\to K^*(K)\ell^+\ell^-$ are considered.
Assuming that the CKM parameters have been determined independently 
of NP and are universal we find then
\be\label{GB/SM}
\frac{\mathcal{B}(B_q\to\mu^+\mu^-)}{\mathcal{B}(B_q\to\mu^+\mu^-)^{\rm SM}}
=\left|\frac{Y_A^q(v)}{\eta_Y Y_0(x_t)}\right|^2, 
\ee
where
\be 
Y_A^q(v)= \eta_Y Y_0(x_t)
-\frac{1}{V_{tb}V^*_{tq}}\frac{\left[\Delta_A^{\mu\bar\mu}(Z^\prime)\right]}{M_{Z^\prime}^2g_\text{SM}^2}
\left[\Delta_R^{qb}(Z^\prime)-\Delta_L^{qb}(Z^\prime)\right]\,
\ee
is generally complex and moreover different for $B_d\to\mu^+\mu^-$ and 
$B_s\to\mu^+\mu^-$  implying violation of the CMFV correlation shown  
in Fig.~\ref{fig:BdvsBs}.
Still  the correlation between ${\mathcal{B}(B_q\to\mu^+\mu^-)^{\rm SM}}$ and 
$\Delta M_q$, when all these observables are calculated directly, could 
offer a useful test of the model.

 In  \cite{Buras:2012jb}  the correlations 
between the following observables have been investigated:
\be\label{Class2} 
\Delta M_s, \quad S_{\psi \phi}, \quad \mathcal{B}(B_s\to\mu^+\mu^-), \quad 
S^s_{\mu\mu}
\ee
in the $B_s$-system and 
\be\label{Class1} 
\Delta M_d, \quad S_{\psi K_S}, \quad \mathcal{B}(B_d\to\mu^+\mu^-), \quad 
S^d_{\mu\mu}
\ee
in $B_d$ system. To this end 
\be\label{DAmumu}
\Delta_A^{\mu\bar\mu}(Z^\prime)=0.5
\ee
 has been chosen,
to be compared with its SM value $\Delta_A^{\mu\bar\mu}(Z)=0.372$. 

Note that for fixed $\Delta_A^{\mu\bar\mu}(Z^\prime)$ the observables in (\ref{Class2}) 
depend only on  two complex variables $\Delta^{bs}_{L,R}(Z^\prime)$ and in 
fact in the LHS, RHS, LRS and ALR scenarios only on $\tilde s_{23}$ and 
$\delta_{23}$. Similarly the observables in (\ref{Class1})
depend on only two complex variables $\Delta^{bd}_{L,R}(Z^\prime)$ and  
in the LHS, RHS, LRS and ALR scenarios only on $\tilde s_{13}$ and 
$\delta_{13}$. As these parameters have been already constrained in 
Step 3, definite correlations between  the observables within each 
set in (\ref{Class2}) and (\ref{Class1}) follow. Once the $U(2)^3$ 
symmetry is imposed correlations between the  sets in (\ref{Class2}) and (\ref{Class1}) are found.
It will be interesting to investigate the impact on these correlations from 
 $b\to s \ell^+\ell^-$ 
and $b\to s\nu\bar\nu$ transitions that we consider in Steps 7 and 9, 
respectively.

It will be useful to present numerical analysis of 
these correlations together with the ones 
resulting from tree-level scalar exchanges and we will first turn our 
attention to the latter exchanges.

\subsubsection{Tree-Level Scalar and Pseudoscalar Exchanges}

\begin{figure}[!tb]
 \centering
\includegraphics[width = 0.35\textwidth]{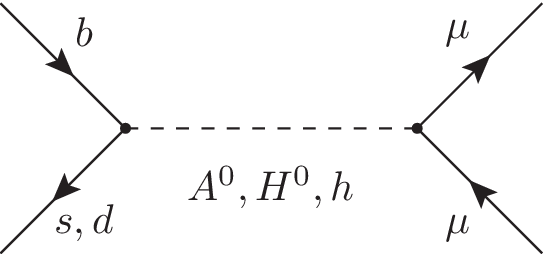}
\caption{\it Tree-level flavour-changing $A^0,H^0,h$ contribution to $\Delta F = 1$ transitions.}\label{fig:FD4}~\\[-2mm]\hrule
\end{figure}

A very detailed analysis of tree-level scalar and pseudoscalar tree-level 
contributions {as shown in Fig.~\ref{fig:FD4}} to decays in question has been performed in \cite{Buras:2013rqa}.
In this case  SM Wilson coefficients remain unchanged but 
the  Wilson coefficients  of scalar and pseudoscalar operators become 
non-zero and are given at $\mu=M_H$ as follows 
\begin{align}
 m_b(M_H)\sin^2\theta_W C_S &= \frac{1}{g_{\text{SM}}^2}\frac{1}{ M_H^2}\frac{\Delta_R^{sb}(H)\Delta_S^{\mu\bar\mu}(H)}{V_{ts}^* V_{tb}},\\
 m_b(M_H)\sin^2\theta_W C_S^\prime &= \frac{1}{g_{\text{SM}}^2}\frac{1}{ M_H^2}\frac{\Delta_L^{sb}(H)\Delta_S^{\mu\bar\mu}(H)}{V_{ts}^* V_{tb}},\\
 m_b(M_H)\sin^2\theta_W C_P &= \frac{1}{g_{\text{SM}}^2}\frac{1}{ M_H^2}\frac{\Delta_R^{sb}(H)\Delta_P^{\mu\bar\mu}(H)}{V_{ts}^* V_{tb}},\\
 m_b(M_H)\sin^2\theta_W C_P^\prime &= \frac{1}{g_{\text{SM}}^2}\frac{1}{ M_H^2}\frac{\Delta_L^{sb}(H)\Delta_P^{\mu\bar\mu}(H)}{V_{ts}^* V_{tb}},
\end{align}
where 
\begin{align}\begin{split}\label{equ:mumuSPLR}
 &\Delta_S^{\mu\bar\mu}(H)= \Delta_R^{\mu\bar\mu}(H)+\Delta_L^{\mu\bar\mu}(H),\\
&\Delta_P^{\mu\bar\mu}(H)= \Delta_R^{\mu\bar\mu}(H)-\Delta_L^{\mu\bar\mu}(H).\end{split}
\end{align}
Here $H$ stands for a scalar or pseudoscalar but if the mass eigenstates 
 has a given CP-parity it is useful to distinguish between a scalar ($H^0$) 
and a pseudoscalar ($A^0$). Then 
\be\label{SP}
\Delta_S^{\mu\bar\mu}(A^0)=0, \quad \Delta_P^{\mu\bar\mu}(H^0)=0
\ee
and only $\Delta_S^{\mu\bar\mu}(H^0)$ and $\Delta_P^{\mu\bar\mu}(A^0)$ 
can be non-vanishing. This is not a general property and in fact in 
the presence of CP-violating effects scalar and pseudoscalars can have 
both couplings. For simplicity, as in \cite{Buras:2013rqa}, we will 
assume (\ref{SP}) to be true.

The crucial property of these couplings following from the hermicity of the 
Hamiltonian is that $\Delta^{\mu\bar\mu}_{S}$ is real and $\Delta^{\mu\bar\mu}_{P}$ purely imaginary. Therefore it is useful to work with 
\be\label{PSEUDO}
\Delta_P^{\mu\bar\mu}(A^0)=i\tilde\Delta_P^{\mu\bar\mu}(A^0).
\ee
where  $\tilde\Delta_P^{\mu\bar\mu}(A^0)$ is real.

It should be emphasized  that in terms of the couplings used in the analysis of $B_{s,d}^0-\bar B_{s,d}^0$ mixings we have generally
\be\label{dictionary}
\Delta_R^{sb}(H)=[\Delta_L^{bs}(H)]^*,\qquad  \Delta_L^{sb}(H)=[\Delta_R^{bs}(H)]^*,
\ee
which should be kept in mind when studying correlations between $\Delta F=1$
and $\Delta F=2$ transitions.

Concerning the values of the $\tilde\Delta_P^{\mu\bar\mu}(H)$ and 
$\Delta_S^{\mu\bar\mu}(H)$ we will set as in \cite{Buras:2013rqa}
\be\label{leptonicset}
\tilde\Delta_P^{\mu\bar\mu}(H)=\pm 0.020\frac{m_b(M_H)}{m_b(m_b)}, \qquad
\Delta_S^{\mu\bar\mu}(H)=0.040\frac{m_b(M_H)}{m_b(m_b)}
\ee
with the latter factor being $0.61$ for $M_H=1\tev$. We show this factor 
explicitly to indicate how the correct scale for $m_b$ affects the allowed 
range for the lepton couplings.
These values assure significant NP effects in 
$B_{s,d}\to\mu^+\mu^-$ while being consistent will all known data.

\subsubsection{Comparison of tree-level $Z'$, pseudoscalar and scalar exchanges}

In Fig.~\ref{fig:BsmuvsSphiZprimeA} we show the correlation between $\overline{\mathcal{B}}(B_{s}\to\mu^+\mu^-)$ and $S_{\psi\phi}$ for
$Z^\prime$ (left panel) and $A^0$ (right panel).
The corresponding plots for the correlation between $S^s_{\mu\mu}$ and 
$S_{\psi\phi}$ and ${\cal A}^{\mu\mu}_{\Delta\Gamma}$ and $S_{\psi\phi}$ are 
shown in Figs.~\ref{fig:SmuvsSphiZprimeA} and~\ref{fig:ADGvsSphiZprimeA}. In Fig.~\ref{fig:BsmuvsSphiH} we show the corresponding 
results for the scalar $H^0$.

\begin{figure}[!tb]
\centering
 \includegraphics[width= 0.45\textwidth]{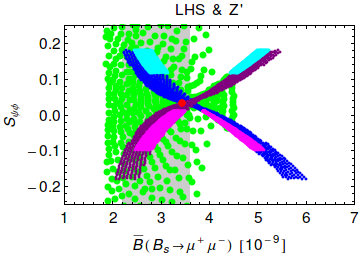}
 \includegraphics[width= 0.45\textwidth]{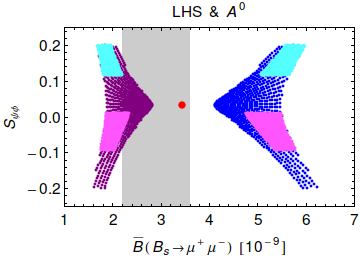}
\caption{\it  $S_{\psi\phi}$ versus $\overline{\mathcal{B}}(B_s\to\mu^+\mu^-)$
for $Z^\prime$ exchange  with $M_{Z^\prime} = 1~$TeV (left) and $A^0$ case with
$M_{A^0} = 1~$TeV (right) in LHS for two oases. The blue and purple
regions are
almost identical for LHS1 and LHS2. The magenta region corresponds to the $U(2)^3$ limit for LHS1 and the cyan region for LHS2. 
The green points in the $Z^\prime$ case indicate the regions that are compatible with $b\to s\ell^+\ell^-$ constraints of 
\cite{Altmannshofer:2012ir}. In the $A^0$ case $b\to s\ell^+\ell^-$ does not give additional constraints.  Gray region:
exp 1$\sigma$ range
$\overline{\mathcal{B}}(B_s\to\mu^+\mu^-) = (2.9\pm0.7)\cdot 10^{-9}$.  Red point: SM central
value.}\label{fig:BsmuvsSphiZprimeA}~\\[-2mm]\hrule
\end{figure}

\begin{figure}[!tb]
 \centering
 \includegraphics[width= 0.45\textwidth]{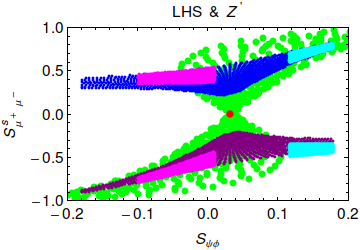}
 \includegraphics[width= 0.45\textwidth]{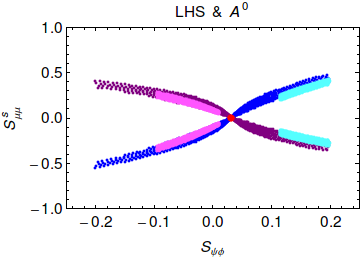}
\caption{\it $S^s_{\mu^+\mu^-}$  versus $S_{\psi\phi}$  in
LHS1 and for $Z^\prime$ (left) and pseudoscalar $A^0$ case (right) both for 1~TeV. The magenta
region corresponds to the $U(2)^3$ limit for LHS1 and the cyan region for LHS2. The green points in the $Z^\prime$ case 
indicate the regions that are compatible with $b\to s\ell^+\ell^-$ constraints. In the $A^0$ case $b\to s\ell^+\ell^-$ does not 
give additional constraints. Red point: SM central
value. }\label{fig:SmuvsSphiZprimeA}~\\[-2mm]\hrule
\end{figure}

\begin{figure}[!tb]
 \centering
\includegraphics[width = 0.45\textwidth]{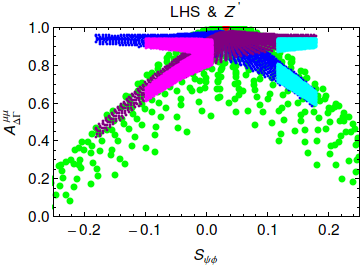}
\includegraphics[width = 0.45\textwidth]{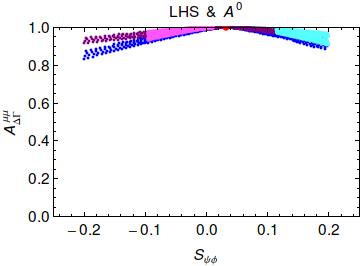}
\caption{\it  $\mathcal{A}^\lambda_{\Delta\Gamma}$ versus $S_{\psi\phi}$  in
LHS1 and for $Z^\prime$ (left) and pseudoscalar $A^0$ case (right)  both for 1~TeV. The magenta
region corresponds to the $U(2)^3$ limit for LHS1 and the cyan region for LHS2.  The green points in the $Z^\prime$ case 
indicate the regions that are compatible with $b\to s\ell^+\ell^-$ constraints. In the $A^0$ case $b\to s\ell^+\ell^-$ does not 
give additional constraints.  Red point: SM central
value. }\label{fig:ADGvsSphiZprimeA}~\\[-2mm]\hrule
\end{figure}

\begin{figure}[!tb]
 \centering
 \includegraphics[width= 0.45\textwidth]{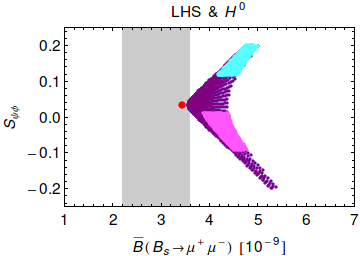}
 \includegraphics[width= 0.45\textwidth]{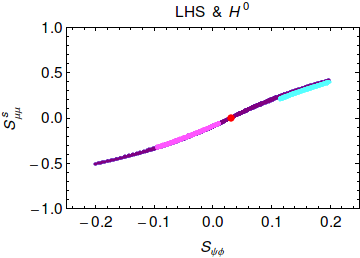}
\includegraphics[width = 0.45\textwidth]{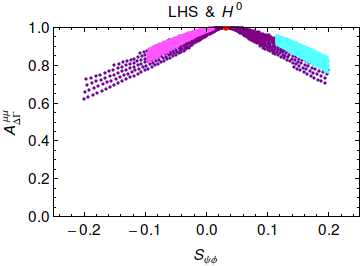}
\caption{\it  $S_{\psi\phi}$ versus $\overline{\mathcal{B}}(B_s\to\mu^+\mu^-)$, $S^s_{\mu^+\mu^-}$  versus $S_{\psi\phi}$  and 
$\mathcal{A}^\lambda_{\Delta\Gamma}$ versus $S_{\psi\phi}$ for scalar $H^0$ case with
$M_{H} = 1~$TeV in LHS1. The two oases (blue and purple) overlap.  The magenta
region corresponds to the $U(2)^3$ limit for LHS1 and the cyan region for LHS2. Red point: SM central
value.}\label{fig:BsmuvsSphiH}~\\[-2mm]\hrule
\end{figure}

The colour coding is as follows:
\begin{itemize}
\item
In the general case {\it blue} and {\it purple} allowed regions correspond to
oases with small and
large $\delta_{23}$, respectively. 
\item
In the $U(2)^3$ symmetry case, the allowed region are shown in {\it magenta} 
and
{\it cyan} for LHS1 and LHS2, respectively, as in this case even in the
$B_s$ system there is dependence on $\vub$ scenario.
These  regions are subregions of the general  blue or purple regions so that
they  cover some parts of them.
\item 
The green points in the $Z^\prime$ case indicate the region that is compatible with constraints from $b\to s\ell^+\ell^-$
transitions. In the scalar and pseudoscalar case the whole oases are compatible with $b\to s\ell^+\ell^-$ (see also
Sec.~\ref{sec:bsllWilson}). 
\end{itemize}

We observe several striking differences between the results for $Z^\prime$, $A^0$ and $H^0$  which allow to distinguish these scenarios from each other:
\begin{itemize}
\item
In the $A^0$ case the asymmetry  $S^s_{\mu^+\mu^-}$ can be zero while this
 is not the case for $Z'$  where
the
requirement of suppression of $\Delta M_s$ directly translates in
$ S^s_{\mu^+\mu^-}$  being non-zero. Consequently in the $Z'$ case the sign of
$ S^s_{\mu^+\mu^-}$  can be used to identify the right oasis. 
This is not possible 
in the case of $A^0$.
\item
On the other hand we observe that in the $A^0$ case
the measurement of $\overline{\mathcal{B}}(B_{s}\to\mu^+\mu^-)$
uniquely chooses the
right oasis. The enhancement of this branching ratio relatively to the
SM chooses the blue oasis
while  suppression 
the purple one. Present data from  LHCb and CMS favour the purple oasis. This distinction is
 not possible in the $Z'$ case. The maximal enhancements and suppressions are comparable in both cases but finding 
$\overline{\mathcal{B}}(B_{s}\to\mu^+\mu^-)$
close to SM value would require in the $A^0$ case either larger
$M_H$ or smaller muon coupling.
\item
Concerning the $H^0$ case, the absence of the interference with the SM 
contribution implies that $\overline{\mathcal{B}}(B_{s}\to\mu^+\mu^-)$
 can only be enhanced in this scenario and
this result is independent of the oasis considered. Thus
finding this branching ratio below its SM value would favour 
 the other two scenarios over scalar one.  The present data from LHCb and CMS indicate that indeed this could be the case. 
But the enhancement is not as
pronounced as in the pseudoscalar case   because in the absence of the 
interference with the SM contribution the correction to the
branching ratio is governed here by the square of the muon coupling and is not 
linearly proportional to it as in the pseudoscalar case.  Therefore in order to exclude this scenario requires significant 
reduction of experimental errors.
\item
 Also CP-asymmetries in the $H^0$ case differ from $Z^\prime$ and $A^0$ cases. Similarly to the branching ratio  there is no dependence on
the oasis considered but more importantly
$S^s_{\mu^+\mu^-}$  can
only increase with increasing $S_{\psi\phi}$.
\item
The correlation between 
$\mathcal{A}^\lambda_{\Delta\Gamma}$ and  $S_{\psi\phi}$ has the same structure 
for $Z^\prime$,  $A^0$ and $H^0$ cases.
We observe that
for  $M_{H}=1\tev$, even for $S_{\psi\phi}$ significantly different from
zero, $\mathcal{A}^\lambda_{\Delta\Gamma}$ does not differ significantly
 from unity in $A^0$ and $H^0$ scenarios. Larger effects for the same 
mass are found in the
$Z^\prime$ case.
\end{itemize}

\begin{figure}[!tb]
\begin{center}
\includegraphics[width=0.46\textwidth]{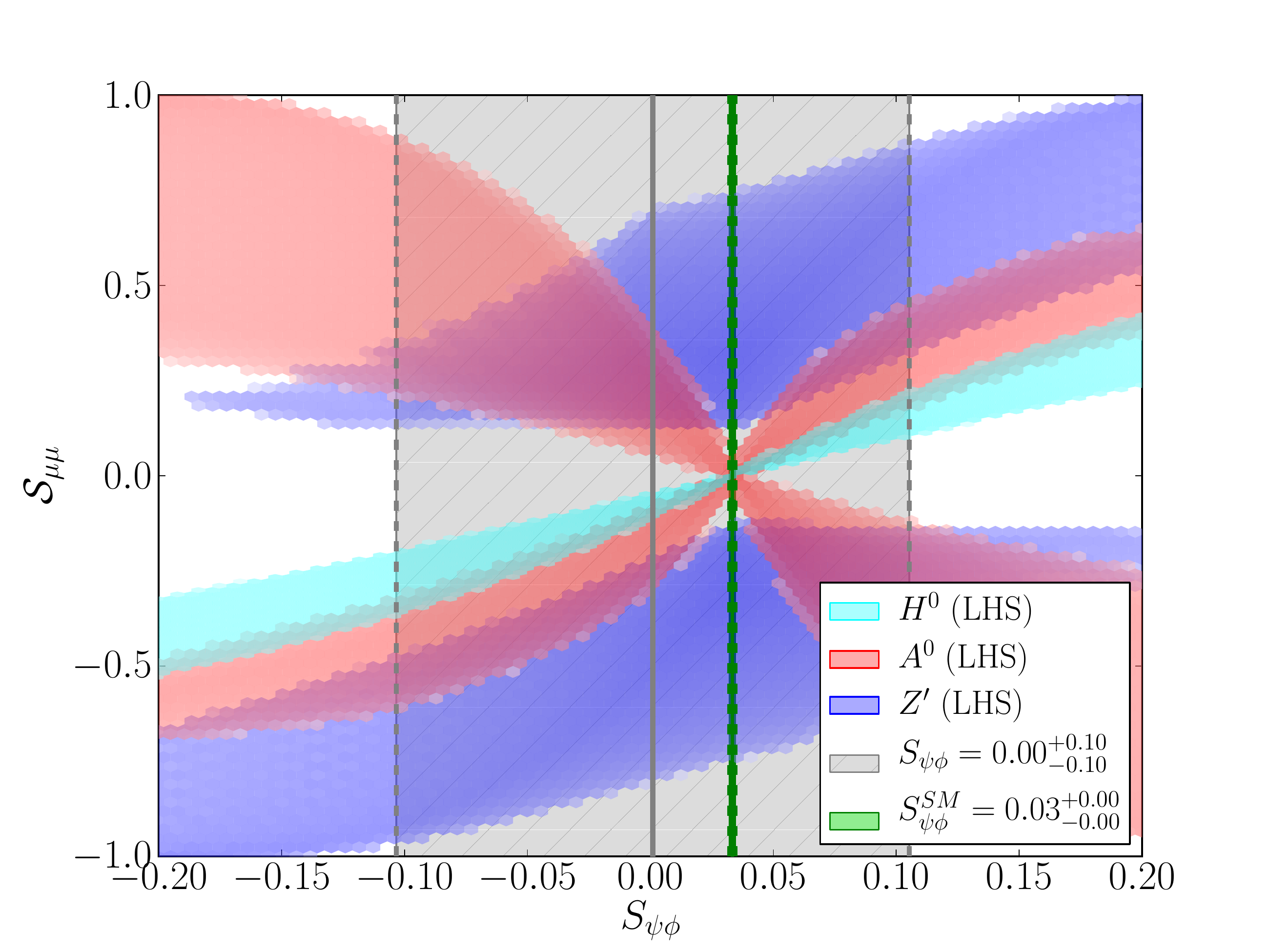}
\includegraphics[width=0.46\textwidth]{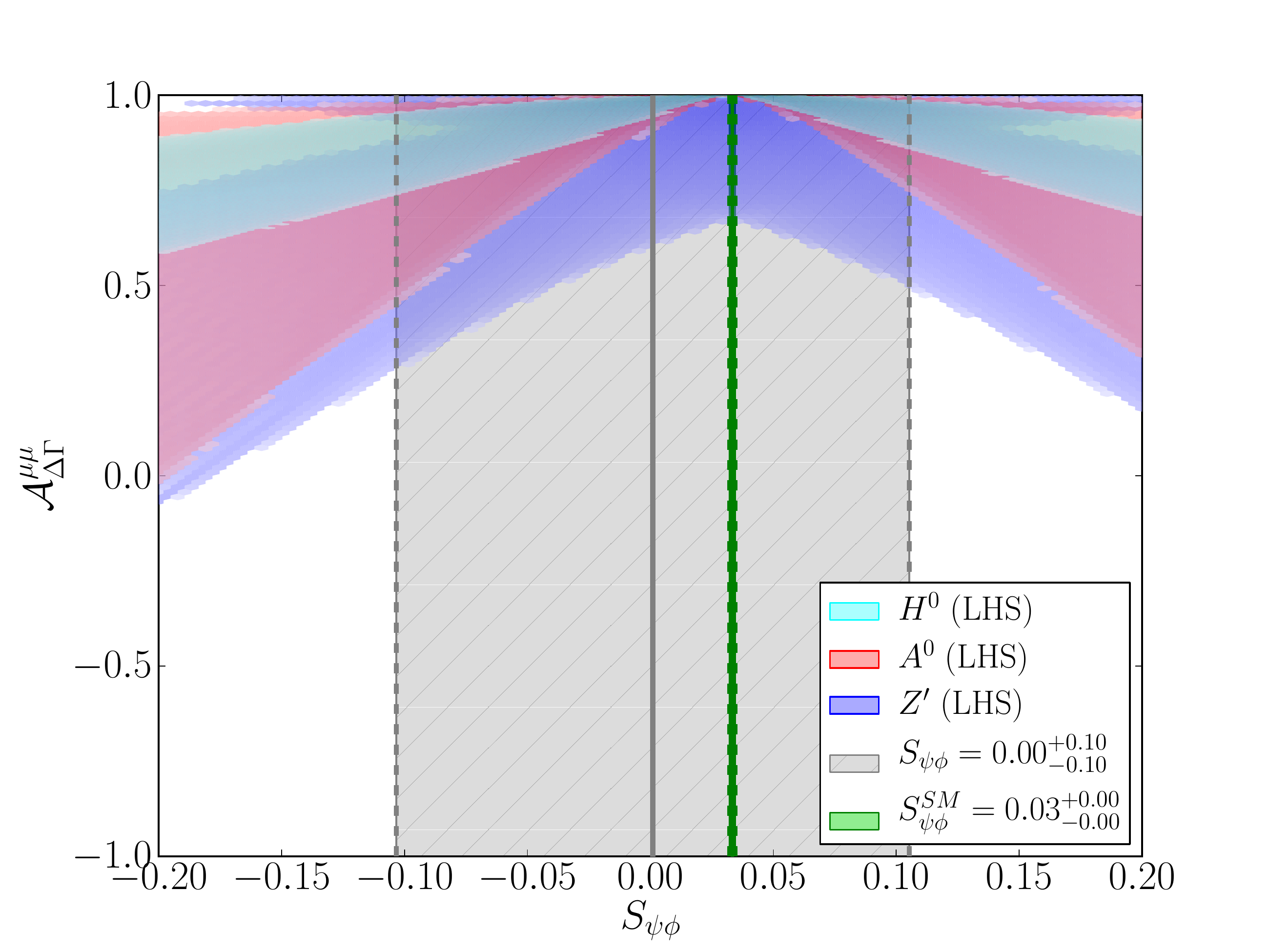}
\includegraphics[width=0.55\textwidth]{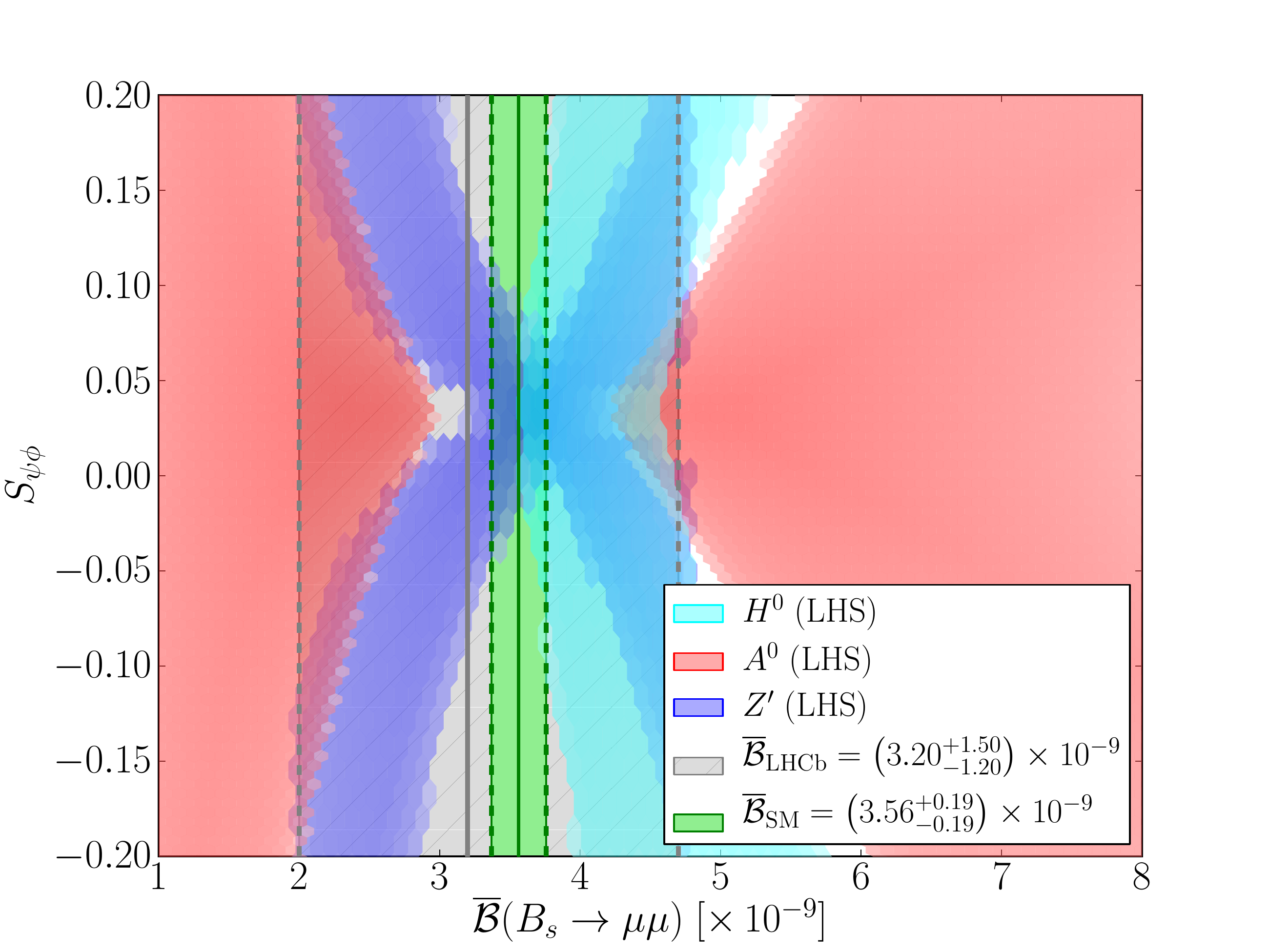}
\caption{\it  Overlay of the correlations for $S_{\mu\mu}^s$ versus $S_{\psi\phi}$ (top left), $A_{\Delta\Gamma}^{\mu\mu}$ versus
$S_{\psi\phi}$ (top right) and $S_{\psi\phi}$ versus $\overline{\mathcal{B}}(B_s\to\mu^+\mu^-)$ (bottom) for tree
level scalar (cyan), pseudoscalar (red) and $Z^\prime$ (blue) exchange (both oases in same colour respectively) in LHS. The lepton
couplings are varied in the ranges $|\Delta_{S,P}^{\mu\mu}(H)| \in [0.012,0.024]$ and $\Delta_A^{\mu\mu}(Z')\in [0.3,0.7]$. From \cite{Buras:2013rqa}.
}\label{fig:grandplot}~\\[-2mm]\hrule
\end{center}
\end{figure}

In Fig.~\ref{fig:grandplot} we summarize our results in the $B_s$ system for tree level $Z^\prime$, $H^0$ and $A^0$  exchanges 
where we also
vary the lepton couplings in a wider range: $|\Delta_{S,P}^{\mu\mu}(H)| \in [0.012,0.024]$ and $\Delta_A^{\mu\mu}(Z')\in 
[0.3,0.7]$.
As explained in \cite{Buras:2013rqa} 
the striking differences between the $A^0$-scenario and $Z'$-scenario
can be traced back to the difference between the phase of the NP correction
to the quantity $P$, defined in (\ref{PP}),
 in these two NP scenarios. As the oasis structure as far as the phase
$\delta_{23}$ is concerned is the same in both scenarios the difference enters
through the muon couplings which are imaginary in the case of $A^0$-scenario but
real in the case of $Z'$.  Taking in addition into
account the sign difference between $Z'$ and pseudoscalar propagator in the
the $b\to s \mu^+\mu^-$ amplitude, which is now not compensated by a hadronic
matrix element, one finds that
\be\label{PP1}
P(Z')=1+ r_{Z'} e^{i \delta_{Z'}}, \qquad P(A^0)=1 +  r_{A^0} e^{i\delta_{A^0}}
\ee
with
\be\label{SHIFT}
 r_{Z'}\approx  r_{A^0}, \qquad    \delta_{Z'}=\delta_{23}-\beta_s, \qquad \delta_{A^0}=\delta_{Z'}-\frac{\pi}{2}.
\ee

Therefore with $\delta_{23}$ of Fig.~\ref{fig:oasesBsLHS1} the phase
$\delta_{Z'}$ is around $90^\circ$ and
$270^\circ$ for the blue and purple oasis, respectively. Correspondingly
$\delta_{A^0}$ is around
$0^\circ$ and $180^0$. This
difference in the phases is at the origin of the differences listed above.
In particular, we understand now why the CP asymmetry $ S^s_{\mu^+\mu^-}$
can vanish in the $A^0$ case, while it was always different from zero in the $Z'$-case. What is interesting is that this difference is just related to the different
particle exchanged: gauge boson and pseudoscalar. We summarize the ranges of
 $\delta_{Z'}$ and  $\delta_{A^0}$ in Table~\ref{tab:PZ}.

Proceeding in an analogous manner for the scalar case we arrive at an important
relation:
\be\label{SZP}
\varphi_S=\delta_{Z'}-\pi,
\ee
where the shift is related to the  sign difference in the $Z'$ and scalar
propagators.
But as seen in (\ref{Rdef})-(\ref{Ssmu}) the three observables given there, all depend
on $2\varphi_S$, implying that from the point of view of these quantities
this shift is irrelevant. As different oases correspond to phases shifted by
$\pi$ this also explains  why in the scalar case the results in different
oases are the same. That the branching ratio can only be enhanced follows
just from the absence of the interference with the SM contributions. In order
to understand the signs in $S_{\mu\mu}^s$ one should note the minus sign in
front of sine in the corresponding formula. Rest follows from (\ref{SZP})
and Table~\ref{tab:PZ}.

\begin{table}[!tb]
\centering
\begin{tabular}{|c||c|c|c|}
\hline
 Oasis  & $\delta_{Z'}$ & $\delta_{A^0}$   \\
\hline
\hline
  \parbox[0pt][1.6em][c]{0cm}{} $B_s$ (blue) & $50^\circ-130^\circ$  & $-40^\circ-(+40^\circ)$   \\
 \parbox[0pt][1.6em][c]{0cm}{}$B_s$ (purple) & $230^\circ-310^\circ$ & $140^\circ-220^\circ$  \\
\hline
 \parbox[0pt][1.6em][c]{0cm}{}$B_d$ (S1) (yellow) & $57^\circ-86^\circ$&  $-33^\circ-(+4^\circ)$   \\
 \parbox[0pt][1.6em][c]{0cm}{} $B_d$ (S1) (green) & $237^\circ-266^\circ$ & $147^\circ-176^\circ$   \\
\parbox[0pt][1.6em][c]{0cm}{}$B_d$ (S2) (yellow) & $103^\circ-125^\circ$&  $13^\circ-35^\circ$   \\
 \parbox[0pt][1.6em][c]{0cm}{} $B_d$ (S2) (green) & $283^\circ-305^\circ$ & $193^\circ-215^\circ$   \\
\hline
 \parbox[0pt][1.6em][c]{0cm}{}$U(2)^3$ (S1) (blue, magenta) & $55^\circ-84^\circ$&  $-35^\circ-(-6^\circ)$   \\
 \parbox[0pt][1.6em][c]{0cm}{} $U(2)^3$ (S1) (purple, magenta) & $235^\circ-264^\circ$ & $145^\circ-174^\circ$   \\
\parbox[0pt][1.6em][c]{0cm}{} $U(2)^3$ (S2) (blue, cyan) & $101^\circ-121^\circ$&  $11^\circ-31^\circ$  \\
 \parbox[0pt][1.6em][c]{0cm}{} $U(2)^3$ (S2) (purple, cyan) & $291^\circ-301^\circ$ & $201^\circ-211^\circ$   \\
\hline
\end{tabular}
\caption{\it Ranges for the values of $\delta_{Z'}$ and $\delta_{A^0}$ as defined in (\ref{PP1}) for the $B_s$ and $B_d$ systems
and various
cases discussed in the text. Also the result for $U(2)^3$ models is shown. 
From \cite{Buras:2013rqa}.}\label{tab:PZ}~\\[-2mm]\hrule
\end{table}

We now turn our attention to the $B_d\to\mu^+\mu^-$ decay. Here we have 
to distinguish between LHS1 and LHS2 scenarios.  Our colour coding is such that
in the general case {\it yellow} and {\it green} allowed regions correspond to
oases with small and
large $\delta_{13}$, respectively. We do not show the impact of the 
 imposition of the $U(2)^3$ symmetry as the resulting reduction of the allowed 
areas amounts typically to $5-10\%$ at most and it is more
transparent not to show it.

\begin{figure}[!tb]
\begin{center}
\includegraphics[width=0.45\textwidth] {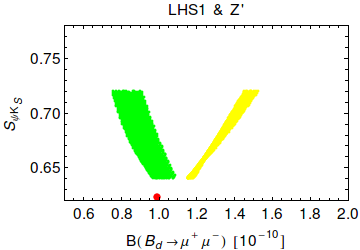}
\includegraphics[width=0.45\textwidth] {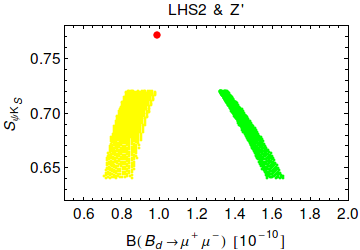}
\caption{\it  $S_{\psi K_S}$ versus   $\mathcal{B}(B_d\to\mu^+\mu^-)$  for $M_{Z^\prime}=1$ TeV in LHS1 (left) and
LHS2 (right) for the yellow and green oases. Red point: SM central
value. }\label{fig:BdmuvsSKSLHS}~\\[-2mm]\hrule
\end{center}
\end{figure}

\begin{figure}[!tb]
\begin{center}
\includegraphics[width=0.45\textwidth] {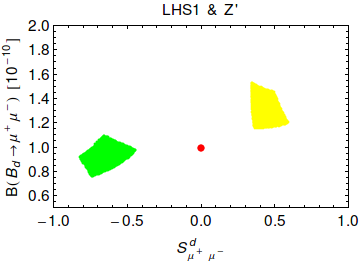}
\includegraphics[width=0.45\textwidth] {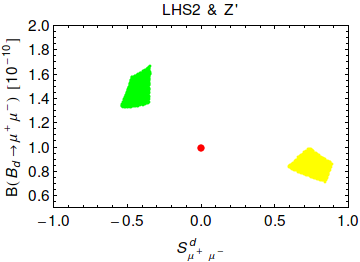}
\caption{\it $\mathcal{B}(B_d\to\mu\bar\mu)$ versus $S_{\mu^+\mu^-}^d$    for $M_{Z^\prime}=1$ TeV in LHS1 (left) and
LHS2 (right).  Red point: SM central
value.}\label{fig:BdmuvsSmudLHS}~\\[-2mm]\hrule
\end{center}
\end{figure}

\begin{figure}[!tb]
\begin{center}
\includegraphics[width=0.45\textwidth] {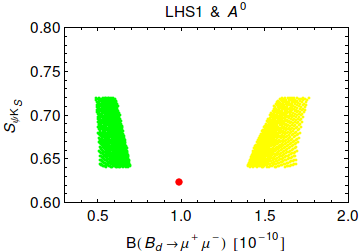}
\includegraphics[width=0.45\textwidth] {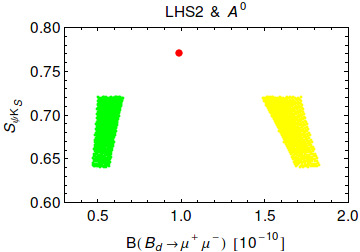}
\caption{\it  $S_{\psi K_S}$ versus   $\mathcal{B}(B_d\to\mu^+\mu^-)$  in $A^0$
scenario for
$M_{H}=1$ TeV in LHS1 (left) and
LHS2 (right) in the yellow and green oases as discussed in the text.   Red point:
SM central
value.}\label{fig:BdmuvsSKSLHSP}~\\[-2mm]\hrule
\end{center}
\end{figure}

\begin{figure}[!tb]
\begin{center}
\includegraphics[width=0.45\textwidth] {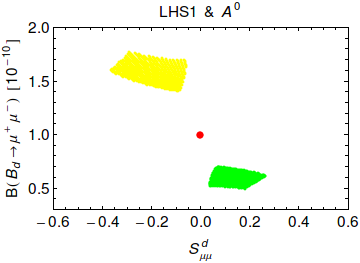}
\includegraphics[width=0.45\textwidth] {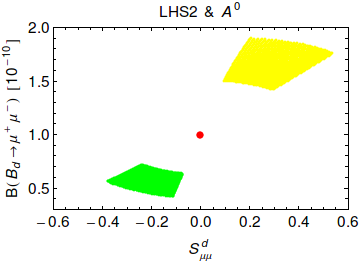}
\caption{\it $\mathcal{B}(B_d\to\mu\bar\mu)$ versus $S_{\mu^+\mu^-}^d$  in $A^0$ case  for $M_{Z^\prime}=1$ TeV in LHS1 (left)
and
LHS2 (right) for the green and yellow oases as discussed in the text.  Red point: SM central
value.}\label{fig:BdmuvsSmudLHSP}~\\[-2mm]\hrule
\end{center}
\end{figure}

\begin{figure}[!tb]
\begin{center}
\includegraphics[width=0.45\textwidth] {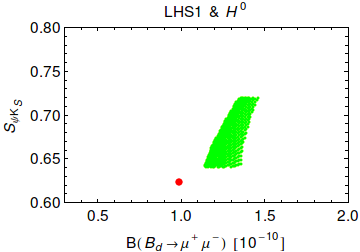}
\includegraphics[width=0.45\textwidth] {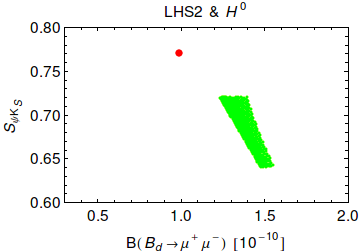}
\caption{\it  $S_{\psi K_S}$ versus   $\mathcal{B}(B_d\to\mu^+\mu^-)$  in $H^0$
scenario for
$M_{H}=1$ TeV in LHS1 (left) and
LHS2 (right) in the yellow and green oases that overlap here. Red point: SM central
value.}\label{fig:BdmuvsSKSLHSS}~\\[-2mm]\hrule
\end{center}
\end{figure}

\begin{figure}[!tb]
\begin{center}
\includegraphics[width=0.45\textwidth] {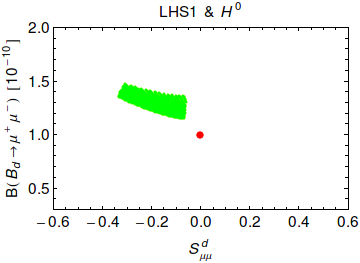}
\includegraphics[width=0.45\textwidth] {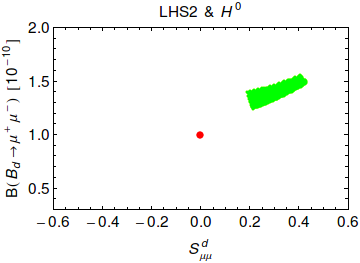}
\caption{\it $\mathcal{B}(B_d\to\mu\bar\mu)$ versus $S_{\mu^+\mu^-}^d$  in $H^0$ case  for $M_{H}=1$ TeV in LHS1 (left) and
LHS2 (right) for the green and yellow oases (they overlap here) as discussed in the text.  Red point: SM central
value.}\label{fig:BdmuvsSmudLHSS}~\\[-2mm]\hrule
\end{center}
\end{figure}

In Figs.~\ref{fig:BdmuvsSKSLHS} and ~\ref{fig:BdmuvsSmudLHS}  we show  $S_{\psi K_S}$ vs  
$\mathcal{B}(B_d\to\mu^+\mu^-)$  and $S_{\mu\mu}^d$ vs  $\mathcal{B}(B_d\to\mu^+\mu^-)$
for 
$Z^\prime$ scenario. The corresponding plots for the $A^0$ and $H^0$ scenarios 
are shown in Figs.~\ref{fig:BdmuvsSKSLHSP}-\ref{fig:BdmuvsSmudLHSS}. 

In order to understand the differences between these two scenarios of NP we
again look at the phase of the correction to $P$ in (\ref{PP1}) which now is given as follows:
\be\label{SHIFTBd}
 r_{Z'}\approx  r_{A^0}, \qquad    \delta_{Z'}=\delta_{13}-\beta, \qquad 
\delta_{A^0}=\delta_{Z'}-\frac{\pi}{2}.
\ee
Note that this time the phase of $V_{td}$ enters the analysis with $\beta\approx 19^\circ$ and $\beta\approx 25^\circ$ for S1 and
S2
scenario of $\vub$, respectively. We find then that in scenario S2 the phase $\delta_{Z'}$ is around
$115^\circ$ and $295^\circ$ for yellow and green oases, respectively.
Correspondingly $\delta_{A^0}$  is around $25^\circ$ and $205^\circ$.
 We summarize the ranges of
 $\delta_{Z'}$ and  $\delta_{A^0}$ in Table~\ref{tab:PZ}.

With
this insight at hand we can easily understand the plots in question
noting that the enhancements and suppressions of $\mathcal{B}(B_d\to\mu^+\mu^-)$ are governed by the cosine of the phase and the signs of $S_{\mu\mu}^d$ by 
the corresponding sines. We leave this exercise to the motivated 
readers and refer 
to \cite{Buras:2013rqa} for a detailed description of the plots. What is interesting is that already the suppressions or enhancements of certain observables 
and correlations or anti-correlations between them could tell us one day 
which of the three NP scenarios if any is favoured by nature.  In fact 
if the present central experimental value for $\mathcal{B}(B_d\to\mu^+\mu^-)$ 
will be confirmed by more precise measurements tree-level $Z^\prime$, $A^0$ and 
$H^0$ exchanges will not be able to describe such data alone when the 
constraints from $\Delta F=2$ transitions are taken into account.

Finally, let us make a few comments on the impact of the imposition of the 
$U(2)^3$ symmetry. The main effect is on $B_s\to\mu^+\mu^-$ and we have
 shown it in all plots above.
Presently  most interesting in this context is the correlation 
between $S_{\psi\phi}$ and
$\mathcal{B}(B_s\to\mu^+\mu^-)$. 
We observe that already the sign of $S_{\psi\phi}$ will decide whether LHS1 or
LHS2 is favoured. Moreover if $\mathcal{B}(B_s\to\mu^+\mu^-)$ will
turned out to be suppressed relatively to the SM then only one oasis 
will survive in each scenario. Comparison with future precise value
of $\vub$ will confirm or rule out this scenario of NP.
These correlations are particular
examples of the correlations in $U(2)^3$ models pointed out in
\cite{Buras:2012sd}. What
is new here is that in a specific model considered by us the $\vub-S_{\psi\phi}$
correlation has now also implications not only for $\mathcal{B}(B_s\to\mu^+\mu^-)$ but also for $S_{\mu\mu}^s$ as seen in other plots. Analogous comments 
can be made in the case of $A^0$ and $H^0$. 

\boldmath
\subsubsection{Dependence of $\Delta F=1$ Transitions on $M_{Z'}$}
\unboldmath
The nominal value of  $M_{Z'}$ in the plots presented in this review is $1\tev$  except for few cases where higher values are considered. The results for 
$M_{Z'}=3\tev$ in 331 models can be found in \cite{Buras:2012dp,Buras:2013dea}.
Here following  \cite{Buras:2012jb,Buras:2013dea} we would like to summarize 
how our results for $\Delta F=1$ transitions can be translated into other 
values of  $M_{Z'}$ in case higher values would be required by the LHC and 
other constraints in a given model. In this translation the lepton couplings 
have to be held fixed.

As presently the constraints on $Z^\prime$ models are dominated by $\Delta F=2$ transitions it turns out  that for a given allowed size of
$\Delta S(B_q)$, NP effects in the one-loop $\Delta F=1$ functions are proportional to $1/M_{Z^\prime}$. That these effects
are only suppressed like  $1/M_{Z^\prime}$  and not like  $1/M^2_{Z^\prime}$ is
the consequence of the increase with $M_{Z^\prime}$ of the allowed values of
the couplings $\Delta_{L,R}^{ij}(Z^\prime)$ extracted from $\Delta F=2$
observables. When NP effects are significantly smaller than the SM 
contribution, only interference between SM and NP contributions matters and 
consequently  this dependence is transfered to the branching ratios.
In
summary, denoting by $\Delta\mathcal{O}^{\rm NP}(M_{Z^\prime}^{(i)})$ NP contributions to a given $\Delta F=1$ observable in
$B_s$ and $B_d$ decays at two ($(i=1,2)$ different values $M_{Z^\prime}^{(i)}$ we have a {\it scaling law}
\be\label{scaling}
\Delta\mathcal{O}^{\rm NP}(M_{Z^\prime}^{(1)})=\frac{M_{Z^\prime}^{(2)}}{M_{Z^\prime}^{(1)}}
\Delta\mathcal{O}^{\rm NP}(M_{Z^\prime}^{(2)}).
\ee

This scaling law is valid in most of observables in $B_s$ and $B_d$ systems 
as NP effects are bounded to be small. In the rare  $K$ decays, like 
$\kpn$ and $\klpn$, where NP contributions for sufficiently low values of  $M_{Z'}$  could be much larger than the 
SM contribution, NP modifications of branching ratios will decrease faster than $1/M_{Z'}$ ($1/M^2_{Z'}$ in the limit of full NP dominance) until NP contributions are sufficiently small so that the  $1/M_{Z'}$ dependence and (\ref{scaling}) is again valid.

Needless to say, when also lepton couplings can be varied in order to 
compensate for the change of $M_{Z'}$, the scaling law could be modified.
In this case the correlations between NP corrections to various 
one-loop functions, derived in  \cite{Buras:2012jb,Buras:2013dea},  are 
helpful in translating our results into the ones obtained for different 
 $M_{Z'}$ and lepton couplings. We refer in particular to
 \cite{Buras:2013dea} where using the data from LEP-II, CMS and ATLAS
the bounds on  $M_{Z'}$  in various 331 models with different lepton couplings 
have been analyzed.

\boldmath
\subsubsection{Flavour Violating SM $Z$ and SM Higgs Boson}
\unboldmath
Let us next look at a possibility that NP will only be detectable 
through modified $Z$ and Higgs couplings. Beginning with
flavour violating $Z$-couplings they 
can be generated in the presence of other neutral gauge bosons and or
new heavy vectorial fermions with $+2/3$ and $-1/3$ electric charges.
RSc is an explicit model of this type \cite{Blanke:2008yr,Buras:2009ka} (see also \cite{delAguila:2011yd}).
Recently, an extensive analysis of flavour violation in the presence
of a vectorial $+2/3$ quark has been presented in
\cite{Botella:2012ju}, where
references to previous literature can be found.

The formalism developed for $Z^\prime$  can be used directly here by setting
\be
M_Z=91.2\gev, \quad \Delta_L^{\nu\bar\nu}(Z)=\Delta_A^{\mu\bar\mu}(Z)=0.372,
\quad  \Delta_V^{\mu\bar\mu}(Z)=-0.028
\ee
The implications of these changes are as follows:
\begin{itemize}
\item
The decrease of the neutral gauge boson mass by an order of magnitude relatively to the nominal value $M_{Z'}=1\tev$ used by us decreases the couplings
$\tilde s_{ij}$ by the same amount without any impact on the phases
$\delta_{ij}$ when the constraints from
$\Delta F=2$ processes are imposed.
\item
As pointed out in \cite{Buras:2012jb} once the parameters $\tilde s_{ij}$
are constrained through $\Delta F=2$ observables
 the decrease of neutral gauge boson mass enhances NP effects in rare $K$
and $B$ decays. This follows from the structure of tree-level contributions
to FCNC processes and is not generally the case when NP contributions are
governed by penguin and box diagrams. 
\item
The latter fact implies that already the present experimental
bounds on $\mathcal{B}(\kpn)$ and $\mathcal{B}(B_{s,d}\to\mu^+\mu^-)$
as well as the data on
$B\to X_s \ell^+\ell^-$, $B\to   K^*\ell^+\ell^-$  and
 $B\to   K\ell^+\ell^-$  decays become more powerful than the $\Delta F=2$
transitions in constraining flavour violating couplings of $Z$ so that
effects in $\Delta F=2$ processes cannot be as large as in $Z'$ case.
\end{itemize}

The patterns of flavour violation through $Z$ in $B_s$, $B_d$ and $K$ 
are strikingly different from each other:
\begin{itemize}
\item
In the $B_s$ system when the constraints from 
$\Delta M_s$  and $S_{\psi\phi}$ are imposed
$\mathcal{B}(B_s\to\mu^+\mu^-)$ is always larger than
its SM value and mostly above the data except in LRS case where NP contributions vanish.  Further constraints follow from $b\to s\ell^+\ell^-$ transitions so 
that one has to conclude that it is
very difficult to suppress
$\Delta M_s$ sufficiently in LHS, LRS and RHS scenarios
without violating the constraints from $b\to s \mu^+\mu^-$
transitions. Thus  we expect 
$\mathcal{B}(B_s\to\mu^+\mu^-)$ to be enhanced over the SM value but 
simultaneously possible tension in $\Delta M_s$ cannot be solved if 
the relevant parameters are like in (\ref{oldf1}).
Future lattice calculations will tell us whether this is indeed a problem.
 Similar conclusions have been reached in
\cite{Altmannshofer:2012ir,Beaujean:2012uj}.  Yet, as demonstrated recently 
in  \cite{Buras:2013qja} by changing the non-perturbative parameters 
agreement with both data on $\Delta F=2$ observables and $B_s\to\mu^+\mu^-$ 
can be obtained and we will summarize this analysis below.
\item
In the $B_d$ system all $\Delta F=2$ constraints can be satisfied. 
We again observe that $\mathcal{B}(B_d\to \mu^+\mu^-)$ can be enhanced 
by almost an order magnitude  and this begins to be a problem for certain 
choices of couplings in view of recent LHCb and CMS data. This is  shown in 
Fig.~\ref{fig:ZBdmuvsSKSLHS} for the LHS1 and LHS2 scenarios. 
Evidently NP effects are much larger than in the $Z^\prime$ case. 
We also show the
results in ALRS1 and ALRS2 scenarios in
which NP effects are smaller than in LHS1 and LHS2 scenarios. 
With improved upper bound on $\mathcal{B}(B_d\to \mu^+\mu^-)$
 LHS1 and LHS2 scenarios could be put into difficulties, while in 
ALRS1 and ALRS2 one could easier satisfy this bound. If such a situation
really took place and NP effects would be observed in this decay, this would
mean that both LH and RH $Z$-couplings in the $B_d$ system would be required
but with opposite sign.
\item
As we will see in Step 8, the effects of flavour violating $Z$ couplings
in $\kpn$ and $\klpn$ can be in principle very large in LHS, RHS and LRS 
scenarios but
they can be bounded by the upper bound on
$K_L\to\mu^+\mu^-$ except for the LR scenarios and the case of purely
imaginary NP contributions in all these scenarios
 where this bound is ineffective. We show in Step 8 in Fig.~\ref{fig:ZKLvsKp}
 few examples  which 
demonstrate that even with the latter constraint taken into account 
flavour violating $Z$ can  have impact on rare $K$ 
decays which is significantly larger than in the $Z^\prime$ case. 
\end{itemize}

\begin{figure}[!tb]
\begin{center}
\includegraphics[width=0.45\textwidth] {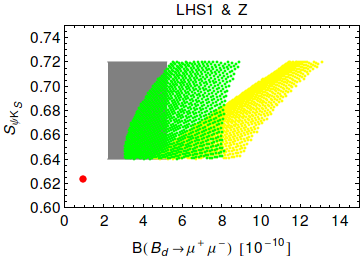}
\includegraphics[width=0.45\textwidth] {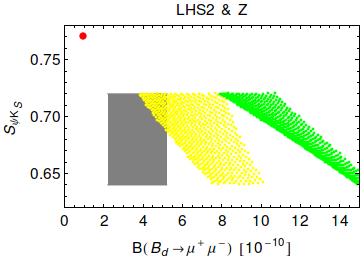}

\includegraphics[width=0.45\textwidth] {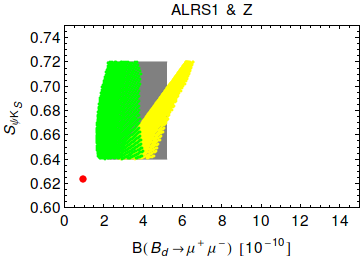}
\includegraphics[width=0.45\textwidth] {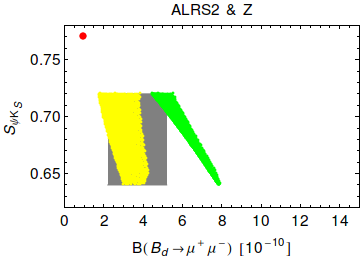}
\caption{\it  $S_{\psi K_S}$ versus   $\mathcal{B}(B_d\to\mu^+\mu^-)$  in LHS1, LHS2 (upper row) and ALRS1, ALRS2 (lower row).
$B_1$: yellow, $B_3$: green.  Red point:
SM central
value. Gray region: $1\sigma$ range of $\mathcal{B}(B_d\to\mu^+\mu^-)=\left(3.6^{+1.6}_{-1.4}\right)\cdot
10^{-10}$ and $2\sigma$ region of $S_{\psi K_S} \in[0.639,0.719]$.}\label{fig:ZBdmuvsSKSLHS}~\\[-2mm]\hrule
\end{center}
\end{figure}

 In summary 
flavour-violating $Z$ couplings in $B_d\to\mu^+\mu^-$ decay,
similarly to $Z'$ couplings in rare $K$ 
decays discussed in Step 8,
could turn out to be an important portal to short distance
scales which cannot be explored
by the LHC. For $B_s\to\mu^+\mu^-$ decay this does not seem to be the case 
any longer.

Concerning the tree-level SM Higgs contributions to FCNCs 
one finds that once the constraints on flavour-violating couplings 
from $\Delta F=2$ observables are imposed,  the smallness of Higgs 
couplings to muons precludes any measurable  effects in $\mathcal{B}(B_d\to\mu^+\mu^-)$ 
and $\mathcal{B}(B_s\to\mu^+\mu^-)$ can be only enhanced by at most 
$8\%$ \cite{Buras:2013rqa}. 
 Still the presence
of such contributions can remove all possible tensions within the SM in
$\Delta F=2$ transitions without being in conflict with constraints from
rare decays.

Similarly to modifications of $Z$ and SM Higgs couplings, also 
  couplings of $W^\pm$ could be modified by NP. There are many papers studying 
  implications of such modifications for FCNC processes. We refer to the 
  recent detailed analysis in \cite{Drobnak:2011aa}, where further references 
 can be found. In particular the constraints on the anomalous $tWb$ interactions turn out to be superior to present direct constraints from top decays 
and production measurements at Tevatron and the LHC.

\boldmath
\subsubsection{Facing the violation of CMFV Relation (\ref{CMFV6})}
\unboldmath
As shown in Fig.~\ref{fig:BdvsBs}
the stringent CMFV relation in (\ref{CMFV6}) appears to be violated by the 
present data. Even if this violation is still not statistically significant 
in view of very inaccurate data on $B_d\to\mu^+\mu^-$ it is of interest to 
see whether tree-level exchanges of $Z^\prime$ and $Z$ could with a certain 
choices of quark and lepton couplings reproduce these data while satisfying 
$\Delta F=2$ constraints and the constraints from 
$B_d\to K^*(K)\mu^+\mu^-$ considered in Step 7. As in the numerical analysis 
presented sofar NP in $\Delta F=2$ processes was governed by (\ref{oldf1}) 
and consequently $C_{B_s}\approx C_{B_d}\approx 0.93$, it is also interesting 
to see what happens when these values are modified.

Such an analysis has been recently performed in
\cite{Buras:2013qja} concentrating on the LHS scenario, which as discussed in 
Step 7 gives a plausible explanation of the $B_d\to K^*(K)\mu^+\mu^-$ data.
Its  outcome 
can be briefly summarized as follows:
\begin{itemize}
\item
The LHS scenario for $Z^\prime$ or $Z$ FCNC couplings  provides
a simple model that allows for the violation of the CMFV relation
between the branching ratios for $B_{d,s}\to \mu^+\mu^-$ and $\Delta M_{s,d}$. 
The plots in Figs.~\ref{fig:BdvsBsLHS} and \ref{fig:ZBdvsBsLHS} for $Z^\prime$ 
and $Z$, respectively,  illustrate this. 
\item
However, to achieve this in the case of $Z^\prime$ the experimental value 
of $\Delta M_s$ must be 
very close to its SM value ($C_{B_s}=1.00\pm0.01$) and $\Delta M_d$ is favoured to be a bit  
{\it larger} than $(\Delta M_d)_{\rm SM}$ ($C_{B_d}=1.04\pm0.01$). $S_{\psi\phi}$ can still deviate 
significantly from its SM value.
\item
In the case of $Z$, both  $\Delta M_s$ and $S_{\psi\phi}$ must be rather close 
to their SM values while $\Delta M_d$ is favoured to be 
{\it smaller} than $(\Delta M_d)_{\rm SM}$ ($C_{B_d}=0.96\pm0.01$).
\end{itemize}
In \cite{Buras:2013qja} details on the dependence of the correlation between 
branching ratios for $B_{s,d}\to\mu^+\mu^-$ and the CP-asymmetries 
$S_{\psi\phi}$ and $S_{\psi K_S}$ on the values of $C_{B_s}$ and $C_{B_d}$ can be 
found. Also the anatomy of the plots in  Figs.~\ref{fig:BdvsBsLHS} and \ref{fig:ZBdvsBsLHS} is presented there.
With the improved data and increased lattice calculations such plots 
will be more informative than presently.

\begin{figure}[!tb]
 \centering
\includegraphics[width = 0.45\textwidth]{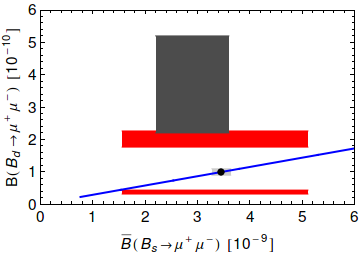}
\includegraphics[width = 0.45\textwidth]{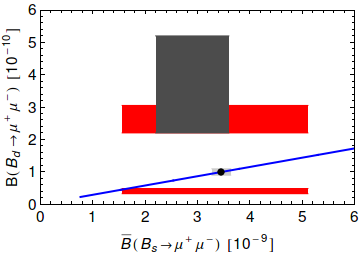}
\caption{ ${\mathcal{B}}(B_{d}\to\mu^+\mu^-)$ versus ${\mathcal{\bar{B}}}(B_{s}\to\mu^+\mu^-)$ in the $Z^\prime$ scenario for $\vub = 0.0034$ (left) and $\vub =
0.0040$ (right) and  $C_{B_d} = 1.04\pm 0.01$, $C_{B_s} = 1.00\pm 0.01$, $\bar{\Delta}_A^{\mu\bar\mu} = 1~\text{TeV}^{-1}$, $0.639\leq
S_{\psi K_s}\leq 0.719$ and $-0.15\leq S_{\psi\phi}\leq 0.15$. SM is represented by the light gray area with black dot and 
 the CMFV prediction by the blue line. Dark gray
 region: Combined exp 1$\sigma$
 range
 $\overline{\mathcal{B}}(B_s\to\mu^+\mu^-) = (2.9\pm0.7)\cdot 10^{-9}$ and $\mathcal{B}(B_d\to\mu^+\mu^-) = (3.6^{+1.6}_{-1.4})\cdot
10^{-10}$.}\label{fig:BdvsBsLHS}~\\[-2mm]\hrule
\end{figure}

\begin{figure}[!tb]
 \centering
\includegraphics[width = 0.45\textwidth]{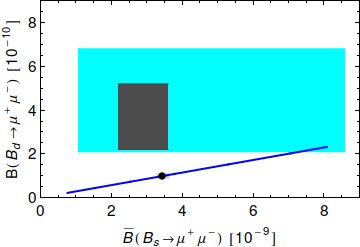}
\includegraphics[width = 0.45\textwidth]{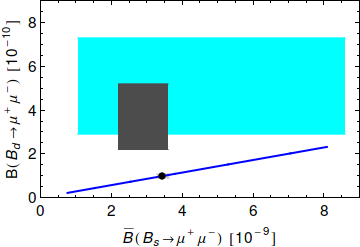}
\caption{ ${\mathcal{B}}(B_{d}\to\mu^+\mu^-)$ versus ${\mathcal{\bar{B}}}(B_{s}\to\mu^+\mu^-)$ in the $Z$-scenario for $\vub = 0.0034$ (left) and $\vub =
0.0040$ (right) and  $C_{B_d} = 0.96\pm 0.01$, $C_{B_s} = 1.00\pm 0.01$, $0.639\leq
S_{\psi K_s}\leq 0.719$ and $-0.15\leq S_{\psi\phi}\leq 0.15$. SM is represented by the light gray area with black dot. Dark gray
 region: Combined exp 1$\sigma$
 range
 $\overline{\mathcal{B}}(B_s\to\mu^+\mu^-) = (2.9\pm0.7)\cdot 10^{-9}$ and $\mathcal{B}(B_d\to\mu^+\mu^-) = (3.6^{+1.6}_{-1.4})\cdot
10^{-10}$.}\label{fig:ZBdvsBsLHS}~\\[-2mm]\hrule
\end{figure}

\boldmath
\subsubsection{$\mathcal{B}(B_s\to\mu^+\mu^-)$ as an Electroweak Precision Test}
\unboldmath
Our review deals dominantly with flavour violation. Yet, in particular NP models  relations between flavour violating and flavour conserving
couplings exist 
so that additional correlations between flavour violating and flavour conserving processes are present. Such correlations can involve on one
hand left-handed 
$Zb\bar b$  and $Zb\bar s$ couplings and on the other hand corresponding 
right-handed 
couplings.  In particular it is known that the measured right-handed $Zb\bar b$ coupling disagrees with its SM value  by  3$\sigma$.
The physics 
responsible for this anomaly can   in some NP models through correlations also have  an impact on FCNC processes. 

Such a correlation has been pointed out first in \cite{Chanowitz:1999jj}, and analyzed in detail 
in the context of MFV in \cite{Haisch:2007ia}.  At that time the information on $Z\to b\bar b$ 
couplings was by far superior to the one from $B_s\to\mu^+\mu^-$ so that 
the bounds on possible deviations of  $Z\to b\bar b$  from their SM values 
implied interesting bounds on FCNC processes, including  $B_s\to\mu^+\mu^-$. 
As pointed out recently in \cite{Guadagnoli:2013mru} the situation is now reversed and  the present data on
 $\mathcal{B}(B_s\to\mu^+\mu^-)$ set already the dominant 
constraints on possible modified flavour diagonal $Z$-boson couplings. In the 
case of MFV models, where significant NP effects are expected only in LH $Z$- couplings, the present bound derived in
\cite{Guadagnoli:2013mru} from  $\mathcal{B}(B_s\to\mu^+\mu^-)$ 
is not much stronger than the one derived from $Z\to b\bar b$. On the other 
hand in generic models with partial compositeness  $\mathcal{B}(B_s\to\mu^+\mu^-)$  sets already now constraint on the RH $Zb\bar b$
coupling that is significantly more stringent than obtained from $Z\to b\bar b$. As a result, in this class of models the present anomaly in
RH $Zb\bar b$ coupling cannot be explained. 
Needless to say, such constraints on diagonal $Zb\bar b$ coupling will become even 
more powerful when the measurement of  $\mathcal{B}(B_s\to\mu^+\mu^-)$ improves  so that this decay will offer electroweak precision tests.

\boldmath
\subsubsection{$B_{s,d}\to\tau^+\tau^-$}
\unboldmath
The leptonic decays $B^0_{s,d}\to\tau^+\tau^-$ could one day play a significant 
role in the tests of NP models. In particular interesting information  on 
the interactions of new particles with the third generation of quarks and 
leptons could be obtained in this manner. In the SM the branching ratios 
in question are enhances by roughly two orders of magnitude over the corresponding decays to the muon pair:
\be
\frac{\mathcal{B}(B_q\to \tau^+\tau^-)}{\mathcal{B}(B_q\to\mu^+\mu^-)}=
\sqrt{1-\frac{4m_\tau^2}{m^2_{B_q}}}\frac{m^2_\tau}{m^2_\mu}\approx 210.
\ee
Tree-level exchange of a neutral SM Higgs with quark flavour violating 
couplings could become important and the same applies to tree-level heavy 
scalar and pseudoscalar exchanges. There are presently no experimental limits 
on these decays, however the interplay with $\Gamma_{12}^s$, and the latest 
measurements of $\Gamma_d/\Gamma_s$ by  LHCb would imply the upper bound 
for branching ratio 
for $B_s^0\to\tau^+\tau^-$ of $3\%$ at $90\%$ C.L. \cite{Dighe:2010nj,Bobeth:2011st}.
Due to significant experimental challenges to observe these decays at the 
LHCb it is then unlikely that  we will benefit from them in this decade and 
we will not discuss them further.

\boldmath
\subsection{Step 5: $B^+\to \tau^+\nu_\tau$}
\unboldmath

\subsubsection{Preliminaries}
We now look at the tree-level decay $B^+ \to \tau^+ \nu$, which was 
the subject of great interest  in the previous dacade as the data from
 BaBar \cite{Aubert:2007xj} and Belle \cite{Ikado:2006un} implied a world average in the ballpark of 
$\mathcal{B}(B^+ \to \tau^+ \nu_\tau)_{\rm exp} = (1.73 \pm 0.35) \times 10^{-4}~$, 
 roughly by a factor of 2 higher than the SM value. Meanwhile, the situation 
changed considerably due to 2012 data from Belle \cite{Adachi:2012mm} so that the present world 
average  that combines BaBar and Belle data reads  \cite{Amhis:2012bh}
\begin{equation} \label{eq:Btaunu_exp}
\mathcal{B}(B^+ \to \tau^+ \nu_\tau)_{\rm exp} = (1.14\pm0.22) 10^{-4}~,
\end{equation}
which is fully consistent with the values quoted in Table~\ref{tab:SMpred} 
with some preference for the inclusive values of $\vub$. 
Yet, the rather large 
experimental error and parametric uncertainties in the SM prediction still
allow in principle for sizable NP contributions. 

In this context one should recall that one of our working assumptions 
was the absence of significant NP contributions to decays governed by 
tree-diagrams. Yet the decay in question could be one of the exceptions 
as it is governed by the smallest element of the CKM matrix $\vub$ and 
its branching ratio is rather small for a tree-level decay. We will therefore 
briefly discuss it in the simplest extensions of the SM. 

The motivation for this study is the sensitivity of this decay to new 
heavy charged gauge bosons and scalars that we did not encounter 
in the previous steps, where neutral gauge bosons and neutral scalars and 
pseudoscalars dominated the scene.

\subsubsection{Standard Model Results}
In the SM $B^+\to \tau^+\nu_\tau$ is
mediated by the $W^\pm$  exchange with the resulting branching ratio  given by
\begin{equation} \label{eq:Btaunu}
\mathcal{B}(B^+ \to \tau^+ \nu_\tau)_{\rm SM} = \frac{G_F^2 m_{B^+} m_\tau^2}{8\pi} \left(1-\frac{m_\tau^2}{m^2_{B^+}} \right)^2 F_{B^+}^2
|V_{ub}|^2 \tau_{B^+}= 6.05~ \vub^2\left(\frac{ F_{B^+}}{185\mev}\right)^2~.
\end{equation}
Evidently this result is subject to significant parametric uncertainties induced in (\ref{eq:Btaunu}) by $F_{B^+}$ and $\vub$. However, recently the 
error on $F_{B^+}$  from lattice QCD decreased significantly so that 
the dominant uncertainty comes from $\vub$. Indeed, as seen in 
Table~\ref{tab:SMpred}, for fixed remaining input parameters, varying $\vub$ 
in the range shown in this table modifies the branching ratio by roughly a 
factor of two.

In the literature
in order to find the SM prediction for this branching ratio one eliminates 
these uncertainties by using $\Delta M_d$,  $\Delta M_d/\Delta M_s$ and 
$S_{\psi K_S}$ \cite{Bona:2009cj,Altmannshofer:2009ne}  
and taking experimental 
values for these three quantities.  To this end $F_{B^+}=F_{B_d}$ is assumed in agreement with lattice values. This strategy has 
a weak point as 
 the experimental 
values of $\Delta M_{d,s}$ used 
in this strategy may not be the ones corresponding to the true value of 
the SM. However, proceeding in this manner one finds 
\cite{Altmannshofer:2009ne}
\begin{equation}\label{eq:BtaunuSM1}
\mathcal{B}(B^+ \to \tau^+ \nu)_{\rm SM}= (0.80 \pm 0.12)\times 10^{-4},
\end{equation}
with a similar result obtained 
by the UTfit collaboration
\cite{Bona:2009cj}.   As seen in
Table~\ref{tab:SMpred} this 
result corresponds to $\vub$ in  the ballpark of $3.6\times 10^{-3}$ and 
 is fully consistent with the data
in~(\ref{eq:Btaunu_exp}).

Unfortunately, the full clarification of a possible
presence of NP in this decay  will have to wait for the
data from SuperKEKB. In the meantime hopefully the error on 
$F^+_B$ from lattice QCD will be further reduced and theoretical advances 
in the determination of  $\vub$ from tree level decays will be 
made allowing us to make a precise prediction 
for this decay without using the experimental value for $\Delta M_d$.

It should be emphasized that for low value of $\vub$ 
the increase of $F_{B^+}$, while enhancing 
the branching ratio in question, would also enhance $\Delta M_d$ which 
in view of our discussion in Step 3 is not favoured by the data. On the 
other hand the increase of $\vub$ while  enhancing 
$\mathcal{B}(B^+ \to \tau^+ \nu)_{\rm SM}$ would also enhance
 $S_{\psi K_S}$  { shifting it away from the data.} This discussion shows clearly 
that before all these parameters will be known significantly more precisely 
than it is the case now, it will be difficult to use this decay for 
the identification of NP. In fact the decays $B_{s,d}\to \mu^+\mu^-$ 
are presently in a much better shape than $B^+ \to \tau^+ \nu$ as 
they are governed by $\vts$, which is presently much better known than 
$\vub$.

In view of this uncertain situation our look at the simplest models 
providing new contributions to this decay will be rather brief.


\subsubsection{CMFV}

To our knowledge $B^+\to\tau^+\nu_\tau$ decay has never been considered in 
CMFV. Here we would like to point out that in this class of models the 
branching ratio for this decay is enhanced (suppressed) for the same (opposite)  sign of the lepton coupling of the new charged gauge boson 
relative to the SM one.  Indeed, the only possibility to modify the SM result up to 
loop corrections in CMFV is through a tree-level exchange of a new charged gauge 
boson, whose flavour interactions with quarks are governed by the CKM 
matrix. In particular the operator structure is the same.

Denoting this gauge boson by $W^\prime$ and the corresponding 
gauge coupling by $\tilde{g}_2$ one has
\be
\frac{\mathcal{B}(B^+\to\tau^+\nu)}{\mathcal{B}(B^+\to\tau^+\nu)^{\rm SM}}=
\left(1+r\frac{\tilde g_2^2}{g^2_2}\frac{M_W^2}{M_{W^\prime}^2}\right)^2,
\ee
where we introduced a factor $r$ allowing a modification 
in the lepton couplings relatively to the SM ones, in particular of its sign. 
Which sign is favoured will be known once the data and SM prediction improve.

If $W^\prime$ with these properties is absent, the branching ratio in this 
framework is not modified with respect to the SM up to loop 
corrections that could involve new particles but are expected to be 
small. A $H^\pm$ exchange generates new operators and is outside this framework.
The same comment applies to gauge bosons with right-handed couplings that we 
will discuss below.

\boldmath
\subsubsection{${\rm 2HDM_{\overline{MFV}}}$}
\unboldmath

Interestingly, when the experimental branching ratio was significantly above 
its SM value, the tension between  theory and experiment in the 
case of $\mathcal{B}(B^+\to\tau^+\nu)$
 increased in the presence of a tree level $H^\pm$
exchange. Indeed such a contribution 
interferes destructively with the $W^\pm$ contribution 
if there are no new sources of CP-violation.
This effect has been calculated  long time ago by 
Hou \cite{Hou:1992sy} and in 
modern times calculated first by Akeroyd and Recksiegel 
\cite{Akeroyd:2003zr}, and later by
Isidori and Paradisi \cite{Isidori:2006pk} in the context of the MSSM. 
The same expression is valid in  ${\rm 2HDM_{\overline{MFV}}}$ framework and 
is given as follows
\cite{Blankenburg:2011ca} 
 \be\label{BP1}
\mathcal{B}(B^+\to\tau^+\nu)_{\rm  2HDM_{\overline{MFV}}}=
{\mathcal{B}(B^+\to\tau^+\nu)_{\rm SM}}
\left[1-\frac{m_B^2}{m^2_{H^\pm}}\frac{\tan^2\beta}{1+(\epsilon_0+\epsilon_1)\tan\beta}
\right]^2.
\ee
In the MSSM  $\epsilon_i$ are  calculable in terms of supersymmetric parameters. In ${\rm 2HDM_{\overline{MFV}}}$ they are just universal parameters that 
can enter other formulae implying correlations between various observables. 
If $\epsilon_i$ are real, positive definite numbers, similarly to MSSM, also 
in this model 
this branching ratio can be strongly suppressed unless 
the choice of model parameters is such that the second term in the parenthesis
is larger than 2. 
Such a
possibility that would necessarily imply a light charged Higgs and large $\tan\beta$ values
seems to be very unlikely in view of the constraints from other 
observables as stressed in the past in the context of MSSM in 
\cite{Antonelli:2008jg} and more recently in the context of the ${\rm 2HDM_{\overline{MFV}}}$ in \cite{Blankenburg:2011ca}.

However,  
Isidori and Blankenburg point out that in ${\rm 2HDM_{\overline{MFV}}}$, 
where $\epsilon_0$ and $\epsilon_1$ are complex numbers 
\be
1+(\epsilon_0+\epsilon_1)\tan\beta\le 0
\ee
is possible provided $\tan\beta$ is large. But then these authors 
find $\mathcal{B}(B\to X_s\gamma)$ to be suppressed relative to the SM 
which is not favoured by the data. 
We will discuss this issue in the next step. 

Let us stress in 
this context that the subscript ``SM'' in (\ref{BP1})  could be misleading 
as what is really meant there, is the formula for this decay in the SM. 
While the SM selects the low ({\it exclusive}) value for $\vub$ in order to 
be in agreement with the experimental value of $S_{\psi K_S}$, the 
 ${\rm 2HDM_{\overline{MFV}}}$ chooses the large ({\it inclusive}) value of 
$\vub$ in order to be consistent with experimental value of $\varepsilon_K$. 
The resulting problem with $S_{\psi K_S}$ is then solved 
as discussed in Step 3 by new phases in $B^0_d-\bar B^0_d$ mixing. But with 
the inclusive value of $\vub$, $\mathcal{B}(B^+\to\tau^+\nu)$ is enhanced 
and as seen in Table~\ref{tab:SMpred} agreement with the data can be obtained.

It appears then that the simplest solution to the possible problem with 
$\mathcal{B}(B^+\to\tau^+\nu)$ in this model is the absence of relevant 
charged Higgs contributions to this decay and sufficiently large value 
of $\vub$.

\subsubsection{Tree-Level Charged Gauge Boson Exchange}

Let us write the effective Hamiltonian for the exchange of a  
charged gauge bosons $W^{\prime +}$ contributing to $B^+ \to \tau^+ \nu_\tau$ as 
follows
\be
{\cal H}_{\rm eff}=C_LO_L+C_RO_R,
\ee
where
\be 
O_L=(\bar b\gamma_\mu P_Lu)(\bar\nu_\tau\gamma^\mu P_L\tau^-), 
\quad
O_R=(\bar b\gamma_\mu P_Ru)(\bar\nu_\tau\gamma^\mu P_L\tau^-)
\ee
and 
\be
C_L=C_L^{\rm SM}+\frac{\Delta_L^{ub*}(W^{\prime +})\Delta_L^{\tau\nu}(W^{\prime +})}{M_{W^{\prime +}}^2}, \quad
C_R= \frac{\Delta_R^{ub*}(W^{\prime +})\Delta_L^{\tau\nu}(W^{\prime +})}{M_{W^{\prime +}}^2}
\ee
with $C_L^{\rm SM}$ having the same structure as the correction from $W^{\prime +}$ with 
\be
\Delta_L^{ub}=\frac{g}{\sqrt{2}}V_{ub}, \qquad \Delta_L^{\nu\tau}=\frac{g}{\sqrt{2}}, \quad \Delta_R^{ub}=0.
\ee
The couplings $\Delta_{L,R}^{ub*}(W^{\prime +})$ could be complex numbers and  contain 
new sources of flavour violation.

Then 
\begin{equation} \label{eq:Btaunu-verygeneral}
\mathcal{B}(B^+ \to \tau^+ \nu_\tau)_{\rm W^{\prime +}} = 
\frac{1}{64\pi}m_{B^+} m_\tau^2 \left(1-\frac{m_\tau^2}{m^2_{B^+}} \right)^2 F_{B^+}^2
\tau_{B^+}|C_R-C_L|^2.
\end{equation}
 Evidently in a model like this it is possible to improve the 
agreement with the data by choosing 
 appropriately the couplings of $W^{\prime +}$.

\subsubsection{Tree-Level  Scalar Exchanges}

We have already discussed such exchanges in the context of  ${\rm 2HDM_{\overline{MFV}}}$. Here we want to mention for completeness that
the decay $B^+\to D^0\tau^+\nu$ being 
sensitive to different couplings of $H^\pm$ can contribute significantly 
to this discussion but form factor uncertainties make this decay less
theoretically clean. A thorough analysis of this decay is presented 
in \cite{Nierste:2008qe} where further references to older papers can be found.

Recently the BABAR collaboration \cite{Lees:2012xj} presented improved analyses
for the ratios
\be
\mathcal{R}(D^{(*)})=\frac{\mathcal{B}(B_d\to D^{(*)}\tau\nu)}{\mathcal{B}(B_d\to D^{(*)}\ell\nu)}
\ee
finding
\be
\mathcal{R}(D)=0.440\pm0.058\pm0.042, \qquad \mathcal{R}(D^*)=0.332\pm0.024\pm0.018
\ee
where the first error is statistical and the second one systematic. These 
results disagree by $2.2\sigma$ and $2.7\sigma$ with the SM, 
respectively \cite{Fajfer:2012vx}
\be
\mathcal{R}_{\rm SM}(D)=0.297\pm0.017, \qquad \mathcal{R}_{\rm SM}(D^*)=0.252\pm0.003~.
\ee
These values update the ones presented first in \cite{Kamenik:2008tj}.

This motivated several theoretical analyses of which we just 
quote four. First the study of these decays in 2HDM of type III \cite{Crivellin:2012ye,Crivellin:2013wna} and in NP models 
with general flavour structure \cite{Fajfer:2012jt}. Moreover in \cite{Ko:2012sv} 2HDM and 3HDM models with the nonminimal flavor violations  originating 
from flavour-dependent gauge interactions have been analyzed.
 It is to be seen whether 
this anomaly remains when the data improve. A recent summary of the situation 
can be found in \cite{Crivellin:2013mba}. In particular 2HDM of type II cannot simultaneously 
describe the data on $\mathcal{R}(D)$ and $\mathcal{R}(D^*)$ but this is 
possible in 2HDM of type III.

In summary it is evident from this discussion that  $B^+ \to \tau^+ \nu_\tau$,  
 $B\to D\tau\nu$ and $B\to D^*\tau\nu$ 
can play a potential role in constraining NP models. Yet, due to the fact 
that the data in the case of $B^+ \to \tau^+ \nu_\tau$   moved significantly towards the SM and because of 
large uncertainty in $\vub$, the identification of a concrete NP at work in this decay
appears to us presently as a big challenge.  The decays $B\to D\tau\nu$ and $B\to D^*\tau\nu$  seem to be more promising but we 
have to wait for improved 
data as well. It looks like in the  SuperKEKB era these three decays taken 
together will be among the stars of flavour physics.

\boldmath
\subsection{Step 6: 
$B\to X_{s}\gamma$ and $B\to K^*\gamma$}
\unboldmath
\subsubsection{Standard Model Results}
The radiative decays in question, in particular $B\to X_s\gamma$, played 
an important role in constraining NP in the last two decades because both 
the experimental data and also the theory have been already in a good
shape for some time. 

The  Hamiltonian  in the SM is given as follows
{\begin{equation} \label{Heff_at_mu}
{\cal H}_{\rm eff}(b\to s\gamma) = - \frac{4 G_{\rm F}}{\sqrt{2}} V_{ts}^* V_{tb}
\left[  C_{7\gamma}(\mu_b) Q_{7\gamma} +  C_{8G}(\mu_b) Q_{8G} \right]\,,
\end{equation}}
where $\mu_b=\ord(m_b)$.
The dipole operators are defined as
\begin{equation}\label{O6B}
Q_{7\gamma}  =  \frac{e}{16\pi^2} m_b \bar{s}_\alpha \sigma^{\mu\nu}
P_R b_\alpha F_{\mu\nu}\,,\qquad            
Q_{8G}     =  \frac{g_s}{16\pi^2} m_b \bar{s}_\alpha \sigma^{\mu\nu}
P_R T^a_{\alpha\beta} b_\beta G^a_{\mu\nu}\,. 
\end{equation}
 While we do not show explicitly the four-quark operators in (\ref{Heff_at_mu}) they are very important for decays considered in this step, in particular 
as far as QCD and electroweak corrections are concerned.

The special role of these decays is that quite generally they  are  loop 
generated processes. As such there are sensitive to NP contributions and 
in contrast to tree-level FCNCs mediated by neutral gauge bosons and 
scalars depend often on the masses and couplings of new heavy fermions. 
But of course new heavy gauge bosons and scalars  contribute to these 
decays in many models as well.
At the CKM-suppressed level, tree-level
$b\to u \bar u s\gamma$ transitions can also contribute but they are small for
the photon energy cut-off $1.6\gev$ usually used 
\cite{Kaminski:2012eb}.

The NNLO QCD calculations of $\mathcal{B}(B\to X_s\gamma)$, that involve a very important 
mixing of dipole operators with current-current operators, have been in the last decade 
at the forefront of perturbative QCD calculations in weak decays. 
The first outcome of these efforts, which included the dominant NNLO 
corrections was already  a rather precise prediction within the SM
\cite{Misiak:2006ab}\footnote{For a historical account of NLO and NNLO corrections to 
this decay see \cite{Buras:2011we}.}
\be\label{bsgth0}
\mathcal{B}(B\to X_s\gamma)_{\rm SM}=(3.15\pm0.23)\times 10^{-4}\,, \qquad (2013)
\ee 
for $E_\gamma\ge 1.6\gev$.
Since then, several new perturbative
contributions have been evaluated
\cite{Czakon:2006ss,Boughezal:2007ny,Asatrian:2006rq,Ewerth:2008nv,Asatrian:2010rq,Ferroglia:2010xe,Misiak:2010tk,Kaminski:2012eb}. Most
recently, the $Q_{1,2}-Q_7$
interference was found in the $m_c=0$ limit \cite{Misiaketal}. An updated 
 NNLO prediction should be available soon.

Also experimentalists made an impressive 
progress in measuring this branching ratio reaching the accuracy of 
$6.4\%$ \cite{Amhis:2012bh}
\be\label{bsgexp}
\mathcal{B}(B\to X_s\gamma)_{\rm exp}=(3.43\pm0.22)\times 10^{-4}\,.
\ee
One expects that in this decade the SuperKEKB 
will reach the accuracy of $3\%$ so that very precise tests of the 
SM and its extensions will be possible.

Comparing the theory with experiment we observe that the experimental value 
is a bit higher than the theory although presently the difference 
amounts to only $1.2\sigma$. However, if the experimental and theoretical errors 
decrease down to $3\%$ without the change in central values we will be 
definitely talking about an anomaly and models in which this branching 
ratio will be enhanced over the SM result will be favoured. Yet, such models 
have to satisfy other constraints as well.

In principle a very sensitive observable to NP CP violating effects is
the direct CP asymmetry in $b\to s\gamma$, i.e. $A_{\rm CP}(b\to s\gamma)$~\cite{Soares:1991te}, because the perturbative contributions
within the SM amount to only $+0.5\%$  \cite{Kagan:1998bh,Kagan:1998ym,Hurth:2003dk}.
Unfortunately, the analysis \cite{Benzke:2010tq} shows that this asymmetry, similar to other direct CP 
asymmetries,  suffers from hadronic uncertainties originating here in the hadronic 
component of the photon. These uncertainties lower the predictive power 
of this observable. Consequently  we do not consider this asymmetry 
as a superstar of flavour physics and will not include it in our investigations.
Similar comments apply to the $B\to X_d\gamma$ decay although CP averaged 
branching ratio could still provide useful results. Yet, we will leave 
this decay from our discussion as well, as the remaining observables considered in our paper are evidently more effective in the search for
NP from the present 
perspective.

 Concerning $B\to V \gamma$ decay we refer first to two fundamental papers 
that include NLO QCD corrections \cite{Bosch:2001gv,Bosch:2004nd}. While the 
branching ratios can already offer useful information, even more
promising is 
the time-dependent
CP asymmetry in   $B\to K^*\gamma$ \cite{Atwood:1997zr,Ball:2006cva,Ball:2006eu}
\begin{equation} \label{eq:SKstargamma}
\frac{\Gamma(\bar B^0(t) \to \bar K^{*0}\gamma) - \Gamma(B^0(t) \to K^{*0}\gamma)}{\Gamma(\bar B^0(t) \to \bar K^{*0}\gamma) + \Gamma(B^0(t)
\to K^{*0}\gamma)} = S_{K^*\gamma} \sin(\Delta M_d t) - C_{K^*\gamma} \cos(\Delta M_d t)~.
\end{equation}
In particular $S_{K^*\gamma}$ offers a very  sensitive probe of 
 right-handed currents. It vanishes
for $C_{7\gamma}^\prime \to 0$ and consequently in the SM being suppressed  by $m_s/m_b$ is very small~\cite{Ball:2006eu}:
\begin{equation}
S_{K^*\gamma}^{\rm SM} = (-2.3 \pm 1.6)\%~.
\label{eq:SKgSM}
\end{equation}

A useful and rather accurate expression for $S_{K^*\gamma}$ has been provided in \cite{Ball:2006cva}
\begin{equation} \label{eq:SKstargamma_NP}
S_{K^*\gamma} \simeq \frac{2}{|C_{7\gamma}|^2 + |C_{7\gamma}^\prime|^2} {\rm Im}\left( e^{-i\phi_d} C_{7\gamma} C_{7\gamma}^\prime\right)~,
\end{equation}
with Wilson coefficients evaluated at $\mu=m_b$ and 
$\sin(\phi_d) = S_{\psi K_S}$.

On the experimental side, while the present value of $S_{K^*\gamma}$ is 
rather inaccurate
\cite{Ushiroda:2006fi,Aubert:2008gy,Asner:2010qj}
\begin{equation}
S_{K^*\gamma}^{\rm exp} = -0.16 \pm 0.22,
\end{equation}
the prospects for accurate  measurements at SuperKEKB are 
 very good \cite{Meadows:2011bk}.

Also isospin asymmetries in $B\to V\gamma$ provide interesting tests 
of the SM and of NP. A detailed recent analysis with references to earlier 
papers can be found in \cite{Lyon:2013gba}. On the experimental side the 
isospin asymmetry in $B\to K^*\gamma$ agrees with the SM, while a $2\sigma$ 
deviation from the SM is found in the case of  $B\to \rho\gamma$ \cite{Amhis:2012bh}.

 \boldmath
\subsubsection{$B \to X_s\gamma$ Beyond  the Standard Model}
\unboldmath
Our discussion of NP contributions to this decay will be very brief. 
The latest review can be found in \cite{Haisch:2008ar}   and 
a detailed analysis of the impact of anomalous $Wtb$ couplings 
has been presented in \cite{Grzadkowski:2008mf}, where further 
references to earlier literature can be found.

As the SM agrees  well with the 
data, NP contributions can be at most in the ballpark of  $20\%$ at the 
level of the branching ratio and they should rather be positive than negative. Consequently this decay will mainly bound 
the parameters of a given extension of the SM. Here we only make a few 
comments.

It is known that $B \to X_s\gamma$ can bound the allowed range of the 
values of charged Higgs ($H^\pm$) mass and of $\tan\beta$ both in 2HDM and 
the MSSM. In 2HDM II the contribution of $H^\pm$ enhances the 
branching ratio and $M_{H^\pm}$ must be larger than $300\gev$ for any 
value of $\tan\beta$. In the MSSM this enhancement can be compensated by 
chargino contributions and the bound is weaker.

As we already stated and discussed in more detail in \cite{Haisch:2008ar} 
the fact that the SM prediction is below the data favours presently 
the models that allow for an enhancement of the branching ratio and 
disfavours those in which only suppression is possible. Table 1 in 
\cite{Haisch:2008ar} is useful in this respect. In particular,
\begin{itemize}
\item
In 2HDM II, Littlest Higgs model without T-parity (LH) and RS $\mathcal{B}(B\to X_s\gamma)$ can only be enhanced and in LHT the enhancement is favoured.
\item
In MFV SUSY GUTs \cite{Altmannshofer:2008vr} and in models with universal 
extra dimensions it can only be suppressed. In particular in the latter 
case lower bound on the compactification scale $1/R$ of $600\gev$ 
can be derived \cite{Agashe:2001xt,Buras:2003mk,Haisch:2007vb,Freitas:2008vh} in this manner.
\item
In more complicated models like MSSM with MFV, general MSSM and 
left-right models both enhancements and suppressions are possible.
\end{itemize}

Another important virtue of this decay is its sensitivity to right-handed (RH) currents. In the case of left-handed (LH) currents the
chirality flip, necessary for $b\to s\gamma$ to occur, can only proceed through the mass of the initial or the final quark. Consequently the
amplitude is proportional to $m_b$ or 
$m_s$. In contrast, when RH currents are present, the chirality flip can take 
place on the internal top quark line resulting in an enhancement factor 
$m_t/m_b$ of the NP contribution relatively to the SM one at the level of the 
amplitude. This is the case of left-right symmetric models in which $B \to X_s\gamma$ has been analyzed by many authors in the past
\cite{Asatrian:1989iu,Asatryan:1990na,Cocolicchio:1988ac,Cho:1993zb,Babu:1993hx,Fujikawa:1993zu,Asatrian:1996as,Bobeth:1999ww,Frank:2010qv,
Guadagnoli:2011id,Blanke:2011ry}. In models 
with heavy fermions ($F$), that couple through RH currents to SM quarks, this 
enhancement, being proportional to $m_F/m_b$ can be very large  \cite{Buras:2011wi} and the 
couplings in question must be strongly suppressed in order to obtain agreement 
with the data. This is for instance the case of gauge flavour models which 
we will briefly describe in Section~\ref{sec:5}. It should be emphasized that
the comments on the $m_t/m_b$ and $m_F/m_b$ enhancements apply also for 
charged and neutral gauge bosons as well as for charged and neutral heavy 
scalars and pseudoscalars.

\boldmath
\subsection{Step 7: $B\to X_s\ell^+\ell^-$ and    $B\to K^*(K)\ell^+\ell^-$}
\unboldmath

\subsubsection{Preliminaries}
While  the branching ratios for $B\to X_s\ell^+\ell^-$ and  $B\to K^*\ell^+\ell^-$
put already significant constraints on NP, the angular observables, 
CP-conserving ones like the well known forward-backward asymmetry 
and CP-violating ones will definitely be useful for distinguishing
various extensions of the SM when the data improve. During the last three years, a number of detailed analyses 
of various CP averaged symmetries ($S_i$) and CP asymmetries ($A_i$) 
provided by the 
angular distributions in the exclusive decay $B\to K^*(\to K\pi)\ell^+\ell^-$
have been performed in 
\cite{Bobeth:2008ij,Egede:2008uy,Altmannshofer:2008dz,Bobeth:2010wg,Altmannshofer:2011gn,DescotesGenon:2011pb,Altmannshofer:2012ir,
Becirevic:2012fy,Bobeth:2011gi,Beaujean:2012uj,DescotesGenon:2012zf,Jager:2012uw}. 
In particular the zeros of some of these 
observables can be accurately predicted. Pioneering experimental analyses 
 performed at BaBar, Belle and Tevatron \cite{Wei:2009zv,Aaltonen:2011ja,Lees:2012tva} provided already
interesting results for the best known forward-backward asymmetry. 
Yet, the recent data from LHCb \cite{Aaij:2011aa,Aaij:2013iag} surpassed the latter ones in 
precision demonstrating 
that the SM is consistent with the present data on the forward-backward 
asymmetry. On the other hand these decays  as we will see below bring 
 new challenges { as the data on $A_i$ and $S_i$  improved last year. Yet 
 in order to reach clear cut conclusions further improvement in 
the data and the reduction of theoretical uncertainties is necessary.
Meanwhile, the present data serve already to bound the parameters in several
 extensions of the SM.}

Compared with previous steps, this one is more challenging as far as the
transparency is concerned. Indeed the effective Hamiltonian for these decays 
involves more local operators and corresponding Wilson coefficients that 
generally are complex quantities. On the other hand the numerous 
symmetries $S_i$ and asymmetries $A_i$ when precisely measured 
will allow one day a detailed insight into the physics behind the 
values of the Wilson coefficients in question. 
In this context it is important to select those $S_i$ and $A_i$ which are particularly useful for the 
tests of NP and are not subject to large form factor uncertainties. While significant progress in this direction has been
already 
done in the literature, a more transparent picture will surely 
emerge 
once the precision on these angular observables will increase with time. 
The  most recent reviews on various optimal strategies for extraction 
of NP from angular observables can be found in \cite{Descotes-Genon:2013vna,Descotes-Genon:2013hba}. Details on these strategies can be
found in 
\cite{Kruger:2005ep,Altmannshofer:2008dz,Egede:2008uy,Bobeth:2010wg,Egede:2010zc,Bobeth:2011gi,Becirevic:2011bp,Bobeth:2012vn,
DescotesGenon:2012zf,Matias:2012xw}.

While it appears from the present perspective that the observables in 
$B_{s,d}\to\mu^+\mu^-$ decays are subject to smaller hadronic uncertainties 
than observables considered here, the strength of $B \to K^{*}\mu^+\mu^-$ 
is not only the presence of several symmetries $S_i$ and asymmetries $A_i$ or 
other constructions like $A_T^i$, $P_i$, $H_T^i$ and alike.
Indeed,  also the presence of an additional variable, the invariant 
mass of the dilepton $(q^2)$, is an important virtue of these decays. 
Studying different observables in different $q^2$ 
bins can indeed one day, as stressed in particular in \cite{Bobeth:2010wg,Bobeth:2011gi,Descotes-Genon:2013vna,Descotes-Genon:2013hba}, not
only help to discover NP, but also to identify it.  The most recent 
study \cite{Descotes-Genon:2013wba} of the so-called {\it primary observables} $P_i$ and $P^\prime_i$ introduced in 
\cite{DescotesGenon:2012zf,Descotes-Genon:2013vna} in the context of the most recent LHCb 
data \cite{Aaij:2013iag,Aaij:2013qta} illustrates this in explicit terms and we 
 will return to these data and the related analyses 
\cite{Descotes-Genon:2013wba,Altmannshofer:2013foa}
 below.

 The story of departures of LHCb data from the SM in the decays in question 
is rather involved but interesting.
In particular 
previous indications for a deviation from SM value of
the isospin asymmetry in $B \to K^{*}\mu^+\mu^-$ decay now disappeared \cite{Aaij:2012cq}. On
the other hand the corresponding asymmetry in $B \to  K \mu^+\mu^-$ decay disagrees 
presently with the SM \cite{Aaij:2012cq}. A recent very detailed analysis of 
the isospin asymmetries in these decays can be found in  \cite{Lyon:2013gba}.

 On the other hand as pointed out in \cite{Descotes-Genon:2013wba,Altmannshofer:2013foa}  and analyzed in detail 
sizable departures from the SM expectations
     in some of the observables $P_i$ or $S_i$ 
are seen in most recent LHCb data \cite{Aaij:2013iag,Aaij:2013qta}.

In order to have a closer look at these issues we need the effective 
Hamiltonian for these decays.  It is given in
(\ref{eq:Heffqll}) with the first 
term given in  (\ref{Heff_at_mu}).
The stars in these decays are the Wilson coefficients  entering this
Hamiltonians. The most important are 
\be\label{WCstars}
 C_{7\gamma},\quad C_9,\quad C_{10},\quad  C^\prime_{7\gamma},
\quad C^\prime_{9},\quad C^\prime_{10}
\ee
where the primed Wilson coefficients correspond to primed operators obtained 
through the replacement $P_L\leftrightarrow P_R$. The scalar and pseudoscalar 
coefficients are more constrained by $B_{s}\to\mu^+\mu^-$ decay but 
we will make few comments on them below.

 The values of the coefficients in (\ref{WCstars}) 
have been calculated in the SM and in its numerous extensions. 
Moreover, they have been constrained in model independent analyses in which 
they have been considered as real or complex parameters. To this end 
the data on $B\to X_s\gamma$, $B\to K^*\gamma$, $B\to X_s\ell^+\ell^-$, 
$B\to K^*\ell^+\ell^-$, $B\to K\ell^+\ell^-$ and $B_s\to\mu^+\mu^-$ have been used. The fact 
that these coefficients enter universally in a number of observables 
allows to obtain correlations between their values. We just refer to selected 
 papers which we found particularly useful for our studies of NP. These 
are  \cite{Altmannshofer:2011gn,Altmannshofer:2012ir,Becirevic:2012fy,Altmannshofer:2013foa}, where  model-independent 
constraints on NP in $b\to s$ transitions have been updated and generalized. 
Further references can be found there and in the text above.

It is useful to consider $B\to X_s\ell^+\ell^-$ decay and 
$B\to K^*\ell^+\ell^-$ in two different regions of the dilepton invariant mass. The low $q^2$ region with 1~GeV$^2 < q^2 < 6$~GeV$^2$, considered already for 
a long time and
the high $q^2$ region with $q^2 > 14.4$~GeV$^2$  which became very relevant 
after theoretical progress made in \cite{Beylich:2011aq}. 
First, in these regions one is not sensitive to the $\bar c c$ resonances. 
Moreover while the branching ratios in the high $q^2$ region are mainly sensitive to NP contributions to the Wilson coefficients
$C_9^{(\prime)}$ and $C_{10}^{(\prime)}$, the branching ratio in the low $q^2$ region {\it also} depends strongly on 
$C_{7\gamma}^{(\prime)}$. Therefore, one expects some correlation between NP 
contributions at low $q^2$ and those in $B\to X_s\gamma$ decay.

In \cite{Altmannshofer:2011gn,Altmannshofer:2012ir} the NP scenarios without important contributions from scalar operators have been
considered. Various analyses show that once the experimental  upper bound on the branching ratio for $B_{s}\to\mu^+\mu^-$ has been taken
into account, the impact of pseudoscalar operators 
$O_P^{(\prime)}$ on $B\to X_s \ell^+\ell^-$ and 
$B\to K^*(K)\ell^+\ell^-$ is minor. However, as stressed in 
\cite{Altmannshofer:2008dz} when lepton mass effects are taken into account 
there is one observable among the many measured in $B\to K^*\ell^+\ell^-$ that 
is sensitive to scalar operators $O_S^{(\prime)}$. This is interesting as 
$B_{s,d}\to\mu^+\mu^-$ decays involve generally both scalar and pseudoscalar 
operators. In this sense angular distribution in $B\to K^*\ell^+\ell^-$ allows 
 to probe the scalar sector of a theory beyond the SM, in a way that is 
theoretically clean and complementary to $B_s\to\mu^+\mu^-$. We refer for 
more details to \cite{Altmannshofer:2008dz}, in particular to Fig.~5 of that 
paper. However, the recent very  improved result  from LHCb and CMS on  $B_s\to\mu^+\mu^-$ in (\ref{LHCb2}) imposed on this 
figure precludes this study
from present 
perspective. 

 While $B\to   K^*\ell^+\ell^-$
 is not as theoretically clean as $B_s\to\mu^+\mu^-$ because of the
presence of form factors, recent advances in lattice calculations \cite{Horgan:2013hoa} give some hopes for improvements. This is also the 
case of  $B\to   K\ell^+\ell^-$, where 
 progress in  lattice calculations of the relevant form factors has been 
reported in  \cite{Bouchard:2013mia,Bouchard:2013eph}. 

As stressed in particular in \cite{Becirevic:2012fy} a
simultaneous
consideration of $B\to   K\ell^+\ell^-$ together with $B_s\to\mu^+\mu^-$  provides
useful tests of extensions of the SM. Indeed, while $B_s\to\mu^+\mu^-$
is sensitive only to the differences  $C_P-C_P'$ and  $C_S-C_S'$, the
decay $B\to   K\ell^+\ell^-$ is sensitive to their sums  $C_P+C_P'$ and  $C_S+C_S'$. A very extensive model independent
analysis of $C_P(C_P')$ and  $C_S(C_S')$ in the context of the data on  $B_s\to\mu^+\mu^-$ and $B\to  K\ell^+\ell^-$ has been
performed in \cite{Becirevic:2012fy}. With improved data a new insight on the 
importance of scalar  and pseudoscalar operators will be possible.

As we already stated above the picture resulting from these analyses is very rich and a brief 
summary of these sometimes numerically challenging analyses is 
a challenge in itself. In what follows we will limit our discussion to a 
number of observations referring to the rich literature for details, in 
particular to 
\cite{Altmannshofer:2011gn,Altmannshofer:2012ir,Becirevic:2012fy,Altmannshofer:2013foa}, as the spirit of these papers fits 
well to our strategies.

\subsubsection{Lessons from Recent Analyses}
The studies of these decays in the SM and its extensions  
have been the subject of numerous analyses almost for the last twenty years
\cite{Bobeth:1999mk,Asatrian:2001de,Asatryan:2001zw,Ghinculov:2003qd,Huber:2005ig,Ligeti:2007sn,Greub:2008cy}.
The most recent studies can be found in
\cite{Bobeth:2008ij,Bobeth:2010wg,Bobeth:2011gi,DescotesGenon:2011yn,Altmannshofer:2011gn,Descotes-Genon:2013wba,Altmannshofer:2013foa,Hambrock:2013zya},
where references to the older papers can be
found. The progress 
in the recent years is the inclusion in these analyses of the data on angular 
observables in 
 $B\to K^*\ell^+\ell^-$. In the simplest case the allowed ranges 
in the space of the real or imaginary parts of a pair of Wilson coefficients, 
or in the complex plane of a single Wilson coefficient are shown. As stressed 
in \cite{Altmannshofer:2011gn} the conclusions drawn from such studies are only 
valid if the chosen Wilson coefficients are indeed the dominant ones in 
a given NP scenario. In fact this is approximately the case in a number 
of models considered in the literature. Few examples are:
\begin{itemize}
\item
 In MFV models with dominance of $Z$ penguins and without new sources of CP violation only the real parts of $C_{7\gamma}$ and $C_{10}$ are
relevant.
\item
In MSSM with MFV and flavour blind phases \cite{Altmannshofer:2008hc}, in effective SUSY with flavour blind phases \cite{Barbieri:2011vn}
and in effective SUSY with a $U(2)^3$ symmetry \cite{Barbieri:2011ci,Barbieri:2011fc}, NP effects in $\Delta B=\Delta S=1$ processes are
dominated by complex contributions to $C_7$ and $C_8$. 
\end{itemize}

The analysis of this type in  \cite{Altmannshofer:2011gn} uses 
the data on 
$B\to K^*\mu^+\mu^-$ at low and high $q^2$, $B\to X_s\ell^+\ell^-$, $B\to X_s\gamma$ and  $B\to K^*\gamma$. The resulting Fig.~2 in that
paper containing 
twelve plots depicts the allowed ranges for various pairs of real and/or 
imaginary parts of chosen Wilson coefficients. While very impressive, such 
plots are rather difficult to digest at first side. Yet the message from 
this analysis is clear. Already present data can exclude sign-flips of 
certain coefficients in certain NP scenarios relative to SM values. Such 
plots will be more informative when the data improve.  

As in many NP models several Wilson coefficients could be affected by new 
contributions,   the authors of \cite{Altmannshofer:2011gn} perform probably 
for the first time a global fit of all Wilson coefficients. In this context 
in addition to the general case, they consider specific examples of NP 
scenarios similar in spirit to the ones introduced in Section~\ref{sec:1}. 
These are the cases of real LH currents, complex LH currents and complex 
RH currents. Again 32 plots resulting from this study shows the complexity 
of such analyses. With improved data such plots will be useful for 
obtaining an insight into the physics involved.  Even if some 
time passed since this analysis has been published  the  
 following observations from this global analysis remain valid:
\begin{itemize}
\item For $C_{7\gamma}$, $C_{9}$ and $C_{10}$ there is little room left
for constructive interference of real NP contributions with the SM.
\item A flipped sign solution with $C_{7\gamma} \simeq -
C_{7\gamma}^\text{SM}$, $C_{9} \simeq - C_{9}^\text{SM}$, and $C_{10}
\simeq - C_{10}^\text{SM}$ is allowed by the data.
\item Sizable imaginary parts for all coefficients are still allowed.
\end{itemize}

A detailed study of CP symmetries and CP asymmetries in concrete BSM scenarios 
can also be found in  \cite{Altmannshofer:2008dz}. In particular it has 
been found that these observables could allow us clear distinction of LHT, 
general 
MSSM and MSSM with flavour blind phases (FBMSSM) not only from SM predictions 
but also among these three scenarios.

 This picture could be modified by the most recent LHCb data  
\cite{Aaij:2013iag,Aaij:2013qta} on angular observables in 
$B_d\to K^*\mu^+\mu^-$ that show significant departures from SM expectations. 
Moreover, new data on the observable $F_L$, 
consistent with LHCb value in \cite{Aaij:2013iag} have been presented by 
CMS \cite{Chatrchyan:2013cda}.
These anomalies in $B_d\to K^*\mu^+\mu^-$ triggered recently 
 two sophisticated analyses \cite{Descotes-Genon:2013wba,Altmannshofer:2013foa} 
with the goal to understand the data and to indicate what type of new physics could be responsible for these departures from the SM. Both analyses point 
toward NP contributions in the 
modified coefficients  $C_{7\gamma}$ and $C_{9}$ with the following shifts with
respect to their SM values: 
\be
C^{\rm NP}_{7\gamma} < 0, \qquad C^{\rm NP}_{9} < 0.
\ee
Other possibilities, in particular involving right-handed currents ($C_9^\prime>0$), have been discussed in \cite{Altmannshofer:2013foa}.  Subsequently several other analyses 
of these data have been presented \cite{Gauld:2013qba,Buras:2013qja,Gauld:2013qja,Beaujean:2013soa,Datta:2013kja,Horgan:2013pva,Buras:2013dea,Richard:2013xfa,Hurth:2013ssa}.
In particular, a recent comprehensive Bayesian analysis of 
the authors of \cite{Beaujean:2012uj,Bobeth:2012vn} in \cite{Beaujean:2013soa} finds that although 
SM works well, if one wants to interpret the data in extensions of the SM
then scenarios in which chirality-flipped operators are included work better 
than the ones without them. In that case they find that the main NP effect is still in $C_9$ 
and in agreement with \cite{Altmannshofer:2013foa} find that in 
the  $C_9-C_9^\prime$ plane the SM point is outside the $2\sigma$ range.

It should be emphasized at this point that these analyses are subject 
to theoretical uncertainties, which have been discussed at length in 
\cite{Khodjamirian:2010vf,Beylich:2011aq,Matias:2012qz,Jager:2012uw,Descotes-Genon:2013wba,Hambrock:2013zya,Hurth:2013ssa} and it remains to be seen whether the observed anomalies are only 
result of statistical fluctuations and/or underestimated error uncertainties. 
This has been in particular emphasized by the authors of  \cite{Beaujean:2013soa} who do 
not think that without significant improvement of the understanding of 
$1/m_b$ corrections and reduction of the uncertainties in hadronic form factors
it will be possible to convincingly demonstrate the presence of NP in 
the decays in question.

Assuming that NP is really at work here we have investigated in \cite{Buras:2013qja} 
whether tree-level $Z^\prime$ and $Z$-exchanges could simultaneously 
explain the  $B_d\to K^*\mu^+\mu^-$ anomalies and the most recent data on 
$B_{s,d}\to\mu^+\mu^-$. In this context we have investigated the correlation 
between these decays and $\Delta F=2$ observables. The outcome of this 
rather extensive analysis for $B_{s,d}\to\mu^+\mu^-$ has been already summarized at the end of Step 4. 
In particular the plots in  Figs.~\ref{fig:BdvsBsLHS} and \ref{fig:ZBdvsBsLHS} 
demonstrate that LHS scenario for $Z^\prime$ or $Z$ FCNC couplings
provides
a simple model that allows for the violation of the CMFV relation
between the branching ratios for $B_{d,s}\to \mu^+\mu^-$ and $\Delta M_{s,d}$.

As far as  the anomalies in $B\to K^*\mu^+\mu^-$ are concerned 
\begin{itemize}
\item
$Z^\prime$ 
with only left-handed couplings is capable of softening the anomalies in 
the observables $F_L$ and $S_5$ in a correlated manner as proposed 
 \cite{Descotes-Genon:2013wba,Altmannshofer:2013foa}. However, a better 
description of the present data is obtained by including also right-handed 
contributions with the RH couplings of approximately the same magnitude 
but opposite sign. This is our ALRS scenario.
We illustrate this in Fig.~\ref{fig:pFLS5LHS}. This is in agreement with the findings in \cite{Altmannshofer:2013foa}.  Several analogous correlations can be found in \cite{Buras:2013qja}. We should emphasize that if $Z^\prime$ is the 
only new particle at scales $\ord(\tev)$ than $C^{\rm NP}_{7\gamma}$ can be 
neglected implying nice correlations shown in  Fig.~\ref{fig:pFLS5LHS}.
\item
The SM $Z$ boson with FCNC couplings to quarks cannot describe
the anomalies in $B\to K^*\mu^+\mu^-$ due to
its small vector coupling to muons. 
\end{itemize}

\begin{figure}[!tb]
 \centering
\includegraphics[width = 0.43\textwidth]{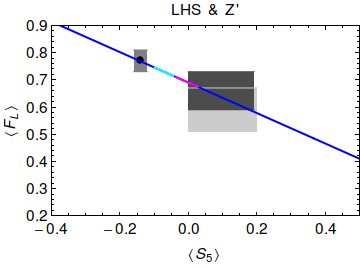}
\includegraphics[width = 0.45\textwidth]{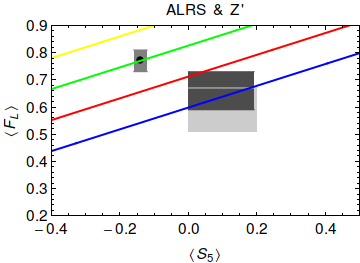}
\caption{Left: $\langle F_L\rangle$ versus $\langle S_5\rangle$ in LHS  where the magenta line corresponds to $C^\text{NP}_9 = -1.6\pm0.3$ and the cyan line to $C^\text{NP}_9 = -0.8\pm0.3$. Right: The same in ALRS 
for different values of $C_9^\text{NP}$: $-2$
(blue), $-1$ (red), $0$ (green) and $1$ (yellow).
The light and dark gray area corresponds to the experimental range for 
$\langle F_L\rangle$ with all data  and only LHCb+CMS data, taken into account,
respectively.  The black point and the
gray box correspond to the SM predictions from \cite{Altmannshofer:2013foa}.}\label{fig:pFLS5LHS}~\\[-2mm]\hrule
\end{figure}
 
In summary, while 
the modification of the Wilson coefficient $C_{7\gamma}$ together with
$C_{9}$ could provide the explanation of the data \cite{Descotes-Genon:2013wba,Altmannshofer:2013foa}, it appears that the most favourite scenario is the 
one with participation of right-handed currents
\cite{Altmannshofer:2013foa,Buras:2013qja,Horgan:2013pva}
\be
C^{\rm NP}_{9} < 0, \qquad C^{\prime}_{9} >0, \qquad   C^{\prime}_{9}\approx -C^{\rm NP}_{9}.
\ee
Yet, the case of NP  present only in the coefficient $C_9$ cannot 
be presently excluded \cite{Descotes-Genon:2013wba,Gauld:2013qba,Buras:2013qja,Gauld:2013qja,Beaujean:2013soa,Buras:2013dea}. Concerning the dynamics,  the favourite physical mechanisms behind these deviations emeraging from these studies is the presence of tree-level $Z^\prime$ exchanges. We will summarize the 
recent results in 331 models \cite{Buras:2013dea} in Section~\ref{sec:331}.

We are looking forward to improved LHCb data in order to see how the story 
of NP in $B\to K^*(K)\mu^+\mu^-$ and $B_{s,d}\to \mu^+\mu^-$ decays evolves with 
time.

\subsubsection{Explicit Bounds on  Wilson Coefficients}\label{sec:bsllWilson}
In the present review we have used the results discussed above to 
constrain the correlations between various observables in models with 
tree-level neutral gauge boson and neutral scalar and pseudoscalar 
exchanges. Such constraints can be found in plots presented in Steps 4 and 9. 
To this end in the case of gauge boson exchanges we use the  bounds 
 from  Figs.~1 and 2 of \cite{Altmannshofer:2012ir}. Approximately these bound can be summarized as follows:\footnote{The
latest updates~\cite{Straub:2013uoa,Altmannshofer:2013oia} show that the recent LHCb measurement of the CP asymmetry
$A_9$~\cite{Aaij:2013iag}
leads to a slightly stronger constraint on the imaginary part of $C_{10}^\prime$: $-1.5\leq \Im(C_{10}^\prime)\leq 1.5$.}
\begin{subequations}\label{equ:ASconstraint}
\begin{align}
 &-2\leq \Re(C_{10}^\prime)\leq 0\,, \quad-2.5\leq \Im(C_{10}^\prime)\leq 2.5\,,\\
&-0.8\leq \Re(C_{10}^\text{NP})\leq 1.8\,,\quad -3\leq \Im(C_{10})\leq 3\,.
\end{align}
\end{subequations}
Especially, 
the LHCb data on
$B\to K^*\mu^+\mu^-$ allow only for  {\it negative} values of the real part
of $C^\prime_{10}$
\be \label{C10C}
\Re( C^\prime_{10}) \le 0
\ee
and this has an impact on  our results in RH and LR scenarios presented in 
Steps 4 and 9.
However for the numerical analysis we use the exact bounds that are smaller than these rectangular bounds. For $C_{10}$~-- relevant
for LHS~--the latter allow a much larger region of parameter space whereas for $C^\prime_{10}$~-- relevant for RHS~-- the approximation
above gives very similar results to the exact bounds in our plots. 
In Figs.~\ref{fig:BsmuvsSphiZprimeA}, \ref{fig:SmuvsSphiZprimeA},
\ref{fig:ADGvsSphiZprimeA}, \ref{fig:BKnuvsBsmu} and \ref{fig:BKstarnuvsBKnu} the green regions in the $Z^\prime$ case are compatible
with the exact bound from \cite{Altmannshofer:2012ir}. The black points in RHS show the excluded regions where the bound in (\ref{C10C})  is violated
which as one can see nearly coincides with the correct bounds (see Figs.~\ref{fig:BKnuvsBsmu} and \ref{fig:BKstarnuvsBKnu}).

Concerning the bounds on the coefficients of scalar operators we quote 
here the bounds derived from the analysis in
\cite{Becirevic:2012fy}. Adjusting their normalization of Wilson coefficients to ours the final result of this paper reads:
\be
m_b|C_S^{(\prime)}|\le 0.7,\qquad   m_b|C_P^{(\prime)}|\le 1.0,
\ee
where the scale in $m_b$ should be the high matching scale.
As demonstrated in \cite{Buras:2013rqa} these bounds do not have presently any impact on
the values of  these coefficients in scenarios with tree-level scalar and 
pseudoscalar exchanges.

In summary this step will definitely bring new insight into short distance 
dynamics during the upgraded analyses of the LHCb and also SuperKEKB will 
play an important role in these studies.

\boldmath
\subsection{Step 8: $\kpn$, $\klpn$ and $K_L\to\mu^+\mu^-$}
\unboldmath

\subsubsection{Preliminaries}
Among the top highlights of flavour physics in this decade
will be the measurements of the branching ratios of two {\it golden} modes
$\kpn$ and $\klpn$. $\kpn$ is CP conserving while $\klpn$ is governed by 
CP violation. Both decays are dominated in the SM and  many of its
extensions by $Z$ penguin diagrams.
It is well known that these decays are theoretically 
very clean and their branching ratios have been calculated within  the SM including NNLO QCD corrections and 
electroweak corrections 
\cite{Buras:2005gr,Buras:2006gb,Brod:2008ss,Brod:2010hi,Buchalla:1997kz}.
Moreover, extensive calculations of isospin breaking effects and 
non-perturbative effects have been done \cite{Isidori:2005xm,Mescia:2007kn}.
 Reviews of these two decays can be found in 
\cite{Buras:2004uu,Isidori:2006yx,Smith:2006qg,Komatsubara:2012pn}. In 
particular in \cite{Buras:2004uu} the status of NP contributions as of 2008 
has been reviewed. A recent short review of  NP signatures in Kaon decays 
can be found in \cite{Blanke:2013goa}.

Assuming that light neutrinos couple only to left-handed currents, the 
general short distance effective Hamiltonian describing both decays  
is given
as follows
\be
\Heff(\nu\nu) =
g_{\text{SM}}^2V_{ts}^\ast 
V_{td} \times \left[ X_{L}(K) (\bar s \gamma^\mu P_L d) 
 +X_{R}(K) (\bar s \gamma^\mu P_R d)\right] \times (\bar \nu \gamma_\nu P_L\nu)\,,
\label{eq:heffKnn}
\ee
where $   g_{\text{SM}}$ is defined in (\ref{gsm}). We have suppressed the 
charm contribution that is represented by $P_c(X)$ below.

The resulting 
branching ratios for the two  
$K \to \pi \nu\bar \nu$ modes 
 can be written generally as
\begin{gather} \label{eq:BRSMKp}   
  \mathcal{B}(K^+\to \pi^+ \nu\bar\nu) = \kappa_+ \left [ \left ( \frac{{\rm Im} X_{\rm eff} }{\lambda^5}
  \right )^2 + \left ( \frac{{\rm Re} X_{\rm eff} }{\lambda^5} 
  - P_c(X)  \right )^2 \right ] \, , \\
\label{eq:BRSMKL} \mathcal{B}( K_L \to \pi^0 \nu\bar\nu) = \kappa_L \left ( \frac{{\rm Im} 
    X_{\rm eff} }{\lambda^5} \right )^2 \, ,
\end{gather}
where \cite{Mescia:2007kn}
\begin{equation}\label{kapp}
\kappa_+=(5.36\pm0.026)\cdot 10^{-11}\,, \quad \kappa_{\rm L}=(2.31\pm0.01)\cdot 10^{-10}
\ee
and \cite{Buras:2005gr,Buras:2006gb,Brod:2008ss,Isidori:2005xm,Mescia:2007kn}.
\be
P_c(X)=0.42\pm0.03.
\end{equation}
The short distance contributions are described by
\be\label{XK}
X_{\rm eff} = V_{ts}^* V_{td} (X_{L}(K) + X_{R}(K))\equiv 
V_{ts}^* V_{td} X(x_t) ( 1 +\xi e^{i\theta}).
\ee
Here
\be\label{XSM}
X_L^{\rm SM}(K)=\eta_X X_0(x_t)=1.464 \pm 0.041,
\ee
results within the SM from $Z$-penguin and box diagrams with
\begin{equation}\label{X0}
X_0(x_t)={\frac{x_t}{8}}\;\left[{\frac{x_t+2}{x_t-1}}
+ {\frac{3 x_t-6}{(x_t -1)^2}}\; \ln x_t\right],
\end{equation}
and $\eta_X=0.994$ for $m_t(m_t)$. 

 It should be remarked that with the definitions of electroweak parameters as 
in Table~\ref{tab:input}, in particular $\sin^2\theta_W$, the electroweak 
corrections to $X_L^{\rm SM}(K)$ are totally negligible \cite{Brod:2010hi} and 
therefore are not exhibited here. {To this end also  $m_t(m_t)$, as discussed in 
the context of the $B_{s,d}\to \mu^+\mu^-$ decays in Step 4, should be used.} That is for $m_t$ only QCD corrections are $\overline{\rm MS}$ renormalized, whereas $m_t$ is on-shell as far as electroweak corrections 
are concerned. See  \cite{Brod:2010hi,Buras:2012ru} for more details.

In order to describe NP contributions
we have introduced the two real parameters $\xi$ and $\theta$ 
that vanish in the SM. 
These formulae 
are in fact  very  general and apply to all extensions of the SM. 
The correlation between the two branching ratios depends generally  on 
two variables $\xi$ and $\theta$ \cite{Buras:2004ub,Buras:2004uu,Buras:2010pz} and measuring 
these branching ratios one day will allow to determine
 $\xi$ and $\theta$
 and compare them with model expectations. We illustrate this in 
Fig.~\ref{fig:Kpinucontour}.

\begin{figure}[!tb]
 \centering
\includegraphics[width = 0.4\textwidth]{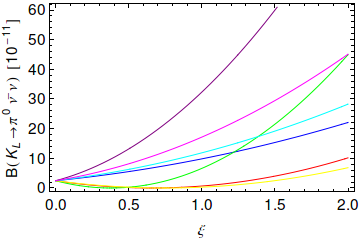}
\hspace{0.3cm}
\includegraphics[width = 0.4\textwidth]{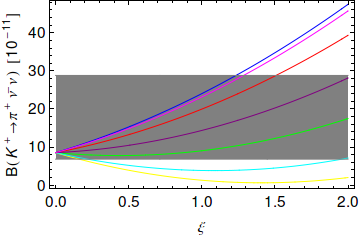}

\includegraphics[width = 0.5\textwidth]{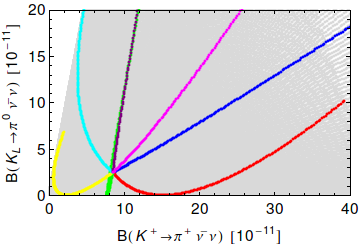}
\caption{\it Top: $K_L \to \pi^0 \nu\bar \nu$ and  $K^+ \to \pi^+ \nu\bar \nu$ as a function of $\xi$ for $\theta = 0$ (blue), 1 (red), 2
(green),
3 (yellow), 4 (cyan), 5 (purple), 6 (magenta). Down: $K_L \to \pi^0 \nu\bar \nu$ vs.  $K^+ \to \pi^+ \nu\bar \nu$
for $\xi\in[0,2]$ and $\theta\in[0,2\pi]$ (light gray) and coloured $\theta$ as before.}\label{fig:Kpinucontour}~\\[-2mm]\hrule
\end{figure}

Unfortunately on the basis of these two 
branching ratios it is not possible to find out how important the 
contributions of right-handed currents are as their effects are hidden in 
a single function $X_{\rm eff}$. In this sense the decays governed 
by $b\to s \nu\bar\nu$ transitions that we will discuss soon 
are superior. Indeed, in this case we have 
three branching ratios to our disposal and one is also sensitive to 
the direction of the spin of $K^*$.

 Experimentally we have \cite{Artamonov:2008qb}
\be\label{EXP1}
\mathcal{B}(\kpn)_\text{exp}=(17.3^{+11.5}_{-10.5})\cdot 10^{-11}\,,
\ee
and the $90\%$ C.L. upper bound \cite{Ahn:2009gb}
\be\label{EXP2}
\mathcal{B}(\klpn)_\text{exp}\le 2.6\cdot 10^{-8}\,.
\ee

The prospects for improved measurements of $\mathcal{B}(\kpn)$ are very good.
One should stress that 
already a measurement of this  branching ratio with an
accuracy of $10\%$ will give us a very important insight into the physics 
at short distance scales. Indeed NA62 experiment at CERN aims at this 
precision and a new experiment at Fermilab (ORKA) should be able to  reach 
the accuracy of $5\%$ which would be truly fantastic. It will take 
longer in the case of $\klpn$ but KOTO experiment at 
J-PARC should provide interesting results in this decade on this 
branching ratio. It should be emphasized  that the combination of 
these two decays is particularly powerful in testing NP.  The future 
prospects for experiments on $K$ decays, in particular $\kpn$ and $\klpn$ 
have been recently reviewed in \cite{Komatsubara:2012pn,E.T.WorcesterfortheORKA:2013cya}.

The decays $K_L\to\pi^0\ell^+\ell^-$ are not as 
theoretically clean as the $K\to\pi\nu\bar\nu$ channels and are less sensitive 
to NP contributions but they probe different operators beyond the SM and 
having accurate branching ratios for them would certainly be useful. 
Further details on this decay can be found in 
\cite{Buchalla:2003sj,Isidori:2004rb,Friot:2004yr,Mescia:2006jd,Prades:2007ud,Buras:1994qa}. As there are no advanced plans to 
measure  these branching ratios in this decade, we will not consider them in what follows. The most recent 
analysis of these decays within $Z^\prime$ models with further references can be 
found in \cite{Buras:2012jb}.

On the other hand the decay $K_L\to\mu^+\mu^-$, even if subject to hadronic 
uncertainties, provides a useful constraint on the extensions of the SM.
We will discuss this decay in this section
 as there are interesting correlations between 
this decay and $\kpn$ which could help to distinguish between various 
NP scenarios.

For  $K_L\to\mu^+\mu^-$ the effective Hamiltonian, suppressing charm contribution and neglecting contributions from scalar operators that are suppressed by 
small $m_{d,s}$, reads
\be
\Heff(\mu\mu) =
-g_{\text{SM}}^2V_{ts}^\ast 
V_{td} \times \left[ Y_{L}(K) (\bar s \gamma^\mu P_L d) 
 +Y_{R}(K) (\bar s \gamma^\mu P_R d)\right] \times (\bar \mu \gamma_\nu P_L\mu)\,.
\label{eq:heffKmumu}
\ee

Only the so-called short distance (SD)
part to a dispersive contribution
to $K_L\to\mu^+\mu^-$ can be reliably calculated. We have then including 
charm contribution
\cite{Buras:2004ub} 
($\lambda=0.226$)
\be
\mathcal{B}(K_L\to\mu^+\mu^-)_{\rm SD} =
 2.08\cdot 10^{-9} 
\left ( \frac{{\rm Re} Y^K_{\rm eff} }{\lambda^5} 
  - \bar P_c(Y)  \right )^2  \,
\ee
where at NNLO \cite{Gorbahn:2006bm}
\be
\bar P_c\left(Y\right) \equiv \left(1-\frac{\lambda^2}{2}\right)P_c\left(Y\right)\,,\qquad
P_c\left(Y\right)=0.113\pm 0.017~.
\ee

The short distance contributions are described by
\be\label{YK}
Y^K_{\rm eff} = V_{ts}^* V_{td} (Y_{L}(K) - Y_{R}(K))
\ee
with
\be
Y_L^{\rm SM}(K) = \eta_Y Y_0(x_t)
\ee
already encountered in $B_{s,d}\to\mu^+\mu^-$ decays and given in 
(\ref{YSM}).
We note 
the minus sign in front of $Y_R$ as opposed to $X_R$ in (\ref{XK}) that 
results from the fact that only the $\gamma_\mu\gamma_5$ part contributes.

The extraction of the short distance
part from the data is subject to considerable uncertainties. The most recent
estimate gives \cite{Isidori:2003ts}
\be\label{eq:KLmm-bound}
\mathcal{B}(K_L\to\mu^+\mu^-)_{\rm SD} \le 2.5 \cdot 10^{-9}\,,
\ee
to be compared with $(0.8\pm0.1)\cdot 10^{-9}$ in the SM 
\cite{Gorbahn:2006bm}.

\subsubsection{Standard Model Results}
The branching ratios for $\kpn$ and $\klpn$ in the SM are given by
\begin{equation}\label{bkpn}
\mathcal{B}(\kpn)=\kappa_+\cdot\left[\left(\frac{\imlt}{\lambda^5}X(x_t)\right)^2+
\left(\frac{\relt}{\lambda^5}X(x_t)-P_c(X)\right)^2
\right]~,
\end{equation}
and
\begin{equation}\label{bklpn}
\mathcal{B}(K_{\rm L}\to\pi^0\nu\bar\nu)=\kappa_{\rm L}\cdot
\left(\frac{\imlt}{\lambda^5}X(x_t)\right)^2.
\end{equation}

The important feature of these expressions is that these two decays 
are described by the same {\it real} function $X(x_t)$. 
The 
present theoretical uncertainties in $\mathcal{B}(\kpn)$ and $\mathcal{B}(\klpn)$ are 
at the level of $2-3\%$ and $1-2\%$, respectively. 
Calculating the branching ratios for the central values 
            of the parameters  in Table~\ref{tab:SMpred}, we find for $\vub = 0.0034$ 
\be\label{SM1}
\mathcal{B}(\kpn)_\text{SM} =8.5\cdot 10^{-11}\,, \quad
\mathcal{B}(\klpn)_\text{SM} =2.5\cdot 10^{-11}\,,
\ee
while for $\vub = 0.0040$ we find
\be\label{SM2}
\mathcal{B}(\kpn)_\text{SM} =8.4\cdot 10^{-11}\,,\quad 
\mathcal{B}(\klpn)_\text{SM} =3.4\cdot 10^{-11}\,.
\ee
We observe that whereas $\mathcal{B}(\kpn)$ is rather insensitive to $\vub$, 
$\mathcal{B}(\klpn)$ increases with increasing $\vub$. The main remaining uncertainty in these branching ratios comes from the $\vcb^4$ dependence and if the 
present value from tree-level decays is used, this uncertainty  amounts to 
roughly $10\%$. As we demonstrated in \cite{Buras:2013raa} this uncertainty within the SM can be decreased significantly with the help of $\varepsilon_K$, 
in particular when the angle $\gamma$ will be known from tree-level decays. 
Therefore, we expect that when the data from NA62 will be available, the 
total uncertainties in both branching ratios will be in the ballpark of $5\%$.

These results  should be compared 
with the experimental values given in (\ref{EXP1}) and (\ref{EXP2}). 
Certainly there is still a significant room left for NP contributions and 
we will now turn our attention to them in the context of simplest extensions 
of the SM.

\subsubsection{CMFV}

In these models $\kpn$ and $\klpn$ are described by a single real 
function $X(v)$ implying a strong correlation between the two branching 
ratios as emphasized in \cite{Buras:2001af}. We show this correlation in  Fig.~\ref{fig:KLvsKpMFV}. 
Thus once the the branching ratio for $\kpn$ will be measured with 
high precision by NA62 and later at Fermilab, we will know also 
precisely the corresponding branching ratio for $\klpn$ that will be universal 
for the full class of CMFV models.

\begin{figure}[!tb]
\begin{center}
\includegraphics[width=0.5\textwidth] {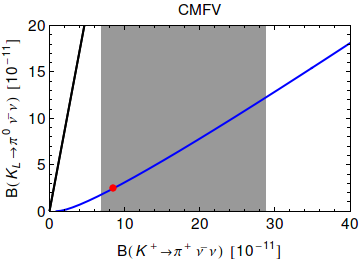}
\caption{\it  $\mathcal{B}(\klpn)$ versus
$\mathcal{B}(\kpn)$ in CMFV. Red point: SM central
value. Gray region:
experimental range of $\mathcal{B}(\kpn)$. { The black line corresponds to the Grossman-Nir bound.} }\label{fig:KLvsKpMFV}~\\[-2mm]\hrule
\end{center}
\end{figure}

\boldmath
\subsubsection{${\rm 2HDM_{\overline{MFV}}}$}
\unboldmath

In this class of models the dominant new contribution comes from charged 
Higgs ($H^\pm$) exchanges in $Z^0$-Penguin diagrams and box diagrams.
While an explicit calculation with present input is missing we do not 
     expect large NP contributions in this scenario.

\subsubsection{Tree-Level Gauge Boson Exchanges}

The contributions of tree-level exchanges to the branching ratios in 
question are known from various studies in $Z^\prime$ models. The new 
feature is the appearance of right-handed current contributions and 
the presence of new flavour violating interactions that can carry new 
CP-violating phases. A very detailed analysis of this simple NP scenario 
has been presented in \cite{Buras:2012jb} and we will summarize 
the most important results of this paper.

The 
branching ratios for the two  
$K \to \pi \nu\bar \nu$ modes are given by  
 (\ref{eq:BRSMKp})--(\ref{XSM}) with  

\be\label{XLK}
X_{\rm L}(K)=\eta_X X_0(x_t)+\frac{\Delta_L^{\nu\bar\nu}(Z')}{g^2_{\rm SM}M_{Z'}^2}
                                       \frac{\Delta_L^{sd}(Z')}{V_{ts}^* V_{td}},
\ee
\be\label{XRK}
X_{\rm R}(K)=\frac{\Delta_L^{\nu\bar\nu}(Z')}{g^2_{\rm SM}M_{Z'}^2}
                                       \frac{\Delta_R^{sd}(Z')}{V_{ts}^* V_{td}},
\ee

As the new $\Delta_{L,R}^{sd}(Z^\prime)$ are complex numbers, these results are 
rather arbitrary.
In a situation like this we have to look for other observables in the $K$ 
system that depend also on these couplings. Here the correlation of $K\to\pi\nu\bar\nu$ 
decays with $\varepsilon_K$ can give insights into  the flavour structure of 
NP contributions and distinguish between models in which NP is dominated 
by left-handed currents or right-handed currents or both left-handed and right-handed currents with similar 
magnitude and phases
 \cite{Blanke:2009pq}. In fact as pointed out in the latter paper a  
 correlation  between $\varepsilon_K$ and $K\to\pi\nu\bar\nu$ 
decays exists that is characteristic for 
all NP frameworks where the
phase in  $\Delta S=2$ amplitudes is the square of the CP-violating 
phase in $\Delta S=1$ FCNC amplitudes. This is 
for instance what happens in the Little Higgs model 
with $T$ parity \cite{Blanke:2006eb}. The introduction of the three 
scenarios for $\Delta_{L,R}$ in Section~\ref{sec:1} was motivated by this work and 
also by \cite{Altmannshofer:2009ne}, where similar scenarios in the context of 
various supersymmetric flavour models have been analyzed. What is novel in our analysis of these scenarios 
is that in the presence of the dominance of NP contributions by 
tree-level exchanges, the correlations in question are particularly 
transparent.

We illustrate this in explicit terms now by considering the set
\be
\varepsilon_K, \quad \kpn, \quad \klpn, \quad K_L\to\mu^+\mu^-
\ee
in the scenarios LHS, RHS and LRS for the $\Delta_{L,R}$ couplings in question. 

The inclusion of $K_L\to\mu^+\mu^-$ in this discussion  leads to interesting 
results. Indeed now
\be
Y_L(K)=Y(x_t)+\frac{\Delta_A^{\mu\bar\mu}(Z^\prime)}{g^2_{\rm SM}M_{Z^\prime}^2}
                                       \frac{\Delta_L^{sd}(Z^\prime)}{V_{ts}^* V_{td}},
\ee
\be
Y_R(K)=\frac{\Delta_A^{\mu\bar\mu}(Z^\prime)}{g^2_{\rm SM}M_{Z^\prime}^2}
                                       \frac{\Delta_R^{sd}(Z^\prime)}{V_{ts}^* V_{td}}.
\ee
We note that up to the lepton couplings NP corrections are the same as in $X_{L,R}(K)$. However, very importantly the function $Y_R(K)$ enters with the 
opposite sign to $X_R(K)$ into the branching ratio for $K_L\to\mu^+\mu^-$ so 
that effectively one has 
\be\label{YAK}
Y_{\rm A}(K)= \eta _Y Y_0(x_t)
+\frac{\left[\Delta_A^{\mu\bar\mu}(Z')\right]}{M_{Z'}^2g_\text{SM}^2}
\left[\frac{\Delta_L^{sd}(Z')-\Delta_R^{sd}(Z')}{V_{ts}^\star V_{td}}\right]\,
\equiv |Y_A(K)|e^{i\theta_Y^K}.
\ee

The minus sign in front of $\Delta_R^{sd}(Z')$ implies an anti-correlation between $\kpn$ and $K_L\to\mu^+\mu^-$ branching ratios noticed already 
 within the RSc scenario in \cite{Blanke:2008yr}.

 We will now summarize the results obtained in \cite{Buras:2012jb}, where 
the leptonic couplings have been chosen to be
\be\label{DAnunu}
\Delta_L^{\nu\bar\nu}(Z^\prime)=\Delta_A^{\mu\bar\mu}(Z^\prime)=0.5~,
\ee
to be compared with its SM value for $Z$ couplings $0.372$.

In our presentation   of particular interest are
the values of the $\delta_{12}$ phase in (\ref{Zprimecouplings}) \cite{Blanke:2009pq} 
\be\label{delta12}
\delta_{12}= n\frac{\pi}{2}, \qquad n=0,1,2,3
\ee
for which NP contributions to $\varepsilon_K$ vanish. As seen in Fig.~\ref{fig:oasesKLHS}
this is only allowed for scenario S2 for which SM agrees well with the data
and NP contributions are not required. In this scenario  $\tilde s_{12}$
can even vanish. In scenario S1, in which NP contributions are required
to reproduce the data,  $\tilde s_{12}$ is bounded from below and
$\delta_{12}$ cannot satisfy~(\ref{delta12}).

\begin{figure}[!tb]
\begin{center}
\includegraphics[width=0.45\textwidth] {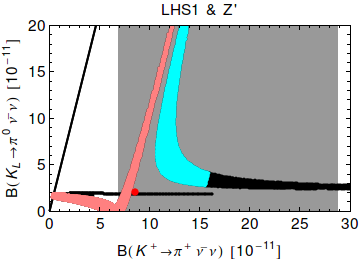}
\includegraphics[width=0.45\textwidth] {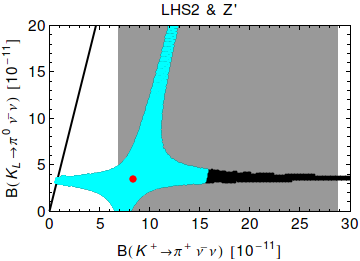}\\
\vspace{0.3cm}
\includegraphics[width=0.45\textwidth] {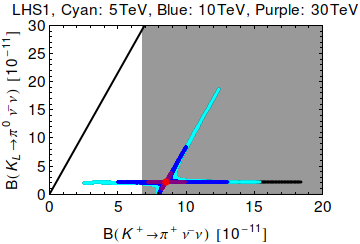}
\includegraphics[width=0.45\textwidth] {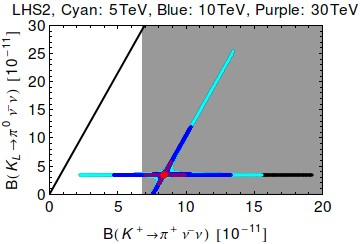}
\caption{\it  $\mathcal{B}(\klpn)$ versus
$\mathcal{B}(\kpn)$ for $M_{Z^\prime} = 1~$TeV (upper panels,  $C_1$: cyan, $C_2$: pink.) and $M_{Z^\prime} = 5~$TeV (cyan),
10~TeV (blue) and 30~TeV (purple) (lower panels) in LHS1 (left) and LHS2
(right).   Black regions are excluded by the upper bound $\mathcal{B}(K_L\to \mu^+\mu^-)\leq 2.5\cdot
10^{-9}$. Red point: SM central
value. Gray region:
experimental range of $\mathcal{B}(\kpn)$.  {The black line corresponds to the Grossman-Nir bound.} 
}\label{fig:KLvsKpLHS}~\\[-2mm]\hrule
\end{center}
\end{figure}

In the upper panels of  Fig.~\ref{fig:KLvsKpLHS} we show  the correlation 
between the branching ratios for $\kpn$ and $\klpn$
in LHS1 and LHS2
 for $M_{Z'}=1\tev$ \cite{Buras:2012jb}. Since only vector currents occur we get the same result for RHS1 and RHS2. We observe the
following pattern of deviations from
the SM expectations:
\begin{itemize}
\item
There are two branches in both scenarios. The difference between LHS1
and LHS2 originates from required NP contributions in LHS1 in order to
agree with the data on $\varepsilon_K$ and the fact that in LHS1 there
are two oases and only one in LHS2.
\item
The horizontal branch in both plots corresponds to $n=0,2$ in (\ref{delta12}), for which
NP contribution to $K\to\pi\nu\bar\nu$ is real and vanishes in the
case of $\klpn$.
\item
The second branch corresponds to $n=1,3$ in (\ref{delta12}), for which NP
contribution is purely imaginary. It is parallel to the Grossman-Nir (GN) bound
\cite{Grossman:1997sk}
that is represented by the solid black line.
\item
The deviations 
from the SM are significantly larger than in the case of
rare $B$ decays. This is a consequence of the weaker constraint from
$\Delta S=2$ processes compared to $\Delta B=2$ and the fact that rare $K$ decays
are stronger suppressed than rare $B$ decays within the SM. 
Yet as seen the largest values corresponding to black areas are 
ruled out through the correlation with $K_L\to\mu^+\mu^-$ as discussed below.
\item
 We observe that even 
at  $M_{Z'}=10\tev$
both branching ratios can still differ by much from SM predictions and for 
 $M_{Z'}\le 20\tev$ NP effects in these decays, in particular 
$\klpn$,  should be detectable in the flavour precision era. 
\end{itemize}

Of particular interest is the correlation  between  $\mathcal{B}(\kpn)$ and  $\mathcal{B}(K_L\to\mu^+\mu^-)$ that we show in  Fig.~\ref{fig:KLmuvsKpLHS}. 
In the case of LHS1 scenario a correlation analogous to this one is found in
the LHT model \cite{Blanke:2009am} but due to fewer free parameters
in $Z'$ model this correlation depends whether oasis $C_1$ or $C_2$ is considered. The horizontal line
in Fig.~\ref{fig:KLmuvsKpLHS} corresponds this time to $n=1,3$ in (\ref{delta12}), for which NP
contribution is purely imaginary, while the other branches correspond
 to $n=0,2$ in (\ref{delta12}), for which
NP contribution to $K\to\pi\nu\bar\nu$ is real.

From Figs.~\ref{fig:KLvsKpLHS} and ~\ref{fig:KLmuvsKpLHS} we obtain the following results:
\begin{itemize}
\item
In the case of the dominance of real NP contributions we find 
for $M_{Z'}=1\tev$
\be\label{UPERBOUND}
\mathcal{B}(\kpn)\le  16\cdot 10^{-11}.
\ee
In this case $\klpn$ is SM-like and $\mathcal{B}(K_L\to\mu^+\mu^-)$ reaches
the upper bound in (\ref{eq:KLmm-bound}). 
\item
In the case of the dominance of  imaginary  NP contributions the bound
on $\mathcal{B}(K_L\to\mu^+\mu^-)$ is ineffective and both
$\mathcal{B}(\kpn)$ and $\mathcal{B}(\klpn)$ can be significantly larger
than the SM predictions and $\mathcal{B}(\kpn)$ can also be larger than its
present experimental central value. We also find that for such large values
the branching ratios are strongly correlated. Inspecting in the LHS2 
scenario when the branch parallel to the GN bound leaves the grey region 
corresponding to the $1\sigma$ region in (\ref{EXP1})
we find a rough upper bound
\be\label{const}
\mathcal{B}(\klpn)\le 85\cdot 10^{-11},
\ee
\end{itemize}
which is much stronger than the present experimental upper bound in 
(\ref{EXP2}).

\begin{figure}[!tb]
\begin{center}
\includegraphics[width=0.45\textwidth] {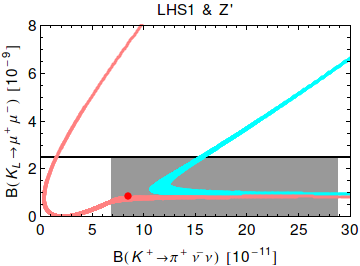}
\includegraphics[width=0.45\textwidth] {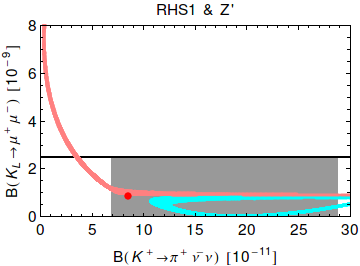}
\caption{\it  $\mathcal{B}(K_L\to\mu^+\mu^-)$ versus  $\mathcal{B}(\kpn)$ for $M_{Z^\prime} = 1~$TeV in LHS1 (left) and RHS1
(right).   $C_1$: cyan, $C_2$: pink. Red
point: SM central
value. Gray region: experimental range of
$\mathcal{B}(\kpn)$ and horizontal black line: upper bound of
$\mathcal{B}(K_L\to\mu^+\mu^-)$.}\label{fig:KLmuvsKpLHS}~\\[-2mm]\hrule
\end{center}
\end{figure}

Finally, in  the right
panel of Fig.~\ref{fig:KLmuvsKpLHS}
we show the correlation  between $\mathcal{B}(\kpn)$ and
 $\mathcal{B}(K_L\to\mu^+\mu^-)$ in
the RHS1 scenario.  Indeed the correlations in both oases
differ from the ones in LHS1.
 This feature is known already from different studies,
in particular in RSc scenario \cite{Blanke:2008yr} and originates in the fact 
that while $\kpn$ is  sensitive to vector couplings, $K_L\to\mu^+\mu^-$
is sensitive to the axial-vector couplings. 
We also note that in the case of the dominance of imaginary NP contributions
corresponding to the horizontal line,  $\mathcal{B}(\kpn)$ and
$\mathcal{B}(\klpn)$ can be large. But otherwise  $\mathcal{B}(\kpn)$ is
suppressed with respect to its SM value and  $\mathcal{B}(\klpn)$ is SM-like.

Finally we also discuss what happens if we exchange the $Z^\prime$ boson with the $Z^0$ boson with flavour violating couplings. Except
for the LR scenario and in case of purely imaginary NP contributions these effects are bounded by $K_L\to\mu^+\mu^-$. In
Fig.~\ref{fig:ZKLvsKp} we show our result for LHS2, RHS2 and LRS2 where the effects can be much larger than in the $Z^\prime$ case.

\begin{figure}[!tb]
\begin{center}
\includegraphics[width=0.45\textwidth] {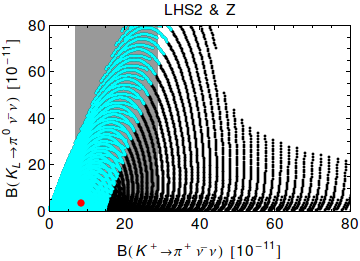}
\includegraphics[width=0.45\textwidth] {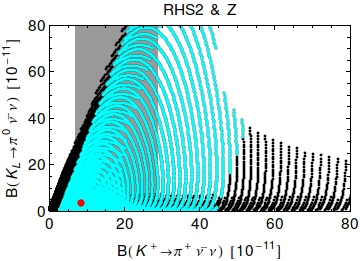}

\includegraphics[width=0.45\textwidth] {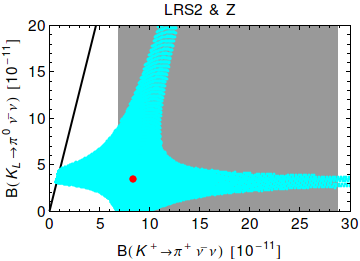}
\caption{\it $\mathcal{B}(\klpn)$ versus
$\mathcal{B}(\kpn)$ in LHS2, RHS2 and LRS2 for $Z^0$ exchange.   Red point: SM central
value. Black regions are excluded by the upper bound $\mathcal{B}(K_L\to \mu^+\mu^-)\leq 2.5\cdot
10^{-9}$. Gray region:
experimental range of $\mathcal{B}(\kpn)$.}\label{fig:ZKLvsKp}~\\[-2mm]\hrule
\end{center}
\end{figure}

\subsubsection{Tree-Level Scalar Exchanges}

If the masses of neutrinos are generated by the couplings to scalars than 
definitely the contributions of these scalars to decays with neutrinos in 
the final state are negligible.
But if the masses of neutrinos are generated by a different mechanism than
coupling to scalars, like in the case of the see-saw mechanism, it is
not a priori obvious that such couplings in some NP scenarios could be
 measurable. Our working assumption in the present paper will be that this is
not the case. Consequently NP
effects of scalars in $\kpn$, $\klpn$ and $b\to s\nu\bar\nu$ transitions considered next 
will be assumed to be negligible in contrast to
$Z'$ models as we have just seen. As demonstrated in \cite{Buras:2013rqa} scalar contributions to  $K_L\to\mu^+\mu^-$ and
$K_L\to\pi^0\ell^+\ell^-$ although in principle larger than for 
 $\kpn$, $\klpn$ and $b\to s\nu\bar\nu$ transitions, are found to be small
and we will not discuss them here.

\boldmath
\subsection{Step 9:
Rare B Decays $B\to X_s\nu\bar\nu$, $B\to K^*\nu\bar\nu$ and 
        $B\to K\nu\bar\nu$}
\unboldmath
\subsubsection{Preliminaries}
The rare decays in question are among the important channels in $B$ physics as 
they allow a transparent study of $Z$ penguin and other electroweak penguin effects in NP scenarios in the absence of dipole operator
contributions and Higgs (scalar) penguin contributions that are often more important than $Z$ contributions in $B\to K^*\ell^+\ell^-$ and
$B_s\to \ell^+\ell^-$ decays 
\cite{Colangelo:1996ay,Buchalla:2000sk,Altmannshofer:2009ma}. However,
their measurements appear to be
even harder than those of the rare $K$ decays just discussed. 
Yet,  SuperKEKB 
should be able to measure them at a satisfactory level.

The inclusive decay $B\to X_s\nu\bar\nu$ is theoretically as clean as 
$K\to\pi\nu\bar\nu$ decays but the parametric uncertainties are a bit
larger. The two exclusive channels are affected by  form factor uncertainties but
in the case of $B\to K^*\nu\bar\nu$ \cite{Altmannshofer:2009ma} 
and $B\to K\nu\bar\nu$ 
\cite{Bartsch:2009qp}
significant progress has been made few years ago. In the latter paper this has been achieved 
by considering simultaneously also $B\to K \ell^+\ell^-$.
Non-perturbative tree level contributions from $B^+\to
\tau^+\nu$ to $B^+\to K^+\nu\bar\nu$ and $B^+\to K^{*+}\nu\bar\nu$ at the 
level of roughly $10\%$ 
have been pointed out \cite{Kamenik:2009kc}.
Therefore the expressions in Eqs.~(\ref{eq:BKnn})--(\ref{eq:Xsnn}) given below, 
as well as the SM results in (\ref{eq:BKnnSM}), refer only to the short-distance contributions 
to these decays. The latter are obtained from the corresponding total rates 
subtracting the reducible long-distance effects pointed out in~\cite{Kamenik:2009kc}.

The general effective Hamiltonian including also right-handed current contributions 
that is used for
the $B \to  \{ X_s, K, K^*\} \nu\bar \nu$ decays is given as follows
\be
\Heff =
 g_{\text{SM}}^2 V_{ts}^\ast 
V_{tb} \times \left[ X_{L}(B_s) (\bar s \gamma^\mu P_L b) 
 +X_{R}(B_s) (\bar s \gamma^\mu P_R b)\right] \times (\bar \nu \gamma_\mu P_L\nu)\,
\label{eq:heffBXsnn}
\ee
and has a very similar structure to the one for $K\to\pi\nu\bar\nu$ decays 
in (\ref{eq:heffKnn}). In particular
 \be
 X_{\rm L}^{\rm SM}(B_s)=X_L^{\rm SM}(K)
 \ee
with $X_L^{\rm SM}(K)$ given in (\ref{XSM}).
Moreover
in models with minimal flavour violation (MFV) there is a striking correlation
 between the branching ratios for $K_L\to\pi^0\nu\bar\nu$ and $B\to X_s\nu\bar\nu$ as also there  the same
 one-loop function
$X(v)$ governs the two processes in question \cite{Buras:2001af}.
This relation is generally
modified in models with non-MFV interactions, in particularly right-handed 
 currents. As we will see below
there are also correlations between $K_L\to\pi^0\nu\bar\nu$, $K^+\to\pi^+ \nu\bar\nu$
and $B\to K^*(\to K\pi)\nu\bar\nu$ that
are useful for the study of various NP scenarios.

The interesting feature of these three $b\to s\nu\bar\nu$ transitions, in particular when 
taken together, is their sensitivity to right-handed currents 
\cite{Colangelo:1996ay,Buchalla:2000sk} studied recently in
\cite{Altmannshofer:2009ma}. 
Following the analysis of the latter paper, 
the branching ratios of the $B \to \{X_s,K, K^*\}\nu\bar \nu$  
modes in the presence of RH currents can be written as follows
 \bea
 \mathcal{B}(B\to K \nu \bar \nu) &=& 
 \mathcal{B}(B\to K \nu \bar \nu)_{\rm SM} \times\left[1 -2\eta \right] \epsilon^2~, \label{eq:BKnn}\\
 \mathcal{B}(B\to K^* \nu \bar \nu) &=& 
 \mathcal{B}(B\to K^* \nu \bar \nu)_{\rm SM}\times\left[1 +1.31\eta \right] \epsilon^2~, \\
 \mathcal{B}(B\to X_s \nu \bar \nu) &=& 
 \mathcal{B}(B\to X_s \nu \bar \nu)_{\rm SM} \times\left[1 + 0.09\eta \right] \epsilon^2~,\label{eq:Xsnn}
 \eea
 where we have introduced the variables 
 \be\label{etaepsilon}
 \epsilon^2 = \frac{ |X_{\rm L}(B_s)|^2 + |X_{\rm R}(B_s)|^2 }{
 |\eta_X X_0(x_t)|^2 }~,  \qquad
 \eta = \frac{ - {\rm Re} \left( X_{\rm L}(B_s) X_{\rm R}^*(B_s)\right) }
{ |X_{\rm L}(B_s)|^2 + |X_{\rm R}(B_s)|^2 }~,
 \ee
with $X_{\rm L,R}$ defined in (\ref{eq:heffBXsnn}). 

We observe that the RH currents signaled here by a non-vanishing $\eta$ enter 
these three branching ratios in a different manner allowing an efficient search 
for the signals of these currents.
Also  the average of the $K^*$ longitudinal polarization fraction $F_L$
 used in the studies of $B\to K^*\ell^+\ell^-$ is a useful variable as
it depends only on  $\eta$:
\be
\label{eq:epseta-FL}
 \langle F_L \rangle = 0.54 \, \frac{(1 + 2 \,\eta)}{(1 + 1.31 \,\eta)}~.
\ee

The experimental bounds~\cite{Barate:2000rc,:2007zk,:2008fr} read
\bea
\mathcal{B}(B\to K \nu \bar \nu)   &<&  1.4 \times 10^{-5}~, \no \\
\mathcal{B}(B\to K^* \nu \bar \nu) &<&  8.0 \times 10^{-5}~, \no \\
\mathcal{B}(B\to X_s \nu \bar \nu)  &<&  6.4 \times 10^{-4}~.
\label{eq:BKnn_exp}
\eea

\subsubsection{Standard Model Results}
In the absence of right-handed currents $\eta=0$ and all three decays 
are fully described by the function $X(x_t)$.
The updated predictions for the SM branching  ratios 
are~\cite{Bartsch:2009qp,Kamenik:2009kc,Altmannshofer:2009ma}
\bea
\mathcal{B}(B\to K \nu \bar \nu)_{\rm SM}   &=& (3.64 \pm 0.47)\times 10^{-6}~, \no \\
\mathcal{B}(B\to K^* \nu \bar \nu)_{\rm SM} &=& (7.2 \pm 1.1)\times 10^{-6}~, \no \\
\mathcal{B}(B\to X_s \nu \bar \nu)_{\rm SM} &=& (2.7 \pm 0.2)\times 10^{-5}~, 
\label{eq:BKnnSM}
\eea


\subsubsection{CMFV}

In this class of models all branching ratios are described as in Step 8 by 
the universal function $X(v)$
\be
X_L(B_s)=X(v), \quad X_R(B_s)=0
\ee
and consequently they are strongly correlated. 
However, most characteristic for this class of models is the correlation 
between the $K\to\pi\nu\bar\nu$ branching ratios and the $b\to s\nu\bar\nu$ 
transitions considered here. This correlation is in particular stringent 
once the CKM parameters have been determined in tree-level decays. We 
show this in Fig.~\ref{fig:bsnunuMFV}.

\begin{figure}[!tb]
\centering
\includegraphics[width = 0.45\textwidth]{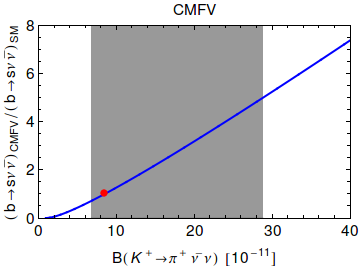}
\includegraphics[width = 0.45\textwidth]{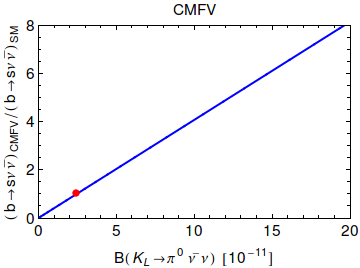}
\caption{\it The ratio $\mathcal{B}(B\to K^{(*)} \nu\bar\nu)_\text{CMFV}/\mathcal{B}(B\to K^{(*)} 
\nu\bar\nu)_\text{SM}=\mathcal{B}(B\to X_s \nu\bar\nu)_\text{CMFV}/\mathcal{B}(B\to X_s \nu\bar\nu)_\text{SM}$ versus
$K^+\to\pi^+\nu\bar\nu$ (left) and $K_L\to\pi^0\nu\bar\nu$(right).
}
 \label{fig:bsnunuMFV}~\\[-2mm]\hrule
\end{figure}

\boldmath
\subsubsection{${\rm 2HDM_{\overline{MFV}}}$}
\unboldmath

 To our knowledge, similarly to the case of $K\to\pi\nu\bar\nu$ decays, 
     no detailed analysis of $b\to s\nu\bar\nu$ transitions exists in the 
     literature. Yet because of tiny couplings of scalar particles to neutrinos 
     such effects could only be relevant at one loop level with charged Higgs 
     contributions at work. We expect these contributions to be small.

\subsubsection{Tree-Level Gauge Boson Exchanges}

Including the SM contribution in this case the couplings $X_{\rm L}$ and $X_{\rm R}$ are giving as 
follows
\be\label{XLB}
 X_{\rm L}(B_q)=\eta_X X_0(x_t)+\left[\frac{\Delta_{L}^{\nu\nu}(Z')}{M_{Z'}^2g^2_{\rm SM}}\right]
\frac{\Delta_{L}^{qb}(Z')}{ V_{tq}^\ast V_{tb}},
\ee
\be\label{XRB}
 X_{\rm R}(B_q)=\left[\frac{\Delta_{L}^{\nu\nu}(Z')}{M_{Z'}^2g^2_{\rm SM}}\right]
\frac{\Delta_{R}^{qb}(Z')}{ V_{tq}^\ast V_{tb}},
\ee

A detailed analysis of these decays has been performed in \cite{Buras:2012jb}. 
We summarize here the most important results of this analysis.

In Fig.~\ref{fig:BXsnuvsBsmuLHS1} (left) we show
$\mathcal{B}(B\to X_s \nu\bar\nu)$ vs
$\mathcal{B}(B_s\to\mu^+\mu^-)$ in LHS1 scenario. This correlation is valid in any oasis due
to the assumed equal sign of the leptonic couplings in (\ref{DAnunu}), 
although, as seen in the plot, the size of NP contribution may depend
on the oasis considered. Significant NP effects are still
possible and suppression of  
$\mathcal{B}(B_s\to\mu^+\mu^-)$ below the SM
value will also imply the suppression of $\mathcal{B}(B\to X_s \nu\bar\nu)$.
If the future data will disagree with this
pattern, the rescue could come from the flip of the signs in $\nu\bar\nu$
or $\mu^+\mu^-$ couplings provided this is allowed by leptonic decays of
$Z'$. As seen on the right of Fig.~\ref{fig:BXsnuvsBsmuLHS1} additional information can come from 
the correlation between
$\mathcal{B}(B\to X_s \nu\bar\nu)$ vs $S_{\psi\phi}$. 

\begin{figure}[!tb]
\centering
\includegraphics[width = 0.45\textwidth]{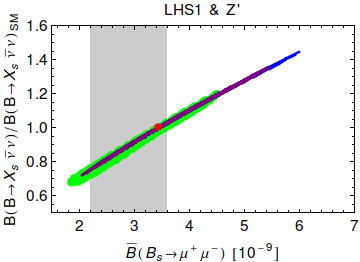}
\includegraphics[width = 0.45\textwidth]{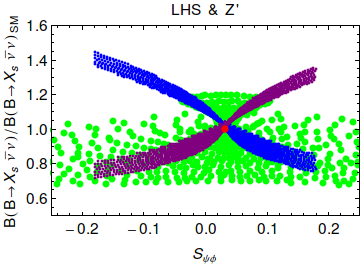}
\caption{\it $\mathcal{B}(B\to X_s \nu\bar\nu)$ versus
$\mathcal{B}(B_s\to\mu^+\mu^-)$ (left) and $\mathcal{B}(B\to X_s \nu\bar\nu)$ versus $S_{\psi\phi}$ (right) in LHS1 for $M_{Z^\prime} =
1~$TeV. The green points 
indicate the regions that are compatible with $b\to s\ell^+\ell^-$ 
constraints. 
}
 \label{fig:BXsnuvsBsmuLHS1}~\\[-2mm]\hrule
\end{figure}

As already emphasized above the decays in question are sensitive to the 
presence of right-handed currents. This is best seen in Fig.~\ref{fig:ep2vseta} 
where we show the
results for all four scenarios considered by us in the $\epsilon-\eta$ plane. 
Indeed a future
determination of $\epsilon$ and 
$\eta$ will tell us whether the nature chooses one of the scenario considered 
by us or a linear combination of them.

As $b\to s\ell^+\ell^-$ transitions have large impact  on the allowed 
size of right-handed currents we show two examples of it in Figs.~\ref{fig:BKnuvsBsmu} and~\ref{fig:BKstarnuvsBKnu}.

\begin{figure}[!tb]
\centering
\includegraphics[width = 0.7\textwidth]{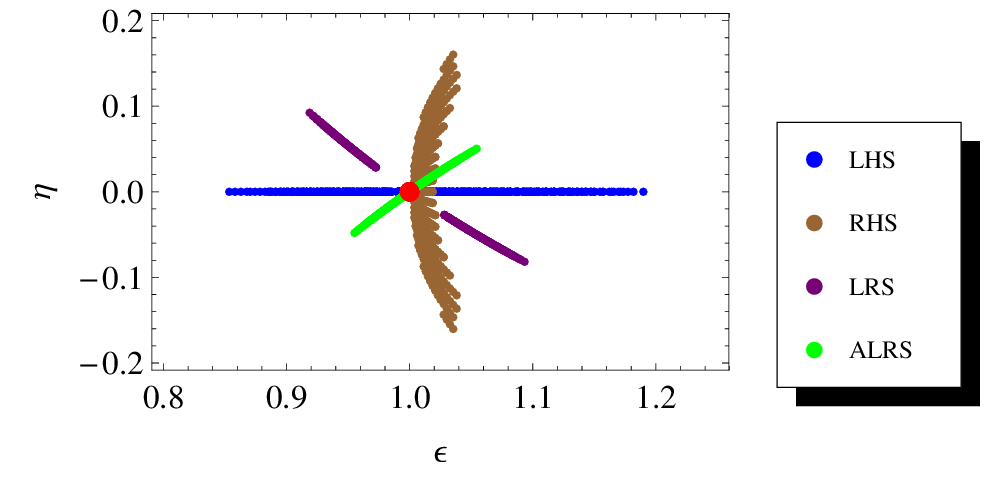}

\caption{\it {  $\eta$ versus $\epsilon$ for scenario LHS1, RHS1, LRS1 and ALRS1.}
}
 \label{fig:ep2vseta}~\\[-2mm]\hrule
\end{figure}

\begin{figure}[!tb]
\centering
\includegraphics[width = 0.45\textwidth]{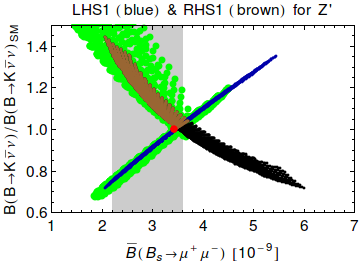}
\includegraphics[width = 0.45\textwidth]{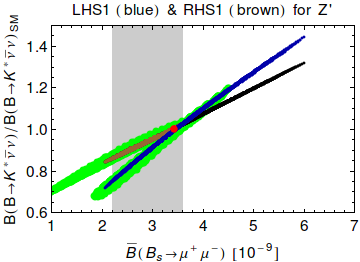}
\caption{\it  $\mathcal{B}(B\to K\nu\bar\nu)$ versus
$\mathcal{B}(B_s\to\mu^+\mu^-)$  (left) and  $\mathcal{B}(B\to K^\star\nu\bar\nu)$ versus
$\mathcal{B}(B_s\to\mu^+\mu^-)$  (right) for
$M_{Z^\prime} = 1~$TeV in LHS1 (blue for both oases $A_{1,3}$) and RHS1 (brown for both oases $A_{1,3}$)). The green points 
indicate the regions that are compatible with $b\to s\ell^+\ell^-$ 
constraints. Black points in RHS show the excluded area due to $b\to 
s\ell^+\ell^-$ transitions explicitly. Gray region: exp 1$\sigma$ range  
$\overline{\mathcal{B}}(B_s\to\mu^+\mu^-) = (2.9\pm 0.7)\cdot 10^{-9}$. Red point: SM central
value. }
 \label{fig:BKnuvsBsmu}~\\[-2mm]\hrule
\end{figure}

\begin{figure}[!tb]
\centering
\includegraphics[width = 0.45\textwidth]{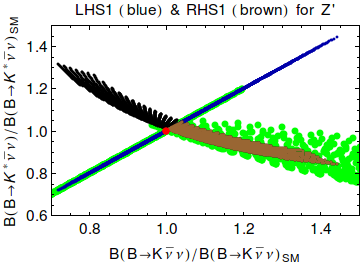}
\includegraphics[width = 0.45\textwidth]{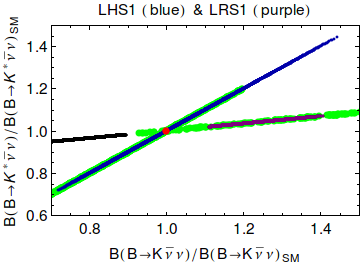}
\caption{\it  $\mathcal{B}(B\to K^\star\nu\bar\nu)$ versus
$\mathcal{B}(B\to K\nu\bar\nu)$  for
$M_{Z^\prime} = 1~$TeV in LHS1 (blue for both oases $A_{1,3}$),  RHS1 (brown for both oases $A_{1,3}$) and LRS1 (purple for both oases
$A_{1,3}$)). The green points 
indicate the regions that are compatible with $b\to s\ell^+\ell^-$ 
constraints. Black points in RHS show the excluded area due to $b\to 
s\ell^+\ell^-$ transitions explicitly. Red point:
SM central value.}
 \label{fig:BKstarnuvsBKnu}~\\[-2mm]\hrule
\end{figure}

\boldmath
\subsection{Step 10: The Ratio $\epe$}
\unboldmath
\subsubsection{Preliminaries}
One of the important actors of the 1990s in flavour physics was the ratio
$\epe$
 that measures the size of the direct CP
violation in $K_L\to\pi\pi$ 
relative to the indirect CP violation described by $\varepsilon_K$. 
In the SM $\varepsilon^\prime$ is governed by QCD penguins but 
receives also an important destructively interfering
 contribution from electroweak
penguins that is generally much more sensitive to NP than the QCD
penguin contribution.

The big challenge in  making predictions for $\epe$ within the SM and 
its extensions is the strong cancellation of 
QCD penguin contributions and electroweak penguin contributions to
 this ratio. In the SM QCD penguins give positive contributions, while the electroweak 
penguins  negative ones.  In order to obtain useful prediction for $\epe$ in the SM the precision on 
the corresponding hadronic parameters
$\bsi$ and $\bei$ should be at least $10\%$. 
Recently significant progress has been made in the case of $\bei$ that is 
relevant for electroweak penguin contribution \cite{Blum:2011ng} but the calculation of 
$\bsi$ is even more important. There are some hopes that 
also this parameter could be known with satisfactory precision in this 
decade \cite{Christ:2009ev,Christ:2013lxa}.
 
This would really be good, as the
calculations of  short distance contributions to this ratio (Wilson 
coefficients of QCD and electroweak penguin operators)  within the SM have been 
 known already 
for twenty years at the NLO level \cite{Buras:1993dy,Ciuchini:1993vr}
and present technology could extend them 
to the NNLO level if necessary. First steps in this direction have been done in 
\cite{Buras:1999st,Gorbahn:2004my}.

 In the { most studied} extensions of the SM, the QCD penguin contributions are not 
modified significantly. On the other hand large NP contributions to electroweak penguins are possible. But they are often correlated with $\klpn$ and $\kpn$ 
decays so that considering $\epe$ and these two decays simultaneously useful 
constraints on model parameters can be derived, again subject to the uncertainties in  $\bsi$ and $\bei$.

The present experimental world average  from 
NA48 \cite{Batley:2002gn}   and 
KTeV \cite{AlaviHarati:2002ye,Worcester:2009qt}, 
\be
\epe=(16.6\pm 2.3)\cdot 10^{-4}~,
\ee
could have an important impact on several extensions of the SM discussed
if $\bsi$ and $\bei$ were known.
An analysis of $\epe$ in the LHT model demonstrates this problem
in explicit terms \cite{Blanke:2007wr}. If one uses $\bsi=\bei=1$ as obtained 
in the large $N$ approach \cite{Bardeen:1986uz,Buras:2014maa}, $(\epe)_{\rm SM}$ is in the ballpark of the experimental data although below it and sizable departures of $\mathcal{B}(\klpn)$ from its SM 
value are not allowed. $\kpn$ being CP conserving and consequently 
not as strongly correlated with $\epe$ as $\klpn$ could still be 
enhanced by $50\%$. On the other hand if $\bsi$ and $\bei$ are different 
from unity and $(\epe)_{\rm SM}$ disagrees with experiment, much more
room for enhancements of rare $K$ decay branching ratios through
NP contributions is available. See also new insight from the recent
analysis in \cite{Buras:2014sba}.
Reviews of $\epe$ can be found in 
\cite{Bertolini:1998vd,Buras:2003zz,Pich:2004ee,Cirigliano:2011ny,Bertolini:2012pu}.

\subsubsection{Basic Formula in the Standard Model}
In the SM ten operators pay the tribute to the $\epe$. These are

{\bf Current--Current:}
\begin{equation}\label{O1} 
Q_1 = (\bar s_{\alpha} u_{\beta})_{V-A}\;(\bar u_{\beta} d_{\alpha})_{V-A}
~~~~~~Q_2 = (\bar su)_{V-A}\;(\bar ud)_{V-A} 
\end{equation}

{\bf QCD--Penguins:}
\begin{equation}\label{O2}
Q_3 = (\bar s d)_{V-A}\sum_{q=u,d,s,c,b}(\bar qq)_{V-A}~~~~~~   
 Q_4 = (\bar s_{\alpha} d_{\beta})_{V-A}\sum_{q=u,d,s,c,b}(\bar q_{\beta} 
       q_{\alpha})_{V-A} 
\end{equation}
\begin{equation}\label{O3}
 Q_5 = (\bar s d)_{V-A} \sum_{q=u,d,s,c,b}(\bar qq)_{V+A}~~~~~  
 Q_6 = (\bar s_{\alpha} d_{\beta})_{V-A}\sum_{q=u,d,s,c,b}
       (\bar q_{\beta} q_{\alpha})_{V+A} 
\end{equation}

{\bf Electroweak Penguins:}
\begin{equation}\label{O4} 
Q_7 = \frac{3}{2}\;(\bar s d)_{V-A}\sum_{q=u,d,s,c,b}e_q\;(\bar qq)_{V+A} 
~~~~~ Q_8 = \frac{3}{2}\;(\bar s_{\alpha} d_{\beta})_{V-A}\sum_{q=u,d,s,c,b}e_q
        (\bar q_{\beta} q_{\alpha})_{V+A}
\end{equation}
\begin{equation}\label{O5} 
 Q_9 = \frac{3}{2}\;(\bar s d)_{V-A}\sum_{q=u,d,s,c,b}e_q(\bar q q)_{V-A}
~~~~~Q_{10} =\frac{3}{2}\;
(\bar s_{\alpha} d_{\beta})_{V-A}\sum_{q=u,d,s,c,b}e_q\;
       (\bar q_{\beta}q_{\alpha})_{V-A} 
\end{equation}
Here, $\alpha,\beta$ denote colours and $e_q$ denotes the electric quark charges reflecting the
electroweak origin of $Q_7,\ldots,Q_{10}$. Finally,
$(\bar sd)_{V-A}\equiv \bar s_\alpha\gamma_\mu(1-\gamma_5) d_\alpha$.

The NLO renormalization group analysis of these operators is rather involved 
\cite{Buras:1993dy,Ciuchini:1993vr} but eventually one can derive an analytic 
formula in terms of the basic one-loop functions \cite{Buras:2003zz}. 
The most recent version of this formula is given as follows \cite{Buras:2014sba}
\be \frac{\varepsilon'}{\varepsilon}= a\IM\lambda^{(K)}_t
\cdot F_{\varepsilon'}(x_t)
\label{epeth}
\ee
where $\lambda^{(K)}_t=V_{td}V_{ts}^*$,  $a=0.92\pm0.02$ and 
\be
F_{\varepsilon'}(x_t) =P_0 + P_X \, X_0(x_t) + 
P_Y \, Y_0(x_t) + P_Z \, Z_0(x_t)+ P_E \, E_0(x_t)~,
\label{FE}
\ee
with the first term dominated by QCD-penguin contributions, the next three 
terms by electroweak penguin contributions and the last term being 
totally negligible. The one-loop functions $X_0$, $Y_0$ and $Z_0$ can be 
found in (\ref{X0}), (\ref{YSM}) and (\ref{ZSM}), respectively.
The coefficients $P_i$ are given in terms of the non-perturbative parameters
$R_6$ and $R_8$ defined in (\ref{RS}) as follows:
\begin{equation}
P_i = r_i^{(0)} + 
r_i^{(6)} R_6 + r_i^{(8)} R_8 \,.
\label{eq:pbePi}
\end{equation}
The coefficients $r_i^{(0)}$, $r_i^{(6)}$ and $r_i^{(8)}$ comprise
information on the Wilson-coefficient functions of the $\Delta S=1$ weak
effective Hamiltonian at the NLO 
and their numerical values can be found in \cite{Buras:2014sba}. These 
numerical values are chosen to satisfy the so-called $\Delta I=1/2$ rule and
emphasize the dominant dependence on the hadronic matrix elements residing in 
 the QCD-penguin operator $Q_6$ and the  electroweak
penguin operator $Q_8$. From Table~1 in  \cite{Buras:2014sba} we find that
for the central value of $\alpha_s(M_Z)=0.1185$ the largest are the coefficients 
$r_0^{(6)}$ and $r_Z^{(8)}$ representing QCD-penguin and electroweak penguin contributions, respectively:
\be
r_0^{(6)}=16.8, \qquad r_Z^{(8)}=-12.6~.
\ee
The fact that these coefficients are of the similar size but having opposite 
signs has been the problem since the end of 1980s when the electroweak penguin contribution 
 increased 
in importance due to the large top-quark mass \cite{Flynn:1989iu,Buchalla:1989we}.

The parameters
$R_6$ and $R_8$ are directly related to the $B$-parameters $\bsi$ and $\bei$ 
representing the hadronic matrix elements of $Q_6$ and $Q_8$, respectively. 
They are defined as
\be\label{RS}
R_6\equiv 1.13\bsi\left[ \frac{114\mev}{m_s(m_c)+m_d(m_c)} \right]^2,
\qquad
R_8\equiv 1.13\bei\left[ \frac{114\mev}{m_s(m_c)+m_d(m_c)} \right]^2,
\ee
where the factor $1.13$ signals the decrease of the value of $m_s$ since 
the analysis in  \cite{Buras:2003zz} has been done. 

A detailed analysis of $\epe$ is clearly beyond this review and we would like 
to make only a few statements. 

In  \cite{Buras:2003zz} it has been found that with $R_8=1.0\pm0.2$  
as obtained 
at that time from lattice QCD, the data could be reproduced within the SM for 
$R_6=1.23\pm0.16$. While in 2003 this value would correspond to $\bsi=1.23$, 
the change in the value of $m_s$ would imply  $\bsi=1.05$, very close to 
the large $N$ value. Now the most recent evaluation of $\bei$  from 
lattice QCD \cite{Blum:2011ng,Blum:2011pu,Blum:2012uk} finds $\bei\approx 0.65$ and thereby implying that $R_8\approx 0.8$. 

 A very recent analysis of $\epe$ in the SM \cite{Buras:2014sba} which uses 
this lattice result finds indeed that for  $\bsi=1.0$  the agreement of 
the SM with the data is good although parametric uncertainties, in 
particular due to $\vub$ and $\vcb$,  allow still for sizable NP 
contributions. Undoubtly we need sufficient precision on  $\bsi$ and these 
two CKM parameters in order to have a clear cut picture of $\epe$.
We are looking forward to the improved values of $\vub$, $\vcb$, $\bsi$ and $\bei$ and 
expect that in the second half of this decade $\epe$ will become again 
an important actor in particle physics. The correlations with $\klpn$ 
and $\kpn$ reanalyzed recently in  \cite{Buras:2014sba} should then 
help us to select favourite NP scenarios in particular if the experimental 
branching ratios for these decays will be known with sufficient accuracy.

\subsection{Step 11: Charm and Top Systems}
\subsubsection{Preliminaries}
Our review is dominated by mixing and decays in $K$, $B_d$ and $B_s$ meson 
systems. In the last two steps we want to emphasize that charm and top physics 
(this step) as well as lepton flavour violation, electric dipole moments and 
$(g-2)_{e,\mu}$ discussed in the next step play important roles in the search 
for new physics. Our discussion will be very brief but we hope that general 
statements and the selected references are still useful for non-experts.
\subsubsection{Charm}
The study of $D$ mesons allows to explore in a unique manner the physics of 
up-type quarks in FCNC processes. This involves $D^0-\bar D^0$ mixing, direct 
and mixing induced CP-violation and rare decays of mesons. Excellent summary 
of the present experimental and theoretical status as well of the future prospects for this field can be found in chapter 4 of
\cite{Bediaga:2012py}. We cannot  add anything new to the information given there but not working recently in 
this field we can provide a number of unbiased statements.

Charm decays have the problem that the intermediate scale of roughly $2\gev$ 
does not allow on the one hand 
to use methods like chiral perturbation theory or large $N$, that are
useful for $K$ physics. On the other hand the methods as heavy quark effective 
theories are not as useful here as in the $B_{s,d}$ systems. Fortunately 
lattice simulation are mostly done around this scale so that the future of this 
field will definitely depend on the progress  made by lattice QCD.

Due to the presence of down quarks in the loop diagrams governing FCNCs within 
the SM, GIM mechanism is very effective so that the short distance part 
of any SM contribution is strongly suppressed. Consequently the background to 
possible NP contributions from this part is significantly smaller than in 
the case of $K$ and $B_{s,d}$ meson systems. This is in particular the case 
of CP violation which is predicted to be tiny in $D$ meson system. Unfortunately large background to NP from hadronic effects make the study of NP effects 
in this system very challenging and even the originally large direct CP 
violation observed by LHCb \cite{Aaij:2011in} could not be uniquely attributed to the signs of 
NP. The recent update shows that the anomaly in question  basically 
disappeared \cite{Aaij:2013bra} but NP could still be hidden under hadronic 
uncertainties.

Yet, the situation could improve in the future and the large amount of theoretical work prompted by these initially exciting LHCb results will definitely 
be very useful when the data improve. It is impossible to review this work which is summarized in \cite{Bediaga:2012py} and we will mention
here only few 
papers that fit very well to the spirit of our review as they discuss
correlations between CP violation in charm decays and other observables 
\cite{Isidori:2011qw,Hochberg:2011ru,Isidori:2012yx}.
These correlations, as in the decays discussed by us in previous steps, 
depend on the model considered, so they may help to identify the NP at 
work. They do not only involve observables in charm system 
like rare decays $D^0\to\phi\gamma$ or 
$D^0\to\mu^+\mu^-$ but also observables measured at high-$p_T$, such as 
$t\bar t$ asymmetries, another highlight from the LHC.

In this context one should mention correlations between $D$ and $K$, which 
could be used to constrain NP effects in $K$ system through the ones in 
charm and vice versa \cite{Blum:2009sk,Gedalia:2012pi}. In particular the universality of CP 
violation in flavour-changing decay processes elaborated in \cite{Gedalia:2012pi}
 allows to predict direct correspondence between NP contributions to 
the direct CP violation in charm and $K_L\to\pi\pi$ represented by $\epe$. 
There is no question about that charm physics will play a significant role 
in the search for NP by  constraining theoretical models and  offering 
complementary information to the one available from  $K$ and $B_{s,d}$ system. 
Yet, from the present perspective clear cut conclusions about the presence 
or absence of relevant NP contributions will be easier to reach by 
studying observables considered by us in previous steps.
\subsubsection{Top Quark}

The heaviest quark, the top quark, played already  a dominant role in our 
review. It governs SM contributions to all observables discussed by us. The 
fact that the SM is doing well indicates that the structure of the CKM matrix 
with three hierarchical top quark couplings to lighter quarks
\be
\vtd\approx 8\times 10^{-3}, \qquad \vts\approx 4\times 10^{-2}, \qquad \vtb\approx 1
\ee
combined with the GIM mechanism represents the flavour properties of the top 
quark well. Yet, as the LHC became a top quark factory, properties of the 
top can be studied also directly, through its production and decay. In the 
latter case FCNC processes like $t\to c\gamma$ can be investigated. It is also 
believed that the top quark is closely related to various aspects of 
electroweak symmetry breaking and the problem of naturalness. Indeed, the top quark having the largest coupling to the Higgs field is the main reason for  the severe fine tuning necessary to keep the Higgs mass close to the electroweak 
scale.

For these reasons we expect that the direct study of top physics, both flavour 
conserving and flavour violating will give us a profound insight into 
short distance dynamics, in particular as hadronic uncertainties at such short 
distance
scales are much smaller than in decays of mesons. The observation of a large 
forward backward asymmetry in $t\bar t$ production at the Tevatron and the 
intensive theoretical studies aiming to explain this phenomenon have shown that 
this type of physics has great potential in constraining various extensions of 
the SM. As this material goes beyond the goals of our review we just wanted 
to emphasize that this is an important field in the search for NP.
A useful collection of articles, which deal with top and flavour physics 
in the LHC era can be found in \cite{Buras:2012ub}.  A detailed study 
of flavour sector with up vector-like quarks including correlations among 
various observables can be found in \cite{Botella:2012ju}.

\boldmath
\subsection{Step 12: Lepton Flavour Violation,  $(g-2)_{\mu,e}$ and EDMs}
\unboldmath
\subsubsection{Preliminaries}

Our review deals dominantly with quark flavour violating processes. Yet 
in the search for NP an important role will also be played by 
\begin{itemize}
\item  Neutrino oscillations, neutrinoless double $\beta$ decay
\item Charged lepton violation
\item Anomalous magnetic moment of the muon  $a_{\mu} =\tfrac{1}{2}(g-2)_\mu$
\item Electric dipole moments of the neutron, atoms and leptons

\end{itemize}

In what follows we will only very briefly discuss these items. 
 Selected reviews of these topics can be found in \cite{Raidal:2008jk,Hewett:2012ns,Jegerlehner:2009ry,Engel:2013lsa,Bernstein:2013hba},
where many references
can be 
found. The study of correlations between LFV, $(g-2)_\mu$ and EDMs in supersymmetric 
flavour models and SUSY GUTS  
can be found in \cite{Altmannshofer:2009ne,Hisano:2009ae,Buras:2010pm,Girrbach:2009uy}. Analogous correlations in models with vector-like leptons have been 
presented in \cite{Falkowski:2013jya} and general expressions for these observables in terms of Wilson coefficients of dimension-six operators can be found in 
 \cite{Crivellin:2013hpa}.

Concerning the first item, the observation of neutrino oscillations is a  clear signal of physics beyond the SM and so far together with
Dark Matter  and the matter-antimatter asymmetry observed in our universe the only  clear sign of  NP. In order to accommodate 
neutrino masses one needs to extend the SM. 
 The most straightforward way is to proceed
in the same manner as for quark and  charged lepton masses  and just introduce three right-handed neutrinos that are singlets
under the SM gauge group anyway. A Dirac mass term is then generated via the usual Higgs coupling $\bar\nu_L Y_{\nu} H \nu_R$. However then
there is also the possibility for a Majorana mass term for the right-handed neutrinos since it is gauge invariant. One would need to
introduce or postulate a further symmetry to forbid this term which is also already an extension of the SM. Furthermore this Majorana mass
term introduces an additional scale $M_R$ and since it is not protected by any symmetry it could be rather high. Then the seesaw mechanism
is at work and can generate light neutrino masses as observed in nature.  Another possibility to get
neutrino masses without
right-handed neutrinos is the introduction of an additional Higgs-triplet field. Either way, the accommodation of neutrino masses requires
an extension of the SM.

In the second and last point from above the interest in the related observables 
is based on the fact that they are suppressed within the SM to 
such a level that any observation of them would clearly signal 
physics beyond the SM. In this respect they differ profoundly from 
all processes discussed by us until now, which suffer from a large 
background coming from the SM and one needs precise theory and precise
experiment to identify NP.
Although  $a_{e,\mu}$ are both flavour- and CP-conserving they also offer powerful probes to test NP.

\boldmath
\subsubsection{Charged Lepton Flavour Violation}
\unboldmath

 The discovery of neutrino oscillations has shown that the individual lepton numbers are not conserved. However, no charged 
 lepton flavour violating decays have been observed to date. 
 In the SM  enriched by  light neutrino masses lepton-flavour violating decays $\ell_j\to\ell_i\gamma$  occur at unobservable
small rates, because the transition amplitudes are suppressed by a factor of $(m_{\nu_j}^2-m_{\nu_i}^2)/M_W^2$. 
On the other hand in many extensions of the SM, like supersymmetric models, littlest Higgs
model with T-parity 
(LHT) or the SM with sequential fourth generation  (SM4) measurable in this decade branching ratios are predicted in particular when the masses of involved new particles are in the LHC reach. However, it should be stressed that in principle LFV can even be sensitive to energy scales as high as $1000\tev$. 
 For a recent analysis within mini-split supersymmetry see   \cite{Altmannshofer:2013lfa}.

The most prominent role in  the LFV studies play the decays
\be
 \mu\to e\gamma,\qquad \tau\to\mu\gamma, \qquad\tau\to e\gamma
\ee
but also the study of decays 
\be
\mu^-\to e^-e^+e^-,\qquad \tau^-\to\mu^-\mu^+\mu^-, \qquad \tau^-\to e^-e^+e^-
\ee
as well as $\mu-e$ conversion in nuclei offer in conjunction with $l_i\to l_j\gamma$ powerful tests of NP.

As our review is dominated by correlations let us just mention how a clear 
cut distinction between supersymmetric models, LHT model and SM4 is possible 
on the basis of these decays.
While it is not possible to distinguish the LHT model from  
         the supersymmetric models on the basis of 
         $\mu\to e\gamma$ alone, it has been  pointed out in 
\cite{Blanke:2007db} that such a distinction can be made 
by measuring any of the 
           ratios $\mathcal{B}(\mu\to 3e)/\mathcal{B}(\mu\to e\gamma)$, 
          $\mathcal{B}(\tau\to 3\mu)/\mathcal{B}(\tau\to \mu\gamma)$, etc. In supersymmetric
          models all these decays are governed by dipole operators so 
         that these ratios are $\ord(\alpha)$ 
\cite{Ellis:2002fe,Arganda:2005ji,Brignole:2004ah,Paradisi:2005tk,Paradisi:2006jp,Paradisi:2005fk,Girrbach:2009uy}.
        In the
        LHT model the LFV decays with three leptons in the final state are
        not governed by dipole operators but by $Z$-penguins and box diagrams
        and the ratios in question turn out to be by almost an order of
        magnitude larger than in supersymmetric models. 
         Other analyses of LFV in the LHT model can be found in 
   \cite{delAguila:2008zu,Goto:2010sn}   and in the MSSM in \cite{Girrbach:2009uy}. In the latter  paper $(g-2)_e$ was used to 
probe  lepton flavour
violating couplings that are correlated with $\tau\to e\gamma$. 

Similarly, as pointed out in \cite{Buras:2010cp}
the pattern of the LFV branching ratios in the SM4 differs significantly from the one encountered in the MSSM, allowing to distinguish these
two models with the help of LFV processes in a transparent manner. 
Also differences from the LHT model were identified.

A detailed analysis of LFV in various extensions of the SM is also motivated by the prospects
in the measurements of LFV processes with much higher sensitivity than
presently available. In particular the MEG experiment at PSI  is 
already testing
 $\mathcal{B}(\mu\to e\gamma)$ at the level of
$\ord(10^{-13})$.  The current upper bound is \cite{Adam:2013mnn}
\begin{align}\label{MEGbound}
 \mathcal{B}(\mu\to e\gamma)\leq 5.7\cdot 10^{-13}\,.
\end{align}
This bound puts also some GUT models under pressure as for example the model discussed in Sec.~\ref{sec:CMM}.
An upgrade for MEG is also already approved \cite{Baldini:2013ke} where they expect to improve the sensitivity down to $6\cdot
10^{-14}$ after three years of running and there is an approved proposal at PSI to do $\mu\to eee$ \cite{Blondel:2013ia}.
The planned accuracy of SuperKEKB of $\ord(10^{-8})$ for
  $\tau\to\mu\gamma$ is also of great interest. This decay can also be studied 
at the LHC. 

An improved upper bound on $\mu-e$ conversion in titanium will also be very important. In this context the dedicated
J-PARC
experiment PRISM/PRIME \cite{Barlow:2011zza} should reach the sensitivity of
$\ord(10^{-18})$, i.\,e. an improvement by six orders of magnitude relative to the present upper bound from SINDRUM-II at PSI
\cite{Kaulard:1998rb}.  Mu2e collaboration will measure $\mu-e$ conversion on aluminium to $6\cdot 10^{-17}$ at 90\% CL around 2020
\cite{Abrams:2012er} which is a factor of $10^4$ better than SINDRUM-II. Another improvement of a factor 10 is planed to { be 
reached
with Project X at Fermilab \cite{Kronfeld:2013uoa}.} In \cite{Cirigliano:2009bz} the model discriminating power of a combined
phenomenological analysis of $\mu \to e \gamma$ and $\mu \to e$
conversion on different { nuclei targets} is discussed. They found that in most cases going from aluminuim to titanium is not very
model-discriminating. A realistic discrimination among models requires a measure
of $\mathcal{B}(\mu\to e,Ti)/\mathcal{B}(\mu\to e,Al)$
 at the level of 5\% or better.

For further detailed review of LFV see \cite{Raidal:2008jk,Feldmann:2011zh,Ibarra:2010zz}. An experimenter's guide for charged LFV  can be found in 
 \cite{Bernstein:2013hba}.

\boldmath
\subsubsection{Anomalous magnetic moments $(g-2)_{\mu,e}$}
\unboldmath
The anomalous magnetic moment of the muon 
\be
a_{\mu}=\frac{(g-2)_\mu}{2}
\ee
 provides an
excellent test for physics beyond the SM.
It can be extracted from the photon-muon vertex function $\Gamma^{\mu}(p^{\prime},p)$
\begin{equation}
\bar{u}(p^{\prime}) \Gamma^{\mu}(p^{\prime},p) u(p)=
\bar{u}(p^{\prime})\left[\gamma^{\mu} F_{V}(q^{2}) + (p+p^{\prime})^{\mu} F_{M}(q^2)\right]u(p)\,,
\end{equation}
with
\begin{equation}
a_{\mu}=-2m_\mu F_{M}(0)\,.
\end{equation}
On the theory side $a_\mu$ receives four dominant contributions:
\begin{equation}\label{amuSM}
a_{\mu}^\text{SM} =a_{\mu}^\text{QED} + a_{\mu}^\text{ew} + a_{\mu}^{\gamma\gamma}+ a_{\mu}^\text{hvp}.
\end{equation}
While the QED \cite{Kinoshita:2004wi,Passera:2006gc,Aoyama:2012wj,Aoyama:2012wk}
and electroweak contributions \cite{Czarnecki:2002nt,Jegerlehner:2009ry} to $a_\mu^\text{SM}$ are known very precisely and the
light--by--light contribution 
  $a_\mu^{\gamma\gamma}$  is currently 
known  with an acceptable  accuracy \cite{Prades:2009tw,Prades:2009qp},
the theoretical uncertainty is dominated by the hadronic vacuum polarization. 
Review of the relevant calculations 
of all these contributions and related extensive analyses can be found in \cite{Jegerlehner:2009ry,Benayoun:2012wc}.

According to the most recent analysis in \cite{Benayoun:2012wc},
the very precise measurement of $a_\mu$ by  the  E821 experiment \cite{Bennett:2006fi} in Brookhaven
differs from its SM prediction by roughly
$4.6\sigma$:
\begin{equation}
a^{\rm{exp}}_{\mu}-a^{\rm{SM}}_{\mu}=(39.4\pm8.5)\times10^{-10},
\label{a-mu}
\end{equation}
where we added various errors discussed in  \cite{Benayoun:2012wc} in quadrature. 

Many models beyond the SM try to explain this discrepancy, especially supersymmetric models were very popular
\cite{Stockinger:2007pe,Marchetti:2008hw,Feroz:2008wr,Nojiri:2008aa,Degrassi:1998es,Heinemeyer:2003dq,Heinemeyer:2004yq}. In SUSY the 
discrepancy could
easily be accommodated for relatively light smuon masses and large $\tan\beta$.
However so far no  light SUSY  particles  have been
discovered. 
Another approach was followed in \cite{Crivellin:2010ty} where the interplay of $(g-2)_{\mu}$ and a soft muon
Yukawa coupling that
is generated radiatively in the MSSM  was studied. 
{With the increased SUSY mass scale the explanation of $(g-2)_\mu$ anomaly 
becomes difficult \cite{Jegerlehner:2012ju}.}

 Of course a new experiment would also be desirable. Fortunately,
the $g-2$ ring at BNL has been disassembled and is on its way to Fermilab for a run around 2016.  The
overall
error should go down by a factor of 2.
 Thus if the central value will remain unchanged the discrepancy with the SM will increas to more than  $8.0\sigma$.

The anomalous magnetic moment of the muon $a_\mu$  is more sensitive 
to lepton flavour conserving NP   than $a_e$ and consequently 
the latter was not as popular as $a_\mu$ in the last decade. However, as 
emphasized in  \cite{Girrbach:2009uy}, the fact that $a_e$ is very precisely 
measured and very precisely calculated within the SM it can also be used 
to probe NP, even if the theory agrees very well with experiment.
Indeed, $a_e$ plays a central role in
QED since its precise measurement provides the best
source of $\alpha_e$  assuming
the validity of QED \cite{Hanneke:2008tm}.  Conversely, one can use a
value of $\alpha_\text{em}$ from a less precise measurement and insert
it into the theory prediction for $a_e$ to probe NP. The most recent calculation yields $a_e  = 1\; 159\; 652\; 182.79 \left(7.71\right)
\times 10^{-12}$ \cite{Aoyama:2007mn},
where the largest uncertainty comes from the second-best measurement of
$\alpha_\text{em}$ which is $ \alpha_\text{em}^{-1} = 137.03599884(91)$ from a Rubidium atom experiment \cite{Clade:2006zz}.
Usually NP contributions to $a_e$ are small
due to the smallness of the electron Yukawa coupling and the 
suppression of the NP scale.  However, multiple flavour changes, resulting effectively in a lepton flavour conserving loop could be
enhanced due to the $\tau$ Yukawa coupling \cite{Girrbach:2009uy}.

\subsubsection{Electric Dipole Moments (EDMs)}
 Even though the experimental sensitivities have improved a lot no EDM of a fundamental particle has been observed so far. 
Nevertheless EDM experiments have already put strong limits on NP models. A permanent EDM of a fundamental particle violates both 
T and P, and  thus~-- assuming CPT symmetry~-- is another way to measure CP violation. 
In the SM the only CP-violating phase of the CKM matrix enters quark EDMs first at three loop (two loop EW + one loop QCD) which
results in negligibly small SM EDMs. 
Consequently EDMs are excellent probes of new CP violating phases of NP models, especially flavour blind phases,  and of strong CP
violation.

A recent review about EDMs can be found in \cite{Engel:2013lsa} which updates the review in \cite{Pospelov:2005pr}. See also \cite{Batell:2012ge}. 
As discussed in
\cite{Engel:2013lsa} by naive dimension analysis  EDMs probe a NP
scale of several TeV. This assumes order one CP-violating phases  $\phi_\text{CP}$ for the electron EDM that arises at one loop 
order:
\begin{align}\label{equ:de}
 d_e\approx e \frac{m_e}{\Lambda^2}\frac{\alpha_e}{4\pi}\sin\phi_\text{CP}\approx
\frac{1}{2}\left(\frac{1~\text{TeV}}{\Lambda}\right)^2\sin\phi_\text{CP} \cdot 10^{-13} e\, \text{fm}~.
\end{align}
 Recently, the upper bound on $d_e$ has been improved by an order of magnitude 
with respect to the previous bound in  \cite{Hudson:2011zz} and reads 
\cite{Baron:2013eja}
\be\label{newde}
|d_e|\le 8.7\cdot 10^{-16} e\,\text{fm}.
\ee
This implies for the CP-violating phase
$|\sin\phi_\text{CP}| 
\lesssim \left(\tfrac{\Lambda}{6~\text{TeV}}\right)^2$. 
The implications of this new bound on MFV have been investigated in \cite{He:2014fva} and other analyses are expected in the near future.

The scale of NP can be even higher for the neutron
and $^{199}$Hg EDMs as they are sensitive to the chromo-magnetic EDM which enters with a factor of $\alpha_s$ rather than the 
fine 
structure
constant $\alpha_e$, pushing the sensitivity closer to 10~TeV. As one can see from (\ref{equ:de}) the sensitivity to the NP 
scale goes
as
$1/\Lambda^2$, whereas in many other cases such as
lepton flavour violation the sensitivity goes as $1/\Lambda^4$. 
Future EDM measurements aim to improve their sensitivity by approximately two orders of magnitude which will then push the mass 
scale
sensitivity into the (20-100)~TeV range.

There are different sources for EDMs.  For hadronic EDMs there is the $\theta$ term of QCD which is very much constrained due to 
the
non-observation of permanent EDMs  of the $^{199}$Hg atom and neutron. Apart from the $\theta$ term, the SM CKM induced EDMs 
would 
be far
smaller in magnitude than the next generation EDM sensitivities.
Consequently, one does not need the same kind of refined hadronic structure computations as one often needs in flavour physics to 
interpret
the EDM results in terms of NP. That being said, the hadronic matrix element problem remains a considerable challenge. 
At 
dimension
six one encounters
several different operators for the first generation fermions that could give rise to EDMs: pure gauge operators $\tilde{G}GG$, four-fermion operators (semi-leptonic and non-leptonic), gauge-higgs operators $\varphi^\dagger\varphi \tilde{G} G$ and gauge-higgs-fermion operators $(\bar Q T^A q_R)\varphi G$.

In experiments one often deals with composite systems and thus nuclear physics is important in determining the EDMs of neutral atoms.
Nuclear structure can also provide an amplifier of atomic EDMs.
In heavier neutral systems there is the shielding of the EDMs of constituents of one charge by those of the other (e.g. protons and
electrons). The transmission of CP violation through a nucleus into an atom must overcome this shielding. Its effectiveness in   doing so
is expressed by a nuclear Schiff moment. In nuclei with asymmetric shapes Schiff moments can be enhanced by two or three orders of
magnitude. For example an octupole deformed 
nuclei such as $^{225}$Ra give enhanced nuclear Schiff moments and, thus, an enhanced atomic EDMs in a diamagnetic system.

Flavour diagonal CP violating phases as needed for electroweak baryogenesis can be strongly constrained by EDMs. In the MSSM, 
for example, this requires rather heavy first and second generation sfermions but at the same time light electroweak 
gauginos below one TeV as well as  a subset of the third
generation sfermions (see  \cite{Morrissey:2012db} for details). { However 
as can be deduced from the plots in \cite{Kozaczuk:2012xv} the improved 
bound on $d_e$ in (\ref{newde}) nearly excludes this possibility. While 
the bino-driven baryogenesis analyzed in \cite{Li:2008ez} is still allowed by
 this new measurements, it further constraints this scenario.}

A new an largely unexplored direction for electroweak 
baryogenesis is flavour non-diagonal CPV that would enter the $B$ or $D$ 
meson
systems \cite{Liu:2011jh,Tulin:2011wi,Cline:2011mm}. Flavour non-diagonal CP violation is far less susceptible to EDM constraints 
than flavour diagonal phases since it arises at multi-loop order. 
In the SM
for example, it is a two-loop effect that involves the one-loop CP-violating penguin operator and  a hadronic loop with two 
$\Delta S=1$ weak interactions.

Finally, let us quote recent studies of EDMs in 2HDM models with flavour 
blind phases \cite{Buras:2010zm,Jung:2013hka} and supersymmetry \cite{Altmannshofer:2013lfa} where further references to 
the rich literature can be found.

\section{Towards Selecting Successful Models}\label{sec:5}
\subsection{Preliminaries}
We have seen in previous sections that considering several theoretically 
clean observables in the context of various extensions of the SM there is 
a chance that we could identify new particles and new forces at very short distance scales that are outside the reach of the LHC. In fact this strategy is 
not new as  most of  elementary particles of the SM have been predicted to exist 
on the basis of low energy data well before their discovery\footnote{ Although 
the non-vanishing neutrino masses came as a surprise and could be regarded as one  of the first  signs of NP beyond the SM.} Moreover, 
this has been achieved by not only the desire to understand the data but simultaneously with the goal to construct a fundamental theory of elementary 
matter and elementary interactions that is predictive and consistent with 
all physics principles we know. Yet, the present situation differs from 
the days when one started to discover first quarks in the following manner. 
Based on time and resources that were required to build the LHC it is 
rather unlikely that a machine probing directly $100-200~\tev$ energy 
scales or short distance scales in the ballpark of a zeptometer ($10^{-21}~$m)
will exist in the first half of this century. Rather a machine as an international linear 
collider with the energy of $1\tev$ will be build in order to study the details 
of physics up to this energy scale. Therefore, the search for new phenomena 
below  $4\times 10^{-20}$~m, that is beyond the LHC, will be in the hands of flavour physics and very rare processes.

There is no question  that the progress in the search for NP at the 
shortest distance scales will require an intensive collaboration of
experimentalists and theorists. In this context  there is the question 
whether top-down or bottom-up approach will turn out to be more efficient 
in reaching this goal. While bottom-up approach using exclusively 
effective theories with basically arbitrary coefficients of local operators 
allowed by symmetries of the SM can provide some insight in what is going 
on, we think that the top-down approach will eventually be more effective 
in the flavour precision era in identifying NP beyond the LHC reach. Yet, 
needless to say 
it would be extremely important to get some directions from direct 
discoveries of new phenomena at the LHC.  This would in particular 
allow the correlations between high energy and low energy observables, which 
is only possible in a top-down approach.

Thus our basic strategy, as already exemplified on previous pages, is to 
look at different models and study different patterns of flavour 
violation in various theories through identification of correlations 
between various observables. The question then arises how to do it 
most efficiently and transparently.

In principle global fits of various observables in a given theory to 
the experimental data appears to be most straightforward. The success or failure 
of a given theory is then decided on the basis of $\chi^2$ or other 
statistical measures. This is clearly a legitimate approach and used 
almost exclusively in the literature. Yet, we think 
that in the first phase of the search for NP a more transparent approach 
could turn out to be more useful. This is what we will present next.

\subsection{DNA-Chart}
As reviewed in \cite{Buras:2010wr,Buras:2012ts} extensive studies of many 
models  allowed to construct various classifications of NP contributions 
in the form of ``DNA'' tables \cite{Altmannshofer:2009ne} and {\it flavour codes}  \cite{Buras:2010wr}. The  ``DNA'' tables in 
\cite{Altmannshofer:2009ne} 
had as a goal to indicate whether in a given theory a value of a 
given observable can differ by a large, moderate or only tiny amount 
from the prediction of the SM.  The {\it flavour codes}  \cite{Buras:2010wr} were more a description of a given model in terms of the presence or absence of left- or right-handed  currents in it and the presence or absence of new CP phases, flavour violating and/or flavour conserving. 

Certainly in both cases there is a room for improvements. In particular in the 
case of the  ``DNA'' tables in  \cite{Altmannshofer:2009ne} we know now that 
in most quark  flavour observables considered there NP effects can be at most by  a factor of $2$ larger than the SM 
contributions. Exceptions are the cases in which 
some branching ratios or asymmetries vanish in the SM. But the particular 
weakness of this approach is the difficulty in depicting the correlations 
between various observables that could be characteristic for a given theory. 
Such correlations are much easier to show on a circle and in 
what follows we would like to formulate this new idea and illustrate it with 
few examples.

{\bf Step 1}

We construct a chart showing different observables, typically 
a branching ratio for a given decay or an asymmetry, like CP-asymmetries 
$S_{\psi K_S}$ and $S_{\psi\phi}$ and quantities $\Delta M_s$, $\Delta M_d$, 
$\varepsilon_K$ and $\varepsilon^\prime$. The important point is to select 
the optimal set of observables which are simple enough so that definite 
predictions in a given theory can be made. 

{\bf Step 2}

In a given theory we calculate the selected observables and investigate 
whether a given observable is enhanced or suppressed relative to the SM 
prediction or is 
basically unchanged. What this means requires a measure, like one or two 
$\sigma$. In the case of asymmetries we will { proceed in the same manner if 
its sign remains unchanged relative to the one in the SM but otherwise we } define the change of its 
sign from $+$ to $-$ as a suppression and the change from  $-$ to $+$ as an
enhancement. For these three situations we will use the following colour 
coding:
\be
{\rm  \colorbox{yellow}{enhancement}}~=~{\rm yellow}, \qquad {\rm \framebox{no~change}}~=~{\rm white}  \qquad {\rm 
\colorbox{black}{\textcolor{white}{\bf suppression}}}~=~{\rm black}
\ee
To this end the predictions within the SM have to be known precisely.

{\bf Step 3}

It is only seldom that a given observable in  a given theory is uniquely 
suppressed or enhanced but frequently two observables are correlated or
uncorrelated, that is the enhancement of one observable implies uniquely
an enhancement (correlation) or suppression (anti-correlation) of another 
observable. It can also happen that no change in the value of a given 
observable implies no change in another observable. There are of course 
other possibilities. The idea then is to connect in our DNA-chart 
a given pair of observables that are correlated with each other
by a line. Absence of a line means that two given observables are 
uncorrelated. In order to distinguish the correlation from anti-correlation 
we will use the following colour coding for the lines in question:
\be
{\rm correlation}~\textcolor{blue}{\Leftrightarrow}~{\rm blue} , \qquad {\rm anti-correlation}~\textcolor{green}{\Leftrightarrow}~{\rm
green}
\ee

We will first make selection of the optimal observables that can be realistically measured in this decade and subsequently we will illustrate 
the DNA-chart on example of few simple  models.                       

 \subsection{Optimal Observables}
On the basis of our presentation in the previous sections we think that one 
should have first a closer look at the following observables.

\be\label{Observables1} 
\varepsilon_K, \quad \Delta M_{s,d}, \quad S_{\psi K_S}, \quad S_{\psi \phi},
\ee

\be\label{Observables2} 
 \kpn, \quad \klpn, \quad \epe,
 \ee

\be\label{Observables3} 
 B_{s,d}\to\mu^+\mu^-, \qquad B\to X_{s}\nu\bar\nu, \quad B\to K^*(K)\nu\bar\nu,
\ee

\be \label{Observables4} 
B\to X_s\gamma, \quad B^+\to \tau^+\nu_\tau~, \quad B\to K^*(K)\mu^+\mu^-,
\ee
 where in the latter case we mean theoretically clean angular observables.
The remaining observables discussed by us will then serve as constraints 
on the model and if measured could also be chosen.

\subsection{Examples of DNA-Charts}
The first DNA-chart which one should in principle construct is the one dictated 
by experiment. This chart will have no correlation lines but will show where 
the SM disagrees with the data and comparing it with DNA-chart specific 
to a given theory will indicate which theories  survived and which have 
been excluded. Unfortunately in view of significant uncertainties in some 
of the SM predictions and rather weak experimental bounds on most interesting 
branching ratios, such an {\it experimental} chart is rather boring at 
present as it is basically white. However, in the second half of this decade 
when LHC and other  machines will provide new data and lattice calculations 
increase their precision it will possible to construct such an experimental DNA chart and we should hope that it will not be completely white.

Here we want to present four examples of DNA-charts.
In Fig.~\ref{fig:CMFVchart} we show the DNA-chart of CMFV and the corresponding chart for $U(2)^3$ models
is shown in Fig.~\ref{fig:U23chart}. 
The DNA-charts representing models with left-handed and right-handed flavour violating
couplings of  $Z$ and $Z^\prime$  can be found in Fig.~\ref{fig:ZPrimechart}.

The interested reader may check that these charts summarize compactly the 
correlations that we discussed in detail at various places in this review.
In particular we observe the following features:
\begin{itemize}
\item
When going from  the DNA-chart of CMFV in  Fig.~\ref{fig:CMFVchart} to the 
one for the $U(2)^3$ models in  Fig.~\ref{fig:U23chart}, the correlations 
between $K$ and $B_{s,d}$ systems are broken as the symmetry is reduced from 
$U(3)^3$ down to $U(2)^3$. The anti-correlation between $S_{\psi\phi}$ and 
$S_{\psi K_S}$ is just the one shown in Fig.~\ref{fig:SvsS}.
\item
As the decays $\kpn$, $\klpn$ and $B\to K\nu\bar\nu$  are only sensitive
to the vector quark currents, they do not change when the couplings are changed from  left-handed to right-handed ones. On the other hand the remaining 
three decays in   Fig.~\ref{fig:ZPrimechart} are sensitive to axial-vector 
couplings implying interchange of enhancements and suppressions when going from 
$L$ to $R$ and also change of correlations to anti-correlations between the 
latter three and the former three decays. Note that the correlation between 
$B_s\to\mu^+\mu^-$  and $B\to K^*\mu^+\mu^-$ does not change as both decays are  sensitive only to axial-vector coupling. 
\item
However, it should be remarked that in order to obtain the correlations or 
anti-correlations in LHS and RHS scenarios it was assumed that the signs 
of the left-handed couplings to neutrinos and the axial-vector couplings 
to muons are the same which does not have to be the case. If they are 
opposite the correlations between the decays with neutrinos and muons in 
the final state change to anti-correlations and vice versa. 
\item
On the other hand due to $SU(2)_L$ symmetry the left-handed $Z^\prime$
 couplings to muons and neutrinos are equal and this implies the relation
\be\label{SU2}
\Delta_{L}^{\nu\bar\nu}(Z')=\frac{\Delta_V^{\mu\bar\mu}(Z')-\Delta_A^{\mu\bar\mu}(Z')}{2}. 
\ee
Therefore, once two of these couplings are determined the third follows uniquely without the freedom mentioned in the previous item.
\item
In the context of the DNA-charts in  Fig.~\ref{fig:ZPrimechart}, the correlations involving $\klpn$ apply only if NP contributions carry some CP-phases. If this is not the case the branching ratio for $\klpn$ will remain unchanged. 
This is evident from our discussion in Step 8 and the plots presented there.
\end{itemize}

In this context let as summarize the following important properties of
 the case of tree-level $Z^\prime$ and $Z$ exchanges when
both LH and RH quark couplings are present which in addition are equal to each 
other (LRS scenario) or differ by sign (ALRS scenario):
\begin{itemize}
\item
In LRS NP contributions to $B_{s,d}\to\mu^+\mu^-$ vanish but not to $\klpn$ 
and $\kpn$.
\item
In ALRS NP contributions to $B_{s,d}\to\mu^+\mu^-$ are non-vanishing and 
this also applies to $B_d\to K^*\mu^+\mu^-$ as seen in the right panel 
of Fig.~\ref{fig:pFLS5LHS}. On the other hand 
they vanish in the case of  $\klpn$, $\kpn$ and $B_d\to K\mu^+\mu^-$
\end{itemize}

\begin{figure}[!tb]
\centering
\includegraphics[width = 0.65\textwidth]{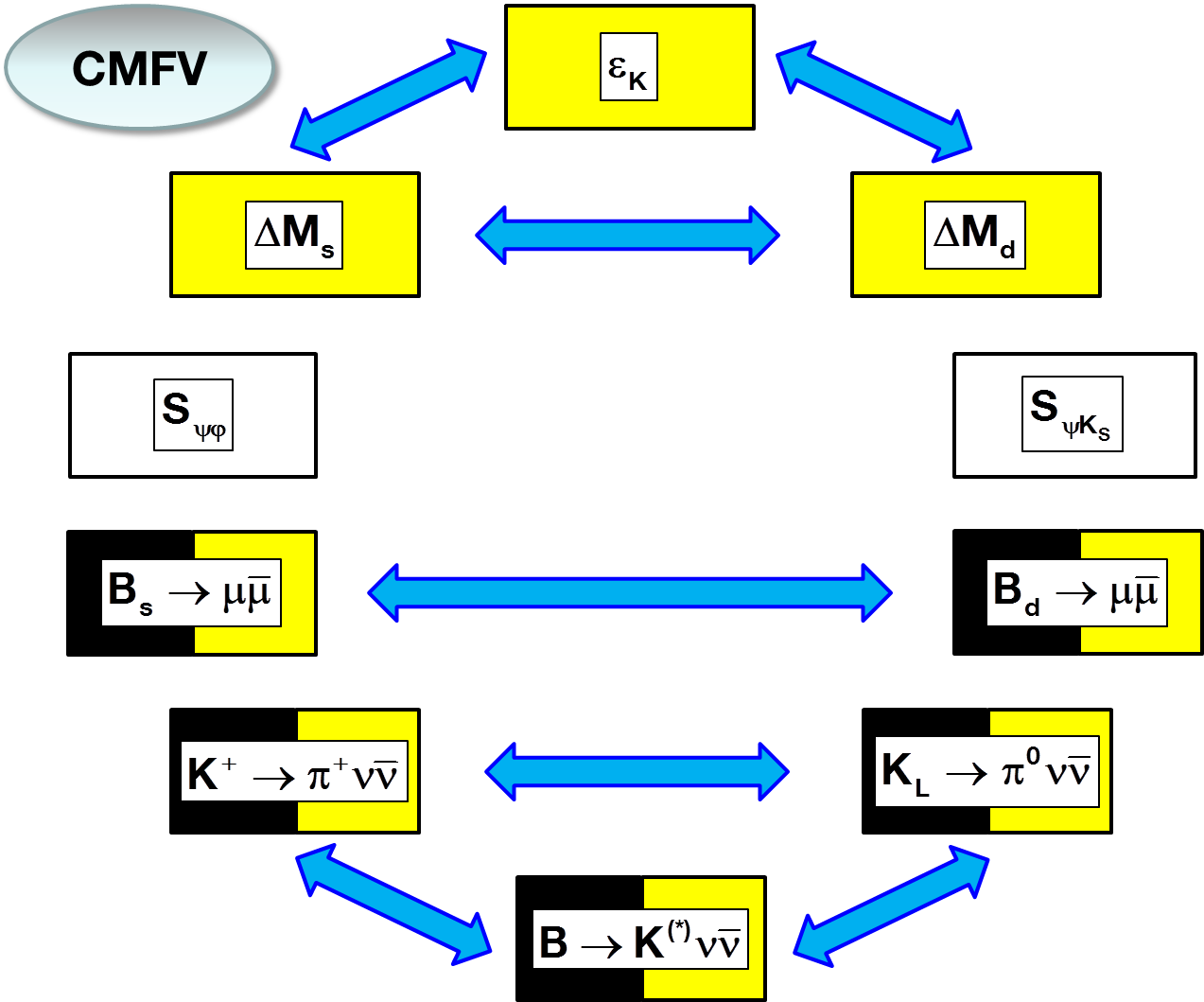}
\caption{\it DNA-chart of CMFV  models. Yellow means   \colorbox{yellow}{enhancement}, black means
\colorbox{black}{\textcolor{white}{\bf suppression}} and white means \protect\framebox{no change}. Blue arrows
\textcolor{blue}{$\Leftrightarrow$}
indicate correlation and green arrows \textcolor{green}{$\Leftrightarrow$} indicate anti-correlation. }
 \label{fig:CMFVchart}~\\[-2mm]\hrule
\end{figure}

\begin{figure}[!tb]
\centering
\includegraphics[width = 0.65\textwidth]{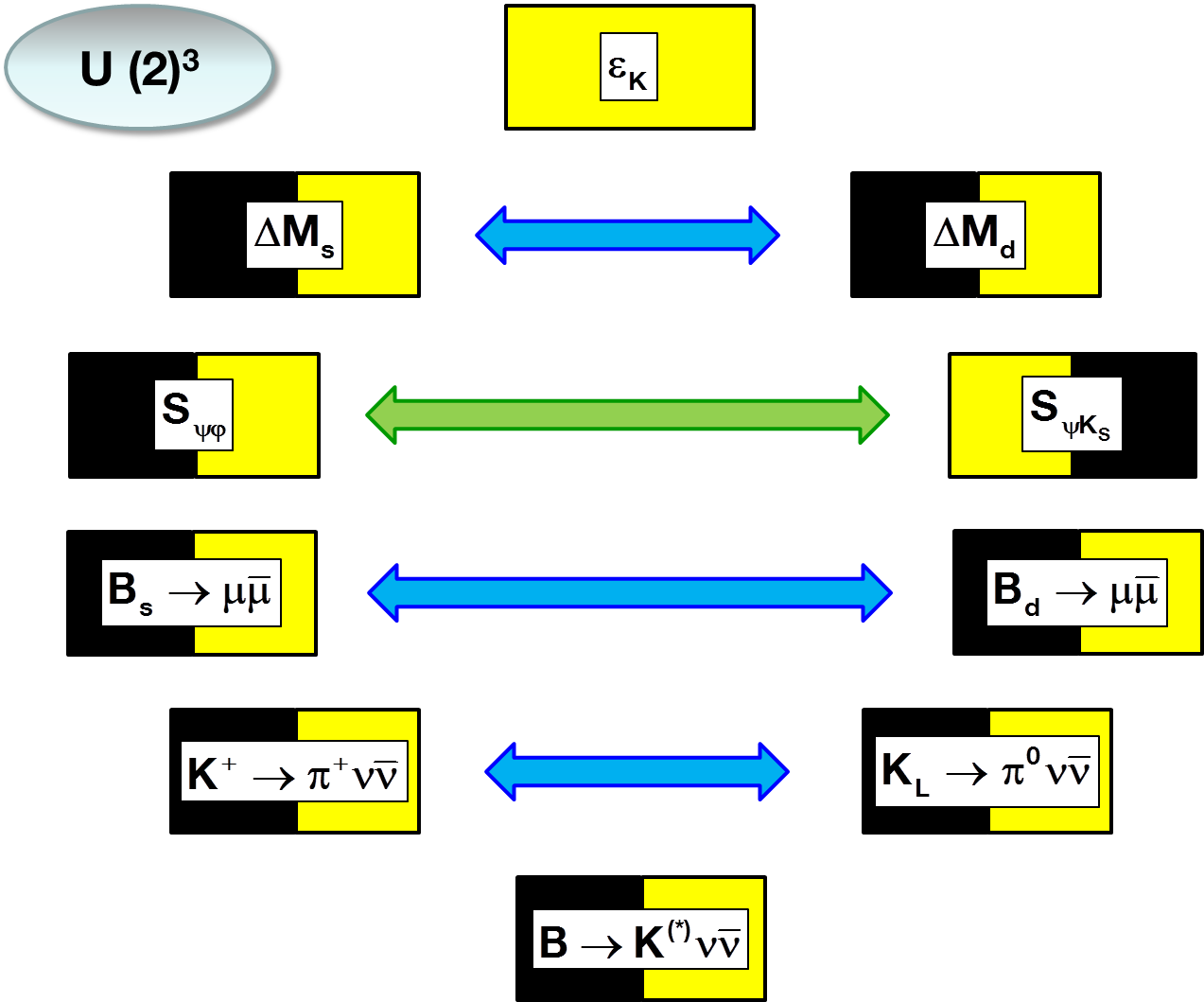}
\caption{\it DNA-chart of $U(2)^3$ models.  Yellow means   \colorbox{yellow}{enhancement}, black means
\colorbox{black}{\textcolor{white}{\bf suppression}} and white means \protect\framebox{no change}. Blue arrows
\textcolor{blue}{$\Leftrightarrow$}
indicate correlation and green arrows \textcolor{green}{$\Leftrightarrow$} indicate anti-correlation. }
 \label{fig:U23chart}~\\[-2mm]\hrule
\end{figure}

\begin{figure}[!tb]
\centering
\includegraphics[width = 0.49\textwidth]{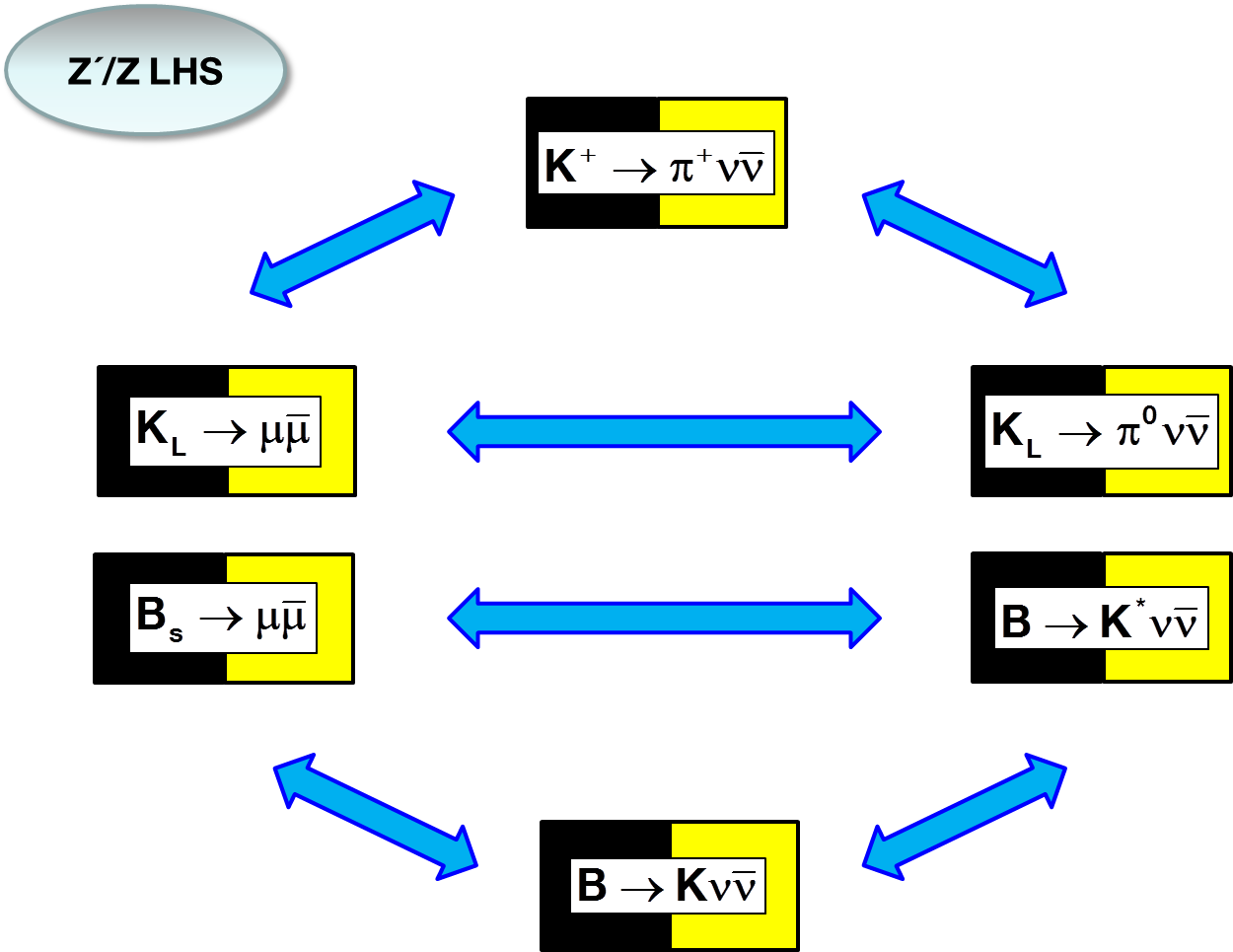}
\includegraphics[width = 0.49\textwidth]{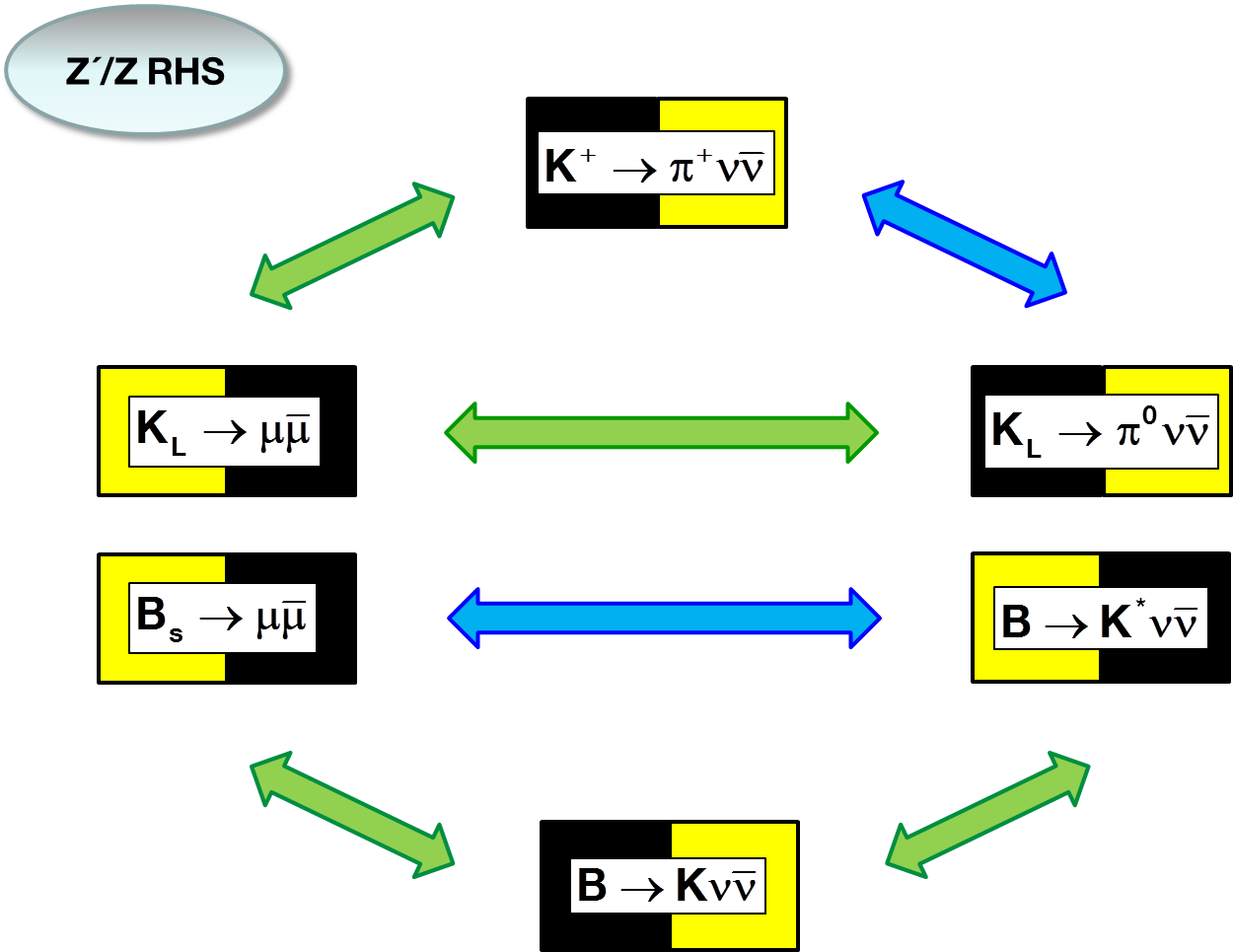}

\caption{\it DNA-charts of $Z^\prime$ models with LH and RH currents.  Yellow means   \colorbox{yellow}{enhancement}, black means
\colorbox{black}{\textcolor{white}{\bf suppression}} and white means \protect\framebox{no change}. Blue arrows
\textcolor{blue}{$\Leftrightarrow$}
indicate correlation and green arrows \textcolor{green}{$\Leftrightarrow$} indicate anti-correlation. }
 \label{fig:ZPrimechart}~\\[-2mm]\hrule
\end{figure}

\subsection{Reviewing concrete models}
The realization of this  strategy in the case of more complicated models 
is more challenging in view of many parameters involved, which often have to 
be determined beyond  flavour physics. However we expect that when 
more data from the LHC and flavour machines around the world will be available 
it will be possible to be more concrete also in the case of these 
more complicated models. Two rather detailed reviews of various patterns 
of flavour violation in a number of favorite and less favorite 
extensions of the SM appeared in \cite{Buras:2010wr,Buras:2012ts}. In view 
of the fact that no totally convincing signs of NP in flavour data has 
been observed since the appearance of the second review, there is no 
point in updating presently these reviews. Basically all these models 
can fit the present data
by adjusting the parameters or increasing the masses of new particles. 
 Therefore we only make a few remarks on some 
of these models 
and indicate in which section of  \cite{Buras:2012ts}
 more details on a given 
model and related references to the original literature can be found.

\subsubsection{331 model}\label{sec:331}

A concrete example for $Z^\prime$ tree-level FCNC discussed in Sec.~\ref{toy} and at various places in Sec.~\ref{sec:4} is a model based on
the gauge group   $SU(3)_C\times SU(3)_L\times U(1)_X$, the so-called 331 model, originally developed in 
\cite{Pisano:1991ee,Frampton:1992wt}. There are different versions of the 331 model characterized by a parameter $\beta$ that
determines the particle content. In
\cite{Buras:2012dp} we consider the $\beta = 1/\sqrt{3}$-model to be called $\overline{331}$ model. Since only left-handed
currents are flavour violating and
effects in $\varepsilon_K$ are rather small it favours inclusive $\vub$ and thus belongs to LHS2. Furthermore also the lepton couplings
are no longer arbitrary but come out automatically from the Lagrangian:  $\Delta_L^{\nu\bar\nu}(Z')=0.14$ and
$\Delta_A^{\mu\bar\mu}(Z')=-0.26$ {for $\beta = 1/\sqrt{3}$}.  For the general $Z^\prime$ scenario
we used $\Delta_L^{\nu\bar\nu}(Z')=0.5$ and $\Delta_A^{\mu\bar\mu}(Z')=0.5$.

In the breaking $SU(3)_L\times U(1)_X\to
SU(2)_L\times U(1)_Y$ to the SM gauge group a new heavy neutral gauge boson $Z^\prime$ appears that mediates FCNC already at tree
level. A nice theoretical feature is that from the requirement of
anomaly cancellation and asymptotic freedom of QCD it follows that one needs $N = 3$ generations. Anomaly cancellation is only possible if
one generation (usually
the
3$^\text{rd}$ is chosen) is treated differently than the other two generations. 

Further studies of the 331 model can  be found in \cite{Liu:1993gy,Diaz:2004fs} where the lepton sector was
analyzed in detail and in \cite{Liu:1994rx,Rodriguez:2004mw,Promberger:2007py} where mixing of neutral mesons as well as a number of rare
$K$ and $B_{d,s}$ decays have been considered. The decay
$b \to s \gamma$ was considered in \cite{Agrawal:1995vp,Promberger:2008xg} and in \cite{Machado:2013jca} also neutral scalar contributions
 were included.

{\bf Flavour structure}

{The  $\overline{331}$ model studied in \cite{Buras:2012dp} has the following fermion content:} 
Left-handed fermions fit in (anti)triplets, while right-handed ones are singlets under $SU(3)_L$.
In the quark sector, the first two generations  fill the two upper components of a triplet, while the third one fills those
of an  anti-triplet; the  third member of the quark (anti)triplet is a new heavy fermion:
\begin{align}
&
 \begin{pmatrix}
                 e\\
-\nu_e\\
\nu_e^c
                \end{pmatrix}_L\,,
\begin{pmatrix}
                 \mu\\
-\nu_\mu\\
\nu_\mu^c
                \end{pmatrix}_L\,,
\begin{pmatrix}
                 \tau\\
-\nu_\tau\\
\nu_\tau^c
                \end{pmatrix}_L\,,\quad\qquad\begin{pmatrix}
           u\\d\\D
          \end{pmatrix}_L\,,
\begin{pmatrix}
 c\\s\\S
\end{pmatrix}\,,\begin{pmatrix}
       b\\-t\\T
      \end{pmatrix}_L\\
&
e_R, \,\mu_R,\, \tau_R,\,\qquad u_R,\, d_R,\, c_R,\, s_R,\, t_R,\, b_R,\,\qquad D_R, S_R, T_R
\end{align}
We need the same number of triplets and anti-triplets due to anomaly cancellation. If one takes into account the three colours of
the quarks we have six triplets and six anti-triplets with this choice. 
Neutral currents mediated by  $Z^\prime$ are affected by the quark mixing because the $Z^\prime$ couplings are generation
non-universal.  
However 
only left-handed  quark
currents are flavour-violating, thus we are left with LHS.  Except for the 
$Z^\prime$ mass the tree-level FCNCs in $B_{d,s}$ and $K$ meson systems depend 
effectively on
2 angles and 2 phases $\tilde s_{23}, \tilde s_{13}, \delta_{1,2}$  such that the $B_d$ sector depends only on $\tilde s_{13},\delta_1$ and the
$B_s$ sector on $\tilde s_{23},\delta_2$.  Then in contrast to the general $Z^\prime$ models, discussed before, the NP parameters in 
$K$ sector are fixed. In particular CP violation is governed there by the phase difference
$\delta_2-\delta_1$. In  more general $Z^\prime$ models  the $K$ sector is  decoupled from $B_{d,s}$ sector.

Concerning phenomenology, it is more restrictive than the one in 
a general $Z^\prime$ model with left-handed couplings and it is of interest to
 investigate how  the 331 models with arbitrary $\beta$ face the new data 
on $B_{s,d}\to \mu^+\mu^-$ and $B_d\to K^*(K)\mu^+\mu^-$
taking into account present constraints from $\Delta F=2$ observables,
low energy precision measurements, LEP-II and the LHC data. Such an analysis 
has been performed in \cite{Buras:2013dea} and we summarize the main results of 
this paper where numerous correlations between various flavour observables can be found.

Studying the
implications of these models  for $\beta=\pm n/\sqrt{3}$ with $n=1,2,3$ we find
that the case $\beta=-\sqrt{3}$
leading to Landau singularities for  $M_{Z^\prime}\approx 4\tev$ can be ruled out when the present constraints on
$Z^\prime$ couplings, in particular from LEP-II, are taken into account.
 For $n=1,2$ interesting results are found for  $M_{Z^\prime}< 4\tev$ with
largest NP effects for $\beta <0$ in   $B_d\to K^*\mu^+\mu^-$ and the ones
in  $B_{s,d}\to\mu^+\mu^-$ for $\beta>0$. As $\RE(C_9^{\rm NP})$  can
reach the values $-0.8$ and  $-0.4$  for $n=2$ and $n=1$, respectively the $B_d\to K^*\mu^+\mu^-$ anomalies can be
softened with
the size depending on
$\Delta M_{s}/(\Delta M_{s})_{\rm SM}$ and the CP-asymmetry $S_{\psi\phi}$.
A correlation between  $\RE(C^{\rm NP}_{9})$ and $\overline{\mathcal{B}}(B_{s}\to\mu^+\mu^-)$, identified for $\beta<0$, implies for 
{\it negative} $\RE(C^{\rm NP}_{9})$ uniquely suppression of $\overline{\mathcal{B}}(B_{s}\to\mu^+\mu^-)$  relative to its SM value which is favoured by 
the data. In turn also
$S_{\psi\phi}< S_{\psi\phi}^{\rm SM}$ is favoured with $S_{\psi\phi}$
having dominantly opposite sign to $S_{\psi\phi}^{\rm SM}$
 and closer to its central experimental value. 
 Another triple correlation is the one between $\RE(C^{\rm NP}_9)$, $\overline{\mathcal{B}}(B_{s}\to\mu^+\mu^-)$ and  $\mathcal{B}(B_d\to K\mu^+\mu^-)$.
 NP effects in $b\to s\nu\bar\nu$ transitions, $\kpn$
and $\klpn$ turn out  to be small. 

We find also that the absence of $B_d\to K^*\mu^+\mu^-$ anomalies in the future
data and confirmation of the suppression of $\overline{\mathcal{B}}(B_{s}\to\mu^+\mu^-)$ relative to its SM value would favour the $\overline{331}$ model ($\beta=1/\sqrt{3}$) summarized in detail above and $M_{Z^\prime}\approx 3\tev$. Assuming lepton universality,
 we find an upper bound  $|C^{\rm NP}_{9}|\le 1.1 (1.4)$ from
LEP-II data for {\it all}  $Z^\prime$ models with only left-handed flavour
violating couplings to quarks when NP contributions to
$\Delta M_s$ at the level of  $10\%(15\%)$  are allowed.

Finally, we refer to a very recent analysis in \cite{Buras:2014yna} in 
which additional effects of $Z-Z'$ mixing and resulting $Z$-mediated FCNCs have been investigated in detail.
We find that these new contributions can indeed be neglected in the case of 
$\Delta F=2$ transitions and decays, like  $B_d\to K^*\mu^+\mu^-$, where they are suppressed by the small vectorial $Z$ coupling to charged 
leptons. However, 
the contributions of tree-level $Z$ exchanges to decays sensitive to 
axial-vector couplings, like $B_{s,d}\to \mu^+\mu^-$ and $B_d\to K\mu^+\mu^-$, 
and those with neutrinos in the final state, like  $b\to s\nu\bar\nu$ transitions, $\kpn$ and $\klpn$ cannot be  generally neglected with size of 
$Z$ contributions depending on $\beta$, $M_{Z^\prime}$ and an additional parameter $\tan\bar\beta$. A detailed summary of these results 
is clearly beyond the scope of this review. We refer to the numerous plots in this paper where 
it can be found how the results on FCNCs in 331 models listed above, in particular correlations between various observables, are modified by these new contributions.  As a byproduct we 
analyzed there for the first time the ratio $\epe$ in 
these models including both $Z^\prime$ and $Z$ contributions.   Our 
analysis of electroweak precision  observables within 331 models demonstrates transparently 
that the interplay of NP effects in electroweak precision observables and those in flavour observables could allow in the future to identify the favourite 
331 model.

\subsubsection{Littlest Higgs Model with T-parity}
As stressed in  Section 3.6 of  \cite{Buras:2012ts} the LHCb data can be 
considered as a relief for this model. 

\begin{itemize}
\item
In this model it was not possible to obtain 
$S_{\psi\phi}$ of $\ord(1)$ and  
 values above 0.3 were rather unlikely. In this model 
also negative values for
$S_{\psi\phi}$ as opposed to  ${\rm 2HDM_{\overline{MFV}}}$ are possible.
\item
Because of new sources of flavour violation 
originating in 
the presence of mirror quarks and new mixing matrices, 
the usual CMFV relations between $K$, $B_d$ and $B_s$ 
systems are violated. This allows to remove 
the $\varepsilon_K-S_{\psi K_S}$ anomaly for both scenarios of $\vub$ and 
also improve agreement with $\Delta M_{s,d}$.
\item
The small value of $S_{\psi\phi}$ from LHCb allows still for sizable 
enhancements of 
$\mathcal{B}(\klpn)$ and $\mathcal{B}(\kpn)$  which would not 
be possible otherwise.
\item
On the other hand rare $B$-decays turn out to be SM-like but still some enhancements 
are possible. 
In particular $\mathcal{B}(B_{s}\to\mu^+\mu^-)$ 
can be enhanced by $30\%$ and a
significant part of this enhancement comes from the T-even sector. The effects 
in $\mathcal{B}(B_{d}\to\mu^+\mu^-)$ can be larger and also suppression is 
possible.
\end{itemize}

\subsubsection{The SM with Sequential Fourth Generation (SM4)}

The LHC data indicate that our nature seems to have only three sequential 
generations of quarks and leptons.
The authors of \cite{Eberhardt:2012gv} performed a statistical analysis including the latest Higgs search results and electroweak
precision observables and concluded that the SM4 is already excluded at roughly 5$\sigma$. 
 Here we mention nevertheless few interesting signatures of this model after the LHCb data 
as far as flavour violation is concerned:
\begin{itemize}
\item
As before the presence of new sources of flavour violations allows to 
remove all existing tensions related to $\Delta F=2$ observables.
\item
The small value of $S_{\psi\phi}$ and the results for
$\mathcal{B}(B_s\to\mu^+\mu^-)$ from LHCb imply now that  $\mathcal{B}(B_d\to\mu^+\mu^-)$ can significantly depart from its  SM
value. On the other hand $\mathcal{B}(B_s\to\mu^+\mu^-)$ is 
SM-like with values below SM prediction being more likely than above it. 
\item
Possible enhancements of $\mathcal{B}(\kpn)$ and $\mathcal{B}(\klpn)$ over the SM3 values are still possible. 
\end{itemize}
More details and references to the original literature can be found in 
Section 3.7 of  \cite{Buras:2012ts}.

\subsubsection{CP conserving 2HDM II}

\begin{figure}[!tb]
 \centering
 \includegraphics[width = 0.4\textwidth]{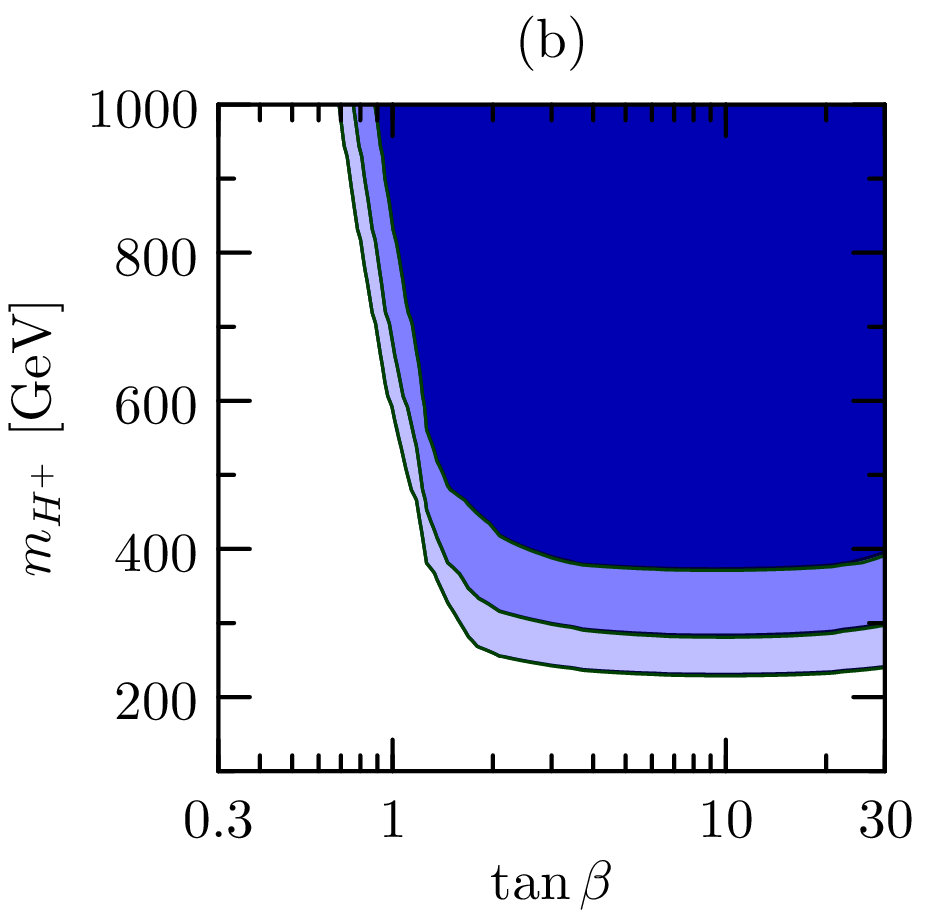}
 \includegraphics[width = 0.55\textwidth]{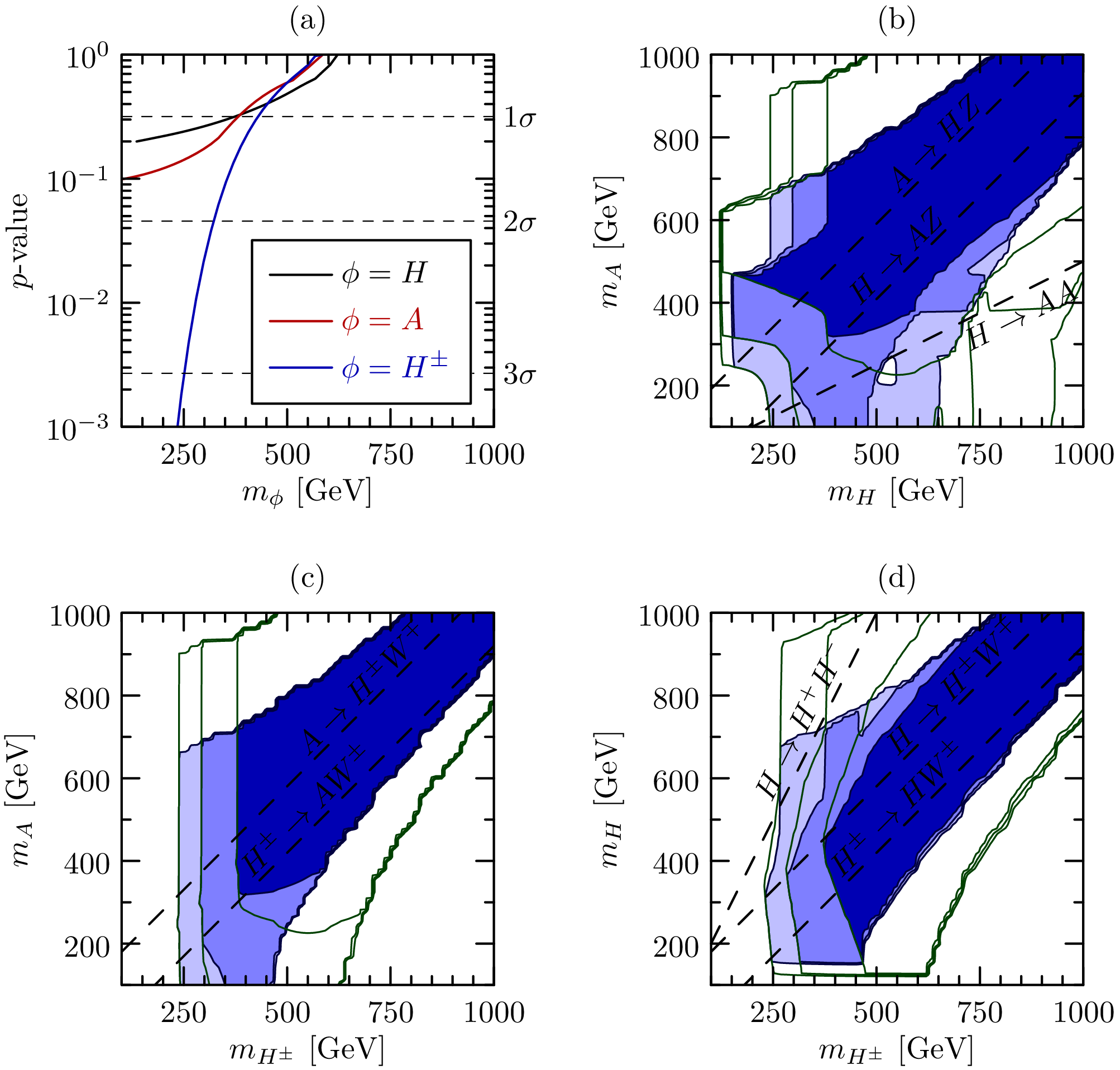}
\caption{\it Allowed regions in the $\tan\beta-m_{H^+}$ plane (left) and mass planes (right). The shaded blue 
areas are the regions allowed at one, two and three standard deviations (dark to light) 
\cite{Eberhardt:2013uba}.}\label{fig:Uli2HDM}~\\[-2mm]\hrule
\end{figure}

The authors of \cite{Eberhardt:2013uba} made a global fit of the CP conserving 2HDM II with a softly broken $Z_2$ symmetry. 
Their analysis includes the experimental constraints from LHC on the mass and signal strength of the Higgs resonance at 126~GeV 
(which is always interpreted as the light CP-even 2HDM Higgs boson $h$), the non-observation of additional Higgs resonances, EWPO 
and flavour data on $B^0-\bar B^0$ mixing and $B\to X_s\gamma$. Furthermore theoretical constraints are taken into account: vacuum 
stability and perturbativity. They find that the parameter region with $\beta-\alpha\approx \frac{\pi}{2}$ where the couplings of 
the light CP-even Higgs boson are SM like is favoured.  In Fig.~\ref{fig:Uli2HDM} (left) the allowed range in the 
$\tan\beta-m_{H^+}$ plane is shown. The lower bound on $m_{H^+}$ of 322~GeV (400~GeV) at 2$\sigma$ (1$\sigma$) for $\tan\beta>1$ 
follows from the constraint from $B\to X_s\gamma$. On the right hand side of Fig.~\ref{fig:Uli2HDM} the allowed mass regions for
$H^0/A^0/H^+$ 
is shown. Flavour and EWP observables exclude scenarios with both $m_H$ and $m_A$ below 300~GeV at $2\sigma$. 

Other recent analyzes of 2HDM II can be found in \cite{Celis:2013rcs,Chiang:2013ixa,Barroso:2013awa,Grinstein:2013npa}. In
\cite{Crivellin:2013mba} it was even stated that 2HDM-II is ruled out by $B\to D(D^*)\tau\nu$ data.  However, it seems to us that such a statement is 
premature as the data could still change in the future and moreover this 
would also imply that the SM is ruled out because the 2HDM-II contains the SM 
in its parameter space in the decoupling limit.

\subsubsection{Supersymmetric Flavour Models (SF)}
None of the 
supersymmetric particles 
has been seen so far. However
one of the important predictions of 
the 
simplest realization of this scenario, the MSSM with $R$-parity, 
is a light Higgs with $m_H\le 130\gev$. The discovery of a Higgs boson at the 
LHC  around 
$125\gev$ could indeed be the first hints for a Higgs of the MSSM but 
it will take some time to verify it. In any case MSSM remains still a viable 
NP scenario at scales $\ord(1\tev)$ although the absence of the discovery 
of supersymmetric particles is rather disappointing.
Similarly the SUSY dreams of large $\mathcal{B}(B_s\to\mu^+\mu^-)$ and $S_{\psi\phi}$ have not been realized at   LHCb and 
CMS. However the data from 
these experiments listed in 
(\ref{LHCb1}), (\ref{LHCb2}) and (\ref{LHCb3}) have certainly an impact on
SUSY predictions. 

In view of a rather rich structure of various SF models analyzed in detail in 
 \cite{Altmannshofer:2009ne} and summarized in Section 3.8 of 
\cite{Buras:2012ts} it is not possible to discuss them adequately here. 
We make only two comments:
\begin{itemize}
\item
 The new 
 data on $\mathcal{B}(B_{s,d}\to\mu^+\mu^-)$ indicate that  there is more room for NP contribution dominated 
by left-handed currents than right-handed currents.
\item
Although the large range of departures from SM expectations 
 found in \cite{Altmannshofer:2009ne} has been significantly narrowed, 
 still significant room for novel SUSY effects is present in quark 
 flavour data. Assuming that SUSY particles will be found, the future 
improved data for $B_{s,d}\to\mu^+\mu^-$ and $S_{\psi\phi}$ as well as
$\gamma$ combined with $\vub$ should help in distinguishing between various 
supersymmetric flavour models.
\end{itemize}

\subsubsection{Supersymmetric SO(10) GUT model}\label{sec:CMM}

Grand Unified Theories open the possibility to transfer the neutrino mixing matrix $U_\text{PMNS}$ to the quark sector and therefore
correlate  leptonic and hadronic observables.
This is accomplished in a
controlled way in a concrete SO(10) SUSY GUT proposed by Chang, Masiero and Murayama (CMM model) where the atmospheric neutrino
mixing
angle induces new $b\to s$ and $\tau\to \mu$ transitions  \cite{Moroi:2000tk,Chang:2002mq}. In \cite{Girrbach:2011an} we have performed a
global analysis in the CMM
model of several flavour processes containing
$\Delta M_s$, $S_{\psi\phi}$, $b\to s\gamma$ and $\tau\to\mu\gamma$ including an extensive renormalization group (RG) analysis to connect
Planck-scale and low-energy parameters. A short
summary of this work can also be found in \cite{Buras:2012ts,Girrbach:2011wt,Nierste:2011na}.

Here we want to shortly summarize the basic features of this model.  At the Planck scale
the flavour symmetry is exact but it is already broken at the SO(10) scale which manifests itself in the appearance of a non-renormalizable
operator in the SO(10) superpotential. The SO(10) symmetry is  broken down to the SM gauge
group via SU(5) and the whole $\mathbf{\bar{5}}$-plet  $\mathbf{5}_i = (d_{Ri}^c,
\,\ell_{Li},\,-\nu_{\ell_i})^T$ and the corresponding supersymmetric partners are then rotated by $U_\text{PMNS}$.
 While at
$M_\text{Pl}$ the
soft masses are still universal, we get a large splitting between the masses  of the 1$^\text{st}$/2$^\text{nd}$ and
3$^\text{rd}$ down-squark and
charged-slepton generation  at the electroweak scale due to RG effects of $y_t$.  The flavour effects in the
CMM model are then mainly determined by the generated mass splitting and the structure of the PMNS matrix.

In \cite{Girrbach:2011an} we used tribimaximal mixing in $U_\text{PMNS}$. However the latest data now show that the reactor neutrino
mixing angle $\theta_{13}\approx 8^\circ$
is indeed non-zero. Consequently whereas effects in \kkm, \bbmd{} and $\mu\to e\gamma$ are very small in the original version of the model,
this  changes when $\theta_{13}\approx 8^\circ$ is taken into account. Now large effects in $\mu\to e\gamma$ are possible. With tribimaximal
mixing large contributions were only
predicted in
observables connecting the 2$^\text{nd}$ and 3$^\text{rd}$ generation. So we focused on $b\to s\gamma$, $\tau\to\mu\gamma$, $\Delta M_s$ and
$S_{\psi\phi}$. Concerning $B_s\to\mu^+\mu^-$, effects are small because the CMM model at low energies appears as a special version of
the
MSSM with small $\tan\beta$ such that this branching ratio stays SM-like.
Another observable that needs further investigation is the Higgs mass which in the CMM model tends to be too small. The
analysis
of \cite{Girrbach:2011an} was done prior to the detection
of the Higgs boson and there we pointed out the Higgs mass could be up to 120~GeV in the
parameter range consistent with flavour observables. An updated analysis of the CMM model however shows that the two new experimental
results, $\theta_{13}\approx 8^\circ$ and $M_H = 126~$GeV, put the CMM model under pressure \cite{NierstePortoroz,NiersteStockel}:
The constraint from $\mathcal{B}(\mu\to e\gamma)$ (see Eq.~(\ref{MEGbound})) supersedes those from $b\to s$ and $\tau\to\mu$ FCNC processes
and requires very heavy
sfermion and gaugino masses ($\approx (8-10)~$TeV).
It is very difficult to
find a range in the parameter space which simultaneously satisfy the Higgs mass constraint and the experimental upper bound on
$\mathcal{B}(\mu\to e\gamma)$. A Higgs mass of  $M_H = 126~$GeV can be accommodated by passing from the MSSM to the NMSSM.

\subsubsection{The Minimal Effective Model with Right-handed Currents: RHMFV}
Few years ago  interest in making another look  
at the right-handed 
 currents in general originated in tensions between inclusive and exclusive
 determinations of the elements of the CKM matrix $|V_{ub}|$ and  $|V_{cb}|$. 
 It could be that these tensions are due to the underestimate of theoretical 
 and/or experimental uncertainties. Yet, as pointed out
 and analyzed in particular in \cite{Crivellin:2009sd,Chen:2008se}, it is a fact 
 that the presence of right-handed  currents could either remove 
 or significantly weaken some of these tensions, especially in the 
 case of $|V_{ub}|$. 

 In \cite{Buras:2010pz} the implications of this idea for other processes 
have been 
investigated in an effective theory approach based on
 a left-right symmetric flavour group
 $SU(3)_L \times SU(3)_R$, commuting with 
an underlying $SU(2)_L \times SU(2)_R \times U(1)_{B-L}$ global symmetry and
broken only by two Yukawa couplings.
The model contains a new 
unitary matrix $\tilde V$  controlling flavour-mixing in the RH sector 
and can be considered as the minimally flavour violating generalization 
to the RH sector. Thus bearing in mind that this model contains non-MFV
interactions from the point of view of the standard MFV hypothesis that
includes only LH charged currents it can be called  RHMFV. Referring to 
\cite{Buras:2010pz} for details, we would like to summarize the 
present status of this model:
\begin{itemize}
\item
In this model it is the high inclusive value 
of $\vub$ that is selected by the model as the true value of this element 
providing simultaneously the explanation of the smaller $\vub$ 
found in SM analysis of exclusive decays and very  high value of $\vub$ 
implied by the previous data for $\mathcal{B}(B^+\to\tau^+\nu_\tau)$. The 
decrease of the latter branching ratio casts some doubts on the explanation 
of the tension between inclusive and exclusive values of $\vub$ by 
right-handed currents but the large experimental error on 
 $\mathcal{B}(B^+\to\tau^+\nu_\tau)$ does not yet exclude this idea.
It could be that the true value of $\vub$ determined in inclusive decays
 is somewhere between its
present central inclusive and exclusive values, like $\vub=3.8\times 10^{-3}$, 
and that the effect of right-handed currents is smaller than previously 
anticipated.
\item
A value like $\vub=3.8\times 10^{-3}$ still implies $\sin 2\beta\approx 0.74$ 
but in this model in the presence of SM-like  $S_{\psi\phi}$ measured by 
LHCb, it is possible due to new phases to achieve the agreement with 
the experimental value of $S_{\psi K_S}$. For  $S_{\psi\phi}=\ord(1)$ this 
would not be possible as stressed in \cite{Buras:2010pz}.
\item
As far as the decays $B_{s,d}\to\mu^+\mu^-$ are concerned, 
already in 2010
the constraint from $B\to X_s \mu^+\mu^-$ precluded 
$\mathcal{B}(B_{s}\to \mu^+\mu^-)$ to be above  
$1\cdot 10^{-8}$.
Moreover NP effects in $B_{d} \to \mu^+\mu^-$ have been  found generally 
to be
smaller than in $B_{s} \to \mu^+\mu^-$. 
But the smallness of $S_{\psi\phi}$ from LHCb modified the  structure of the RH 
matrix and  one should expect that the opposite is true in accordance with 
the room left for NP in $B_{d} \to \mu^+\mu^-$ by  the LHCb data. But to be sure a more detailed numerical analysis is required.
\end{itemize}

There are other interesting consequences of this NP scenario that can be 
found in \cite{Buras:2010pz} and \cite{Crivellin:2011ba} even if some of them will be modified due 
to changes in the structure of the RH matrix. It looks 
like RHMFV could still remain a useful framework when more precise 
experimental data for observables just mentioned
 will be available in the second half of this 
decade.

\subsubsection{A Randall-Sundrum Model with Custodial Protection}
Models with a warped extra dimension first proposed by Randall and 
Sundrum provide a geometrical explanation of the hierarchy between the 
Planck scale and the EW scale. Moreover, when the SM quarks and leptons 
are allowed to propagate in the fifth dimension (bulk), these models naturally 
generate the hierarchies in the fermion masses and mixing angles through different localization of the fermions in the bulk. 

In order to avoid problems
with electroweak precision tests (EWPT) and FCNC processes, the gauge group 
is
generally larger than the SM gauge group \cite{Agashe:2003zs,Csaki:2003zu}:
\be
G_\text{RSc}=SU(3)_c\times SU(2)_L\times SU(2)_R\times U(1)_X
\ee
 and similarly to the LHT model
new heavy gauge bosons, KK gauge bosons, are present. Moreover, a special 
choice of fermion representation protects the left-handed flavour conserving 
couplings in order to agree with the data, in particular in the case of
 $Z\to b\bar b$
\cite{Agashe:2006at}.

The increased symmetry provides 
a custodial protection also for left-handed 
flavour violating couplings of $Z$ to down-quarks and to corresponding right-handed couplings to up-quarks \cite{Blanke:2008zb,Blanke:2008yr,Buras:2009ka}. 
We will call this model RSc to indicate the custodial protection. 
 Detailed analyses of electroweak precision
tests and FCNCs in a RS model without and with 
custodial 
protection can also be found in \cite{Casagrande:2008hr,Bauer:2008xb,Bauer:2009cf}. 

The different placing of fermions in the bulk generates non-universal 
couplings of fermions to KK gauge bosons and $Z$ and after the rotation to mass eigenstates induces FCNC transitions at the tree-level. As we discussed tree-level 
FCNCs due to exchanges of a single gauge boson $Z^\prime$ or $Z$, it is 
instructive to emphasize the differences between our examples and the RS 
scenario. These are:
\begin{itemize}
\item
First of all there are several new heavy gauge bosons. 
The lightest new gauge bosons 
are the KK--gluons, the KK-photon and the 
electroweak KK gauge bosons $W^\pm_H$, $W^{\prime\pm}$, $Z_H$ and $Z^\prime$,
all with masses $M_{KK}$ at least as large as $2-3\tev$ as required by the consistency 
with the EWPT \cite{Agashe:2003zs,Csaki:2003zu,Agashe:2006at}.
\item
While in our simple examples a given gauge boson was the dominant NP 
effect in $K$, $B_s$ and $B_d$ systems, the situation in RSc is different. 
NP in $\varepsilon_K$ is dominated by KK gluons, $B^0_{s,d}-\bar B^0_{s,d}$ 
systems by KK gluons and KK weak gauge bosons, while rare $K$ and $B_{s,d}$ 
decays by right-handed flavour-violating couplings of $Z$ to down-quarks.
Therefore the correlations between $\Delta F=2$ and $\Delta F=1$ 
observables found in our simple scenarios are absent here.
\item
Yet, the problematic KK gluon contributions to $\varepsilon_K$, 
requiring some fine-tuning of the parameters have an indirect 
impact on other observables as the space of parameters is severely 
reduced. Moreover, the fact that RSc has a goal to explain 
the masses and mixing angles implies as mentioned below
 some correlations between 
different meson systems which were absent in our examples.
\end{itemize}

A very extensive analysis of FCNCs has been presented prior to the LHCb 
data in \cite{Blanke:2008zb,Blanke:2008yr}. 
The branching ratios for $B_{s,d}\to \mu^+\mu^-$ and 
             $B\to X_{s,d}\nu\bar\nu$ have been found to 
             be  SM-like: the maximal enhancements
            of these branching ratios amount to $15\%$. This is clearly 
            consistent with the present  LHCb and CMS  data
 but the situation may 
            change this year. An anti-correlation in the size of NP effects 
has been found between $S_{\psi\phi}$ and rare $K$ decays precluding, 
similar to the LHT model, visible effects in the latter in the presence of a
large  $S_{\psi\phi}$. The smallness of  $S_{\psi\phi}$ are good news for
 rare $K$ decays in the RSc framework as now sizable enhancements of 
branching ratios for $\kpn$ and $\klpn$ are allowed.

So far so good. In addition to $\varepsilon_K$ large NP contributions in the RS framework that
require some tunings of parameters in order to be in agreement with the experimental data have been found in $\varepsilon'/\varepsilon$
\cite{Gedalia:2009ws,Bauer:2009cf}. 
Moreover it appears that the  fine tuning in this ratio is not necessarily consistent with the one required in the
case of $\varepsilon_K$.
As far as transitions dominated by dipole operators are concerned some fine tuning of NP contributions to EDMs 
\cite{Agashe:2004cp,Iltan:2007sc} and 
$\mathcal{B}(\mu\rightarrow  e\gamma)$ \cite{Agashe:2006iy,Davidson:2007si,Agashe:2009tu} is required. 
After the recent data from the MEG experiment at PSI \cite{Adam:2013mnn}
this is in particular the case of $\mathcal{B}(\mu\to e\gamma)$
when considered in conjunction with $\mathcal{B}(\mu\to 3e)$ \cite{Csaki:2010aj}. Sizable contributions are possible also to the $b\to
s\gamma$ transition. However as they affect mostly the chirality-flipped Wilson coefficient $C'_7$, $\mathcal{B}(B\to X_s\gamma)$ remains in
good agreement with the data \cite{Blanke:2012tv,Agashe:2004cp,Agashe:2008uz}.

It appears then that this scenario, unless extended by some  flavour symmetry, does not look like a favorite one 
for NP around few TeV scales. On the other hand 
 many of the ideas and
concepts that characterize most of the physics discussed in the context 
of RS scenario  
do not rely on the assumption
of additional dimensions and
as indicated by AdS/CFT correspondence
 we can regard RS models
as a mere computational tool for certain 
strongly coupled theories. Therefore in spite of some tensions in this NP 
scenario the techniques developed in the last decade will 
certainly play an important role in the phenomenology if
a new strong dynamics will show up at the LHC after its upgrade.

\subsubsection{Composite Higgs and Partial Compositeness}
This brings us to the idea which still has not been ruled out in 
spite of the discovery of a boson that looks like the Higgs boson of the 
SM. The severe fine-tuning problem which this model faces can still be 
avoided if the Higgs boson is a composite object. Then the question arises 
how in such a model fermion masses can be generated without at the same 
time violating the stringent bounds of FCNCs. The most popular mechanism 
to achieve this goal is an old 4D idea which is known as partial compositeness~\cite{Kaplan:1991dc}. In this NP scenario SM fermions couple
linearly to heavy 
composite fermions with the same quantum numbers. The SM
fermion masses are then  generated in a seesaw-like manner and the mass eigenstates are superpositions of elementary and composite fields.
Light quarks 
are dominantly elementary while the degree of compositeness is 
large for the top quark. 

This idea for
explaining the fermion mass hierarchies by hierarchical composite-elementary mixings, already used in RS scenario discussed previously,
leads to a suppression of \mbox{FCNCs} even if the strong sector is completely flavour-anarchic
\cite{Grossman:1999ra,Huber:2000ie,Gherghetta:2000qt}. Yet, as we have seen in the 5D 
setting even this mechanism is not powerful enough to satisfy the bounds 
from FCNCs without some degree of fine-tuning for the masses of KK gluons, 
 represented here by the resonances of the strong sector, in the reach of the 
LHC \cite{Agashe:2004cp,Csaki:2008zd,Blanke:2008zb}. 
For this reason, various mechanisms have been suggested to further suppress flavour violation. One idea is to impose a flavour symmetry
under which the strong sector is invariant and which is only broken by the composite-elementary mixings
\cite{Cacciapaglia:2007fw,Barbieri:2008zt,Redi:2011zi,Barbieri:2012uh,Redi:2012uj}. Alternative solutions 
include flavour symmetries broken also in the strong sector \cite{Fitzpatrick:2007sa,Santiago:2008vq,Csaki:2008eh}. Also an extension of the
(flavour-blind) global symmetry of the strong sector has been proposed in  \cite{Bauer:2011ah}.

In addition as we have seen in the case of RSc, protection mechanisms 
have to be invoked to satisfy electroweak precision tests, in 
particular related to the $T$ parameter,
\cite{Agashe:2003zs,Csaki:2003zu} that requires the extension of the gauge group. 
In the 4D setting this  means that
the strong sector should be invariant under the custodial symmetry $SU(2)_L\times SU(2)_R\times U(1)_X$. Moreover, the presence of
heavy vectorial composite fermions that mix with the SM fermions and the 
presence of new heavy vector resonances implies modifications of $Z$ couplings 
leading to unacceptable $Z$ coupling to left-handed $b$ quarks and tree-level FCNCs mediated by $Z$. As already discussed in the context of
RS a particular choice of fermion representation allows to remove these problems both for 
$Z\to b\bar b$ \cite{Agashe:2006at} and also FCNCs \cite{Blanke:2008zb,Blanke:2008yr,Buras:2009ka}.
In the 4D setting this is equivalent to making the strong sector (approximately) invariant under a discrete symmetry \cite{Agashe:2006at}.

The important point to be made here, emphasized also recently by Straub \cite{Straub:2013zca}, is that 
the resulting pattern of FCNCs mediated by $Z$ will generally depend on
\begin{itemize}
\item
The flavour symmetry imposed on the strong sector admitting also 
the case of an  anarchic strong sector,
\item
Choice of the fermion representations to satisfy the bounds on $Z$ 
couplings.
\end{itemize}

A simple 4D effective framework to study the phenomenology of these different 
possibilities is  given by the two-site approach proposed in 
\cite{Contino:2006nn}. In this framework, one considers only one set of fermion resonances with heavy Dirac masses as well as spin-1
resonances associated to the global symmetry $SU(3)_c\times SU(2)_L\times SU(2)_R\times U(1)_X$ which 
can be considered as new ``heavy gauge bosons''. This approach can be viewed as a truncation of 5D warped (RS) models, taking into account
only the lightest set of KK states. This approximation has already been used in \cite{Blanke:2008zb,Blanke:2008yr,Buras:2009ka} in the
context of RSc as discussed above and 
is particularly justified in the case when FCNCs appear already at tree-level.

In the language of 4D strongly coupled theories 
the RSc scenario discussed previously is custodially 
protected flavour-anarchic model where the left-handed quarks couple to 
a single composite fermion. In such a framework NP effects in rare $K$ and $B_{s,d}$ 
decays as analyzed in \cite{Blanke:2008zb,Blanke:2008yr,Buras:2009ka} are 
full dominated by RH $Z$-couplings and the pattern of flavour violation 
with implied correlations is described by  the right DNA chart in Fig.~\ref{fig:ZPrimechart}.

However, there are other possibilities \cite{Straub:2013zca}. In a custodially protected flavour-anarchic model, where the left-handed up-
and down-type quarks couple to 
two different composite fermions rare $K$ and $B_{s,d}$ 
decays  are 
full dominated by LH $Z$-couplings. The pattern of flavour violation 
with implied correlations is summarized by the  left DNA chart in Fig.~\ref{fig:ZPrimechart}. Indeed 
the results for this scenario in Fig.~4 in \cite{Straub:2013zca} can easily be understood 
on the basis of the DNA-chart in Fig.~\ref{fig:ZPrimechart}.

Next one can consider a model with partial compositeness in 
which the strong sector possesses 
$U(2)^3$ flavour symmetry \cite{Barbieri:2012uh,Barbieri:2012tu}, minimally broken by the composite-elementary mixings of right-handed
quarks. In 
this  case as already discussed at length by us and 
also seen in Fig.~4 of \cite{Straub:2013zca}  the pattern of flavour violation with 
implied correlations is summarized by the DNA-chart in Fig.~\ref{fig:U23chart}.
Useful set of references to models with partial compositeness can be found in 
\cite{Straub:2013zca}.

\subsubsection{Gauged Flavour Models}

In these models \cite{Grinstein:2010ve,Feldmann:2010yp,Guadagnoli:2011id} 
a MFV-like ansatz is   implemented in the context of maximal gauge flavour (MGF) symmetries: in the limit
of vanishing Yukawa interactions these gauge symmetries are the largest non-Abelian ones allowed by the Lagrangian
of the model. The particle spectrum is enriched by
new heavy gauge bosons, carrying neither colour nor electric charges, and exotic fermions
to cancel anomalies. Furthermore, the new exotic fermions give rise to the SM fermion
masses through a seesaw mechanism, in a way similar to how the light left-handed (LH) neutrinos
obtain masses by the heavy RH ones. 

Even if this approach has some similarities to the usual MFV description, the presence of 
flavour-violating neutral gauge bosons and exotic fermions introduces modifications of the SM couplings and
tends to lead to dangerous contributions to FCNC processes mediated by the new heavy particles.

In \cite{Buras:2011wi}  a detailed analysis
of $\Delta F=2$ observables 
and of $B\to X_s\gamma$ in the framework of a specific 
MGF model of 
Grinstein {\it et al.}  \cite{Grinstein:2010ve} 
including all relevant contributions has been presented.
The number of parameters in this model is much smaller than in
some of the extensions of the SM discussed above and therefore it is not obvious
that the present tensions on the flavour data can be removed or at least softened. Therefore it is of interest to summarize the status of this model 
in the light of the discussions of FCNCs in the previous sections. The 
situation is as follows:
\begin{itemize}
\item
After imposition of the constraint from $\varepsilon_K$
only small deviations from the SM values of $S_{\psi K_s}$ and $S_{\psi\phi}$ 
are allowed. While at the time of our analysis in \cite{Buras:2011wi} this appeared as a possible 
problem, this result is fully consistent with present LHCb data.
Consequently  this model selects the scenario with 
exclusive (small) value of $|V_{ub}|$.  
\item
The structure of correlations between $\Delta F=2$ observables 
is very similar to models with CMFV and represented by the DNA-chart in 
 Fig.~\ref{fig:CMFVchart}. In particular $|\varepsilon_K|$ is enhanced without modifying $S_{\psi K_S}$. Moreover, $\Delta M_{d}$ and $\Delta M_{s}$  
are strongly 
correlated in this model with $\varepsilon_K$  and  the enhancement 
of the latter implies the enhancement of $\Delta M_{s,d}$. In fact 
the $\varepsilon_K-\Delta M_{s,d}$ tension discussed at length in Step 3 
of our review has been pointed out in  \cite{Buras:2011wi}. Thus the 
future of this model depends on the values of $\vcb$ and a of 
number of non-perturbative parameters as analyzed in \cite{Buras:2013raa}
\item
However, the  main problem of this scenario in 2011,  the branching ratio 
for $B^+\to\tau^+\nu_\tau$, that in this  model  is in the ballpark 
of $0.7\times 10^{-4}$, softened significantly in view of the 2012 
data from Belle. 
\end{itemize}

\subsubsection{New Vectorlike Fermions: a Minimal Theory of Fermion Masses}\label{sec:vectorlike}
We end the review of NP models by summarizing the results obtained within a model with vectorlike fermions based on
\cite{Buras:2011ph,Buras:2013td} that can be seen as a Minimal Theory of
Fermion Masses (MTFM). The idea is to explain SM fermion masses and mixings by their dynamical mixing with new heavy vectorlike
fermion $F$. Very simplified the Lagrangian has the following form: $\mathcal{L}\propto m \bar f F + M \bar F F + \lambda h F F$,
where $M$ denotes the heavy mass scale, $m$ characterizes the mixing and $\lambda$ is a Yukawa coupling. Thus the light fermions
have an admixture of heavy fermions with explicit mass terms.

This mass generation mechanism bears some similarities to the one in models with  partial compositeness and gauge flavour models just discussed. 
 As in this model the Higgs couples only to vectorlike heavy fermions but not to chiral fermions of the SM, 
that SM Yukawas arise solely through mixing. We reduce the number of parameters such that
it is still possible to reproduce the SM Yukawa couplings and that at the same time
flavour violation is suppressed. In this way we can identify the minimal FCNC effects.
A central formula is the leading order expression for the SM quark masses
\begin{align}
m_{ij}^X = v \varepsilon_i^Q \varepsilon_j^X \lambda_{ij}^X\,,\qquad (X = U,D)\,,\qquad \varepsilon_i^{Q,U,D} =
\frac{m_i^{Q,U,D}}{M_i^{Q,U,D}} \,.
\end{align}
In \cite{Buras:2011ph} the heavy Yukawa couplings $\lambda^{U,D}$ have been assumed to be anarchical $\mathcal{O}(1)$ real 
numbers which allowed a first look at the phenomenological implications. In \cite{Buras:2013td} the so called TUM 
(Trivially Unitary Model) was studied in more detail. We assumed universality of heavy masses $M_i^Q = M_i^U = M_i^D = M$ and 
unitary Yukawa matrices. With this the flavour structure simplified considerably. Furthermore 
we concentrated on flavour violation in the down sector  and thus set $\lambda^U = \mathds{1}$. After fitting the SM quark masses 
and the CKM matrix  we are left with only four new real parameters and no new phases: $M,\,\varepsilon_3^Q,\, 
s_{13}^d,\, s_{23}^d$. The latter two parameters are angles of $\lambda^D$ (the third angle is fixed by the fitting procedure) 
and from fitting $m_t$ it follows that $0.8\leq \varepsilon_3^Q\leq 1$.

\begin{figure}[!tb]
\centering
\includegraphics[width = 0.38\textwidth]{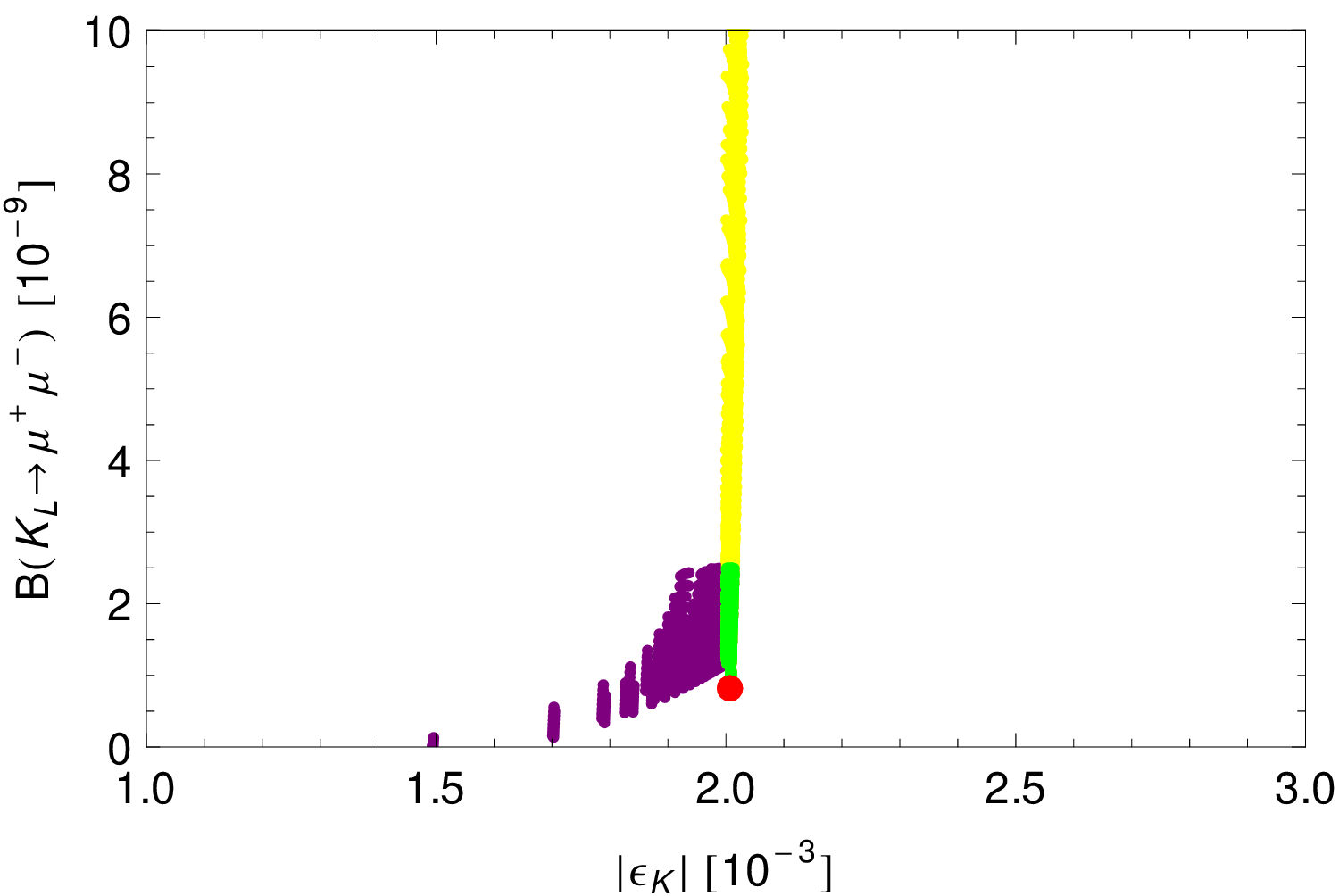}
\hspace{0.5cm}
\includegraphics[width = 0.38\textwidth]{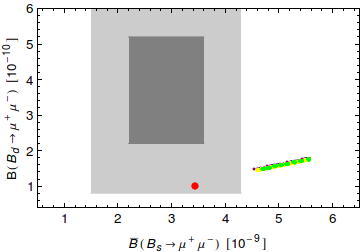}
\caption{\it $\mathcal{B}(K_L\to\mu^+\mu^-)$ vs. $|\varepsilon_K|$ and  $\mathcal{B}(B_d\to\mu^+\mu^-)$  vs. 
$\overline{\mathcal{B}}(B_s\to\mu^+\mu^-)$ for $M = 3~$TeV and $|V_{ub}| = 0.0037$. Green points are
compatible with both bounds for $|\varepsilon_K|$ (\protect\ref{C3}) and $\mathcal{B}(K_L\to\mu^+\mu^-)$ 
(\protect\ref{eq:KLmm-bound}), yellow is
only compatible with  $|\varepsilon_K|$ and purple only with $\mathcal{B}(K_L\to\mu^+\mu^-)$. The red
point corresponds to the SM central value.  The dark/light gray range shows the overlap of the $1\sigma/2\sigma$ experimental 
values of
$\mathcal{B}(B_d\to\mu^+\mu^-)$  vs. 
$\overline{\mathcal{B}}(B_s\to\mu^+\mu^-)$. }\label{fig:TUM}~\\[-2mm]\hrule
\end{figure}

The new contributions to FCNC processes are dominated
by tree-level flavour violating $Z$ couplings to quarks.  The simplest version of the MTFM, the TUM, is capable of describing the 
known quark mass spectrum and the elements of the CKM matrix favouring $|V_{ub} | \approx 0.0037$. Since there are no new phases in the TUM
$S_{\psi K_S}$ stays SM-like and thus the large inclusive value of $|V_{ub}|$ is disfavored. Although effects in $\varepsilon_K$ 
can in principle be large, the effects are bounded by $ \mathcal{B}(K_L\to\mu^+\mu^-)_{\rm SD} \le 2.5 \cdot 10^{-9}$.
For a $|V_{ub}|$ in between exclusive and inclusive value it is still possible to find regions in the parameter space that satisfy Eq.~(\ref{C3})
and~(\ref{eq:KLmm-bound}) but then the prediction of the model is that $S_{\psi K_S}\approx 0.72$ which is by $2\sigma$ 
higher than its present experimental central value. In Fig.~\ref{fig:TUM} (left) we show the correlation 
$\mathcal{B}(K_L\to\mu^+\mu^-)$ vs. $|\varepsilon_K|$ for $M = 3~$TeV where only the green points satisfy (\ref{eq:KLmm-bound}) and 
(\ref{C3}) simultaneously. In the TUM effects in $B_{s,d}$ mixings are negligible  and the pattern of deviations from 
SM predictions in rare $B$ decays is 
CMFV-like as can be see 
on the right hand side of Fig.~\ref{fig:TUM}. However   
$\mathcal{B}(B_{s,d}\to\mu^+\mu^-)$ are uniquely enhanced  over their SM 
values. For $M=3~$TeV these enhancements amount to at least $35\%$ 
and can be as large as  a factor of two. With increasing $M$  the enhancements  decrease. However they remain 
sufficiently large for 
$M\le 5~$TeV to be detected in the flavour precision era.
Also effects in $K\to \pi\nu\bar\nu$ transitions are enhanced by a similar amount.
 
At the time when our paper was published there was a hope that the 
enhancement of $\mathcal{B}(B_{s}\to\mu^+\mu^-)$ uniquely predicted by 
the model would be confirmed by the improved data. As seen on  the right hand side of Fig.~\ref{fig:TUM} the most recent data from 
 LHCb and CMS do not support this prediction and either the value of $M$ 
has to be increased or the TUM has to be made less trivial.

\section{Summary and Shopping List}\label{sec:sum}
Our review of strategies for the identification of New Physics through 
quark flavour violating processes is approaching the end. In the spirit 
of our previous reviews \cite{Buras:2009if,Buras:2010wr,Buras:2012ts} we 
have addressed the question how in principle one could identify NP with 
the help of quark flavour violating processes. In contrast to  \cite{Buras:2009if,Buras:2010wr} we 
have concentrated on the simplest extensions of the SM, describing more 
complicated ones only in the final part of this review. These simple 
constructions are helpful in identifying certain patterns of flavour violation.
In particular
correlations between various observables characteristic for these scenarios 
can distinguish between them. These features are exposed
compactly by the DNA-charts in Figs.~\ref{fig:CMFVchart}-\ref{fig:ZPrimechart}. 
Our extensive study of models in which flavour violation is 
governed by tree-level exchanges of gauge bosons, scalars and pseudoscalars with different couplings exemplified by  LHS, RHS, LRS and ALRS scenarios 
shows that future measurements can tell us which one of them is favoured by nature.

However we are aware of the fact that these simple scenarios are not fully 
representative for more complicated models in which a collection of 
several new particles and a number of new parameters can wash out  
various correlations identified by us. This is in particular the case 
of models in which FCNCs appear first at one-loop level and the FCNC amplitudes
depend on the masses of exchanged gauge bosons, fermions  and scalars and 
their couplings to SM particles. In CMFV, MFV at large and models with 
$U(2)^3$ some general pattern of flavour violation can still be identified. 
But this is much harder in the case of models with non-MFV contributions. 

Our review shows that without some concrete signs of NP in high energy collisions 
at the LHC a successful execution of the whole strategy presented in this review will  be challenging. On the other hand  with many observables accurately measured some 
picture of the physics beyond the LHC scales could in principle emerge from 
flavour physics and rare processes  alone. Yet, there is still a hope that the second half 
of this decade will bring the discoveries of new particles at the LHC and 
this would give us some concrete directions for the next steps through flavour physics 
that would allow us to get a better indirect insight into the physics at short distance scales outside the reach of the LHC.  

We end our review with a short shopping list which involves only quark flavour observables:

\begin{itemize}
\item
Precise values of all non-perturbative parameters relevant for $\Delta F=2$ 
transitions from lattice QCD. This means also hadronic matrix elements 
of new operators outside the framework of CMFV. In fact this will be 
the progress made in the coming years when most of the experiments will 
sharpen their tools for the second half of this decade.
\item
Precise determinations of CKM parameters from tree-level decays. This goal 
will be predominantly addressed by SuperKEKB but in the case of the 
angle $\gamma$, LHCb will provide a very important contribution.
\item
Precise values of $S_{\psi K_S}$ and $S_{\psi\phi}$ together with 
improved understanding of hadronic uncertainties represented by 
QCD penguins. 
\item
Precise measurements of $\mathcal{B}(B_s\to\mu^+\mu^-)$ and $\mathcal{B}(B_d\to\mu^+\mu^-)$. It is important that both branching are measured as this in 
the interplay with $\Delta M_s$ and $\Delta M_d$  and precise values of 
$\hat B_{B_s}$ and $\hat B_{B_d}$ would provide a powerful test of CMFV. 
It is evident from our presentation that the observables related to 
the time dependent rate would by far enrich these studies.
\item
Precise measurements of the multitude of angular observables in 
 $B\to K(K^*)\ell^+\ell^-$ accompanied by improved form factors can still 
provide important information about NP.  In particular it will be 
important to clarify the anomalies observed recently by the LHCb experiment 
as discussed in Step 7 of our strategy. 
\item
Precise measurements of $\mathcal{B}(\kpn)$ and $\mathcal{B}(\klpn)$. 
The first messages will come from NA62 and then hopefully from J-Parc and 
ORKA.
\item
Precise measurements of the branching ratios for the trio $B\to X_s\nu\bar\nu$, 
$B\to K^*\nu\bar\nu$ and $B\to K\nu\bar\nu$. These decays are in 
the hands of SuperKEKB.
\item 
Precise determination of  $\mathcal{B}(B^+\to\tau^+\nu_\tau)$, again in 
the hands of SuperKEKB.
\item
Precise measurement of  $\mathcal{B}(B\to X_s\gamma)$.
\item
Precise lattice results for the  parameters $\bsi$ and $\bei$ entering 
the evaluation of $\epe$.
\end{itemize}

A special role will be played by charm physics as it allows us to learn 
more about flavour physics in the up-quark sector. But the future of this 
field will depend on the progress on reduction of the hadronic uncertainties.

Next a very important role in the search for NP, as discussed in Step 12, will be played by 
lepton flavour violating decays, EDMs and $(g-2)_{e,\mu}$. But this is another 
story  and we discussed these topics only very briefly in
our review.

Finally a crucial role in these investigations will be played by theorists, 
both in the case of inventing new ideas for identifying new physics and 
constructing new extensions of the Standard Model with fewer parameters 
and thereby more predictive.

In any case this decade is expected to bring a big step forward in the 
search for new particles and new forces and we should hope that one day 
the collaboration of experimentalists and theorists will enable us to 
get some insight into the  Zeptouniverse.

\section*{Acknowledgements}

 We would like to thank all our collaborators for exciting time we had 
together while exploring different avenues beyond the Standard Model. In connection with this review
 we thank in particular Michael Ramsey-Musolf for  illuminating discussions about EDMs and Wolfgang Altmannshofer for detailed comments on  $b\to
s\ell^+\ell^-$. We also thank Bob Bernstein, Monika Blanke, Christoph Bobeth, Gerhard Buchalla, Svjetlana Fajfer, Mikolaj Misiak, David Straub and Cecilia Tarantino for useful informations.
This research was dominantly financed and done in the context of the ERC Advanced Grant project ``FLAVOUR'' (267104). It was also partially supported by the 
DFG cluster of excellence ``Origin and Structure of the Universe''.

\bibliographystyle{JHEP}
\bibliography{allrefs}

\providecommand{\href}[2]{#2}\begingroup\raggedright\begin{thebibliography}{100}

\bibitem{Cabibbo:1963yz}
N.~Cabibbo, {\it {Unitary Symmetry and Leptonic Decays}},  {\em Phys. Rev.
  Lett.} {\bf 10} (1963) 531--532.

\bibitem{Kobayashi:1973fv}
M.~Kobayashi and T.~Maskawa, {\it {CP Violation in the Renormalizable Theory of
  Weak Interaction}},  {\em Prog. Theor. Phys.} {\bf 49} (1973) 652--657.

\bibitem{Pontecorvo:1957qd}
B.~Pontecorvo, {\it {Inverse beta processes and nonconservation of lepton
  charge}},  {\em Sov. Phys. JETP} {\bf 7} (1958) 172--173.

\bibitem{Maki:1962mu}
Z.~Maki, M.~Nakagawa, and S.~Sakata, {\it {Remarks on the unified model of
  elementary particles}},  {\em Prog. Theor. Phys.} {\bf 28} (1962) 870.

\bibitem{Glashow:1970gm}
S.~L. Glashow, J.~Iliopoulos, and L.~Maiani, {\it {Weak Interactions with
  Lepton-Hadron Symmetry}},  {\em Phys. Rev.} {\bf D2} (1970) 1285--1292.

\bibitem{Buchalla:2008jp}
M.~Artuso {\em et.~al.}, {\it {$B$, $D$ and $K$ decays}},  {\em Eur. Phys. J.}
  {\bf C57} (2008) 309--492, [\href{http://xxx.lanl.gov/abs/0801.1833}{{\tt
  arXiv:0801.1833}}].

\bibitem{Raidal:2008jk}
M.~Raidal {\em et.~al.}, {\it {Flavour physics of leptons and dipole moments}},
   {\em Eur. Phys. J.} {\bf C57} (2008) 13--182,
  [\href{http://xxx.lanl.gov/abs/0801.1826}{{\tt arXiv:0801.1826}}].

\bibitem{Antonelli:2009ws}
M.~Antonelli, D.~M. Asner, D.~A. Bauer, T.~G. Becher, M.~Beneke, {\em et.~al.},
  {\it {Flavor Physics in the Quark Sector}},  {\em Phys.Rept.} {\bf 494}
  (2010) 197--414, [\href{http://xxx.lanl.gov/abs/0907.5386}{{\tt
  arXiv:0907.5386}}].

\bibitem{Bona:2007qt}
{\bf SuperB Collaboration} Collaboration, M.~Bona {\em et.~al.}, {\it {SuperB:
  A High-Luminosity Asymmetric $e^+ e^-$ Super Flavor Factory. Conceptual
  Design Report}},  \href{http://xxx.lanl.gov/abs/0709.0451}{{\tt
  arXiv:0709.0451}}.

\bibitem{Browder:2008em}
T.~E. Browder, T.~Gershon, D.~Pirjol, A.~Soni, and J.~Zupan, {\it {New Physics
  at a Super Flavor Factory}},  {\em Rev.Mod.Phys.} {\bf 81} (2009) 1887--1941,
  [\href{http://xxx.lanl.gov/abs/0802.3201}{{\tt arXiv:0802.3201}}].

\bibitem{Adeva:2009ny}
{\bf The LHCb Collaboration} Collaboration, B.~Adeva {\em et.~al.}, {\it
  {Roadmap for selected key measurements of LHCb}},
  \href{http://xxx.lanl.gov/abs/0912.4179}{{\tt arXiv:0912.4179}}.

\bibitem{Buras:2009if}
A.~J. Buras, {\it {Flavour Theory: 2009}},  {\em PoS} {\bf EPS-HEP2009} (2009)
  024, [\href{http://xxx.lanl.gov/abs/0910.1032}{{\tt arXiv:0910.1032}}].

\bibitem{Isidori:2010kg}
G.~Isidori, Y.~Nir, and G.~Perez, {\it {Flavor Physics Constraints for Physics
  Beyond the Standard Model}},  {\em Ann.Rev.Nucl.Part.Sci.} {\bf 60} (2010)
  355, [\href{http://xxx.lanl.gov/abs/1002.0900}{{\tt arXiv:1002.0900}}].

\bibitem{Fleischer:2010qb}
R.~Fleischer, {\it {B Physics in the LHC Era: Selected Topics}},  {\em
  Nucl.Phys.Proc.Suppl.} {\bf 209} (2010) 3--8,
  [\href{http://xxx.lanl.gov/abs/1010.0496}{{\tt arXiv:1010.0496}}].

\bibitem{Nir:2010jr}
Y.~Nir, {\it {Flavour physics and CP violation}},
  \href{http://xxx.lanl.gov/abs/1010.2666}{{\tt arXiv:1010.2666}}.

\bibitem{Hurth:2010tk}
T.~Hurth and M.~Nakao, {\it {Radiative and Electroweak Penguin Decays of B
  Mesons}},  {\em Ann.Rev.Nucl.Part.Sci.} {\bf 60} (2010) 645--677,
  [\href{http://xxx.lanl.gov/abs/1005.1224}{{\tt arXiv:1005.1224}}].

\bibitem{Buras:2010wr}
A.~J. Buras, {\it {Minimal flavour violation and beyond: Towards a flavour code
  for short distance dynamics}},  {\em Acta Phys.Polon.} {\bf B41} (2010)
  2487--2561, [\href{http://xxx.lanl.gov/abs/1012.1447}{{\tt
  arXiv:1012.1447}}].

\bibitem{Ciuchini:2011ca}
M.~Ciuchini and A.~Stocchi, {\it {Physics Opportunities at the Next Generation
  of Precision Flavor Physics}},  {\em Ann.Rev.Nucl.Part.Sci.} {\bf 61} (2011)
  491--517, [\href{http://xxx.lanl.gov/abs/1110.3920}{{\tt arXiv:1110.3920}}].

\bibitem{Meadows:2011bk}
B.~Meadows, M.~Blanke, A.~Stocchi, A.~Drutskoy, A.~Cervelli, {\em et.~al.},
  {\it {The impact of SuperB on flavour physics}},
  \href{http://xxx.lanl.gov/abs/1109.5028}{{\tt arXiv:1109.5028}}.

\bibitem{Buras:2012ts}
A.~J. Buras and J.~Girrbach, {\it {BSM models facing the recent LHCb data: A
  First look}},  {\em Acta Phys.Polon.} {\bf B43} (2012) 1427,
  [\href{http://xxx.lanl.gov/abs/1204.5064}{{\tt arXiv:1204.5064}}].

\bibitem{Borissov:2013yha}
G.~Borissov, R.~Fleischer, and M.-H. Schune, {\it {Rare Decays and CP Violation
  in the $B_s$ System}},  {\em Ann.Rev.Nucl.Part.Sci.} {\bf 63} (2013)
  205--235, [\href{http://xxx.lanl.gov/abs/1303.5575}{{\tt arXiv:1303.5575}}].

\bibitem{Bediaga:2012py}
{\bf LHCb Collaboration} Collaboration, R.~Aaij {\em et.~al.}, {\it
  {Implications of LHCb measurements and future prospects}},  {\em EPJ C} {\bf
  73} (2013) 2373, [\href{http://xxx.lanl.gov/abs/1208.3355}{{\tt
  arXiv:1208.3355}}].

\bibitem{Hewett:2012ns}
J.~Hewett, H.~Weerts, R.~Brock, J.~Butler, B.~Casey, {\em et.~al.}, {\it
  {Fundamental Physics at the Intensity Frontier}},
  \href{http://xxx.lanl.gov/abs/1205.2671}{{\tt arXiv:1205.2671}}.

\bibitem{Hurth:2012vp}
T.~Hurth and F.~Mahmoudi, {\it {New physics search with flavour in the LHC
  era}},  {\em Rev.Mod.Phys.} {\bf 85} (2013) 795,
  [\href{http://xxx.lanl.gov/abs/1211.6453}{{\tt arXiv:1211.6453}}].

\bibitem{Stone:2012yr}
S.~Stone, {\it {New physics from ﬂavour}},  {\em PoS} {\bf ICHEP2012} (2013)
  033, [\href{http://xxx.lanl.gov/abs/1212.6374}{{\tt arXiv:1212.6374}}].

\bibitem{Isidori:2013ez}
G.~Isidori, {\it {Flavor physics and CP violation}},
  \href{http://xxx.lanl.gov/abs/1302.0661}{{\tt arXiv:1302.0661}}.

\bibitem{Kronfeld:2013uoa}
A.~S. Kronfeld, R.~S. Tschirhart, U.~Al-Binni, W.~Altmannshofer,
  C.~Ankenbrandt, {\em et.~al.}, {\it {Project X: Physics Opportunities}},
  \href{http://xxx.lanl.gov/abs/1306.5009}{{\tt arXiv:1306.5009}}.

\bibitem{Cirigliano:2013lpa}
V.~Cirigliano and M.~J. Ramsey-Musolf, {\it {Low Energy Probes of Physics
  Beyond the Standard Model}},  {\em Prog.Part.Nucl.Phys.} {\bf 71} (2013)
  2--20, [\href{http://xxx.lanl.gov/abs/1304.0017}{{\tt arXiv:1304.0017}}].

\bibitem{Charles:2013aka}
J.~Charles, S.~Descotes-Genon, Z.~Ligeti, S.~Monteil, M.~Papucci, {\em
  et.~al.}, {\it {Future sensitivity to new physics in $B_d$, $B_s$ and $K$
  mixings}},  \href{http://xxx.lanl.gov/abs/1309.2293}{{\tt arXiv:1309.2293}}.

\bibitem{Butler:2013kdw}
{\bf Quark Flavor Physics Working Group} Collaboration, J.~Butler {\em
  et.~al.}, {\it {Report of the Quark Flavor Physics Working Group}},
  \href{http://xxx.lanl.gov/abs/1311.1076}{{\tt arXiv:1311.1076}}.

\bibitem{Branco:1999fs}
G.~C. Branco, L.~Lavoura, and J.~P. Silva, {\it {CP Violation}},  {\em
  Int.Ser.Monogr.Phys.} {\bf 103} (1999) 1--536.

\bibitem{Bigi:2000yz}
I.~I. Bigi and A.~Sanda, {\it {CP violation}},  {\em
  Camb.Monogr.Part.Phys.Nucl.Phys.Cosmol.} {\bf 9} (2000) 1--382.

\bibitem{Lunghi:2008aa}
E.~Lunghi and A.~Soni, {\it {Possible Indications of New Physics in
  $B_d$-mixing and in $\sin(2 \beta)$ Determinations}},  {\em Phys. Lett.} {\bf
  B666} (2008) 162--165, [\href{http://xxx.lanl.gov/abs/0803.4340}{{\tt
  arXiv:0803.4340}}].

\bibitem{Buras:2008nn}
A.~J. Buras and D.~Guadagnoli, {\it {Correlations among new CP violating
  effects in $\Delta F = 2$ observables}},  {\em Phys. Rev.} {\bf D78} (2008)
  033005, [\href{http://xxx.lanl.gov/abs/0805.3887}{{\tt arXiv:0805.3887}}].

\bibitem{Altmannshofer:2009ne}
W.~Altmannshofer, A.~J. Buras, S.~Gori, P.~Paradisi, and D.~M. Straub, {\it
  {Anatomy and Phenomenology of FCNC and CPV Effects in SUSY Theories}},  {\em
  Nucl.Phys.} {\bf B830} (2010) 17--94,
  [\href{http://xxx.lanl.gov/abs/0909.1333}{{\tt arXiv:0909.1333}}].

\bibitem{Blanke:2009pq}
M.~Blanke, {\it {Insights from the Interplay of $K\rightarrow \pi
  \nu\overline{\nu}$ and $\epsilon_K$ on the New Physics Flavour Structure}},
  {\em Acta Phys.Polon.} {\bf B41} (2010) 127,
  [\href{http://xxx.lanl.gov/abs/0904.2528}{{\tt arXiv:0904.2528}}].

\bibitem{Altmannshofer:2011gn}
W.~Altmannshofer, P.~Paradisi, and D.~M. Straub, {\it {Model-Independent
  Constraints on New Physics in $b\to s\gamma$ Transitions}},  {\em JHEP} {\bf
  1204} (2012) 008, [\href{http://xxx.lanl.gov/abs/1111.1257}{{\tt
  arXiv:1111.1257}}].

\bibitem{Altmannshofer:2012ir}
W.~Altmannshofer and D.~M. Straub, {\it {Cornering New Physics in $b\to s$
  Transitions}},  {\em JHEP} {\bf 1208} (2012) 121,
  [\href{http://xxx.lanl.gov/abs/1206.0273}{{\tt arXiv:1206.0273}}].

\bibitem{Buras:2012sd}
A.~J. Buras and J.~Girrbach, {\it {On the Correlations between Flavour
  Observables in Minimal $U(2)^3$ Models}},  {\em JHEP} {\bf 1301} (2013) 007,
  [\href{http://xxx.lanl.gov/abs/1206.3878}{{\tt arXiv:1206.3878}}].

\bibitem{Buras:2012jb}
A.~J. Buras, F.~De~Fazio, and J.~Girrbach, {\it {The Anatomy of Z' and Z with
  Flavour Changing Neutral Currents in the Flavour Precision Era}},  {\em JHEP}
  {\bf 1302} (2013) 116, [\href{http://xxx.lanl.gov/abs/1211.1896}{{\tt
  arXiv:1211.1896}}].

\bibitem{Buras:2012dp}
A.~J. Buras, F.~De~Fazio, J.~Girrbach, and M.~V. Carlucci, {\it {The Anatomy of
  Quark Flavour Observables in 331 Models in the Flavour Precision Era}},  {\em
  JHEP} {\bf 1302} (2013) 023, [\href{http://xxx.lanl.gov/abs/1211.1237}{{\tt
  arXiv:1211.1237}}].

\bibitem{Buras:2013td}
A.~J. Buras, J.~Girrbach, and R.~Ziegler, {\it {Particle-Antiparticle Mixing,
  CP Violation and Rare K and B Decays in a Minimal Theory of Fermion Masses}},
   {\em JHEP} {\bf 1304} (2013) 168,
  [\href{http://xxx.lanl.gov/abs/1301.5498}{{\tt arXiv:1301.5498}}].

\bibitem{Buras:2013uqa}
A.~J. Buras, R.~Fleischer, J.~Girrbach, and R.~Knegjens, {\it {Probing New
  Physics with the $B_s\to\mu^+\mu^-$ Time-Dependent Rate}},  {\em JHEP} {\bf
  1307} (2013) 77, [\href{http://xxx.lanl.gov/abs/1303.3820}{{\tt
  arXiv:1303.3820}}].

\bibitem{Buras:2013rqa}
A.~J. Buras, F.~De~Fazio, J.~Girrbach, R.~Knegjens, and M.~Nagai, {\it {The
  Anatomy of Neutral Scalars with FCNCs in the Flavour Precision Era}},  {\em
  JHEP} {\bf 1306} (2013) 111, [\href{http://xxx.lanl.gov/abs/1303.3723}{{\tt
  arXiv:1303.3723}}].

\bibitem{Buras:2013raa}
A.~J. Buras and J.~Girrbach, {\it {Stringent Tests of Constrained Minimal
  Flavour Violation through $\Delta F=2$ Transitions}},  {\em The European
  Physical Journal C} {\bf 9} (73) 2013,
  [\href{http://xxx.lanl.gov/abs/1304.6835}{{\tt arXiv:1304.6835}}].

\bibitem{Buras:2013qja}
A.~J. Buras and J.~Girrbach, {\it {Left-handed Z' and Z FCNC quark couplings
  facing new $b \to s \mu^+ \mu^-$ data}},  {\em JHEP} {\bf 1312} (2013) 009,
  [\href{http://xxx.lanl.gov/abs/1309.2466}{{\tt arXiv:1309.2466}}].

\bibitem{Buras:2013dea}
A.~J. Buras, F.~De~Fazio, and J.~Girrbach, {\it {331 models facing new $b \to
  s\mu^+ \mu^-$ data}},  {\em JHEP} {\bf 1402} (2014) 112,
  [\href{http://xxx.lanl.gov/abs/1311.6729}{{\tt arXiv:1311.6729}}].

\bibitem{Blanke:2006eb}
M.~Blanke {\em et.~al.}, {\it {Rare and CP-violating $K$ and $B$ decays in the
  Littlest Higgs model with T-parity}},  {\em JHEP} {\bf 01} (2007) 066,
  [\href{http://xxx.lanl.gov/abs/hep-ph/0610298}{{\tt hep-ph/0610298}}].

\bibitem{Blanke:2008yr}
M.~Blanke, A.~J. Buras, B.~Duling, K.~Gemmler, and S.~Gori, {\it {Rare K and B
  Decays in a Warped Extra Dimension with Custodial Protection}},  {\em JHEP}
  {\bf 03} (2009) 108, [\href{http://xxx.lanl.gov/abs/0812.3803}{{\tt
  arXiv:0812.3803}}].

\bibitem{Buchalla:1995vs}
G.~Buchalla, A.~J. Buras, and M.~E. Lautenbacher, {\it {Weak decays beyond
  leading logarithms}},  {\em Rev.Mod.Phys.} {\bf 68} (1996) 1125--1144,
  [\href{http://xxx.lanl.gov/abs/hep-ph/9512380}{{\tt hep-ph/9512380}}].

\bibitem{Buras:1998raa}
A.~J. Buras, {\it {Weak Hamiltonian, CP violation and rare decays}},
  \href{http://xxx.lanl.gov/abs/hep-ph/9806471}{{\tt hep-ph/9806471}}. In
  'Probing the Standard Model of Particle Interactions', F.David and R. Gupta,
  eds., 1998, Elsevier Science B.V.

\bibitem{Buras:2011we}
A.~J. Buras, {\it {Climbing NLO and NNLO Summits of Weak Decays}},
  \href{http://xxx.lanl.gov/abs/1102.5650}{{\tt arXiv:1102.5650}}.

\bibitem{Buras:2000dm}
A.~J. Buras, P.~Gambino, M.~Gorbahn, S.~Jager, and L.~Silvestrini, {\it
  Universal unitarity triangle and physics beyond the standard model},  {\em
  Phys. Lett.} {\bf B500} (2001) 161--167,
  [\href{http://xxx.lanl.gov/abs/hep-ph/0007085}{{\tt hep-ph/0007085}}].

\bibitem{Buras:2003jf}
A.~J. Buras, {\it Minimal flavor violation},  {\em Acta Phys. Polon.} {\bf B34}
  (2003) 5615--5668, [\href{http://xxx.lanl.gov/abs/hep-ph/0310208}{{\tt
  hep-ph/0310208}}].

\bibitem{Blanke:2006ig}
M.~Blanke, A.~J. Buras, D.~Guadagnoli, and C.~Tarantino, {\it {Minimal Flavour
  Violation Waiting for Precise Measurements of $\Delta M_s$, $S_{\psi \phi}$,
  $A^s_\text{SL}$, $|V_{ub}|$, $\gamma$ and $B^0_{s,d} \to \mu^+ \mu^-$}},
  {\em JHEP} {\bf 10} (2006) 003,
  [\href{http://xxx.lanl.gov/abs/hep-ph/0604057}{{\tt hep-ph/0604057}}].

\bibitem{Inami:1980fz}
T.~Inami and C.~Lim, {\it {Effects of Superheavy Quarks and Leptons in
  Low-Energy Weak Processes $K_L\to\mu^+\mu^-$, $K^+\to\pi^+\nu\bar\nu$ and
  $K^0-\bar K^0$}},  {\em Prog.Theor.Phys.} {\bf 65} (1981) 297.

\bibitem{Buchalla:1990qz}
G.~Buchalla, A.~J. Buras, and M.~K. Harlander, {\it Penguin box expansion:
  Flavor changing neutral current processes and a heavy top quark},  {\em Nucl.
  Phys.} {\bf B349} (1991) 1--47.

\bibitem{Buras:1994ec}
A.~J. Buras, M.~E. Lautenbacher, and G.~Ostermaier, {\it {Waiting for the top
  quark mass, $K^+ \to \pi^+ \nu\bar\nu$, $B_s^0 - \bar B_s^0$ mixing and CP
  asymmetries in $B$ decays}},  {\em Phys. Rev.} {\bf D50} (1994) 3433--3446,
  [\href{http://xxx.lanl.gov/abs/hep-ph/9403384}{{\tt hep-ph/9403384}}].

\bibitem{Buras:2000xq}
A.~J. Buras and R.~Buras, {\it {A Lower bound on $\sin$ 2 beta from minimal
  flavor violation}},  {\em Phys.Lett.} {\bf B501} (2001) 223--230,
  [\href{http://xxx.lanl.gov/abs/hep-ph/0008273}{{\tt hep-ph/0008273}}].

\bibitem{Blanke:2006yh}
M.~Blanke and A.~J. Buras, {\it {Lower bounds on $\Delta M_{s,d}$ from
  constrained minimal flavour violation}},  {\em JHEP} {\bf 0705} (2007) 061,
  [\href{http://xxx.lanl.gov/abs/hep-ph/0610037}{{\tt hep-ph/0610037}}].

\bibitem{Buras:2003td}
A.~J. Buras, {\it {Relations between $\Delta M_{s,d}$ and $B_{s,d} \to \mu^+
  \mu^-$ in models with minimal flavour violation}},  {\em Phys. Lett.} {\bf
  B566} (2003) 115--119, [\href{http://xxx.lanl.gov/abs/hep-ph/0303060}{{\tt
  hep-ph/0303060}}].

\bibitem{Carrasco:2013zta}
N.~Carrasco, M.~Ciuchini, P.~Dimopoulos, R.~Frezzotti, V.~Gimenez, {\em
  et.~al.}, {\it {B-physics from Nf=2 tmQCD: the Standard Model and beyond}},
  \href{http://xxx.lanl.gov/abs/1308.1851}{{\tt arXiv:1308.1851}}.

\bibitem{D'Ambrosio:2002ex}
G.~D'Ambrosio, G.~F. Giudice, G.~Isidori, and A.~Strumia, {\it {Minimal flavour
  violation: An effective field theory approach}},  {\em Nucl. Phys.} {\bf
  B645} (2002) 155--187, [\href{http://xxx.lanl.gov/abs/hep-ph/0207036}{{\tt
  hep-ph/0207036}}].

\bibitem{Feldmann:2006jk}
T.~Feldmann and T.~Mannel, {\it {Minimal Flavour Violation and Beyond}},  {\em
  JHEP} {\bf 0702} (2007) 067,
  [\href{http://xxx.lanl.gov/abs/hep-ph/0611095}{{\tt hep-ph/0611095}}].

\bibitem{Colangelo:2008qp}
G.~Colangelo, E.~Nikolidakis, and C.~Smith, {\it {Supersymmetric models with
  minimal flavour violation and their running}},  {\em Eur.Phys.J.} {\bf C59}
  (2009) 75--98, [\href{http://xxx.lanl.gov/abs/0807.0801}{{\tt
  arXiv:0807.0801}}].

\bibitem{Paradisi:2008qh}
P.~Paradisi, M.~Ratz, R.~Schieren, and C.~Simonetto, {\it {Running minimal
  flavor violation}},  {\em Phys.Lett.} {\bf B668} (2008) 202--209,
  [\href{http://xxx.lanl.gov/abs/0805.3989}{{\tt arXiv:0805.3989}}].

\bibitem{Mercolli:2009ns}
L.~Mercolli and C.~Smith, {\it {EDM constraints on flavored CP-violating
  phases}},  {\em Nucl.Phys.} {\bf B817} (2009) 1--24,
  [\href{http://xxx.lanl.gov/abs/0902.1949}{{\tt arXiv:0902.1949}}].

\bibitem{Feldmann:2009dc}
T.~Feldmann, M.~Jung, and T.~Mannel, {\it {Sequential Flavour Symmetry
  Breaking}},  {\em Phys.Rev.} {\bf D80} (2009) 033003,
  [\href{http://xxx.lanl.gov/abs/0906.1523}{{\tt arXiv:0906.1523}}].

\bibitem{Kagan:2009bn}
A.~L. Kagan, G.~Perez, T.~Volansky, and J.~Zupan, {\it {General Minimal Flavor
  Violation}},  {\em Phys.Rev.} {\bf D80} (2009) 076002,
  [\href{http://xxx.lanl.gov/abs/0903.1794}{{\tt arXiv:0903.1794}}].

\bibitem{Paradisi:2009ey}
P.~Paradisi and D.~M. Straub, {\it {The SUSY CP Problem and the MFV
  Principle}},  {\em Phys.Lett.} {\bf B684} (2010) 147--153,
  [\href{http://xxx.lanl.gov/abs/0906.4551}{{\tt arXiv:0906.4551}}].

\bibitem{Isidori:2010gz}
G.~Isidori, {\it {B Physics in the LHC Era}},
  \href{http://xxx.lanl.gov/abs/1001.3431}{{\tt arXiv:1001.3431}}.

\bibitem{Hurth:2008jc}
T.~Hurth, G.~Isidori, J.~F. Kamenik, and F.~Mescia, {\it {Constraints on New
  Physics in MFV models: A Model-independent analysis of $\Delta F = $1
  processes}},  {\em Nucl. Phys.} {\bf B808} (2009) 326--346,
  [\href{http://xxx.lanl.gov/abs/0807.5039}{{\tt arXiv:0807.5039}}].

\bibitem{Isidori:2012ts}
G.~Isidori and D.~M. Straub, {\it {Minimal Flavour Violation and Beyond}},
  {\em Eur.Phys.J.} {\bf C72} (2012) 2103,
  [\href{http://xxx.lanl.gov/abs/1202.0464}{{\tt arXiv:1202.0464}}].

\bibitem{Baek:1998yn}
S.~Baek and P.~Ko, {\it {Probing SUSY induced CP violations at B factories}},
  {\em Phys.Rev.Lett.} {\bf 83} (1999) 488--491,
  [\href{http://xxx.lanl.gov/abs/hep-ph/9812229}{{\tt hep-ph/9812229}}].

\bibitem{Baek:1999qy}
S.~Baek and P.~Ko, {\it {Effects of supersymmetric CP violating phases on $B\to
  X(s) lepton^+ lepton^-$ and $\epsilon(K)$}},  {\em Phys.Lett.} {\bf B462}
  (1999) 95--102, [\href{http://xxx.lanl.gov/abs/hep-ph/9904283}{{\tt
  hep-ph/9904283}}].

\bibitem{Bartl:2001wc}
A.~Bartl, T.~Gajdosik, E.~Lunghi, A.~Masiero, W.~Porod, {\em et.~al.}, {\it
  {General flavor blind MSSM and CP violation}},  {\em Phys.Rev.} {\bf D64}
  (2001) 076009, [\href{http://xxx.lanl.gov/abs/hep-ph/0103324}{{\tt
  hep-ph/0103324}}].

\bibitem{Ellis:2007kb}
J.~Ellis, J.~S. Lee, and A.~Pilaftsis, {\it {B-Meson Observables in the
  Maximally CP-Violating MSSM with Minimal Flavour Violation}},  {\em Phys.
  Rev.} {\bf D76} (2007) 115011, [\href{http://xxx.lanl.gov/abs/0708.2079}{{\tt
  arXiv:0708.2079}}].

\bibitem{Altmannshofer:2008hc}
W.~Altmannshofer, A.~Buras, and P.~Paradisi, {\it {Low Energy Probes of CP
  Violation in a Flavor Blind MSSM}},  {\em Phys.Lett.} {\bf B669} (2008)
  239--245, [\href{http://xxx.lanl.gov/abs/0808.0707}{{\tt arXiv:0808.0707}}].

\bibitem{Pich:2009sp}
A.~Pich and P.~Tuzon, {\it {Yukawa Alignment in the Two-Higgs-Doublet Model}},
  {\em Phys.Rev.} {\bf D80} (2009) 091702,
  [\href{http://xxx.lanl.gov/abs/0908.1554}{{\tt arXiv:0908.1554}}].

\bibitem{Blum:2010mj}
K.~Blum, Y.~Hochberg, and Y.~Nir, {\it {Implications of large dimuon CP
  asymmetry in $B_{d,s}$ decays on minimal flavor violation with low tan
  $\beta$}},  {\em JHEP} {\bf 1009} (2010) 035,
  [\href{http://xxx.lanl.gov/abs/1007.1872}{{\tt arXiv:1007.1872}}].

\bibitem{Dobrescu:2010rh}
B.~A. Dobrescu, P.~J. Fox, and A.~Martin, {\it {CP violation in $B_s$ mixing
  from heavy Higgs exchange}},  {\em Phys.Rev.Lett.} {\bf 105} (2010) 041801,
  [\href{http://xxx.lanl.gov/abs/1005.4238}{{\tt arXiv:1005.4238}}].

\bibitem{Altmannshofer:2011iv}
W.~Altmannshofer and M.~Carena, {\it {B Meson Mixing in Effective Theories of
  Supersymmetric Higgs Bosons}},  {\em Phys.Rev.} {\bf D85} (2012) 075006,
  [\href{http://xxx.lanl.gov/abs/1110.0843}{{\tt arXiv:1110.0843}}].

\bibitem{Altmannshofer:2011rm}
W.~Altmannshofer, M.~Carena, S.~Gori, and A.~de~la Puente, {\it {Signals of CP
  Violation Beyond the MSSM in Higgs and Flavor Physics}},  {\em Phys.Rev.}
  {\bf D84} (2011) 095027, [\href{http://xxx.lanl.gov/abs/1107.3814}{{\tt
  arXiv:1107.3814}}].

\bibitem{Buras:2010mh}
A.~J. Buras, M.~V. Carlucci, S.~Gori, and G.~Isidori, {\it {Higgs-mediated
  FCNCs: Natural Flavour Conservation vs. Minimal Flavour Violation}},  {\em
  JHEP} {\bf 1010} (2010) 009, [\href{http://xxx.lanl.gov/abs/1005.5310}{{\tt
  arXiv:1005.5310}}].

\bibitem{Branco:2011iw}
G.~Branco, P.~Ferreira, L.~Lavoura, M.~Rebelo, M.~Sher, {\em et.~al.}, {\it
  {Theory and phenomenology of two-Higgs-doublet models}},  {\em Phys.Rept.}
  {\bf 516} (2012) 1--102, [\href{http://xxx.lanl.gov/abs/1106.0034}{{\tt
  arXiv:1106.0034}}].

\bibitem{Barbieri:2011ci}
R.~Barbieri, G.~Isidori, J.~Jones-Perez, P.~Lodone, and D.~M. Straub, {\it
  {U(2) and Minimal Flavour Violation in Supersymmetry}},  {\em Eur.Phys.J.}
  {\bf C71} (2011) 1725, [\href{http://xxx.lanl.gov/abs/1105.2296}{{\tt
  arXiv:1105.2296}}].

\bibitem{Barbieri:2011fc}
R.~Barbieri, P.~Campli, G.~Isidori, F.~Sala, and D.~M. Straub, {\it {B-decay
  CP-asymmetries in SUSY with a $U(2)^3$ flavour symmetry}},  {\em Eur.Phys.J.}
  {\bf C71} (2011) 1812, [\href{http://xxx.lanl.gov/abs/1108.5125}{{\tt
  arXiv:1108.5125}}].

\bibitem{Barbieri:2012uh}
R.~Barbieri, D.~Buttazzo, F.~Sala, and D.~M. Straub, {\it {Flavour physics from
  an approximate $U(2)^3$ symmetry}},  {\em JHEP} {\bf 1207} (2012) 181,
  [\href{http://xxx.lanl.gov/abs/1203.4218}{{\tt arXiv:1203.4218}}].

\bibitem{Barbieri:2012bh}
R.~Barbieri, D.~Buttazzo, F.~Sala, and D.~M. Straub, {\it {Less Minimal Flavour
  Violation}},  {\em JHEP} {\bf 1210} (2012) 040,
  [\href{http://xxx.lanl.gov/abs/1206.1327}{{\tt arXiv:1206.1327}}].

\bibitem{Crivellin:2011fb}
A.~Crivellin, L.~Hofer, and U.~Nierste, {\it {The MSSM with a Softly Broken
  $U(2)^3$ Flavor Symmetry}},  {\em PoS} {\bf EPS-HEP2011} (2011) 145,
  [\href{http://xxx.lanl.gov/abs/1111.0246}{{\tt arXiv:1111.0246}}].

\bibitem{Crivellin:2011sj}
A.~Crivellin, L.~Hofer, U.~Nierste, and D.~Scherer, {\it {Phenomenological
  consequences of radiative flavor violation in the MSSM}},  {\em Phys.Rev.}
  {\bf D84} (2011) 035030, [\href{http://xxx.lanl.gov/abs/1105.2818}{{\tt
  arXiv:1105.2818}}].

\bibitem{Crivellin:2008mq}
A.~Crivellin and U.~Nierste, {\it {Supersymmetric renormalisation of the CKM
  matrix and new constraints on the squark mass matrices}},  {\em Phys.Rev.}
  {\bf D79} (2009) 035018, [\href{http://xxx.lanl.gov/abs/0810.1613}{{\tt
  arXiv:0810.1613}}].

\bibitem{Buras:2012fs}
A.~J. Buras and J.~Girrbach, {\it {Complete NLO QCD Corrections for Tree Level
  Delta F = 2 FCNC Processes}},  {\em JHEP} {\bf 1203} (2012) 052,
  [\href{http://xxx.lanl.gov/abs/1201.1302}{{\tt arXiv:1201.1302}}].

\bibitem{Buras:2012gm}
A.~J. Buras and J.~Girrbach, {\it {Completing NLO QCD Corrections for Tree
  Level Non-Leptonic $\Delta F = 1$ Decays Beyond the Standard Model}},
  \href{http://xxx.lanl.gov/abs/1201.2563}{{\tt arXiv:1201.2563}}.

\bibitem{Ligeti:2006pm}
Z.~Ligeti, M.~Papucci, and G.~Perez, {\it {Implications of the measurement of
  the $B_s^0 - \bar B_s^0$ mass difference}},  {\em Phys. Rev. Lett} {\bf 97}
  (2006) 101801, [\href{http://xxx.lanl.gov/abs/hep-ph/0604112}{{\tt
  hep-ph/0604112}}].

\bibitem{Buchalla:1994tr}
G.~Buchalla and A.~J. Buras, {\it $\sin2\beta$ from $k \to \pi \nu\bar\nu$},
  {\em Phys. Lett.} {\bf B333} (1994) 221--227,
  [\href{http://xxx.lanl.gov/abs/hep-ph/9405259}{{\tt hep-ph/9405259}}].

\bibitem{Buras:2001af}
A.~J. Buras and R.~Fleischer, {\it {Bounds on the unitarity triangle,
  $\sin2\beta$ and $K \to\pi \nu\bar\nu$ decays in models with minimal flavor
  violation}},  {\em Phys. Rev.} {\bf D64} (2001) 115010,
  [\href{http://xxx.lanl.gov/abs/hep-ph/0104238}{{\tt hep-ph/0104238}}].

\bibitem{Wolfenstein:1983yz}
L.~Wolfenstein, {\it {Parametrization of the Kobayashi-Maskawa Matrix}},  {\em
  Phys.Rev.Lett.} {\bf 51} (1983) 1945.

\bibitem{Charles:2004jd}
{\bf CKMfitter Group} Collaboration, J.~Charles {\em et.~al.}, {\it {CP
  violation and the CKM matrix: Assessing the impact of the asymmetric $B$
  factories}},  {\em Eur.Phys.J.} {\bf C41} (2005) 1--131,
  [\href{http://xxx.lanl.gov/abs/hep-ph/0406184}{{\tt hep-ph/0406184}}].
  http://www.ckmfitter.in2p3.fr.

\bibitem{Bona:2005eu}
{\bf UTfit} Collaboration, M.~Bona {\em et.~al.}, {\it {The UTfit collaboration
  report on the status of the unitarity triangle beyond the standard model. I:
  Model- independent analysis and minimal flavour violation}},  {\em JHEP} {\bf
  03} (2006) 080, [\href{http://xxx.lanl.gov/abs/hep-ph/0509219}{{\tt
  hep-ph/0509219}}]. http://www.utfit.org/UTfit/ResultsSummer2013PostEPS.

\bibitem{Eigen:2013cv}
G.~Eigen, G.~Dubois-Felsmann, D.~G. Hitlin, and F.~C. Porter, {\it {Global CKM
  Fits with the Scan Method}},  {\em PoS} {\bf ICHEP2012} (2013) 320,
  [\href{http://xxx.lanl.gov/abs/1301.5867}{{\tt arXiv:1301.5867}}].

\bibitem{Nakamura:2010zzi}
{\bf Particle Data Group} Collaboration, K.~Nakamura {\em et.~al.}, {\it
  {Review of particle physics}},  {\em J.Phys.G} {\bf G37} (2010) 075021.

\bibitem{Beringer:1900zz}
{\bf Particle Data Group} Collaboration, J.~Beringer {\em et.~al.}, {\it
  {Review of Particle Physics (RPP)}},  {\em Phys.Rev.} {\bf D86} (2012)
  010001.

\bibitem{Aoki:2013ldr}
S.~Aoki, Y.~Aoki, C.~Bernard, T.~Blum, G.~Colangelo, {\em et.~al.}, {\it
  {Review of lattice results concerning low energy particle physics}},
  \href{http://xxx.lanl.gov/abs/1310.8555}{{\tt arXiv:1310.8555}}.

\bibitem{Dowdall:2013tga}
{\bf HPQCD Collaboration} Collaboration, R.~Dowdall, C.~Davies, R.~Horgan,
  C.~Monahan, and J.~Shigemitsu, {\it {B-meson decay constants from improved
  lattice NRQCD and physical u, d, s and c sea quarks}},  {\em Phys.Rev.Lett.}
  {\bf 110} (2013) 222003, [\href{http://xxx.lanl.gov/abs/1302.2644}{{\tt
  arXiv:1302.2644}}].

\bibitem{Chetyrkin:2009fv}
K.~Chetyrkin, J.~Kuhn, A.~Maier, P.~Maierhofer, P.~Marquard, {\em et.~al.},
  {\it {Charm and Bottom Quark Masses: An Update}},  {\em Phys.Rev.} {\bf D80}
  (2009) 074010, [\href{http://xxx.lanl.gov/abs/0907.2110}{{\tt
  arXiv:0907.2110}}].

\bibitem{Laiho:2009eu}
J.~Laiho, E.~Lunghi, and R.~S. Van~de Water, {\it {Lattice QCD inputs to the
  CKM unitarity triangle analysis}},  {\em Phys. Rev.} {\bf D81} (2010) 034503,
  [\href{http://xxx.lanl.gov/abs/0910.2928}{{\tt arXiv:0910.2928}}]. Updates
  available on {\tt http://latticeaverages.org/}.

\bibitem{Allison:2008xk}
{\bf HPQCD Collaboration} Collaboration, I.~Allison {\em et.~al.}, {\it
  {High-Precision Charm-Quark Mass from Current-Current Correlators in Lattice
  and Continuum QCD}},  {\em Phys.Rev.} {\bf D78} (2008) 054513,
  [\href{http://xxx.lanl.gov/abs/0805.2999}{{\tt arXiv:0805.2999}}].

\bibitem{Buras:1990fn}
A.~J. Buras, M.~Jamin, and P.~H. Weisz, {\it {Leading and next-to-leading QCD
  corrections to $\varepsilon$ parameter and $B^0-\bar{B}^0$ mixing in the
  presence of a heavy top quark}},  {\em Nucl. Phys.} {\bf B347} (1990)
  491--536.

\bibitem{Urban:1997gw}
J.~Urban, F.~Krauss, U.~Jentschura, and G.~Soff, {\it {Next-to-leading order
  QCD corrections for the $B^0 - \bar B^0$ mixing with an extended Higgs
  sector}},  {\em Nucl. Phys.} {\bf B523} (1998) 40--58,
  [\href{http://xxx.lanl.gov/abs/hep-ph/9710245}{{\tt hep-ph/9710245}}].

\bibitem{Aaltonen:2012ra}
{\bf CDF Collaboration, D0 Collaboration} Collaboration, T.~Aaltonen {\em
  et.~al.}, {\it {Combination of the top-quark mass measurements from the
  Tevatron collider}},  {\em Phys.Rev.} {\bf D86} (2012) 092003,
  [\href{http://xxx.lanl.gov/abs/1207.1069}{{\tt arXiv:1207.1069}}].

\bibitem{Amhis:2012bh}
{\bf Heavy Flavor Averaging Group} Collaboration, Y.~Amhis {\em et.~al.}, {\it
  {Averages of B-Hadron, C-Hadron, and tau-lepton properties as of early
  2012}},  \href{http://xxx.lanl.gov/abs/1207.1158}{{\tt arXiv:1207.1158}}.
  http://www.slac.stanford.edu/xorg/hfag.

\bibitem{Buras:2010pza}
A.~J. Buras, D.~Guadagnoli, and G.~Isidori, {\it {On $\epsilon_K$ beyond lowest
  order in the Operator Product Expansion}},  {\em Phys.Lett.} {\bf B688}
  (2010) 309--313, [\href{http://xxx.lanl.gov/abs/1002.3612}{{\tt
  arXiv:1002.3612}}].

\bibitem{Brod:2011ty}
J.~Brod and M.~Gorbahn, {\it {Next-to-Next-to-Leading-Order Charm-Quark
  Contribution to the CP Violation Parameter $\varepsilon_K$ and $\Delta
  M_K$}},  {\em Phys.Rev.Lett.} {\bf 108} (2012) 121801,
  [\href{http://xxx.lanl.gov/abs/1108.2036}{{\tt arXiv:1108.2036}}].

\bibitem{Brod:2010mj}
J.~Brod and M.~Gorbahn, {\it {$\epsilon_K$ at Next-to-Next-to-Leading Order:
  The Charm-Top-Quark Contribution}},  {\em Phys.Rev.} {\bf D82} (2010) 094026,
  [\href{http://xxx.lanl.gov/abs/1007.0684}{{\tt arXiv:1007.0684}}].

\bibitem{Gambino:2013rza}
P.~Gambino and C.~Schwanda, {\it {Inclusive semileptonic fits, heavy quark
  masses, and $V_{cb}$}},  \href{http://xxx.lanl.gov/abs/1307.4551}{{\tt
  arXiv:1307.4551}}.

\bibitem{Ricciardi:2013cda}
G.~Ricciardi, {\it {Determination of the CKM matrix elements |V(xb)|}},  {\em
  Mod.Phys.Lett.} {\bf A28} (2013) 1330016,
  [\href{http://xxx.lanl.gov/abs/1305.2844}{{\tt arXiv:1305.2844}}].

\bibitem{Bailey:2014tva}
J.~A. Bailey, A.~Bazavov, C.~Bernard, C.~Bouchard, C.~DeTar, {\em et.~al.},
  {\it {Update of $|V_{cb}|$ from the $\bar{B}\to D^*\ell\bar{\nu}$ form factor
  at zero recoil with three-flavor lattice QCD}},
  \href{http://xxx.lanl.gov/abs/1403.0635}{{\tt arXiv:1403.0635}}.

\bibitem{Buras:2014sba}
A.~J. Buras, F.~De~Fazio, and J.~Girrbach, {\it {$\Delta I=1/2$ Rule,
  $\varepsilon'/\varepsilon$ and $K\to\pi\nu\bar\nu$ in Z'(Z) and G' Models
  with FCNC Quark Couplings}},  \href{http://xxx.lanl.gov/abs/1404.3824}{{\tt
  arXiv:1404.3824}}.

\bibitem{Buras:2002yj}
A.~J. Buras, F.~Parodi, and A.~Stocchi, {\it {The CKM matrix and the unitarity
  triangle: Another look}},  {\em JHEP} {\bf 0301} (2003) 029,
  [\href{http://xxx.lanl.gov/abs/hep-ph/0207101}{{\tt hep-ph/0207101}}].

\bibitem{Tarantino:2012mq}
C.~Tarantino, {\it {Flavor Lattice QCD in the Precision Era}},  {\em PoS} {\bf
  ICHEP2012} (2013) 023, [\href{http://xxx.lanl.gov/abs/1210.0474}{{\tt
  arXiv:1210.0474}}].

\bibitem{Davies:2012qf}
C.~Davies, {\it {Standard Model Heavy Flavor physics on the Lattice}},  {\em
  PoS} {\bf LATTICE2011} (2011) 019,
  [\href{http://xxx.lanl.gov/abs/1203.3862}{{\tt arXiv:1203.3862}}].

\bibitem{Gamiz:2013waa}
E.~G\'amiz, {\it {Flavour physics from lattice QCD}},  {\em PoS} {\bf
  ConfinementX} (2012) 241, [\href{http://xxx.lanl.gov/abs/1303.3971}{{\tt
  arXiv:1303.3971}}].

\bibitem{Sachrajda:2013fxa}
{\bf RBC-UKQCD} Collaboration, C.~T. Sachrajda, {\it {Prospects for Lattice
  Calculations of Rare Kaon Decay Amplitudes}},  {\em PoS} {\bf KAON13} (2013)
  019.

\bibitem{Christ:2013lxa}
N.~Christ, {\it {Nonleptonic Kaon Decays from Lattice QCD}},  {\em PoS} {\bf
  KAON13} (2013) 029.

\bibitem{Buras:2001ra}
A.~J. Buras, S.~Jager, and J.~Urban, {\it {Master formulae for $\Delta F=2$ NLO
  QCD factors in the standard model and beyond}},  {\em Nucl.Phys.} {\bf B605}
  (2001) 600--624, [\href{http://xxx.lanl.gov/abs/hep-ph/0102316}{{\tt
  hep-ph/0102316}}].

\bibitem{Boyle:2012qb}
{\bf RBC and UKQCD Collaborations} Collaboration, P.~Boyle, N.~Garron, and
  R.~Hudspith, {\it {Neutral kaon mixing beyond the standard model with $n_f =
  2+1$ chiral fermions}},  {\em Phys.Rev.} {\bf D86} (2012) 054028,
  [\href{http://xxx.lanl.gov/abs/1206.5737}{{\tt arXiv:1206.5737}}].

\bibitem{Bertone:2012cu}
{\bf Collaboration ETM} Collaboration, V.~Bertone {\em et.~al.}, {\it {Kaon
  Mixing Beyond the SM from Nf=2 tmQCD and model independent constraints from
  the UTA}},  {\em JHEP} {\bf 1303} (2013) 089,
  [\href{http://xxx.lanl.gov/abs/1207.1287}{{\tt arXiv:1207.1287}}].

\bibitem{Bouchard:2011xj}
C.~Bouchard, E.~Freeland, C.~Bernard, A.~El-Khadra, E.~Gamiz, {\em et.~al.},
  {\it {Neutral $B$ mixing from $2+1$ flavor lattice-QCD: the Standard Model
  and beyond}},  {\em PoS} {\bf LATTICE2011} (2011) 274,
  [\href{http://xxx.lanl.gov/abs/1112.5642}{{\tt arXiv:1112.5642}}].

\bibitem{Bae:2013tca}
{\bf SWME Collaboration} Collaboration, T.~Bae {\em et.~al.}, {\it {Neutral
  kaon mixing from new physics: matrix elements in $N_f=2+1$ QCD}},
  \href{http://xxx.lanl.gov/abs/1309.2040}{{\tt arXiv:1309.2040}}.

\bibitem{Blanke:2011ry}
M.~Blanke, A.~J. Buras, K.~Gemmler, and T.~Heidsieck, {\it {$\Delta F = 2$
  observables and $B\to X_q\gamma$ in the Left-Right Asymmetric Model: Higgs
  particles striking back}},  {\em JHEP} {\bf 1203} (2012) 024,
  [\href{http://xxx.lanl.gov/abs/1111.5014}{{\tt arXiv:1111.5014}}].

\bibitem{LHCb-CONF-2013-006}
{\it Improved constraints on $\gamma$ from $b^\pm\to dk^\pm$ decays including
  first results on 2012 data}, . Linked to LHCb-ANA-2013-012.

\bibitem{Fleischer:2010ib}
R.~Fleischer and R.~Knegjens, {\it {In Pursuit of New Physics with $B^0_s\to
  K^+K^-$}},  {\em Eur.Phys.J.} {\bf C71} (2011) 1532,
  [\href{http://xxx.lanl.gov/abs/1011.1096}{{\tt arXiv:1011.1096}}].

\bibitem{Aaij:2013zfa}
{\bf LHCb collaboration} Collaboration, R.~Aaij {\em et.~al.}, {\it
  {Measurement of the CKM angle gamma from a combination of $B\to Dh$
  analyses}},  \href{http://xxx.lanl.gov/abs/1305.2050}{{\tt arXiv:1305.2050}}.

\bibitem{Buras:2011wi}
A.~J. Buras, M.~V. Carlucci, L.~Merlo, and E.~Stamou, {\it {Phenomenology of a
  Gauged $SU(3)^3$ Flavour Model}},  {\em JHEP} {\bf 1203} (2012) 088,
  [\href{http://xxx.lanl.gov/abs/1112.4477}{{\tt arXiv:1112.4477}}].

\bibitem{Buras:1985yx}
A.~J. Buras and J.-M. G\'erard, {\it {$1/N$ Expansion for Kaons}},  {\em
  Nucl.Phys.} {\bf B264} (1986) 371.

\bibitem{Bardeen:1987vg}
W.~A. Bardeen, A.~J. Buras, and J.-M. G\'erard, {\it {The B Parameter Beyond
  the Leading Order of 1/N Expansion}},  {\em Phys.Lett.} {\bf B211} (1988)
  343.

\bibitem{Gerard:2010jt}
J.-M. G\'erard, {\it {An upper bound on the Kaon B-parameter and ${\rm
  Re}(\epsilon_K)$}},  {\em JHEP} {\bf 1102} (2011) 075,
  [\href{http://xxx.lanl.gov/abs/1012.2026}{{\tt arXiv:1012.2026}}].

\bibitem{Buras:2014maa}
A.~J. Buras, J.-M. Gerard, and W.~A. Bardeen, {\it {Large $N$ Approach to Kaon
  Decays and Mixing 28 Years Later: $\Delta I = 1/2$ Rule, $\hat B_K$ and
  $\Delta M_K$}},  \href{http://xxx.lanl.gov/abs/1401.1385}{{\tt
  arXiv:1401.1385}}.

\bibitem{Buras:2010zm}
A.~J. Buras, G.~Isidori, and P.~Paradisi, {\it {EDMs versus CPV in $B_{s,d}$
  mixing in two Higgs doublet models with MFV}},  {\em Phys.Lett.} {\bf B694}
  (2011) 402--409, [\href{http://xxx.lanl.gov/abs/1007.5291}{{\tt
  arXiv:1007.5291}}].

\bibitem{Buras:2012xxx}
A.~J. Buras, J.~Girrbach, and M.~Nagai, ``{${\rm 2HDM_{\overline{MFV}}}$ Facing
  Recent LHCb Data}.'' {\it, unpublished}.

\bibitem{Guadagnoli:2013mru}
D.~Guadagnoli and G.~Isidori, {\it {BR($B_s\to \mu^+\mu^-$) as an electroweak
  precision test}},  {\em Phys.Lett.} {\bf B724} (2013) 63--67,
  [\href{http://xxx.lanl.gov/abs/1302.3909}{{\tt arXiv:1302.3909}}].

\bibitem{Altmannshofer:2013oia}
W.~Altmannshofer, {\it {The $B_s$ → $µ^+ µ^−$ and $B_d$ → $µ^+ µ^−$
  Decays: Standard Model and Beyond}},  {\em PoS} {\bf Beauty2013} (2013) 024,
  [\href{http://xxx.lanl.gov/abs/1306.0022}{{\tt arXiv:1306.0022}}].

\bibitem{Isidori:2002qe}
G.~Isidori and A.~Retico, {\it {$B_{s,d}\to \ell^+ \ell^-$ and $K_L\to \ell^+
  \ell^-$ in SUSY models with non-minimal sources of flavour mixing}},  {\em
  JHEP} {\bf 09} (2002) 063,
  [\href{http://xxx.lanl.gov/abs/hep-ph/0208159}{{\tt hep-ph/0208159}}].

\bibitem{Bobeth:2001jm}
C.~Bobeth, A.~J. Buras, F.~Kruger, and J.~Urban, {\it Qcd corrections to $\bar
  b \to x_{d,s}\nu\bar\nu$, $\bar b_{d,s}\to\ell^+\ell^-$, $k \to\pi\nu\bar\nu$
  and $k_l\to\mu^+\mu^-$ in the mssm},  {\em Nucl. Phys.} {\bf B630} (2002)
  87--131, [\href{http://xxx.lanl.gov/abs/hep-ph/0112305}{{\tt
  hep-ph/0112305}}].

\bibitem{Dedes:2008iw}
A.~Dedes, J.~Rosiek, and P.~Tanedo, {\it {Complete One-Loop MSSM Predictions
  for $B\to$ lepton lepton' at the Tevatron and LHC}},  {\em Phys.Rev.} {\bf
  D79} (2009) 055006, [\href{http://xxx.lanl.gov/abs/0812.4320}{{\tt
  arXiv:0812.4320}}].

\bibitem{DescotesGenon:2011pb}
S.~Descotes-Genon, J.~Matias, and J.~Virto, {\it {An analysis of $B_{d,s}$
  mixing angles in presence of New Physics and an update of $B_s \to K^{0*}
  \bar K^{0*}$}},  {\em Phys.Rev.} {\bf D85} (2012) 034010,
  [\href{http://xxx.lanl.gov/abs/1111.4882}{{\tt arXiv:1111.4882}}].

\bibitem{deBruyn:2012wj}
K.~De~Bruyn, R.~Fleischer, R.~Knegjens, P.~Koppenburg, M.~Merk, {\em et.~al.},
  {\it {Branching Ratio Measurements of $B_s$ Decays}},  {\em Phys.Rev.} {\bf
  D86} (2012) 014027, [\href{http://xxx.lanl.gov/abs/1204.1735}{{\tt
  arXiv:1204.1735}}].

\bibitem{deBruyn:2012wk}
K.~De~Bruyn, R.~Fleischer, R.~Knegjens, P.~Koppenburg, M.~Merk, {\em et.~al.},
  {\it {Probing New Physics via the $B^0_s\to \mu^+\mu^-$ Effective Lifetime}},
   {\em Phys.Rev.Lett.} {\bf 109} (2012) 041801,
  [\href{http://xxx.lanl.gov/abs/1204.1737}{{\tt arXiv:1204.1737}}].

\bibitem{Fleischer:2012fy}
R.~Fleischer, {\it {On Branching Ratios of $B_s$ Decays and the Search for New
  Physics in $B^0_s\to \mu^+\mu^-$}},  {\em Nucl.Phys.Proc.Suppl.} {\bf
  241-242} (2013) 135--140, [\href{http://xxx.lanl.gov/abs/1208.2843}{{\tt
  arXiv:1208.2843}}].

\bibitem{Buchalla:1998ba}
G.~Buchalla and A.~J. Buras, {\it {The rare decays $K\to\pi \nu\bar\nu$, $B\to
  X \nu\bar\nu$ and $B\to \ell^+\ell^-$: An Update}},  {\em Nucl.Phys.} {\bf
  B548} (1999) 309--327, [\href{http://xxx.lanl.gov/abs/hep-ph/9901288}{{\tt
  hep-ph/9901288}}].

\bibitem{Misiak:1999yg}
M.~Misiak and J.~Urban, {\it {QCD corrections to FCNC decays mediated by Z
  penguins and W boxes}},  {\em Phys.Lett.} {\bf B451} (1999) 161--169,
  [\href{http://xxx.lanl.gov/abs/hep-ph/9901278}{{\tt hep-ph/9901278}}].

\bibitem{Buchalla:1997kz}
G.~Buchalla and A.~J. Buras, {\it Two-loop large-$m_t$ electroweak corrections
  to $k \to\pi\nu\bar\nu$ for arbitrary higgs boson mass},  {\em Phys. Rev.}
  {\bf D57} (1998) 216--223,
  [\href{http://xxx.lanl.gov/abs/hep-ph/9707243}{{\tt hep-ph/9707243}}].

\bibitem{Bobeth:2003at}
C.~Bobeth, P.~Gambino, M.~Gorbahn, and U.~Haisch, {\it {Complete NNLO QCD
  analysis of $\bar B \to X_s \ell^+ \ell^-$ and higher order electroweak
  effects}},  {\em JHEP} {\bf 04} (2004) 071,
  [\href{http://xxx.lanl.gov/abs/hep-ph/0312090}{{\tt hep-ph/0312090}}].

\bibitem{Huber:2005ig}
T.~Huber, E.~Lunghi, M.~Misiak, and D.~Wyler, {\it {Electromagnetic logarithms
  in $\bar B\to X(s) l^+ l^-$}},  {\em Nucl.Phys.} {\bf B740} (2006) 105--137,
  [\href{http://xxx.lanl.gov/abs/hep-ph/0512066}{{\tt hep-ph/0512066}}].

\bibitem{Misiak:2011bf}
M.~Misiak, {\it {Rare $B$-Meson Decays}},
  \href{http://xxx.lanl.gov/abs/1112.5978}{{\tt arXiv:1112.5978}}.

\bibitem{Buras:2012ru}
A.~J. Buras, J.~Girrbach, D.~Guadagnoli, and G.~Isidori, {\it {On the Standard
  Model prediction for BR($B_{s,d} \to \mu^+\mu^-)$}},  {\em Eur.Phys.J.} {\bf
  C72} (2012) 2172, [\href{http://xxx.lanl.gov/abs/1208.0934}{{\tt
  arXiv:1208.0934}}].

\bibitem{Bobeth:2013tba}
C.~Bobeth, M.~Gorbahn, and E.~Stamou, {\it {Electroweak Corrections to $B_{s,d}
  \to \ell^+ \ell^-$}},  \href{http://xxx.lanl.gov/abs/1311.1348}{{\tt
  arXiv:1311.1348}}.

\bibitem{Hermann:2013kca}
T.~Hermann, M.~Misiak, and M.~Steinhauser, {\it {Three-loop QCD corrections to
  $B_{s}\to \ell^+ \ell^-$}},  \href{http://xxx.lanl.gov/abs/1311.1347}{{\tt
  arXiv:1311.1347}}.

\bibitem{Bobeth:2013uxa}
C.~Bobeth, M.~Gorbahn, T.~Hermann, M.~Misiak, E.~Stamou, {\em et.~al.}, {\it
  {$B_{s,d}\to \ell^+ \ell^-$ in the Standard Model}},
  \href{http://xxx.lanl.gov/abs/1311.0903}{{\tt arXiv:1311.0903}}.

\bibitem{Brod:2010hi}
J.~Brod, M.~Gorbahn, and E.~Stamou, {\it {Two-Loop Electroweak Corrections for
  the $K \to \pi \nu \bar{nu}$ Decays}},  {\em Phys.Rev.} {\bf D83} (2011)
  034030, [\href{http://xxx.lanl.gov/abs/1009.0947}{{\tt arXiv:1009.0947}}].

\bibitem{Aaij:2013aka}
{\bf LHCb collaboration} Collaboration, R.~Aaij {\em et.~al.}, {\it
  {Measurement of the $B^0_s \to \mu^+ \mu^-$ branching fraction and search for
  $B^0 \to \mu^+ \mu^-$ decays at the LHCb experiment}},  {\em Phys.Rev.Lett.}
  {\bf 111} (2013) 101805, [\href{http://xxx.lanl.gov/abs/1307.5024}{{\tt
  arXiv:1307.5024}}].

\bibitem{Chatrchyan:2013bka}
{\bf CMS Collaboration} Collaboration, S.~Chatrchyan {\em et.~al.}, {\it
  {Measurement of the $B_s \to \mu\mu$ branching fraction and search for $B_0
  \to \mu\mu$ with the CMS Experiment}},  {\em Phys.Rev.Lett.} {\bf 111} (2013)
  101804, [\href{http://xxx.lanl.gov/abs/1307.5025}{{\tt arXiv:1307.5025}}].

\bibitem{CMS-PAS-BPH-13-007}
{\it Combination of results on the rare decays $b\to \mu^+\mu^-$ from the cms
  and lhcb experiments},  Tech. Rep. CMS-PAS-BPH-13-007, CERN, Geneva, 2013.

\bibitem{Logan:2000iv}
H.~E. Logan and U.~Nierste, {\it {$B_{s,d} \to \ell^+ \ell^-$ in a two Higgs
  doublet model}},  {\em Nucl.Phys.} {\bf B586} (2000) 39--55,
  [\href{http://xxx.lanl.gov/abs/hep-ph/0004139}{{\tt hep-ph/0004139}}].

\bibitem{Altmannshofer:2012ks}
W.~Altmannshofer, M.~Carena, N.~R. Shah, and F.~Yu, {\it {Indirect Probes of
  the MSSM after the Higgs Discovery}},  {\em JHEP} {\bf 1301} (2013) 160,
  [\href{http://xxx.lanl.gov/abs/1211.1976}{{\tt arXiv:1211.1976}}].

\bibitem{Buras:2009ka}
A.~J. Buras, B.~Duling, and S.~Gori, {\it {The Impact of Kaluza-Klein Fermions
  on Standard Model Fermion Couplings in a RS Model with Custodial
  Protection}},  {\em JHEP} {\bf 0909} (2009) 076,
  [\href{http://xxx.lanl.gov/abs/0905.2318}{{\tt arXiv:0905.2318}}].

\bibitem{delAguila:2011yd}
F.~del Aguila, J.~de~Blas, P.~Langacker, and M.~Perez-Victoria, {\it {Impact of
  extra particles on indirect Z' limits}},  {\em Phys.Rev.} {\bf D84} (2011)
  015015, [\href{http://xxx.lanl.gov/abs/1104.5512}{{\tt arXiv:1104.5512}}].

\bibitem{Botella:2012ju}
F.~Botella, G.~Branco, and M.~Nebot, {\it {The Hunt for New Physics in the
  Flavour Sector with up vector-like quarks}},  {\em JHEP} {\bf 1212} (2012)
  040, [\href{http://xxx.lanl.gov/abs/1207.4440}{{\tt arXiv:1207.4440}}].

\bibitem{Beaujean:2012uj}
F.~Beaujean, C.~Bobeth, D.~van Dyk, and C.~Wacker, {\it {Bayesian Fit of
  Exclusive $b \to s \bar\ell\ell$ Decays: The Standard Model Operator Basis}},
   {\em JHEP} {\bf 1208} (2012) 030,
  [\href{http://xxx.lanl.gov/abs/1205.1838}{{\tt arXiv:1205.1838}}].

\bibitem{Drobnak:2011aa}
J.~Drobnak, S.~Fajfer, and J.~F. Kamenik, {\it {Probing anomalous tWb
  interactions with rare B decays}},  {\em Nucl.Phys.} {\bf B855} (2012)
  82--99, [\href{http://xxx.lanl.gov/abs/1109.2357}{{\tt arXiv:1109.2357}}].

\bibitem{Chanowitz:1999jj}
M.~S. Chanowitz, {\it {The $Z\to \bar b b$ decay asymmetry and flavor changing
  neutral currents}},  \href{http://xxx.lanl.gov/abs/hep-ph/9905478}{{\tt
  hep-ph/9905478}}.

\bibitem{Haisch:2007ia}
U.~Haisch and A.~Weiler, {\it {Determining the Sign of the $Z$ Penguin
  Amplitude}},  {\em Phys. Rev.} {\bf D76} (2007) 074027,
  [\href{http://xxx.lanl.gov/abs/0706.2054}{{\tt arXiv:0706.2054}}].

\bibitem{Dighe:2010nj}
A.~Dighe, A.~Kundu, and S.~Nandi, {\it {Enhanced $B_s - \bar{B}_s$ lifetime
  difference and anomalous like-sign dimuon charge asymmetry from new physics
  in $B_s\to \tau^+ \tau^-$}},  {\em Phys.Rev.} {\bf D82} (2010) 031502,
  [\href{http://xxx.lanl.gov/abs/1005.4051}{{\tt arXiv:1005.4051}}].

\bibitem{Bobeth:2011st}
C.~Bobeth and U.~Haisch, {\it {New Physics in $\Gamma_12^s$: ($\bar{s}
  b$)$(\bar{\tau} \tau)$ Operators}},  {\em Acta Phys.Polon.} {\bf B44} (2013)
  127--176, [\href{http://xxx.lanl.gov/abs/1109.1826}{{\tt arXiv:1109.1826}}].

\bibitem{Aubert:2007xj}
{\bf BABAR Collaboration} Collaboration, B.~Aubert {\em et.~al.}, {\it {A
  Search for $B^{+} \to \tau^{+} \nu$ with Hadronic $B$ tags}},  {\em
  Phys.Rev.} {\bf D77} (2008) 011107,
  [\href{http://xxx.lanl.gov/abs/0708.2260}{{\tt arXiv:0708.2260}}].

\bibitem{Ikado:2006un}
{\bf Belle Collaboration} Collaboration, K.~Ikado {\em et.~al.}, {\it {Evidence
  of the Purely Leptonic Decay $B^- \to\tau^-\bar\nu_\tau$}},  {\em
  Phys.Rev.Lett.} {\bf 97} (2006) 251802,
  [\href{http://xxx.lanl.gov/abs/hep-ex/0604018}{{\tt hep-ex/0604018}}].

\bibitem{Adachi:2012mm}
{\bf Belle Collaboration} Collaboration, I.~Adachi {\em et.~al.}, {\it
  {Measurement of $B^- \to \tau^- \bar{\nu}_\tau$ with a Hadronic Tagging
  Method Using the Full Data Sample of Belle}},  {\em Phys.Rev.Lett.} {\bf 110}
  (2013) 131801, [\href{http://xxx.lanl.gov/abs/1208.4678}{{\tt
  arXiv:1208.4678}}].

\bibitem{Bona:2009cj}
{\bf UTfit Collaboration} Collaboration, M.~Bona {\em et.~al.}, {\it {An
  Improved Standard Model Prediction Of $BR(B\to \tau \nu)$ And Its
  Implications For New Physics}},  {\em Phys.Lett.} {\bf B687} (2010) 61--69,
  [\href{http://xxx.lanl.gov/abs/0908.3470}{{\tt arXiv:0908.3470}}].

\bibitem{Hou:1992sy}
W.-S. Hou, {\it {Enhanced charged Higgs boson effects in $B^-\to\tau \bar\nu,
  \mu \bar\nu$ and $b\to tau\bar\nu +X$}},  {\em Phys.Rev.} {\bf D48} (1993)
  2342--2344.

\bibitem{Akeroyd:2003zr}
A.~Akeroyd and S.~Recksiegel, {\it {The Effect of $H^\pm$ on $B^\pm\to\tau^\pm
  \nu_\tau$ and $B^\pm\to \mu^\pm\nu_\mu$}},  {\em J.Phys.G} {\bf G29} (2003)
  2311--2317, [\href{http://xxx.lanl.gov/abs/hep-ph/0306037}{{\tt
  hep-ph/0306037}}].

\bibitem{Isidori:2006pk}
G.~Isidori and P.~Paradisi, {\it {Hints of large tan(beta) in flavour
  physics}},  {\em Phys. Lett.} {\bf B639} (2006) 499--507,
  [\href{http://xxx.lanl.gov/abs/hep-ph/0605012}{{\tt hep-ph/0605012}}].

\bibitem{Blankenburg:2011ca}
G.~Blankenburg and G.~Isidori, {\it {$B \to \tau \nu$ in two-Higgs doublet
  models with MFV}},  {\em Eur.Phys.J.Plus} {\bf 127} (2012) 85,
  [\href{http://xxx.lanl.gov/abs/1107.1216}{{\tt arXiv:1107.1216}}].

\bibitem{Antonelli:2008jg}
{\bf FlaviaNet Working Group on Kaon Decays} Collaboration, M.~Antonelli {\em
  et.~al.}, {\it {Precision tests of the Standard Model with leptonic and
  semileptonic kaon decays}},  \href{http://xxx.lanl.gov/abs/0801.1817}{{\tt
  arXiv:0801.1817}}. {Updates available on {\tt
  http://www.lnf.infn.it/wg/vus/}.}

\bibitem{Nierste:2008qe}
U.~Nierste, S.~Trine, and S.~Westhoff, {\it {Charged-Higgs effects in a new
  $B\to D\tau\nu$ differential decay distribution}},  {\em Phys.Rev.} {\bf D78}
  (2008) 015006, [\href{http://xxx.lanl.gov/abs/0801.4938}{{\tt
  arXiv:0801.4938}}].

\bibitem{Lees:2012xj}
{\bf BaBar Collaboration} Collaboration, J.~Lees {\em et.~al.}, {\it {Evidence
  for an excess of $\bar{B} \to D^{(*)} \tau^-\bar{\nu}_\tau$ decays}},  {\em
  Phys.Rev.Lett.} {\bf 109} (2012) 101802,
  [\href{http://xxx.lanl.gov/abs/1205.5442}{{\tt arXiv:1205.5442}}].

\bibitem{Fajfer:2012vx}
S.~Fajfer, J.~F. Kamenik, and I.~Nisandzic, {\it {On the $B \to D^* \tau \bar
  \nu_{\tau}$ Sensitivity to New Physics}},  {\em Phys.Rev.} {\bf D85} (2012)
  094025, [\href{http://xxx.lanl.gov/abs/1203.2654}{{\tt arXiv:1203.2654}}].

\bibitem{Kamenik:2008tj}
J.~F. Kamenik and F.~Mescia, {\it {$B\to D\tau\nu$ Branching Ratios:
  Opportunity for Lattice QCD and Hadron Colliders}},  {\em Phys.Rev.} {\bf
  D78} (2008) 014003, [\href{http://xxx.lanl.gov/abs/0802.3790}{{\tt
  arXiv:0802.3790}}].

\bibitem{Crivellin:2012ye}
A.~Crivellin, C.~Greub, and A.~Kokulu, {\it {Explaining $B\to D\tau\nu$, $B\to
  D^*\tau\nu$ and $B\to \tau\nu$ in a 2HDM of type III}},  {\em Phys.Rev.} {\bf
  D86} (2012) 054014, [\href{http://xxx.lanl.gov/abs/1206.2634}{{\tt
  arXiv:1206.2634}}].

\bibitem{Crivellin:2013wna}
A.~Crivellin, A.~Kokulu, and C.~Greub, {\it {Flavor-phenomenology of
  two-Higgs-doublet models with generic Yukawa structure}},  {\em Phys.Rev.}
  {\bf D87} (2013) 094031, [\href{http://xxx.lanl.gov/abs/1303.5877}{{\tt
  arXiv:1303.5877}}].

\bibitem{Fajfer:2012jt}
S.~Fajfer, J.~F. Kamenik, I.~Nisandzic, and J.~Zupan, {\it {Implications of
  Lepton Flavor Universality Violations in B Decays}},  {\em Phys.Rev.Lett.}
  {\bf 109} (2012) 161801, [\href{http://xxx.lanl.gov/abs/1206.1872}{{\tt
  arXiv:1206.1872}}].

\bibitem{Ko:2012sv}
P.~Ko, Y.~Omura, and C.~Yu, {\it {$B\to D^(*) \tau \nu$ and $B\to \tau \nu$ in
  chiral U(1)' models with flavored multi Higgs doublets}},  {\em JHEP} {\bf
  1303} (2013) 151, [\href{http://xxx.lanl.gov/abs/1212.4607}{{\tt
  arXiv:1212.4607}}].

\bibitem{Crivellin:2013mba}
A.~Crivellin, C.~Greub, and A.~Kokulu, {\it {Flavour-violation in
  two-Higgs-doublet models}},  \href{http://xxx.lanl.gov/abs/1309.4806}{{\tt
  arXiv:1309.4806}}.

\bibitem{Kaminski:2012eb}
M.~Kaminski, M.~Misiak, and M.~Poradzinski, {\it {Tree-level contributions to
  $B\to X_s \gamma$}},  {\em Phys.Rev.} {\bf D86} (2012) 094004,
  [\href{http://xxx.lanl.gov/abs/1209.0965}{{\tt arXiv:1209.0965}}].

\bibitem{Misiak:2006ab}
M.~Misiak and M.~Steinhauser, {\it {NNLO QCD corrections to the $\bar B \to
  X(s) \gamma$ matrix elements using interpolation in $m_c$}},  {\em
  Nucl.Phys.} {\bf B764} (2007) 62--82,
  [\href{http://xxx.lanl.gov/abs/hep-ph/0609241}{{\tt hep-ph/0609241}}].

\bibitem{Czakon:2006ss}
M.~Czakon, U.~Haisch, and M.~Misiak, {\it {Four-Loop Anomalous Dimensions for
  Radiative Flavour-Changing Decays}},  {\em JHEP} {\bf 0703} (2007) 008,
  [\href{http://xxx.lanl.gov/abs/hep-ph/0612329}{{\tt hep-ph/0612329}}].

\bibitem{Boughezal:2007ny}
R.~Boughezal, M.~Czakon, and T.~Schutzmeier, {\it {NNLO fermionic corrections
  to the charm quark mass dependent matrix elements in $\bar B\to X_s
  \gamma$}},  {\em JHEP} {\bf 0709} (2007) 072,
  [\href{http://xxx.lanl.gov/abs/0707.3090}{{\tt arXiv:0707.3090}}].

\bibitem{Asatrian:2006rq}
H.~Asatrian, T.~Ewerth, H.~Gabrielyan, and C.~Greub, {\it {Charm quark mass
  dependence of the electromagnetic dipole operator contribution to $\bar B\to
  X_s \gamma$ at $O(\alpha^2_s)$}},  {\em Phys.Lett.} {\bf B647} (2007)
  173--178, [\href{http://xxx.lanl.gov/abs/hep-ph/0611123}{{\tt
  hep-ph/0611123}}].

\bibitem{Ewerth:2008nv}
T.~Ewerth, {\it {Fermionic corrections to the interference of the electro- and
  chromomagnetic dipole operators in $\bar B\to X_s \gamma$ at
  $O(\alpha^2_s)$}},  {\em Phys.Lett.} {\bf B669} (2008) 167--172,
  [\href{http://xxx.lanl.gov/abs/0805.3911}{{\tt arXiv:0805.3911}}].

\bibitem{Asatrian:2010rq}
H.~Asatrian, T.~Ewerth, A.~Ferroglia, C.~Greub, and G.~Ossola, {\it {Complete
  $(O_7,O_8)$ contribution to $ B\to X_s \gamma$ at order $\alpha^2_s$}},  {\em
  Phys.Rev.} {\bf D82} (2010) 074006,
  [\href{http://xxx.lanl.gov/abs/1005.5587}{{\tt arXiv:1005.5587}}].

\bibitem{Ferroglia:2010xe}
A.~Ferroglia and U.~Haisch, {\it {Chromomagnetic Dipole-Operator Corrections in
  $\bar{B} \to X_{s\gamma}$ at $O(\beta_0 \alpha_s^2)$}},  {\em Phys.Rev.} {\bf
  D82} (2010) 094012, [\href{http://xxx.lanl.gov/abs/1009.2144}{{\tt
  arXiv:1009.2144}}].

\bibitem{Misiak:2010tk}
M.~Misiak and M.~Poradzinski, {\it {Completing the Calculation of BLM
  corrections to $\bar B \to X_s \gamma$}},  {\em Phys.Rev.} {\bf D83} (2011)
  014024, [\href{http://xxx.lanl.gov/abs/1009.5685}{{\tt arXiv:1009.5685}}].

\bibitem{Misiaketal}
M.~Czakon, P.~Fiedler, T.~Huber, M.~Misiak, T.~Schutzmeier, and M.~Steinhauser,
  ``{to be published}.'' {2013}.

\bibitem{Soares:1991te}
J.~M. Soares, {\it {CP violation in radiative b decays}},  {\em Nucl.Phys.}
  {\bf B367} (1991) 575--590.

\bibitem{Kagan:1998bh}
A.~L. Kagan and M.~Neubert, {\it {Direct CP violation in $B\to X_s \gamma$
  decays as a signature of new physics}},  {\em Phys.Rev.} {\bf D58} (1998)
  094012, [\href{http://xxx.lanl.gov/abs/hep-ph/9803368}{{\tt
  hep-ph/9803368}}].

\bibitem{Kagan:1998ym}
A.~L. Kagan and M.~Neubert, {\it {QCD anatomy of $B\to X_s\gamma$ decays}},
  {\em Eur.Phys.J.} {\bf C7} (1999) 5--27,
  [\href{http://xxx.lanl.gov/abs/hep-ph/9805303}{{\tt hep-ph/9805303}}].

\bibitem{Hurth:2003dk}
T.~Hurth, E.~Lunghi, and W.~Porod, {\it {Untagged $\bar B\to X_{s+d} \gamma$ CP
  asymmetry as a probe for new physics}},  {\em Nucl.Phys.} {\bf B704} (2005)
  56--74, [\href{http://xxx.lanl.gov/abs/hep-ph/0312260}{{\tt
  hep-ph/0312260}}].

\bibitem{Benzke:2010tq}
M.~Benzke, S.~J. Lee, M.~Neubert, and G.~Paz, {\it {Long-Distance Dominance of
  the CP Asymmetry in $B\to X_{s,d}\gamma$ Decays}},  {\em Phys.Rev.Lett.} {\bf
  106} (2011) 141801, [\href{http://xxx.lanl.gov/abs/1012.3167}{{\tt
  arXiv:1012.3167}}].

\bibitem{Bosch:2001gv}
S.~W. Bosch and G.~Buchalla, {\it {The Radiative decays $B\to V \gamma$ at
  next-to-leading order in QCD}},  {\em Nucl.Phys.} {\bf B621} (2002) 459--478,
  [\href{http://xxx.lanl.gov/abs/hep-ph/0106081}{{\tt hep-ph/0106081}}].

\bibitem{Bosch:2004nd}
S.~W. Bosch and G.~Buchalla, {\it {Constraining the unitarity triangle with
  $B\to V\gamma$ }},  {\em JHEP} {\bf 0501} (2005) 035,
  [\href{http://xxx.lanl.gov/abs/hep-ph/0408231}{{\tt hep-ph/0408231}}].

\bibitem{Atwood:1997zr}
D.~Atwood, M.~Gronau, and A.~Soni, {\it {Mixing induced CP asymmetries in
  radiative B decays in and beyond the standard model}},  {\em Phys.Rev.Lett.}
  {\bf 79} (1997) 185--188, [\href{http://xxx.lanl.gov/abs/hep-ph/9704272}{{\tt
  hep-ph/9704272}}].

\bibitem{Ball:2006cva}
P.~Ball and R.~Zwicky, {\it {Time-dependent CP Asymmetry in $B\to K^* \gamma$
  as a (Quasi) Null Test of the Standard Model}},  {\em Phys.Lett.} {\bf B642}
  (2006) 478--486, [\href{http://xxx.lanl.gov/abs/hep-ph/0609037}{{\tt
  hep-ph/0609037}}].

\bibitem{Ball:2006eu}
P.~Ball, G.~W. Jones, and R.~Zwicky, {\it {$B\to V \gamma$ beyond QCD
  factorisation}},  {\em Phys.Rev.} {\bf D75} (2007) 054004,
  [\href{http://xxx.lanl.gov/abs/hep-ph/0612081}{{\tt hep-ph/0612081}}].

\bibitem{Ushiroda:2006fi}
{\bf Belle Collaboration} Collaboration, Y.~Ushiroda {\em et.~al.}, {\it
  {Time-Dependent CP Asymmetries in $B^0\to K^0_s \pi^0 \gamma$ transitions}},
  {\em Phys.Rev.} {\bf D74} (2006) 111104,
  [\href{http://xxx.lanl.gov/abs/hep-ex/0608017}{{\tt hep-ex/0608017}}].

\bibitem{Aubert:2008gy}
{\bf BABAR Collaboration} Collaboration, B.~Aubert {\em et.~al.}, {\it
  {Measurement of Time-Dependent CP Asymmetry in $B^0 \to K^0_{S} \pi^0 \gamma$
  Decays}},  {\em Phys.Rev.} {\bf D78} (2008) 071102,
  [\href{http://xxx.lanl.gov/abs/0807.3103}{{\tt arXiv:0807.3103}}].

\bibitem{Asner:2010qj}
{\bf Heavy Flavor Averaging Group} Collaboration, D.~Asner {\em et.~al.}, {\it
  {Averages of $b$-hadron, $c$-hadron, and $\tau$-lepton Properties}},
  \href{http://xxx.lanl.gov/abs/1010.1589}{{\tt arXiv:1010.1589}}. Long author
  list - awaiting processing.

\bibitem{Lyon:2013gba}
J.~Lyon and R.~Zwicky, {\it {Isospin asymmetries in $B\to (K^*,\rho) \gamma/
  l^+ l^-$ and $B \to K l^+ l^-$ in and beyond the Standard Model}},  {\em
  Phys.Rev.} {\bf D88} (2013) 094004,
  [\href{http://xxx.lanl.gov/abs/1305.4797}{{\tt arXiv:1305.4797}}].

\bibitem{Haisch:2008ar}
U.~Haisch, {\it {$\bar{B} \to X_{s} \gamma$: Standard Model and Beyond}},
  \href{http://xxx.lanl.gov/abs/0805.2141}{{\tt arXiv:0805.2141}}.

\bibitem{Grzadkowski:2008mf}
B.~Grzadkowski and M.~Misiak, {\it {Anomalous Wtb coupling effects in the weak
  radiative B-meson decay}},  {\em Phys.Rev.} {\bf D78} (2008) 077501,
  [\href{http://xxx.lanl.gov/abs/0802.1413}{{\tt arXiv:0802.1413}}].

\bibitem{Altmannshofer:2008vr}
W.~Altmannshofer, D.~Guadagnoli, S.~Raby, and D.~M. Straub, {\it {SUSY GUTs
  with Yukawa unification: A Go/no-go study using FCNC processes}},  {\em Phys.
  Lett.} {\bf B668} (2008) 385--391,
  [\href{http://xxx.lanl.gov/abs/0801.4363}{{\tt arXiv:0801.4363}}].

\bibitem{Agashe:2001xt}
K.~Agashe, N.~G. Deshpande, and G.~H. Wu, {\it Universal extra dimensions and
  $b \to s \gamma$},  {\em Phys. Lett.} {\bf B514} (2001) 309--314,
  [\href{http://xxx.lanl.gov/abs/hep-ph/0105084}{{\tt hep-ph/0105084}}].

\bibitem{Buras:2003mk}
A.~J. Buras, A.~Poschenrieder, M.~Spranger, and A.~Weiler, {\it The impact of
  universal extra dimensions on $b \to x_s \gamma$, $b \to x_s$gluon, $b \to
  x_s \mu^+ \mu^-$, $k_l \to\pi^0 e^+ e^-$, and $\varepsilon'/\varepsilon$},
  {\em Nucl. Phys.} {\bf B678} (2004) 455--490,
  [\href{http://xxx.lanl.gov/abs/hep-ph/0306158}{{\tt hep-ph/0306158}}].

\bibitem{Haisch:2007vb}
U.~Haisch and A.~Weiler, {\it {Bound on minimal universal extra dimensions from
  $\bar{B}\to X_s\gamma$}},  {\em Phys.Rev.} {\bf D76} (2007) 034014,
  [\href{http://xxx.lanl.gov/abs/hep-ph/0703064}{{\tt hep-ph/0703064}}].

\bibitem{Freitas:2008vh}
A.~Freitas and U.~Haisch, {\it {$\bar{B}\to X_s \gamma$ in two universal extra
  dimensions}},  {\em Phys.Rev.} {\bf D77} (2008) 093008,
  [\href{http://xxx.lanl.gov/abs/0801.4346}{{\tt arXiv:0801.4346}}].

\bibitem{Asatrian:1989iu}
G.~Asatrian and A.~Ionnisian, {\it {RARE B MESON DECAYS IN $SU(2)_L \times
  SU(2)_R \times U(1)$ MODEL}},  {\em Mod.Phys.Lett.} {\bf A5} (1990)
  1089--1096.

\bibitem{Asatryan:1990na}
G.~Asatryan and A.~Ioannisyan, {\it {The $b\to s \gamma$ decay in $SU(2)_L
  \times SU(2)_R \times U(1)$ model. (In Russian)}},  {\em Sov.J.Nucl.Phys.}
  {\bf 51} (1990) 858--860.

\bibitem{Cocolicchio:1988ac}
D.~Cocolicchio, G.~Costa, G.~L. Fogli, J.~Kim, and A.~Masiero, {\it {RARE $B$
  DECAYS IN LEFT-RIGHT SYMMETRIC MODELS}},  {\em Phys.Rev.} {\bf D40} (1989)
  1477.

\bibitem{Cho:1993zb}
P.~L. Cho and M.~Misiak, {\it {$b\to s \gamma$ decay in $SU(2)_L \times SU(2)_R
  \times U(1)$ extensions of the Standard Model}},  {\em Phys.Rev.} {\bf D49}
  (1994) 5894--5903, [\href{http://xxx.lanl.gov/abs/hep-ph/9310332}{{\tt
  hep-ph/9310332}}].

\bibitem{Babu:1993hx}
K.~Babu, K.~Fujikawa, and A.~Yamada, {\it {Constraints on left-right symmetric
  models from the process $b\to s \gamma$}},  {\em Phys.Lett.} {\bf B333}
  (1994) 196--201, [\href{http://xxx.lanl.gov/abs/hep-ph/9312315}{{\tt
  hep-ph/9312315}}].

\bibitem{Fujikawa:1993zu}
K.~Fujikawa and A.~Yamada, {\it {Test of the chiral structure of the top -
  bottom charged current by the process $b\to s \gamma$}},  {\em Phys.Rev.}
  {\bf D49} (1994) 5890--5893.

\bibitem{Asatrian:1996as}
G.~Asatrian and A.~Ioannisian, {\it {CP violation in the decay $b\to s \gamma$
  in the left-right symmetric model}},  {\em Phys.Rev.} {\bf D54} (1996)
  5642--5646, [\href{http://xxx.lanl.gov/abs/hep-ph/9603318}{{\tt
  hep-ph/9603318}}].

\bibitem{Bobeth:1999ww}
C.~Bobeth, M.~Misiak, and J.~Urban, {\it {Matching conditions for $b\to s
  \gamma$ and $b\to s g$ in extensions of the standard model}},  {\em
  Nucl.Phys.} {\bf B567} (2000) 153--185,
  [\href{http://xxx.lanl.gov/abs/hep-ph/9904413}{{\tt hep-ph/9904413}}].

\bibitem{Frank:2010qv}
M.~Frank, A.~Hayreter, and I.~Turan, {\it {B Decays in an Asymmetric Left-Right
  Model}},  {\em Phys.Rev.} {\bf D82} (2010) 033012,
  [\href{http://xxx.lanl.gov/abs/1005.3074}{{\tt arXiv:1005.3074}}].

\bibitem{Guadagnoli:2011id}
D.~Guadagnoli, R.~N. Mohapatra, and I.~Sung, {\it {Gauged Flavor Group with
  Left-Right Symmetry}},  {\em JHEP} {\bf 1104} (2011) 093,
  [\href{http://xxx.lanl.gov/abs/1103.4170}{{\tt arXiv:1103.4170}}].

\bibitem{Bobeth:2008ij}
C.~Bobeth, G.~Hiller, and G.~Piranishvili, {\it {CP Asymmetries in bar $B \to
  \bar{K}^* (\to \bar{K} \pi) \bar{\ell} \ell$ and Untagged $\bar{B}_s$, $B_s
  \to \phi (\to K^{+} K^-) \bar{\ell} \ell$ Decays at NLO}},  {\em JHEP} {\bf
  0807} (2008) 106, [\href{http://xxx.lanl.gov/abs/0805.2525}{{\tt
  arXiv:0805.2525}}].

\bibitem{Egede:2008uy}
U.~Egede, T.~Hurth, J.~Matias, M.~Ramon, and W.~Reece, {\it {New observables in
  the decay mode $\bar B_d\to\bar K^{*0} l^+ l^-$}},  {\em JHEP} {\bf 0811}
  (2008) 032, [\href{http://xxx.lanl.gov/abs/0807.2589}{{\tt
  arXiv:0807.2589}}].

\bibitem{Altmannshofer:2008dz}
W.~Altmannshofer, P.~Ball, A.~Bharucha, A.~J. Buras, D.~M. Straub, {\em
  et.~al.}, {\it {Symmetries and Asymmetries of $B \to K^{*} \mu^{+} \mu^{-}$
  Decays in the Standard Model and Beyond}},  {\em JHEP} {\bf 0901} (2009) 019,
  [\href{http://xxx.lanl.gov/abs/0811.1214}{{\tt arXiv:0811.1214}}].

\bibitem{Bobeth:2010wg}
C.~Bobeth, G.~Hiller, and D.~van Dyk, {\it {The Benefits of $\bar{B} \to
  \bar{K}^* l^+ l^-$ Decays at Low Recoil}},  {\em JHEP} {\bf 1007} (2010) 098,
  [\href{http://xxx.lanl.gov/abs/1006.5013}{{\tt arXiv:1006.5013}}].

\bibitem{Becirevic:2012fy}
D.~Becirevic, N.~Kosnik, F.~Mescia, and E.~Schneider, {\it {Complementarity of
  the constraints on New Physics from $B_s\to\mu^+\mu^-$ and from $B\to K
  \ell^+\ell^-$ decays}},  {\em Phys.Rev.} {\bf D86} (2012) 034034,
  [\href{http://xxx.lanl.gov/abs/1205.5811}{{\tt arXiv:1205.5811}}].

\bibitem{Bobeth:2011gi}
C.~Bobeth, G.~Hiller, and D.~van Dyk, {\it {More Benefits of Semileptonic Rare
  B Decays at Low Recoil: CP Violation}},  {\em JHEP} {\bf 1107} (2011) 067,
  [\href{http://xxx.lanl.gov/abs/1105.0376}{{\tt arXiv:1105.0376}}].

\bibitem{DescotesGenon:2012zf}
S.~Descotes-Genon, J.~Matias, M.~Ramon, and J.~Virto, {\it {Implications from
  clean observables for the binned analysis of $B\to K*\mu^+\mu^-$ at large
  recoil}},  {\em JHEP} {\bf 1301} (2013) 048,
  [\href{http://xxx.lanl.gov/abs/1207.2753}{{\tt arXiv:1207.2753}}].

\bibitem{Jager:2012uw}
S.~Jager and J.~M. Camalich, {\it {On $B\to V \ell \ell$ at small dilepton
  invariant mass, power corrections, and new physics}},  {\em JHEP} {\bf 1305}
  (2013) 043, [\href{http://xxx.lanl.gov/abs/1212.2263}{{\tt
  arXiv:1212.2263}}].

\bibitem{Wei:2009zv}
{\bf BELLE Collaboration} Collaboration, J.-T. Wei {\em et.~al.}, {\it
  {Measurement of the Differential Branching Fraction and Forward-Backword
  Asymmetry for $B\to K^{(*)}l^+l^-$}},  {\em Phys.Rev.Lett.} {\bf 103} (2009)
  171801, [\href{http://xxx.lanl.gov/abs/0904.0770}{{\tt arXiv:0904.0770}}].

\bibitem{Aaltonen:2011ja}
{\bf CDF Collaboration} Collaboration, T.~Aaltonen {\em et.~al.}, {\it
  {Measurements of the Angular Distributions in the Decays $B \to K^{(*)} \mu^+
  \mu^-$ at CDF}},  {\em Phys.Rev.Lett.} {\bf 108} (2012) 081807,
  [\href{http://xxx.lanl.gov/abs/1108.0695}{{\tt arXiv:1108.0695}}].

\bibitem{Lees:2012tva}
{\bf BaBar Collaboration} Collaboration, J.~Lees {\em et.~al.}, {\it
  {Measurement of Branching Fractions and Rate Asymmetries in the Rare Decays
  $B \to K^{(*)} l^+ l^-$}},  {\em Phys.Rev.} {\bf D86} (2012) 032012,
  [\href{http://xxx.lanl.gov/abs/1204.3933}{{\tt arXiv:1204.3933}}].

\bibitem{Aaij:2011aa}
{\bf LHCb Collaboration} Collaboration, R.~Aaij {\em et.~al.}, {\it
  {Differential branching fraction and angular analysis of the decay $B^{0} \to
  K^{*0} \mu^+ \mu^-$}},  {\em Phys.Rev.Lett.} {\bf 108} (2012) 181806,
  [\href{http://xxx.lanl.gov/abs/1112.3515}{{\tt arXiv:1112.3515}}].

\bibitem{Aaij:2013iag}
{\bf LHCb Collaboration} Collaboration, R.~Aaij {\em et.~al.}, {\it
  {Differential branching fraction and angular analysis of the decay $B^{0} \to
  K^{*0} \mu^{+}\mu^{-}$}},  {\em JHEP} {\bf 1308} (2013) 131,
  [\href{http://xxx.lanl.gov/abs/1304.6325}{{\tt arXiv:1304.6325}}].

\bibitem{Descotes-Genon:2013vna}
S.~Descotes-Genon, T.~Hurth, J.~Matias, and J.~Virto, {\it {Optimizing the
  basis of ${B} \to {K}^{*}\ell^+ \ell^-$ observables in the full kinematic
  range}},  {\em JHEP} {\bf 1305} (2013) 137,
  [\href{http://xxx.lanl.gov/abs/1303.5794}{{\tt arXiv:1303.5794}}].

\bibitem{Descotes-Genon:2013hba}
S.~Descotes-Genon, T.~Hurth, J.~Matias, and J.~Virto, {\it {$B\to K^*\ell\ell$:
  The New Frontier of New Physics searches in Flavor}},
  \href{http://xxx.lanl.gov/abs/1305.4808}{{\tt arXiv:1305.4808}}.

\bibitem{Kruger:2005ep}
F.~Kruger and J.~Matias, {\it {Probing new physics via the transverse
  amplitudes of $B^0\to K*0 (K^- \pi^+) l^+l^-$ at large recoil}},  {\em
  Phys.Rev.} {\bf D71} (2005) 094009,
  [\href{http://xxx.lanl.gov/abs/hep-ph/0502060}{{\tt hep-ph/0502060}}].

\bibitem{Egede:2010zc}
U.~Egede, T.~Hurth, J.~Matias, M.~Ramon, and W.~Reece, {\it {New physics reach
  of the decay mode $\bar{B} \to \bar{K}^{*0}\ell^+\ell^-$}},  {\em JHEP} {\bf
  1010} (2010) 056, [\href{http://xxx.lanl.gov/abs/1005.0571}{{\tt
  arXiv:1005.0571}}].

\bibitem{Becirevic:2011bp}
D.~Becirevic and E.~Schneider, {\it {On transverse asymmetries in $B\to K^*
  l^+l^-$}},  {\em Nucl.Phys.} {\bf B854} (2012) 321--339,
  [\href{http://xxx.lanl.gov/abs/1106.3283}{{\tt arXiv:1106.3283}}].

\bibitem{Bobeth:2012vn}
C.~Bobeth, G.~Hiller, and D.~van Dyk, {\it {General Analysis of $\bar{B} \to
  \bar{K}^{(*)}\ell^+ \ell^-$ Decays at Low Recoil}},  {\em Phys.Rev.} {\bf
  D87} (2013) 034016, [\href{http://xxx.lanl.gov/abs/1212.2321}{{\tt
  arXiv:1212.2321}}].

\bibitem{Matias:2012xw}
J.~Matias, F.~Mescia, M.~Ramon, and J.~Virto, {\it {Complete Anatomy of
  $\bar{B}_d \to \bar{K}^{* 0} (\to; K \pi)l^+l^-$ and its angular
  distribution}},  {\em JHEP} {\bf 1204} (2012) 104,
  [\href{http://xxx.lanl.gov/abs/1202.4266}{{\tt arXiv:1202.4266}}].

\bibitem{Descotes-Genon:2013wba}
S.~Descotes-Genon, J.~Matias, and J.~Virto, {\it {Understanding the $B\to
  K^*\mu^+\mu^-$ Anomaly}},  {\em Phys. Rev. D 88,} {\bf 074002} (2013)
  [\href{http://xxx.lanl.gov/abs/1307.5683}{{\tt arXiv:1307.5683}}].

\bibitem{Aaij:2013qta}
{\bf LHCb collaboration} Collaboration, R.~Aaij {\em et.~al.}, {\it
  {Measurement of form-factor independent observables in the decay $B^{0} \to
  K^{*0} \mu^+ \mu^-$}},  {\em Phys.Rev.Lett.} {\bf 111} (2013) 191801,
  [\href{http://xxx.lanl.gov/abs/1308.1707}{{\tt arXiv:1308.1707}}].

\bibitem{Altmannshofer:2013foa}
W.~Altmannshofer and D.~M. Straub, {\it {New physics in $B \to
  K^*{\mu}{\mu}$?}},  \href{http://xxx.lanl.gov/abs/1308.1501}{{\tt
  arXiv:1308.1501}}.

\bibitem{Aaij:2012cq}
{\bf LHCb Collaboration} Collaboration, R.~Aaij {\em et.~al.}, {\it
  {Measurement of the isospin asymmetry in $B \to K^{(*)}\mu^+\mu^-$ decays}},
  {\em JHEP} {\bf 1207} (2012) 133,
  [\href{http://xxx.lanl.gov/abs/1205.3422}{{\tt arXiv:1205.3422}}].

\bibitem{Beylich:2011aq}
M.~Beylich, G.~Buchalla, and T.~Feldmann, {\it {Theory of $B\to K^{(*)}l^+l^-$
  decays at high $q^2$: OPE and quark-hadron duality}},  {\em Eur.Phys.J.} {\bf
  C71} (2011) 1635, [\href{http://xxx.lanl.gov/abs/1101.5118}{{\tt
  arXiv:1101.5118}}].

\bibitem{Horgan:2013hoa}
R.~R. Horgan, Z.~Liu, S.~Meinel, and M.~Wingate, {\it {Lattice QCD calculation
  of form factors describing the rare decays $B \to K^* \ell^+ \ell^-$ and $B_s
  \to \phi \ell^+ \ell^-$}},  {\em Phys.Rev.} {\bf D89} (2014) 094501,
  [\href{http://xxx.lanl.gov/abs/1310.3722}{{\tt arXiv:1310.3722}}].

\bibitem{Bouchard:2013mia}
C.~Bouchard, G.~P. Lepage, C.~Monahan, H.~Na, and J.~Shigemitsu, {\it {Standard
  Model predictions for $B\to Kll$ with form factors from lattice QCD}},  {\em
  Phys. Rev. Lett. 111,} {\bf 162002} (2013)
  [\href{http://xxx.lanl.gov/abs/1306.0434}{{\tt arXiv:1306.0434}}].

\bibitem{Bouchard:2013eph}
C.~Bouchard, G.~P. Lepage, C.~Monahan, H.~Na, and J.~Shigemitsu, {\it {Rare
  decay $B\to Kll$ form factors from lattice QCD}},  {\em Phys. Rev. D 88,}
  {\bf 054509} (2013) 054509, [\href{http://xxx.lanl.gov/abs/1306.2384}{{\tt
  arXiv:1306.2384}}].

\bibitem{Bobeth:1999mk}
C.~Bobeth, M.~Misiak, and J.~Urban, {\it {Photonic penguins at two loops and
  $m_t$-dependence of $BR(B\to X_s \ell^+ \ell^-)$}},  {\em Nucl. Phys.} {\bf
  B574} (2000) 291--330, [\href{http://xxx.lanl.gov/abs/hep-ph/9910220}{{\tt
  hep-ph/9910220}}].

\bibitem{Asatrian:2001de}
H.~Asatrian, H.~Asatrian, C.~Greub, and M.~Walker, {\it {Two loop virtual
  corrections to $B\to X(s) l^+ l^-$ in the standard model}},  {\em Phys.Lett.}
  {\bf B507} (2001) 162--172,
  [\href{http://xxx.lanl.gov/abs/hep-ph/0103087}{{\tt hep-ph/0103087}}].

\bibitem{Asatryan:2001zw}
H.~H. Asatryan, H.~M. Asatrian, C.~Greub, and M.~Walker, {\it {Calculation of
  two loop virtual corrections to $b\to s l^+ l^-$ in the standard model}},
  {\em Phys. Rev.} {\bf D65} (2002) 074004,
  [\href{http://xxx.lanl.gov/abs/hep-ph/0109140}{{\tt hep-ph/0109140}}].

\bibitem{Ghinculov:2003qd}
A.~Ghinculov, T.~Hurth, G.~Isidori, and Y.~P. Yao, {\it {The rare decay B -->
  X/s l+ l- to NNLL precision for arbitrary dilepton invariant mass}},  {\em
  Nucl. Phys.} {\bf B685} (2004) 351--392,
  [\href{http://xxx.lanl.gov/abs/hep-ph/0312128}{{\tt hep-ph/0312128}}].

\bibitem{Ligeti:2007sn}
Z.~Ligeti and F.~J. Tackmann, {\it {Precise predictions for $B\to X(s) l^+ l^-$
  in the large $q^{**2}$ region}},  {\em Phys.Lett.} {\bf B653} (2007)
  404--410, [\href{http://xxx.lanl.gov/abs/0707.1694}{{\tt arXiv:0707.1694}}].

\bibitem{Greub:2008cy}
C.~Greub, V.~Pilipp, and C.~Schupbach, {\it {Analytic calculation of two-loop
  QCD corrections to $b \to s l^+ l^-$ in the high $q^{**2}$ region}},  {\em
  JHEP} {\bf 0812} (2008) 040, [\href{http://xxx.lanl.gov/abs/0810.4077}{{\tt
  arXiv:0810.4077}}].

\bibitem{DescotesGenon:2011yn}
S.~Descotes-Genon, D.~Ghosh, J.~Matias, and M.~Ramon, {\it {Exploring New
  Physics in the $C_7-C_7^\prime$ plane}},  {\em JHEP} {\bf 1106} (2011) 099,
  [\href{http://xxx.lanl.gov/abs/1104.3342}{{\tt arXiv:1104.3342}}].

\bibitem{Hambrock:2013zya}
C.~Hambrock, G.~Hiller, S.~Schacht, and R.~Zwicky, {\it {$B\to K^*$ Form
  Factors from Flavor Data to QCD and Back}},
  \href{http://xxx.lanl.gov/abs/1308.4379}{{\tt arXiv:1308.4379}}.

\bibitem{Barbieri:2011vn}
R.~Barbieri, P.~Lodone, and D.~M. Straub, {\it {CP Violation in Supersymmetry
  with Effective Minimal Flavour Violation}},  {\em JHEP} {\bf 1105} (2011)
  049, [\href{http://xxx.lanl.gov/abs/1102.0726}{{\tt arXiv:1102.0726}}].

\bibitem{Chatrchyan:2013cda}
{\bf CMS Collaboration} Collaboration, S.~Chatrchyan {\em et.~al.}, {\it
  {Angular analysis and branching fraction measurement of the decay $B^0 \to
  K^{*0} \mu^+\mu^-$}},  {\em Phys.Lett.} {\bf B727} (2013) 77--100,
  [\href{http://xxx.lanl.gov/abs/1308.3409}{{\tt arXiv:1308.3409}}].

\bibitem{Gauld:2013qba}
R.~Gauld, F.~Goertz, and U.~Haisch, {\it {On minimal $Z'$ explanations of the
  $B\to K^*\mu^+\mu^-$ anomaly}},
  \href{http://xxx.lanl.gov/abs/1308.1959}{{\tt arXiv:1308.1959}}.

\bibitem{Gauld:2013qja}
R.~Gauld, F.~Goertz, and U.~Haisch, {\it {An explicit Z'-boson explanation of
  the $B \to K^* \mu^+ \mu^-$ anomaly}},  {\em JHEP} {\bf 1401} (2014) 069,
  [\href{http://xxx.lanl.gov/abs/1310.1082}{{\tt arXiv:1310.1082}}].

\bibitem{Beaujean:2013soa}
F.~Beaujean, C.~Bobeth, and D.~van Dyk, {\it {Comprehensive Bayesian Analysis
  of Rare (Semi)leptonic and Radiative B Decays}},
  \href{http://xxx.lanl.gov/abs/1310.2478}{{\tt arXiv:1310.2478}}.

\bibitem{Datta:2013kja}
A.~Datta, M.~Duraisamy, and D.~Ghosh, {\it {Explaining the $B \to K^\ast \mu^+
  \mu^-$ anomaly with scalar interactions}},
  \href{http://xxx.lanl.gov/abs/1310.1937}{{\tt arXiv:1310.1937}}.

\bibitem{Horgan:2013pva}
R.~R. Horgan, Z.~Liu, S.~Meinel, and M.~Wingate, {\it {Calculation of $B^0 \to
  K^{*0} \mu^+ \mu^-$ and $B_s^0 \to \phi \mu^+ \mu^-$ observables using form
  factors from lattice QCD}},  \href{http://xxx.lanl.gov/abs/1310.3887}{{\tt
  arXiv:1310.3887}}.

\bibitem{Richard:2013xfa}
F.~Richard, {\it {A $Z^\prime$ interpretation of $B_d\to K^*\mu^+\mu^-$ data
  and consequences for high energy colliders}},
  \href{http://xxx.lanl.gov/abs/1312.2467}{{\tt arXiv:1312.2467}}.

\bibitem{Hurth:2013ssa}
T.~Hurth and F.~Mahmoudi, {\it {On the LHCb anomaly in $B\to K^*l^+l^-$}},
  \href{http://xxx.lanl.gov/abs/1312.5267}{{\tt arXiv:1312.5267}}.

\bibitem{Khodjamirian:2010vf}
A.~Khodjamirian, T.~Mannel, A.~Pivovarov, and Y.-M. Wang, {\it {Charm-loop
  effect in $B \to K^{(*)} \ell^{+} \ell^{-}$ and $B\to K^*\gamma$}},  {\em
  JHEP} {\bf 1009} (2010) 089, [\href{http://xxx.lanl.gov/abs/1006.4945}{{\tt
  arXiv:1006.4945}}].

\bibitem{Matias:2012qz}
J.~Matias, {\it {On the S-wave pollution of $B\to K* l^+l-$ observables}},
  {\em Phys.Rev.} {\bf D86} (2012) 094024,
  [\href{http://xxx.lanl.gov/abs/1209.1525}{{\tt arXiv:1209.1525}}].

\bibitem{Straub:2013uoa}
D.~M. Straub, {\it {Constraints on new physics from rare (semi-)leptonic B
  decays}},  \href{http://xxx.lanl.gov/abs/1305.5704}{{\tt arXiv:1305.5704}}.

\bibitem{Buras:2005gr}
A.~J. Buras, M.~Gorbahn, U.~Haisch, and U.~Nierste, {\it {The rare decay $K^+
  \to \pi^+ \nu \bar\nu$ at the next-to-next-to-leading order in QCD}},  {\em
  Phys. Rev. Lett.} {\bf 95} (2005) 261805,
  [\href{http://xxx.lanl.gov/abs/hep-ph/0508165}{{\tt hep-ph/0508165}}].

\bibitem{Buras:2006gb}
A.~J. Buras, M.~Gorbahn, U.~Haisch, and U.~Nierste, {\it {Charm quark
  contribution to $K^+ \to \pi^+ \nu \bar\nu$ at next-to-next-to-leading
  order}},  {\em JHEP} {\bf 11} (2006) 002,
  [\href{http://xxx.lanl.gov/abs/hep-ph/0603079}{{\tt hep-ph/0603079}}].

\bibitem{Brod:2008ss}
J.~Brod and M.~Gorbahn, {\it {Electroweak Corrections to the Charm Quark
  Contribution to $K^+ \to \pi^+ \nu \bar\nu$}},  {\em Phys. Rev.} {\bf D78}
  (2008) 034006, [\href{http://xxx.lanl.gov/abs/0805.4119}{{\tt
  arXiv:0805.4119}}].

\bibitem{Isidori:2005xm}
G.~Isidori, F.~Mescia, and C.~Smith, {\it {Light-quark loops in $K
  \to\pi\nu\bar\nu$}},  {\em Nucl. Phys.} {\bf B718} (2005) 319--338,
  [\href{http://xxx.lanl.gov/abs/hep-ph/0503107}{{\tt hep-ph/0503107}}].

\bibitem{Mescia:2007kn}
F.~Mescia and C.~Smith, {\it {Improved estimates of rare K decay
  matrix-elements from $K_{\ell3}$ decays}},  {\em Phys. Rev.} {\bf D76} (2007)
  034017, [\href{http://xxx.lanl.gov/abs/0705.2025}{{\tt arXiv:0705.2025}}].

\bibitem{Buras:2004uu}
A.~J. Buras, F.~Schwab, and S.~Uhlig, {\it {Waiting for precise measurements of
  $K^{+} \to \pi^{+} \nu \bar{\nu}$ and $K_{L} \to \pi^0 \nu \bar{\nu}$}},
  {\em Rev. Mod. Phys.} {\bf 80} (2008) 965--1007,
  [\href{http://xxx.lanl.gov/abs/hep-ph/0405132}{{\tt hep-ph/0405132}}].

\bibitem{Isidori:2006yx}
G.~Isidori, {\it {Flavor Physics with light quarks and leptons}},  {\em eConf}
  {\bf C060409} (2006) 035, [\href{http://xxx.lanl.gov/abs/hep-ph/0606047}{{\tt
  hep-ph/0606047}}].

\bibitem{Smith:2006qg}
C.~Smith, {\it {Theory review on rare K decays: Standard model and beyond}},
  \href{http://xxx.lanl.gov/abs/hep-ph/0608343}{{\tt hep-ph/0608343}}.

\bibitem{Komatsubara:2012pn}
T.~Komatsubara, {\it {Experiments with K-Meson Decays}},  {\em
  Prog.Part.Nucl.Phys.} {\bf 67} (2012) 995--1018,
  [\href{http://xxx.lanl.gov/abs/1203.6437}{{\tt arXiv:1203.6437}}].

\bibitem{Blanke:2013goa}
M.~Blanke, {\it {New Physics Signatures in Kaon Decays}},  {\em PoS} {\bf
  KAON13} (2013) 010, [\href{http://xxx.lanl.gov/abs/1305.5671}{{\tt
  arXiv:1305.5671}}].

\bibitem{Buras:2004ub}
A.~J. Buras, R.~Fleischer, S.~Recksiegel, and F.~Schwab, {\it {Anatomy of
  prominent $B$ and $K$ decays and signatures of CP-violating new physics in
  the electroweak penguin sector}},  {\em Nucl. Phys.} {\bf B697} (2004)
  133--206, [\href{http://xxx.lanl.gov/abs/hep-ph/0402112}{{\tt
  hep-ph/0402112}}].

\bibitem{Buras:2010pz}
A.~J. Buras, K.~Gemmler, and G.~Isidori, {\it {Quark flavour mixing with
  right-handed currents: an effective theory approach}},  {\em Nucl.Phys.} {\bf
  B843} (2011) 107--142, [\href{http://xxx.lanl.gov/abs/1007.1993}{{\tt
  arXiv:1007.1993}}].

\bibitem{Artamonov:2008qb}
{\bf E949} Collaboration, A.~V. Artamonov {\em et.~al.}, {\it {New measurement
  of the $K^{+} \to \pi^{+} \nu \bar{\nu}$ branching ratio}},  {\em Phys. Rev.
  Lett.} {\bf 101} (2008) 191802,
  [\href{http://xxx.lanl.gov/abs/0808.2459}{{\tt arXiv:0808.2459}}].

\bibitem{Ahn:2009gb}
{\bf E391a Collaboration} Collaboration, J.~Ahn {\em et.~al.}, {\it
  {Experimental study of the decay $K^0_L\to\pi^0\nu \bar\nu$}},  {\em
  Phys.Rev.} {\bf D81} (2010) 072004,
  [\href{http://xxx.lanl.gov/abs/0911.4789}{{\tt arXiv:0911.4789}}].

\bibitem{E.T.WorcesterfortheORKA:2013cya}
{\bf ORKA} Collaboration, E.~Worcester, {\it {ORKA, The Golden Kaon Experiment:
  Precision measurement of $K^+ \to \pi^+ \nu \bar{\nu}$ and other rare
  processes}},  {\em PoS} {\bf KAON13} (2013) 035,
  [\href{http://xxx.lanl.gov/abs/1305.7245}{{\tt arXiv:1305.7245}}].

\bibitem{Buchalla:2003sj}
G.~Buchalla, G.~D'Ambrosio, and G.~Isidori, {\it {Extracting short-distance
  physics from $K_{L,S} \to \pi^0 e^+ e^-$ decays}},  {\em Nucl. Phys.} {\bf
  B672} (2003) 387--408, [\href{http://xxx.lanl.gov/abs/hep-ph/0308008}{{\tt
  hep-ph/0308008}}].

\bibitem{Isidori:2004rb}
G.~Isidori, C.~Smith, and R.~Unterdorfer, {\it {The rare decay $K_L \to \pi^0
  \mu^+\mu^-$ within the SM}},  {\em Eur. Phys. J.} {\bf C36} (2004) 57--66,
  [\href{http://xxx.lanl.gov/abs/hep-ph/0404127}{{\tt hep-ph/0404127}}].

\bibitem{Friot:2004yr}
S.~Friot, D.~Greynat, and E.~De~Rafael, {\it {Rare kaon decays revisited}},
  {\em Phys. Lett.} {\bf B595} (2004) 301--308,
  [\href{http://xxx.lanl.gov/abs/hep-ph/0404136}{{\tt hep-ph/0404136}}].

\bibitem{Mescia:2006jd}
F.~Mescia, C.~Smith, and S.~Trine, {\it {$K_L\to\pi^0 e^+e^-$ and
  $K_L\to\pi^0\mu^+\mu^-$: A binary star on the stage of flavor physics}},
  {\em JHEP} {\bf 08} (2006) 088,
  [\href{http://xxx.lanl.gov/abs/hep-ph/0606081}{{\tt hep-ph/0606081}}].

\bibitem{Prades:2007ud}
J.~Prades, {\it {ChPT Progress on Non-Leptonic and Radiative Kaon Decays}},
  {\em PoS} {\bf KAON} (2008) 022,
  [\href{http://xxx.lanl.gov/abs/0707.1789}{{\tt arXiv:0707.1789}}].

\bibitem{Buras:1994qa}
A.~J. Buras, M.~E. Lautenbacher, M.~Misiak, and M.~Munz, {\it {Direct CP
  violation in $K_L \to\pi^0 e^+ e^-$ beyond leading logarithms}},  {\em Nucl.
  Phys.} {\bf B423} (1994) 349--383,
  [\href{http://xxx.lanl.gov/abs/hep-ph/9402347}{{\tt hep-ph/9402347}}].

\bibitem{Gorbahn:2006bm}
M.~Gorbahn and U.~Haisch, {\it {Charm quark contribution to $K_L \to \mu^+
  \mu^-$ at next-to-next-to-leading order}},  {\em Phys. Rev. Lett.} {\bf 97}
  (2006) 122002, [\href{http://xxx.lanl.gov/abs/hep-ph/0605203}{{\tt
  hep-ph/0605203}}].

\bibitem{Isidori:2003ts}
G.~Isidori and R.~Unterdorfer, {\it {On the short-distance constraints from
  $K_{L,S} \to \mu^+ \mu^-$ }},  {\em JHEP} {\bf 01} (2004) 009,
  [\href{http://xxx.lanl.gov/abs/hep-ph/0311084}{{\tt hep-ph/0311084}}].

\bibitem{Grossman:1997sk}
Y.~Grossman and Y.~Nir, {\it {$K_L\to\pi^0\nu\bar\nu$ beyond the standard
  model}},  {\em Phys. Lett.} {\bf B398} (1997) 163--168,
  [\href{http://xxx.lanl.gov/abs/hep-ph/9701313}{{\tt hep-ph/9701313}}].

\bibitem{Blanke:2009am}
M.~Blanke, A.~J. Buras, B.~Duling, S.~Recksiegel, and C.~Tarantino, {\it {FCNC
  Processes in the Littlest Higgs Model with T-Parity: a 2009 Look}},  {\em
  Acta Phys.Polon.} {\bf B41} (2010) 657--683,
  [\href{http://xxx.lanl.gov/abs/0906.5454}{{\tt arXiv:0906.5454}}].

\bibitem{Colangelo:1996ay}
P.~Colangelo, F.~De~Fazio, P.~Santorelli, and E.~Scrimieri, {\it {Rare $B \to
  K^{(*)} \nu\bar\nu$ decays at $B$ factories}},  {\em Phys.Lett.} {\bf B395}
  (1997) 339--344, [\href{http://xxx.lanl.gov/abs/hep-ph/9610297}{{\tt
  hep-ph/9610297}}].

\bibitem{Buchalla:2000sk}
G.~Buchalla, G.~Hiller, and G.~Isidori, {\it {Phenomenology of non-standard Z
  couplings in exclusive semileptonic $b\to s$ transitions}},  {\em Phys. Rev.}
  {\bf D63} (2001) 014015, [\href{http://xxx.lanl.gov/abs/hep-ph/0006136}{{\tt
  hep-ph/0006136}}].

\bibitem{Altmannshofer:2009ma}
W.~Altmannshofer, A.~J. Buras, D.~M. Straub, and M.~Wick, {\it {New strategies
  for New Physics search in $B \to K^{*} \nu \bar{\nu}$, $B \to K \nu
  \bar{\nu}$ and $B \to X_{s} \nu \bar{\nu}$ decays}},  {\em JHEP} {\bf 04}
  (2009) 022, [\href{http://xxx.lanl.gov/abs/0902.0160}{{\tt
  arXiv:0902.0160}}].

\bibitem{Bartsch:2009qp}
M.~Bartsch, M.~Beylich, G.~Buchalla, and D.-N. Gao, {\it {Precision Flavour
  Physics with $B\to K \nu \bar\nu$ and $B\to K l^+ l^-$}},  {\em JHEP} {\bf
  0911} (2009) 011, [\href{http://xxx.lanl.gov/abs/0909.1512}{{\tt
  arXiv:0909.1512}}].

\bibitem{Kamenik:2009kc}
J.~F. Kamenik and C.~Smith, {\it {Tree-level contributions to the rare decays
  $B^+ \to \pi^+ \nu \bar\nu, B^+\to K^+ \nu\bar\nu$, and $B^+\to
  K^{*+}\nu\bar\nu$ in the Standard Model}},  {\em Phys.Lett.} {\bf B680}
  (2009) 471--475, [\href{http://xxx.lanl.gov/abs/0908.1174}{{\tt
  arXiv:0908.1174}}].

\bibitem{Barate:2000rc}
{\bf ALEPH} Collaboration, R.~Barate {\em et.~al.}, {\it {Measurements of $BR(b
  \to \tau^- \bar\nu_\tau X)$ and $BR(b \to \tau^- \bar\nu_\tau D^{*\pm} X)$
  and upper limits on $BR(B^- \to \tau^- \bar\nu_\tau)$ and $BR(b \to \nu
  \bar\nu)$}},  {\em Eur. Phys. J.} {\bf C19} (2001) 213--227,
  [\href{http://xxx.lanl.gov/abs/hep-ex/0010022}{{\tt hep-ex/0010022}}].

\bibitem{:2007zk}
{\bf BELLE} Collaboration, K.~F. Chen {\em et.~al.}, {\it {Search for $B\to
  h^{(*)} \nu \bar\nu$ Decays at Belle}},  {\em Phys. Rev. Lett.} {\bf 99}
  (2007) 221802, [\href{http://xxx.lanl.gov/abs/0707.0138}{{\tt
  arXiv:0707.0138}}].

\bibitem{:2008fr}
{\bf BABAR} Collaboration, B.~Aubert {\em et.~al.}, {\it {Search for $B\to K^*
  \nu \bar\nu$ decays}},  {\em Phys. Rev.} {\bf D78} (2008) 072007,
  [\href{http://xxx.lanl.gov/abs/0808.1338}{{\tt arXiv:0808.1338}}].

\bibitem{Blum:2011ng}
T.~Blum, P.~Boyle, N.~Christ, N.~Garron, E.~Goode, {\em et.~al.}, {\it {The
  $K\to(\pi\pi)_{I=2}$ Decay Amplitude from Lattice QCD}},  {\em
  Phys.Rev.Lett.} {\bf 108} (2012) 141601,
  [\href{http://xxx.lanl.gov/abs/1111.1699}{{\tt arXiv:1111.1699}}].

\bibitem{Christ:2009ev}
{\bf RBC Collaboration, UKQCD Collaboration} Collaboration, N.~H. Christ, {\it
  {Theoretical strategies for $\epsilon'/\epsilon$}},  {\em PoS} {\bf KAON09}
  (2009) 027, [\href{http://xxx.lanl.gov/abs/0912.2917}{{\tt
  arXiv:0912.2917}}].

\bibitem{Buras:1993dy}
A.~J. Buras, M.~Jamin, and M.~E. Lautenbacher, {\it The anatomy of
  $\varepsilon'/ \varepsilon$ beyond leading logarithms with improved hadronic
  matrix elements},  {\em Nucl. Phys.} {\bf B408} (1993) 209--285,
  [\href{http://xxx.lanl.gov/abs/hep-ph/9303284}{{\tt hep-ph/9303284}}].

\bibitem{Ciuchini:1993vr}
M.~Ciuchini, E.~Franco, G.~Martinelli, and L.~Reina, {\it {The $\Delta S = 1$
  effective Hamiltonian including next-to-leading order QCD and QED
  corrections}},  {\em Nucl.Phys.} {\bf B415} (1994) 403--462,
  [\href{http://xxx.lanl.gov/abs/hep-ph/9304257}{{\tt hep-ph/9304257}}].

\bibitem{Buras:1999st}
A.~J. Buras, P.~Gambino, and U.~A. Haisch, {\it Electroweak penguin
  contributions to non-leptonic $\delta f = 1$ decays at nnlo},  {\em Nucl.
  Phys.} {\bf B570} (2000) 117--154,
  [\href{http://xxx.lanl.gov/abs/hep-ph/9911250}{{\tt hep-ph/9911250}}].

\bibitem{Gorbahn:2004my}
M.~Gorbahn and U.~Haisch, {\it {Effective Hamiltonian for non-leptonic $|\Delta
  F| = 1$ decays at NNLO in QCD}},  {\em Nucl.Phys.} {\bf B713} (2005)
  291--332, [\href{http://xxx.lanl.gov/abs/hep-ph/0411071}{{\tt
  hep-ph/0411071}}].

\bibitem{Batley:2002gn}
{\bf NA48 Collaboration} Collaboration, J.~Batley {\em et.~al.}, {\it {A
  Precision measurement of direct CP violation in the decay of neutral kaons
  into two pions}},  {\em Phys.Lett.} {\bf B544} (2002) 97--112,
  [\href{http://xxx.lanl.gov/abs/hep-ex/0208009}{{\tt hep-ex/0208009}}].

\bibitem{AlaviHarati:2002ye}
{\bf KTeV Collaboration} Collaboration, A.~Alavi-Harati {\em et.~al.}, {\it
  {Measurements of direct CP violation, CPT symmetry, and other parameters in
  the neutral kaon system}},  {\em Phys.Rev.} {\bf D67} (2003) 012005,
  [\href{http://xxx.lanl.gov/abs/hep-ex/0208007}{{\tt hep-ex/0208007}}].

\bibitem{Worcester:2009qt}
{\bf KTeV Collaboration} Collaboration, E.~Worcester, {\it {The Final
  Measurement of $\varepsilon'/\varepsilon$ from KTeV}},
  \href{http://xxx.lanl.gov/abs/0909.2555}{{\tt arXiv:0909.2555}}.

\bibitem{Blanke:2007wr}
M.~Blanke, A.~J. Buras, S.~Recksiegel, C.~Tarantino, and S.~Uhlig, {\it
  {Correlations between $\varepsilon'/\varepsilon$ and Rare $K$ Decays in the
  Littlest Higgs Model with T-Parity}},  {\em JHEP} {\bf 06} (2007) 082,
  [\href{http://xxx.lanl.gov/abs/0704.3329}{{\tt 0704.3329}}].

\bibitem{Bardeen:1986uz}
W.~A. Bardeen, A.~J. Buras, and J.-M. G\'erard, {\it {The $K\to\pi \pi$ Decays
  in the Large N Limit: Quark Evolution}},  {\em Nucl.Phys.} {\bf B293} (1987)
  787.

\bibitem{Bertolini:1998vd}
S.~Bertolini, M.~Fabbrichesi, and J.~O. Eeg, {\it {Theory of the CP violating
  parameter $\epsilon'/\epsilon$}},  {\em Rev.Mod.Phys.} {\bf 72} (2000)
  65--93, [\href{http://xxx.lanl.gov/abs/hep-ph/9802405}{{\tt
  hep-ph/9802405}}].

\bibitem{Buras:2003zz}
A.~J. Buras and M.~Jamin, {\it $\varepsilon'/\varepsilon$ at the nlo: 10 years
  later},  {\em JHEP} {\bf 01} (2004) 048,
  [\href{http://xxx.lanl.gov/abs/hep-ph/0306217}{{\tt hep-ph/0306217}}].

\bibitem{Pich:2004ee}
A.~Pich, {\it {$\varepsilon'/\varepsilon$ in the standard model: Theoretical
  update}},  \href{http://xxx.lanl.gov/abs/hep-ph/0410215}{{\tt
  hep-ph/0410215}}.

\bibitem{Cirigliano:2011ny}
V.~Cirigliano, G.~Ecker, H.~Neufeld, A.~Pich, and J.~Portoles, {\it {Kaon
  Decays in the Standard Model}},  {\em Rev.Mod.Phys.} {\bf 84} (2012) 399,
  [\href{http://xxx.lanl.gov/abs/1107.6001}{{\tt arXiv:1107.6001}}].

\bibitem{Bertolini:2012pu}
S.~Bertolini, J.~O. Eeg, A.~Maiezza, and F.~Nesti, {\it {New physics in
  $\epsilon'$ from gluomagnetic contributions and limits on Left-Right
  symmetry}},  {\em Phys.Rev.} {\bf D86} (2012) 095013,
  [\href{http://xxx.lanl.gov/abs/1206.0668}{{\tt arXiv:1206.0668}}].

\bibitem{Flynn:1989iu}
J.~M. Flynn and L.~Randall, {\it {The Electromagnetic Penguin Contribution to
  $\varepsilon^\prime / \varepsilon$ for Large Top Quark Mass}},  {\em
  Phys.Lett.} {\bf B224} (1989) 221.

\bibitem{Buchalla:1989we}
G.~Buchalla, A.~J. Buras, and M.~K. Harlander, {\it The anatomy of
  $\varepsilon' / \varepsilon$ in the standard model},  {\em Nucl. Phys.} {\bf
  B337} (1990) 313--362.

\bibitem{Blum:2011pu}
T.~Blum, P.~Boyle, N.~Christ, N.~Garron, E.~Goode, {\em et.~al.}, {\it {$K\to
  \pi\pi$ Decay amplitudes from Lattice QCD}},  {\em Phys.Rev.} {\bf D84}
  (2011) 114503, [\href{http://xxx.lanl.gov/abs/1106.2714}{{\tt
  arXiv:1106.2714}}].

\bibitem{Blum:2012uk}
T.~Blum, P.~Boyle, N.~Christ, N.~Garron, E.~Goode, {\em et.~al.}, {\it {Lattice
  determination of the $K \to (\pi\pi)_{I=2}$ Decay Amplitude $A_2$}},  {\em
  Phys.Rev.} {\bf D86} (2012) 074513,
  [\href{http://xxx.lanl.gov/abs/1206.5142}{{\tt arXiv:1206.5142}}].

\bibitem{Aaij:2011in}
{\bf LHCb Collaboration} Collaboration, R.~Aaij {\em et.~al.}, {\it {Evidence
  for CP violation in time-integrated $D^0 \to h^-h^+$ decay rates}},  {\em
  Phys.Rev.Lett.} {\bf 108} (2012) 111602,
  [\href{http://xxx.lanl.gov/abs/1112.0938}{{\tt arXiv:1112.0938}}].

\bibitem{Aaij:2013bra}
{\bf LHCb collaboration} Collaboration, R.~Aaij {\em et.~al.}, {\it {Search for
  direct $CP$ violation in $D^0 \rightarrow h^- h^+$ modes using semileptonic
  $B$ decays}},  {\em Phys.Lett.} {\bf B723} (2013) 33--43,
  [\href{http://xxx.lanl.gov/abs/1303.2614}{{\tt arXiv:1303.2614}}].

\bibitem{Isidori:2011qw}
G.~Isidori, J.~F. Kamenik, Z.~Ligeti, and G.~Perez, {\it {Implications of the
  LHCb Evidence for Charm CP Violation}},  {\em Phys.Lett.} {\bf B711} (2012)
  46--51, [\href{http://xxx.lanl.gov/abs/1111.4987}{{\tt arXiv:1111.4987}}].

\bibitem{Hochberg:2011ru}
Y.~Hochberg and Y.~Nir, {\it {Relating direct CP violation in D decays and the
  forward-backward asymmetry in $t\bar t$ production}},  {\em Phys.Rev.Lett.}
  {\bf 108} (2012) 261601, [\href{http://xxx.lanl.gov/abs/1112.5268}{{\tt
  arXiv:1112.5268}}].

\bibitem{Isidori:2012yx}
G.~Isidori and J.~F. Kamenik, {\it {Shedding light on CP violation in the charm
  system via $D\to V \gamma$ decays}},  {\em Phys.Rev.Lett.} {\bf 109} (2012)
  171801, [\href{http://xxx.lanl.gov/abs/1205.3164}{{\tt arXiv:1205.3164}}].

\bibitem{Blum:2009sk}
K.~Blum, Y.~Grossman, Y.~Nir, and G.~Perez, {\it {Combining $K^0 - \bar K^0$
  mixing and $D^0 -\bar D^0$ mixing to constrain the flavor structure of new
  physics}},  {\em Phys.Rev.Lett.} {\bf 102} (2009) 211802,
  [\href{http://xxx.lanl.gov/abs/0903.2118}{{\tt arXiv:0903.2118}}].

\bibitem{Gedalia:2012pi}
O.~Gedalia, J.~F. Kamenik, Z.~Ligeti, and G.~Perez, {\it {On the Universality
  of CP Violation in $\Delta F = 1$ Processes}},  {\em Phys.Lett.} {\bf B714}
  (2012) 55--61, [\href{http://xxx.lanl.gov/abs/1202.5038}{{\tt
  arXiv:1202.5038}}].

\bibitem{Buras:2012ub}
A.~J. Buras, G.~Perez, T.~A. Schwarz, and T.~M. Tait, {\it {Top and flavour
  physics in the LHC era}},  {\em Eur.Phys.J.} {\bf C72} (2012) 2105.

\bibitem{Jegerlehner:2009ry}
F.~Jegerlehner and A.~Nyffeler, {\it {The Muon $g-2$}},  {\em Phys. Rept.} {\bf
  477} (2009) 1--110, [\href{http://xxx.lanl.gov/abs/0902.3360}{{\tt
  arXiv:0902.3360}}].

\bibitem{Engel:2013lsa}
J.~Engel, M.~J. Ramsey-Musolf, and U.~van Kolck, {\it {Electric Dipole Moments
  of Nucleons, Nuclei, and Atoms: The Standard Model and Beyond}},  {\em
  Prog.Part.Nucl.Phys.} {\bf 71} (2013) 21--74,
  [\href{http://xxx.lanl.gov/abs/1303.2371}{{\tt arXiv:1303.2371}}].

\bibitem{Bernstein:2013hba}
R.~H. Bernstein and P.~S. Cooper, {\it {Charged Lepton Flavor Violation: An
  Experimenter's Guide}},  {\em Phys.Rept.} (2013)
  [\href{http://xxx.lanl.gov/abs/1307.5787}{{\tt arXiv:1307.5787}}].

\bibitem{Hisano:2009ae}
J.~Hisano, M.~Nagai, P.~Paradisi, and Y.~Shimizu, {\it {Waiting for $\mu\to
  e\gamma$ from the MEG experiment}},  {\em JHEP} {\bf 0912} (2009) 030,
  [\href{http://xxx.lanl.gov/abs/0904.2080}{{\tt arXiv:0904.2080}}].

\bibitem{Buras:2010pm}
A.~J. Buras, M.~Nagai, and P.~Paradisi, {\it {Footprints of SUSY GUTs in
  Flavour Physics}},  {\em JHEP} {\bf 1105} (2011) 005,
  [\href{http://xxx.lanl.gov/abs/1011.4853}{{\tt arXiv:1011.4853}}].

\bibitem{Girrbach:2009uy}
J.~Girrbach, S.~Mertens, U.~Nierste, and S.~Wiesenfeldt, {\it {Lepton flavour
  violation in the MSSM}},  {\em JHEP} {\bf 05} (2010) 026,
  [\href{http://xxx.lanl.gov/abs/0910.2663}{{\tt arXiv:0910.2663}}].

\bibitem{Falkowski:2013jya}
A.~Falkowski, D.~M. Straub, and A.~Vicente, {\it {Vector-like leptons: Higgs
  decays and collider phenomenology}},
  \href{http://xxx.lanl.gov/abs/1312.5329}{{\tt arXiv:1312.5329}}.

\bibitem{Crivellin:2013hpa}
A.~Crivellin, S.~Najjari, and J.~Rosiek, {\it {Lepton Flavor Violation in the
  Standard Model with general Dimension-Six Operators}},
  \href{http://xxx.lanl.gov/abs/1312.0634}{{\tt arXiv:1312.0634}}.

\bibitem{Altmannshofer:2013lfa}
W.~Altmannshofer, R.~Harnik, and J.~Zupan, {\it {Low Energy Probes of PeV Scale
  Sfermions}},  {\em JHEP} {\bf 1311} (2013) 202,
  [\href{http://xxx.lanl.gov/abs/1308.3653}{{\tt arXiv:1308.3653}}].

\bibitem{Blanke:2007db}
M.~Blanke, A.~J. Buras, B.~Duling, A.~Poschenrieder, and C.~Tarantino, {\it
  {Charged Lepton Flavour Violation and $(g-2)_\mu$ in the Littlest Higgs Model
  with T-Parity: a clear Distinction from Supersymmetry}},  {\em JHEP} {\bf 05}
  (2007) 013, [\href{http://xxx.lanl.gov/abs/hep-ph/0702136}{{\tt
  hep-ph/0702136}}].

\bibitem{Ellis:2002fe}
J.~R. Ellis, J.~Hisano, M.~Raidal, and Y.~Shimizu, {\it {A new parametrization
  of the seesaw mechanism and applications in supersymmetric models}},  {\em
  Phys. Rev.} {\bf D66} (2002) 115013,
  [\href{http://xxx.lanl.gov/abs/hep-ph/0206110}{{\tt hep-ph/0206110}}].

\bibitem{Arganda:2005ji}
E.~Arganda and M.~J. Herrero, {\it {Testing supersymmetry with lepton flavor
  violating tau and mu decays}},  {\em Phys. Rev.} {\bf D73} (2006) 055003,
  [\href{http://xxx.lanl.gov/abs/hep-ph/0510405}{{\tt hep-ph/0510405}}].

\bibitem{Brignole:2004ah}
A.~Brignole and A.~Rossi, {\it {Anatomy and phenomenology of $\mu \tau$ lepton
  flavour violation in the MSSM}},  {\em Nucl. Phys.} {\bf B701} (2004) 3--53,
  [\href{http://xxx.lanl.gov/abs/hep-ph/0404211}{{\tt hep-ph/0404211}}].

\bibitem{Paradisi:2005tk}
P.~Paradisi, {\it {Higgs-mediated $\tau\to\mu$ and $\tau\to e$ transitions in
  II Higgs doublet model and supersymmetry}},  {\em JHEP} {\bf 02} (2006) 050,
  [\href{http://xxx.lanl.gov/abs/hep-ph/0508054}{{\tt hep-ph/0508054}}].

\bibitem{Paradisi:2006jp}
P.~Paradisi, {\it {Higgs-mediated $e \to \mu$ transitions in II Higgs doublet
  model and supersymmetry}},  {\em JHEP} {\bf 08} (2006) 047,
  [\href{http://xxx.lanl.gov/abs/hep-ph/0601100}{{\tt hep-ph/0601100}}].

\bibitem{Paradisi:2005fk}
P.~Paradisi, {\it {Constraints on SUSY lepton flavour violation by rare
  processes}},  {\em JHEP} {\bf 10} (2005) 006,
  [\href{http://xxx.lanl.gov/abs/hep-ph/0505046}{{\tt hep-ph/0505046}}].

\bibitem{delAguila:2008zu}
F.~del Aguila, J.~I. Illana, and M.~D. Jenkins, {\it {Precise limits from
  lepton flavour violating processes on the Littlest Higgs model with
  T-parity}},  {\em JHEP} {\bf 01} (2009) 080,
  [\href{http://xxx.lanl.gov/abs/0811.2891}{{\tt arXiv:0811.2891}}].

\bibitem{Goto:2010sn}
T.~Goto, Y.~Okada, and Y.~Yamamoto, {\it {Tau and muon lepton flavor violations
  in the littlest Higgs model with T-parity}},  {\em Phys.Rev.} {\bf D83}
  (2011) 053011, [\href{http://xxx.lanl.gov/abs/1012.4385}{{\tt
  arXiv:1012.4385}}].

\bibitem{Buras:2010cp}
A.~J. Buras, B.~Duling, T.~Feldmann, T.~Heidsieck, and C.~Promberger, {\it
  {Lepton Flavour Violation in the Presence of a Fourth Generation of Quarks
  and Leptons}},  {\em JHEP} {\bf 1009} (2010) 104,
  [\href{http://xxx.lanl.gov/abs/1006.5356}{{\tt arXiv:1006.5356}}].

\bibitem{Adam:2013mnn}
{\bf MEG Collaboration} Collaboration, J.~Adam {\em et.~al.}, {\it {New
  constraint on the existence of the $\mu^+\to e^+ \gamma$ decay}},  {\em
  Phys.Rev.Lett.} {\bf 110} (2013) 201801,
  [\href{http://xxx.lanl.gov/abs/1303.0754}{{\tt arXiv:1303.0754}}].

\bibitem{Baldini:2013ke}
A.~Baldini, F.~Cei, C.~Cerri, S.~Dussoni, L.~Galli, {\em et.~al.}, {\it {MEG
  Upgrade Proposal}},  \href{http://xxx.lanl.gov/abs/1301.7225}{{\tt
  arXiv:1301.7225}}.

\bibitem{Blondel:2013ia}
A.~Blondel, A.~Bravar, M.~Pohl, S.~Bachmann, N.~Berger, {\em et.~al.}, {\it
  {Research Proposal for an Experiment to Search for the Decay $\mu\to eee$}},
  \href{http://xxx.lanl.gov/abs/1301.6113}{{\tt arXiv:1301.6113}}.

\bibitem{Barlow:2011zza}
R.~Barlow, {\it {The PRISM/PRIME project}},  {\em Nucl.Phys.Proc.Suppl.} {\bf
  218} (2011) 44--49.

\bibitem{Kaulard:1998rb}
{\bf SINDRUM II Collaboration} Collaboration, J.~Kaulard {\em et.~al.}, {\it
  {Improved limit on the branching ratio of $\mu \to e$ conversion on
  titanium}},  {\em Phys.Lett.} {\bf B422} (1998) 334--338.

\bibitem{Abrams:2012er}
{\bf Mu2e Collaboration} Collaboration, R.~Abrams {\em et.~al.}, {\it {Mu2e
  Conceptual Design Report}},  \href{http://xxx.lanl.gov/abs/1211.7019}{{\tt
  arXiv:1211.7019}}.

\bibitem{Cirigliano:2009bz}
V.~Cirigliano, R.~Kitano, Y.~Okada, and P.~Tuzon, {\it {On the model
  discriminating power of $\mu \to e$ conversion in nuclei}},  {\em Phys.Rev.}
  {\bf D80} (2009) 013002, [\href{http://xxx.lanl.gov/abs/0904.0957}{{\tt
  arXiv:0904.0957}}].

\bibitem{Feldmann:2011zh}
T.~Feldmann, {\it {Lepton Flavour Violation Theory}},  {\em PoS} {\bf
  BEAUTY2011} (2011) 017, [\href{http://xxx.lanl.gov/abs/1105.2139}{{\tt
  arXiv:1105.2139}}].

\bibitem{Ibarra:2010zz}
A.~Ibarra, {\it {Neutrino physics and lepton flavour violation: A theoretical
  overview}},  {\em Nuovo Cim.} {\bf C033N5} (2010) 67--75.

\bibitem{Kinoshita:2004wi}
T.~Kinoshita and M.~Nio, {\it {Improved $\alpha^4$ term of the muon anomalous
  magnetic moment}},  {\em Phys.Rev.} {\bf D70} (2004) 113001,
  [\href{http://xxx.lanl.gov/abs/hep-ph/0402206}{{\tt hep-ph/0402206}}].

\bibitem{Passera:2006gc}
M.~Passera, {\it {Precise mass-dependent QED contributions to leptonic g-2 at
  order $\alpha^2$ and $\alpha^3$}},  {\em Phys.Rev.} {\bf D75} (2007) 013002,
  [\href{http://xxx.lanl.gov/abs/hep-ph/0606174}{{\tt hep-ph/0606174}}].

\bibitem{Aoyama:2012wj}
T.~Aoyama, M.~Hayakawa, T.~Kinoshita, and M.~Nio, {\it {Tenth-Order QED
  Contribution to the Electron g-2 and an Improved Value of the Fine Structure
  Constant}},  {\em Phys.Rev.Lett.} {\bf 109} (2012) 111807,
  [\href{http://xxx.lanl.gov/abs/1205.5368}{{\tt arXiv:1205.5368}}].

\bibitem{Aoyama:2012wk}
T.~Aoyama, M.~Hayakawa, T.~Kinoshita, and M.~Nio, {\it {Complete Tenth-Order
  QED Contribution to the Muon g-2}},  {\em Phys.Rev.Lett.} {\bf 109} (2012)
  111808, [\href{http://xxx.lanl.gov/abs/1205.5370}{{\tt arXiv:1205.5370}}].

\bibitem{Czarnecki:2002nt}
A.~Czarnecki, W.~J. Marciano, and A.~Vainshtein, {\it {Refinements in
  electroweak contributions to the muon anomalous magnetic moment}},  {\em
  Phys.Rev.} {\bf D67} (2003) 073006,
  [\href{http://xxx.lanl.gov/abs/hep-ph/0212229}{{\tt hep-ph/0212229}}].

\bibitem{Prades:2009tw}
J.~Prades, E.~de~Rafael, and A.~Vainshtein, {\it {Hadronic Light-by-Light
  Scattering Contribution to the Muon Anomalous Magnetic Moment}},
  \href{http://xxx.lanl.gov/abs/0901.0306}{{\tt arXiv:0901.0306}}.

\bibitem{Prades:2009qp}
J.~Prades, {\it {Standard Model Prediction of the Muon Anomalous Magnetic
  Moment}},  {\em Acta Phys.Polon.Supp.} {\bf 3} (2010) 75--86,
  [\href{http://xxx.lanl.gov/abs/0909.2546}{{\tt arXiv:0909.2546}}].

\bibitem{Benayoun:2012wc}
M.~Benayoun, P.~David, L.~DelBuono, and F.~Jegerlehner, {\it {An Update of the
  HLS Estimate of the Muon g-2}},  {\em Eur.Phys.J.} {\bf C73} (2013) 2453,
  [\href{http://xxx.lanl.gov/abs/1210.7184}{{\tt arXiv:1210.7184}}].

\bibitem{Bennett:2006fi}
{\bf Muon G-2} Collaboration, G.~W. Bennett {\em et.~al.}, {\it {Final report
  of the muon E821 anomalous magnetic moment measurement at BNL}},  {\em Phys.
  Rev.} {\bf D73} (2006) 072003,
  [\href{http://xxx.lanl.gov/abs/hep-ex/0602035}{{\tt hep-ex/0602035}}].

\bibitem{Stockinger:2007pe}
D.~Stockinger, {\it {$(g-2)_{\mu}$ and supersymmetry: Status and prospects}},
  \href{http://xxx.lanl.gov/abs/0710.2429}{{\tt arXiv:0710.2429}}.

\bibitem{Marchetti:2008hw}
S.~Marchetti, S.~Mertens, U.~Nierste, and D.~Stockinger, {\it
  {$\tan\beta$-enhanced supersymmetric corrections to the anomalous magnetic
  moment of the muon}},  {\em Phys.Rev.} {\bf D79} (2009) 013010,
  [\href{http://xxx.lanl.gov/abs/0808.1530}{{\tt arXiv:0808.1530}}].

\bibitem{Feroz:2008wr}
F.~Feroz, B.~C. Allanach, M.~Hobson, S.~S. AbdusSalam, R.~Trotta, {\em
  et.~al.}, {\it {Bayesian Selection of $sign(\mu)$ within mSUGRA in Global
  Fits Including WMAP5 Results}},  {\em JHEP} {\bf 0810} (2008) 064,
  [\href{http://xxx.lanl.gov/abs/0807.4512}{{\tt arXiv:0807.4512}}].

\bibitem{Nojiri:2008aa}
M.~Nojiri, T.~Plehn, G.~Polesello, J.~M. Alexander, B.~Allanach, {\em et.~al.},
  {\it {Physics Beyond the Standard Model: Supersymmetry}},
  \href{http://xxx.lanl.gov/abs/0802.3672}{{\tt arXiv:0802.3672}}.

\bibitem{Degrassi:1998es}
G.~Degrassi and G.~Giudice, {\it {QED logarithms in the electroweak corrections
  to the muon anomalous magnetic moment}},  {\em Phys.Rev.} {\bf D58} (1998)
  053007, [\href{http://xxx.lanl.gov/abs/hep-ph/9803384}{{\tt
  hep-ph/9803384}}].

\bibitem{Heinemeyer:2003dq}
S.~Heinemeyer, D.~Stockinger, and G.~Weiglein, {\it {Two loop SUSY corrections
  to the anomalous magnetic moment of the muon}},  {\em Nucl.Phys.} {\bf B690}
  (2004) 62--80, [\href{http://xxx.lanl.gov/abs/hep-ph/0312264}{{\tt
  hep-ph/0312264}}].

\bibitem{Heinemeyer:2004yq}
S.~Heinemeyer, D.~Stockinger, and G.~Weiglein, {\it {Electroweak and
  supersymmetric two-loop corrections to $(g-2)_{\mu}$}},  {\em Nucl.Phys.}
  {\bf B699} (2004) 103--123,
  [\href{http://xxx.lanl.gov/abs/hep-ph/0405255}{{\tt hep-ph/0405255}}].

\bibitem{Crivellin:2010ty}
A.~Crivellin, J.~Girrbach, and U.~Nierste, {\it {Yukawa coupling and anomalous
  magnetic moment of the muon: an update for the LHC era}},  {\em Phys.Rev.}
  {\bf D83} (2011) 055009, [\href{http://xxx.lanl.gov/abs/1010.4485}{{\tt
  arXiv:1010.4485}}].

\bibitem{Jegerlehner:2012ju}
F.~Jegerlehner, {\it {Implications of low and high energy measurements on SUSY
  models}},  {\em Frascati Phys.Ser.} {\bf 54} (2012) 42--51,
  [\href{http://xxx.lanl.gov/abs/1203.0806}{{\tt arXiv:1203.0806}}].

\bibitem{Hanneke:2008tm}
D.~Hanneke, S.~Fogwell, and G.~Gabrielse, {\it {New Measurement of the Electron
  Magnetic Moment and the Fine Structure Constant}},  {\em Phys.Rev.Lett.} {\bf
  100} (2008) 120801, [\href{http://xxx.lanl.gov/abs/0801.1134}{{\tt
  arXiv:0801.1134}}].

\bibitem{Aoyama:2007mn}
T.~Aoyama, M.~Hayakawa, T.~Kinoshita, and M.~Nio, {\it {Revised value of the
  eighth-order QED contribution to the anomalous magnetic moment of the
  electron}},  {\em Phys.Rev.} {\bf D77} (2008) 053012,
  [\href{http://xxx.lanl.gov/abs/0712.2607}{{\tt arXiv:0712.2607}}].

\bibitem{Clade:2006zz}
P.~Clade, E.~de~Mirandes, M.~Cadoret, S.~Guellati-Khelifa, C.~Schwob, {\em
  et.~al.}, {\it {Determination of the Fine Structure Constant Based on Bloch
  Oscillations of Ultracold Atoms in a Vertical Optical Lattice}},  {\em
  Phys.Rev.Lett.} {\bf 96} (2006) 033001.

\bibitem{Pospelov:2005pr}
M.~Pospelov and A.~Ritz, {\it {Electric dipole moments as probes of new
  physics}},  {\em Annals Phys.} {\bf 318} (2005) 119--169,
  [\href{http://xxx.lanl.gov/abs/hep-ph/0504231}{{\tt hep-ph/0504231}}].

\bibitem{Batell:2012ge}
B.~Batell, {\it {Flavor-diagonal CP violation}},  {\em Eur.Phys.J.} {\bf C72}
  (2012) 2127.

\bibitem{Hudson:2011zz}
J.~Hudson, D.~Kara, I.~Smallman, B.~Sauer, M.~Tarbutt, {\em et.~al.}, {\it
  {Improved measurement of the shape of the electron}},  {\em Nature} {\bf 473}
  (2011) 493--496.

\bibitem{Baron:2013eja}
{\bf ACME Collaboration} Collaboration, J.~Baron {\em et.~al.}, {\it {Order of
  Magnitude Smaller Limit on the Electric Dipole Moment of the Electron}},
  {\em Science} (2013) [\href{http://xxx.lanl.gov/abs/1310.7534}{{\tt
  arXiv:1310.7534}}].

\bibitem{He:2014fva}
X.-G. He, C.-J. Lee, S.-F. Li, and J.~Tandean, {\it {A Large Electron EDM and
  Minimal Flavor Violation}},  \href{http://xxx.lanl.gov/abs/1401.2615}{{\tt
  arXiv:1401.2615}}.

\bibitem{Morrissey:2012db}
D.~E. Morrissey and M.~J. Ramsey-Musolf, {\it {Electroweak baryogenesis}},
  {\em New J.Phys.} {\bf 14} (2012) 125003,
  [\href{http://xxx.lanl.gov/abs/1206.2942}{{\tt arXiv:1206.2942}}].

\bibitem{Kozaczuk:2012xv}
J.~Kozaczuk, S.~Profumo, M.~J. Ramsey-Musolf, and C.~L. Wainwright, {\it
  {Supersymmetric Electroweak Baryogenesis Via Resonant Sfermion Sources}},
  {\em Phys.Rev.} {\bf D86} (2012) 096001,
  [\href{http://xxx.lanl.gov/abs/1206.4100}{{\tt arXiv:1206.4100}}].

\bibitem{Li:2008ez}
Y.~Li, S.~Profumo, and M.~Ramsey-Musolf, {\it {Bino-driven Electroweak
  Baryogenesis with highly suppressed Electric Dipole Moments}},  {\em
  Phys.Lett.} {\bf B673} (2009) 95--100,
  [\href{http://xxx.lanl.gov/abs/0811.1987}{{\tt arXiv:0811.1987}}].

\bibitem{Liu:2011jh}
T.~Liu, M.~J. Ramsey-Musolf, and J.~Shu, {\it {Electroweak Beautygenesis: From
  $b \to s$ CP-violation to the Cosmic Baryon Asymmetry}},  {\em
  Phys.Rev.Lett.} {\bf 108} (2012) 221301,
  [\href{http://xxx.lanl.gov/abs/1109.4145}{{\tt arXiv:1109.4145}}].

\bibitem{Tulin:2011wi}
S.~Tulin and P.~Winslow, {\it {Anomalous $B$ meson mixing and baryogenesis}},
  {\em Phys.Rev.} {\bf D84} (2011) 034013,
  [\href{http://xxx.lanl.gov/abs/1105.2848}{{\tt arXiv:1105.2848}}].

\bibitem{Cline:2011mm}
J.~M. Cline, K.~Kainulainen, and M.~Trott, {\it {Electroweak Baryogenesis in
  Two Higgs Doublet Models and $B$ meson anomalies}},  {\em JHEP} {\bf 1111}
  (2011) 089, [\href{http://xxx.lanl.gov/abs/1107.3559}{{\tt
  arXiv:1107.3559}}].

\bibitem{Jung:2013hka}
M.~Jung and A.~Pich, {\it {Electric Dipole Moments in Two-Higgs-Doublet
  Models}},  \href{http://xxx.lanl.gov/abs/1308.6283}{{\tt arXiv:1308.6283}}.

\bibitem{Pisano:1991ee}
F.~Pisano and V.~Pleitez, {\it {An SU(3) x U(1) model for electroweak
  interactions}},  {\em Phys.Rev.} {\bf D46} (1992) 410--417,
  [\href{http://xxx.lanl.gov/abs/hep-ph/9206242}{{\tt hep-ph/9206242}}].

\bibitem{Frampton:1992wt}
P.~H. Frampton, {\it {Chiral dilepton model and the flavor question}},  {\em
  Phys. Rev. Lett.} {\bf 69} (1992) 2889--2891.

\bibitem{Liu:1993gy}
J.~T. Liu and D.~Ng, {\it {Lepton flavor changing processes and CP violation in
  the 331 model}},  {\em Phys.Rev.} {\bf D50} (1994) 548--557,
  [\href{http://xxx.lanl.gov/abs/hep-ph/9401228}{{\tt hep-ph/9401228}}].

\bibitem{Diaz:2004fs}
R.~A. Diaz, R.~Martinez, and F.~Ochoa, {\it {SU(3)(c) x SU(3)(L) x U(1)(X)
  models for beta arbitrary and families with mirror fermions}},  {\em
  Phys.Rev.} {\bf D72} (2005) 035018,
  [\href{http://xxx.lanl.gov/abs/hep-ph/0411263}{{\tt hep-ph/0411263}}].

\bibitem{Liu:1994rx}
J.~T. Liu, {\it {Generation nonuniversality and flavor changing neutral
  currents in the 331 model}},  {\em Phys.Rev.} {\bf D50} (1994) 542--547,
  [\href{http://xxx.lanl.gov/abs/hep-ph/9312312}{{\tt hep-ph/9312312}}].

\bibitem{Rodriguez:2004mw}
J.~A. Rodriguez and M.~Sher, {\it {FCNC and rare B decays in 3-3-1 models}},
  {\em Phys.Rev.} {\bf D70} (2004) 117702,
  [\href{http://xxx.lanl.gov/abs/hep-ph/0407248}{{\tt hep-ph/0407248}}].

\bibitem{Promberger:2007py}
C.~Promberger, S.~Schatt, and F.~Schwab, {\it {Flavor Changing Neutral Current
  Effects and CP Violation in the Minimal 3-3-1 Model}},  {\em Phys.Rev.} {\bf
  D75} (2007) 115007, [\href{http://xxx.lanl.gov/abs/hep-ph/0702169}{{\tt
  hep-ph/0702169}}].

\bibitem{Agrawal:1995vp}
J.~Agrawal, P.~H. Frampton, and J.~T. Liu, {\it {The Decay $b \to s \gamma$ in
  the 3-3-1 model}},  {\em Int.J.Mod.Phys.} {\bf A11} (1996) 2263--2280,
  [\href{http://xxx.lanl.gov/abs/hep-ph/9502353}{{\tt hep-ph/9502353}}].

\bibitem{Promberger:2008xg}
C.~Promberger, S.~Schatt, F.~Schwab, and S.~Uhlig, {\it {Bounding the Minimal
  331 Model through the Decay $B \to X_s \gamma$}},  {\em Phys.Rev.} {\bf D77}
  (2008) 115022, [\href{http://xxx.lanl.gov/abs/0802.0949}{{\tt
  arXiv:0802.0949}}].

\bibitem{Machado:2013jca}
A.~Machado, J.~Montero, and V.~Pleitez, {\it {FCNC in the minimal 3-3-1 model
  revisited}},  {\em Phys.Rev.} {\bf D88} (2013) 113002,
  [\href{http://xxx.lanl.gov/abs/1305.1921}{{\tt arXiv:1305.1921}}].

\bibitem{Buras:2014yna}
A.~J. Buras, F.~De~Fazio, and J.~Girrbach-Noe, {\it {Z-Z' Mixing and Z-Mediated
  FCNCs in $SU(3)_C \times SU(3)_L \times U(1)_X$ Models}},
  \href{http://xxx.lanl.gov/abs/1405.3850}{{\tt arXiv:1405.3850}}.

\bibitem{Eberhardt:2012gv}
O.~Eberhardt, G.~Herbert, H.~Lacker, A.~Lenz, A.~Menzel, {\em et.~al.}, {\it
  {Impact of a Higgs boson at a mass of 126 GeV on the standard model with
  three and four fermion generations}},  {\em Phys.Rev.Lett.} {\bf 109} (2012)
  241802, [\href{http://xxx.lanl.gov/abs/1209.1101}{{\tt arXiv:1209.1101}}].

\bibitem{Eberhardt:2013uba}
O.~Eberhardt, U.~Nierste, and M.~Wiebusch, {\it {Status of the
  two-Higgs-doublet model of type II}},
  \href{http://xxx.lanl.gov/abs/1305.1649}{{\tt arXiv:1305.1649}}.

\bibitem{Celis:2013rcs}
A.~Celis, V.~Ilisie, and A.~Pich, {\it {LHC constraints on two-Higgs doublet
  models}},  {\em JHEP} {\bf 1307} (2013) 053,
  [\href{http://xxx.lanl.gov/abs/1302.4022}{{\tt arXiv:1302.4022}}].

\bibitem{Chiang:2013ixa}
C.-W. Chiang and K.~Yagyu, {\it {Implications of Higgs boson search data on the
  two-Higgs doublet models with a softly broken $Z_2$ symmetry}},  {\em JHEP}
  {\bf 1307} (2013) 160, [\href{http://xxx.lanl.gov/abs/1303.0168}{{\tt
  arXiv:1303.0168}}].

\bibitem{Barroso:2013awa}
A.~Barroso, P.~Ferreira, I.~Ivanov, and R.~Santos, {\it {Metastability bounds
  on the two Higgs doublet model}},  {\em JHEP} {\bf 1306} (2013) 045,
  [\href{http://xxx.lanl.gov/abs/1303.5098}{{\tt arXiv:1303.5098}}].

\bibitem{Grinstein:2013npa}
B.~Grinstein and P.~Uttayarat, {\it {Carving Out Parameter Space in Type-II Two
  Higgs Doublets Model}},  {\em JHEP} {\bf 1306} (2013) 094,
  [\href{http://xxx.lanl.gov/abs/1304.0028}{{\tt arXiv:1304.0028}}].

\bibitem{Moroi:2000tk}
T.~Moroi, {\it {CP violation in $B_d \to \phi K_S$ in SUSY GUT with
  right-handed neutrinos}},  {\em Phys.Lett.} {\bf B493} (2000) 366--374,
  [\href{http://xxx.lanl.gov/abs/hep-ph/0007328}{{\tt hep-ph/0007328}}].

\bibitem{Chang:2002mq}
D.~Chang, A.~Masiero, and H.~Murayama, {\it {Neutrino mixing and large CP
  violation in $B$ physics}},  {\em Phys.Rev.} {\bf D67} (2003) 075013,
  [\href{http://xxx.lanl.gov/abs/hep-ph/0205111}{{\tt hep-ph/0205111}}].

\bibitem{Girrbach:2011an}
J.~Girrbach, S.~Jager, M.~Knopf, W.~Martens, U.~Nierste, {\em et.~al.}, {\it
  {Flavor Physics in an SO(10) Grand Unified Model}},  {\em JHEP} {\bf 1106}
  (2011) 044, [\href{http://xxx.lanl.gov/abs/1101.6047}{{\tt
  arXiv:1101.6047}}].

\bibitem{Girrbach:2011wt}
J.~Girrbach, {\it {Flavour Physics in an SO(10) Grand Unified Model}},  {\em
  PoS} {\bf EPS-HEP2011} (2011) 183,
  [\href{http://xxx.lanl.gov/abs/1108.4852}{{\tt arXiv:1108.4852}}].

\bibitem{Nierste:2011na}
U.~Nierste, {\it {Flavour physics, supersymmetry and grand unification}},
  \href{http://xxx.lanl.gov/abs/1107.0621}{{\tt arXiv:1107.0621}}.

\bibitem{NierstePortoroz}
U.~Nierste, ``Impact of the higgs discovery on two models of new physics.''
  Talk given at Portoroz, 14.-18. April 2013.

\bibitem{NiersteStockel}
U.~Nierste and J.~Stockel, ``{work in progress}.'' {2013}.

\bibitem{Crivellin:2009sd}
A.~Crivellin, {\it {Effects of right-handed charged currents on the
  determinations of $|V_{ub}|$ and $|V_{cb}|$}},  {\em Phys.Rev.} {\bf D81}
  (2010) 031301, [\href{http://xxx.lanl.gov/abs/0907.2461}{{\tt
  arXiv:0907.2461}}].

\bibitem{Chen:2008se}
C.-H. Chen and S.-h. Nam, {\it {Left-right mixing on leptonic and semileptonic
  $b\to u$ decays}},  {\em Phys.Lett.} {\bf B666} (2008) 462--466,
  [\href{http://xxx.lanl.gov/abs/0807.0896}{{\tt arXiv:0807.0896}}].

\bibitem{Crivellin:2011ba}
A.~Crivellin and L.~Mercolli, {\it {$B\to X_d \gamma$ and constraints on new
  physics}},  {\em Phys.Rev.} {\bf D84} (2011) 114005,
  [\href{http://xxx.lanl.gov/abs/1106.5499}{{\tt arXiv:1106.5499}}].

\bibitem{Agashe:2003zs}
K.~Agashe, A.~Delgado, M.~J. May, and R.~Sundrum, {\it Rs1, custodial isospin
  and precision tests},  {\em JHEP} {\bf 08} (2003) 050,
  [\href{http://xxx.lanl.gov/abs/hep-ph/0308036}{{\tt hep-ph/0308036}}].

\bibitem{Csaki:2003zu}
C.~Csaki, C.~Grojean, L.~Pilo, and J.~Terning, {\it Towards a realistic model
  of higgsless electroweak symmetry breaking},  {\em Phys. Rev. Lett.} {\bf 92}
  (2004) 101802, [\href{http://xxx.lanl.gov/abs/hep-ph/0308038}{{\tt
  hep-ph/0308038}}].

\bibitem{Agashe:2006at}
K.~Agashe, R.~Contino, L.~Da~Rold, and A.~Pomarol, {\it {A custodial symmetry
  for $Z b \bar b$}},  {\em Phys. Lett.} {\bf B641} (2006) 62--66,
  [\href{http://xxx.lanl.gov/abs/hep-ph/0605341}{{\tt hep-ph/0605341}}].

\bibitem{Blanke:2008zb}
M.~Blanke, A.~J. Buras, B.~Duling, S.~Gori, and A.~Weiler, {\it {$\Delta F=2$
  Observables and Fine-Tuning in a Warped Extra Dimension with Custodial
  Protection}},  {\em JHEP} {\bf 03} (2009) 001,
  [\href{http://xxx.lanl.gov/abs/0809.1073}{{\tt arXiv:0809.1073}}].

\bibitem{Casagrande:2008hr}
S.~Casagrande, F.~Goertz, U.~Haisch, M.~Neubert, and T.~Pfoh, {\it {Flavor
  Physics in the Randall-Sundrum Model: I. Theoretical Setup and Electroweak
  Precision Tests}},  {\em JHEP} {\bf 10} (2008) 094,
  [\href{http://xxx.lanl.gov/abs/0807.4937}{{\tt arXiv:0807.4937}}].

\bibitem{Bauer:2008xb}
M.~Bauer, S.~Casagrande, L.~Grunder, U.~Haisch, and M.~Neubert, {\it {Little
  Randall-Sundrum models: $\varepsilon_K$ strikes again}},  {\em Phys.Rev.}
  {\bf D79} (2009) 076001, [\href{http://xxx.lanl.gov/abs/0811.3678}{{\tt
  arXiv:0811.3678}}].

\bibitem{Bauer:2009cf}
M.~Bauer, S.~Casagrande, U.~Haisch, and M.~Neubert, {\it {Flavor Physics in the
  Randall-Sundrum Model: II. Tree-Level Weak-Interaction Processes}},  {\em
  JHEP} {\bf 1009} (2010) 017, [\href{http://xxx.lanl.gov/abs/0912.1625}{{\tt
  arXiv:0912.1625}}].

\bibitem{Gedalia:2009ws}
O.~Gedalia, G.~Isidori, and G.~Perez, {\it {Combining Direct and Indirect Kaon
  CP Violation to Constrain the Warped KK Scale}},  {\em Phys.Lett.} {\bf B682}
  (2009) 200--206, [\href{http://xxx.lanl.gov/abs/0905.3264}{{\tt
  arXiv:0905.3264}}].

\bibitem{Agashe:2004cp}
K.~Agashe, G.~Perez, and A.~Soni, {\it Flavor structure of warped extra
  dimension models},  {\em Phys. Rev.} {\bf D71} (2005) 016002,
  [\href{http://xxx.lanl.gov/abs/hep-ph/0408134}{{\tt hep-ph/0408134}}].

\bibitem{Iltan:2007sc}
E.~O. Iltan, {\it {The effects of lepton KK modes on the lepton electric dipole
  moments in the Randall Sundrum scenario}},  {\em Eur. Phys. J.} {\bf C54}
  (2008) 583--590, [\href{http://xxx.lanl.gov/abs/0708.3765}{{\tt
  arXiv:0708.3765}}].

\bibitem{Agashe:2006iy}
K.~Agashe, A.~E. Blechman, and F.~Petriello, {\it {Probing the Randall-Sundrum
  geometric origin of flavor with lepton flavor violation}},  {\em Phys. Rev.}
  {\bf D74} (2006) 053011, [\href{http://xxx.lanl.gov/abs/hep-ph/0606021}{{\tt
  hep-ph/0606021}}].

\bibitem{Davidson:2007si}
S.~Davidson, G.~Isidori, and S.~Uhlig, {\it {Solving the flavour problem with
  hierarchical fermion wave functions}},  {\em Phys. Lett.} {\bf B663} (2008)
  73--79, [\href{http://xxx.lanl.gov/abs/0711.3376}{{\tt arXiv:0711.3376}}].

\bibitem{Agashe:2009tu}
K.~Agashe, {\it {Relaxing Constraints from Lepton Flavor Violation in 5D
  Flavorful Theories}},  {\em Phys.Rev.} {\bf D80} (2009) 115020,
  [\href{http://xxx.lanl.gov/abs/0902.2400}{{\tt arXiv:0902.2400}}].

\bibitem{Csaki:2010aj}
C.~Csaki, Y.~Grossman, P.~Tanedo, and Y.~Tsai, {\it {Warped penguin diagrams}},
   {\em Phys.Rev.} {\bf D83} (2011) 073002,
  [\href{http://xxx.lanl.gov/abs/1004.2037}{{\tt arXiv:1004.2037}}].

\bibitem{Blanke:2012tv}
M.~Blanke, B.~Shakya, P.~Tanedo, and Y.~Tsai, {\it {The Birds and the $B$s in
  RS: The $b \to s \gamma$ penguin in a warped extra dimension}},  {\em JHEP}
  {\bf 1208} (2012) 038, [\href{http://xxx.lanl.gov/abs/1203.6650}{{\tt
  arXiv:1203.6650}}].

\bibitem{Agashe:2008uz}
K.~Agashe, A.~Azatov, and L.~Zhu, {\it {Flavor Violation Tests of
  Warped/Composite SM in the Two-Site Approach}},  {\em Phys. Rev.} {\bf D79}
  (2009) 056006, [\href{http://xxx.lanl.gov/abs/0810.1016}{{\tt
  arXiv:0810.1016}}].

\bibitem{Kaplan:1991dc}
D.~B. Kaplan, {\it {Flavor at SSC energies: A New mechanism for dynamically
  generated fermion masses}},  {\em Nucl.Phys.} {\bf B365} (1991) 259--278.

\bibitem{Grossman:1999ra}
Y.~Grossman and M.~Neubert, {\it Neutrino masses and mixings in
  non-factorizable geometry},  {\em Phys. Lett.} {\bf B474} (2000) 361--371,
  [\href{http://xxx.lanl.gov/abs/hep-ph/9912408}{{\tt hep-ph/9912408}}].

\bibitem{Huber:2000ie}
S.~J. Huber and Q.~Shafi, {\it Fermion masses, mixings and proton decay in a
  randall-sundrum model},  {\em Phys. Lett.} {\bf B498} (2001) 256--262,
  [\href{http://xxx.lanl.gov/abs/hep-ph/0010195}{{\tt hep-ph/0010195}}].

\bibitem{Gherghetta:2000qt}
T.~Gherghetta and A.~Pomarol, {\it Bulk fields and supersymmetry in a slice of
  ads},  {\em Nucl. Phys.} {\bf B586} (2000) 141--162,
  [\href{http://xxx.lanl.gov/abs/hep-ph/0003129}{{\tt hep-ph/0003129}}].

\bibitem{Csaki:2008zd}
C.~Csaki, A.~Falkowski, and A.~Weiler, {\it {The Flavor of the Composite
  Pseudo-Goldstone Higgs}},  {\em JHEP} {\bf 09} (2008) 008,
  [\href{http://xxx.lanl.gov/abs/0804.1954}{{\tt arXiv:0804.1954}}].

\bibitem{Cacciapaglia:2007fw}
G.~Cacciapaglia {\em et.~al.}, {\it {A GIM Mechanism from Extra Dimensions}},
  {\em JHEP} {\bf 04} (2008) 006,
  [\href{http://xxx.lanl.gov/abs/0709.1714}{{\tt arXiv:0709.1714}}].

\bibitem{Barbieri:2008zt}
R.~Barbieri, G.~Isidori, and D.~Pappadopulo, {\it {Composite fermions in
  Electroweak Symmetry Breaking}},  {\em JHEP} {\bf 0902} (2009) 029,
  [\href{http://xxx.lanl.gov/abs/0811.2888}{{\tt arXiv:0811.2888}}].

\bibitem{Redi:2011zi}
M.~Redi and A.~Weiler, {\it {Flavor and CP Invariant Composite Higgs Models}},
  {\em JHEP} {\bf 1111} (2011) 108,
  [\href{http://xxx.lanl.gov/abs/1106.6357}{{\tt arXiv:1106.6357}}].

\bibitem{Redi:2012uj}
M.~Redi, {\it {Composite MFV and Beyond}},  {\em Eur.Phys.J.} {\bf C72} (2012)
  2030, [\href{http://xxx.lanl.gov/abs/1203.4220}{{\tt arXiv:1203.4220}}].

\bibitem{Fitzpatrick:2007sa}
A.~L. Fitzpatrick, G.~Perez, and L.~Randall, {\it {Flavor anarchy in a
  Randall-Sundrum model with 5D minimal flavor violation and a low Kaluza-Klein
  scale}},  {\em Phys.Rev.Lett.} {\bf 100} (2008) 171604,
  [\href{http://xxx.lanl.gov/abs/0710.1869}{{\tt arXiv:0710.1869}}].

\bibitem{Santiago:2008vq}
J.~Santiago, {\it {Minimal Flavor Protection: A New Flavor Paradigm in Warped
  Models}},  {\em JHEP} {\bf 12} (2008) 046,
  [\href{http://xxx.lanl.gov/abs/0806.1230}{{\tt arXiv:0806.1230}}].

\bibitem{Csaki:2008eh}
C.~Csaki, A.~Falkowski, and A.~Weiler, {\it {A Simple Flavor Protection for
  RS}},  {\em Phys.Rev.} {\bf D80} (2009) 016001,
  [\href{http://xxx.lanl.gov/abs/0806.3757}{{\tt arXiv:0806.3757}}].

\bibitem{Bauer:2011ah}
M.~Bauer, R.~Malm, and M.~Neubert, {\it {A Solution to the Flavor Problem of
  Warped Extra-Dimension Models}},  {\em Phys.Rev.Lett.} {\bf 108} (2012)
  081603, [\href{http://xxx.lanl.gov/abs/1110.0471}{{\tt arXiv:1110.0471}}].

\bibitem{Straub:2013zca}
D.~M. Straub, {\it {Anatomy of flavour-changing Z couplings in models with
  partial compositeness}},  {\em JHEP} {\bf 1308} (2013) 108,
  [\href{http://xxx.lanl.gov/abs/1302.4651}{{\tt arXiv:1302.4651}}].

\bibitem{Contino:2006nn}
R.~Contino, T.~Kramer, M.~Son, and R.~Sundrum, {\it Warped/composite
  phenomenology simplified},  {\em JHEP} {\bf 05} (2007) 074,
  [\href{http://xxx.lanl.gov/abs/hep-ph/0612180}{{\tt hep-ph/0612180}}].

\bibitem{Barbieri:2012tu}
R.~Barbieri, D.~Buttazzo, F.~Sala, D.~M. Straub, and A.~Tesi, {\it {A 125 GeV
  composite Higgs boson versus flavour and electroweak precision tests}},  {\em
  JHEP} {\bf 1305} (2013) 069, [\href{http://xxx.lanl.gov/abs/1211.5085}{{\tt
  arXiv:1211.5085}}].

\bibitem{Grinstein:2010ve}
B.~Grinstein, M.~Redi, and G.~Villadoro, {\it {Low Scale Flavor Gauge
  Symmetries}},  {\em JHEP} {\bf 1011} (2010) 067,
  [\href{http://xxx.lanl.gov/abs/1009.2049}{{\tt arXiv:1009.2049}}].

\bibitem{Feldmann:2010yp}
T.~Feldmann, {\it {See-Saw Masses for Quarks and Leptons in SU(5)}},  {\em
  JHEP} {\bf 1104} (2011) 043, [\href{http://xxx.lanl.gov/abs/1010.2116}{{\tt
  arXiv:1010.2116}}].

\bibitem{Buras:2011ph}
A.~J. Buras, C.~Grojean, S.~Pokorski, and R.~Ziegler, {\it {FCNC Effects in a
  Minimal Theory of Fermion Masses}},  {\em JHEP} {\bf 1108} (2011) 028,
  [\href{http://xxx.lanl.gov/abs/1105.3725}{{\tt arXiv:1105.3725}}].

\end{thebibliography}\endgroup
\end{document}